# Coexisting ferroelectric and antiferroelectric phases in dipole ordered substances. Material properties and possible applications.

V. M. Ishchuk

Science & Technology Center "Reaktivelektron" of the National Academy of Sciences of Ukraine, Donetsk, 81046, Ukraine

V. L. Sobolev

Department of Physics, South Dakota School of Mines & Technology, Rapid City, SD 57701

**Abstract**

Comprehensive review of investigations of substances with a small difference in the energies of the ferroelectric and antiferroelectric types of dipole ordering is presented. Detailed analysis of the stability of homogeneous phases and conditions for existence of the inhomogeneous state of the substance containing domains of the coexisting ferroelectric and antiferroelectric phases is presented. It is shown that the interaction of locally separated domains of the ferroelectric and antiferroelectric phases stabilizes an inhomogeneous state of these materials. The analysis of physical phenomena caused by the presence of domains of the coexisting ferroelectric and antiferroelectric phases in different dipole ordered substances is presented. Peculiarities of behavior of the systems with the inhomogeneous state of coexisting phases under the action of changing external thermodynamic parameters (temperature, pressure, electric field, and chemical composition) are analyzed and their importance for applications of these materials is discussed.

**Content**

# 1. Introduction

Condensed matter physics often deals with inhomogeneous states of the substances containing domains of two (or even more) coexisting phases. For instance, in the substances characterized by the first order phase transition there is an interval of thermodynamic parameters (hysteresis region) within which domains of both phases coexist. This hysteresis region is located between the lability boundaries of each of the phases (participating in the phase transition) in the diagram of phase states of the substance. An initiation of the stable state out of the metastable one occurs by the nucleation, and within a certain interval of thermodynamic parameters the domains of these two phases coexist. It was routinely considered that the presence of the interphase boundary (the boundary separating domains of the coexisting phases) leads to an increase in the free energy of the system in question. In other words, the energy of the interphase boundary is positive (though this fact has not been proved so far).

However, a situation when the said rule appeared to be violated is well-known at present. The state of the type II superconductors placed in an external magnetic field is one of the examples. Within a certain field interval (between the first and the second critical fields) the normal (resistive) and superconducting phases coexist [1, 2]. The inhomogeneous state of the type II superconductors is caused by the negative surface energy of the boundary between these two phases. However, such a phenomenon is observed only in a magnetic field and is a consequence of the redistribution of energy between the sample and the field: the energy of the inhomogeneous system decreases, whereas the density of the magnetic field energy rises over the whole space. As a result the energy of the system "superconductor – magnetic field" increases, and the inhomogeneous state of the superconductor is maintained due to the energy of the source of electric current creating the magnetic field. The evidence of such rise in total energy is the fact that the superconductor regains its uniform state when the magnetic field source is switched off (i.e. in the absence of the field).

A natural question arises during the consideration of the above example: whether there exist the substances able to create the field, which will lead to the effects observed in the type II superconductors (i.e. the inhomogeneous state)? This question is to be answered as follows. Such substances do exist – these are magnetic and electric dipole ordered materials at the temperatures below the Curie point. The inhomogeneous state of coexisting ferromagnetic and antiferromagnetic domains is present in magnetic materials and the inhomogeneous state of coexisting domains of the ferroelectric (FE) and antiferroelectric (AFE) phases is present in the second type of substances.

Such inhomogeneous ordered states are advantageous from the viewpoint of energy. Spontaneous appearance of the order parameter leads to an advent of the field conjugated to the order parameter (the scattering or demagnetizing field). Here it should be noted that such field do not penetrate ferroelectrics: it is shielded by the charges leaking on the surface, however, for our consideration this is not essential. Among the mentioned substances, there exist such compounds for which the two-phase state of domains of the coexisting phases may exist under certain conditions (certain values of thermodynamic parameters, i.e. temperature, field intensity, pressure etc.). These are the substances, which undergo the first order phase transition between the antiferromagnetic and ferromagnetic phases or the transition between the antiferroelectric and ferroelectric phases. Therefore, it seems very interesting to consider the influence of the field generated by the order parameter of the ferromagnetic or ferroelectric phase on the general state of the substance. For definiteness, let us dwell on physical systems characterized by the phase transition between the antiferroelectric and ferroelectric phases in dipole ordered state.

Such systems are of a great interest form the viewpoint of material science especially for development of new functional materials. Ferroelectrics (for the most part it pertains to ferroelectric piezoelectric ceramics) have been used for a long time and they have no alternatives in some applications. This circumstance determined a comprehensive scrutiny of these substances form the points of view of both fundamental and applied physics.

Antiferroelectrics are studied to a lesser extent than ferroelectrics. As of yet their scope of applications is narrow − mainly, it covers materials for special purposes (largely the development of materials for military applications). A growing interest to the substances with the antiferroelectric order has been manifested in recent years. Primarily this interest is devoted to the compounds in which the phase transition between the antiferroelectric and ferroelectric phases takes place under the action of external influences (such as electric field, hydrostatic and uniaxial pressure, heat.

Why materials with the phase transitions between the antiferroelectric and ferroelectric states are so attractive for practical applications? Different types of dipole ordering determine the fundamental difference in physical parameters (characteristics) of these states. A significant change in the parameters of the substance takes place in the process of phase transition. These changes can be used in different type of sensors and actuating devices. The changes in parameters during the phase transformation may also be useful in smooth tuning of devices and for control of their parameters.

The main disadvantage of all well studied to date substances with phase transformations between the antiferroelectric and ferroelectric states is the abrupt nature of this transition (as a consequence of the first order phase transition). This circumstance prevents the smooth change of the physical parameters of these substances. Such disadvantage is eliminated in the compounds with the inhomogeneous state of the coexisting ferroelectric and antiferroelectric phases.

Present review contains results of investigations of the nature of the inhomogeneous state of the coexisting ferroelectric and antiferroelectric phases and behavior of materials in which such inhomogeneous structures are realized. The material of this review is presented in such a way that in all cases the consideration of theoretical models is immediately followed by the comparison with experimental data available in the literature.

The second chapter devoted to the analysis of results of studies of the relative stability of ferroelectric and antiferroelectric phases in the substances of interest.

In the third chapter it is shown that the stable (from the viewpoint of energy) inhomogeneous quasi-two-phase state may be present in a substance under fulfilment of the two conditions. First, the difference in free energies of the antiferroelectric (AFE) and ferroelectric (FE) phases has to be small within a wide interval of changes of external parameters. Second, the interphase interaction of domains of the coexisting phases has to be taken into account. The uniform state of any phase (AFE or FE) will be a metastable state at these values of thermodynamic parameters. The consideration is made in the frames of the phenomenological Landau-Devonshire approach to phase transitions in ferroelectric substances. The relations between the phenomenological parameters in the expression for the thermodynamic potential determining the stability the two-phase state are established.

Results of transmission electron microscopy studies that confirm the existence of the inhomogeneous two-phase state are also presented.

The fourth chapter devoted to consideration of substances with wide hysteretic regions in the diagram of phase states. The inhomogeneous states discussed in previous chapter are present within these hysteretic regions. The factors providing the possibility of realization of such stable two-phase states are found in the frames of the Landau-Devonshire approach to phase transitions in ferroelectric materials. The discussion of the phase diagrams of these substances subjected to the action of external factors of different physical nature is presented.

This chapter also contains experimental results supporting theoretical conclusions. The diagrams of the phase states for the $PbZr_{1-y}Ti_yO_3$ (PZT) based solid solutions are presented. These phase diagrams demonstrate the presence of extended hysteresis regions of the antiferroelectric to ferroelectric (AFE-FE) phase transformations. The "composition-temperature" (Y-T), "pressure-temperature" (P-T), and "temperature-electric field" (T-E) phase diagrams for the above-mentioned solid solutions are obtained and discussed.

The fifth chapter covers the discussion of the AFE-FE transition in the presence of a DC electric field. It is shown that the inhomogeneous state of coexisting domains of the FE and AFE phases is realized in the bulk of the substance. Its geometrical properties are analogous to those of the intermediate state in antiferromagnets or superconductors. However, the nature of such a state is quite different: it is the interaction between the domains of the coexisting phases.

The chapter number six contains a consideration of the processes that occur in the vicinity of the interphase boundaries separating the coexisting domains of the FE and AFE phases in the inhomogeneous state of the substance. Local mechanical stresses arising in the said region are shown to lead to the local decomposition of the solid solution and to the processes of long-duration relaxation. The experimental results obtained on different substances in which the difference in the free energies of the FE and AFE states is relatively small are presented.

The seventh chapter covers the results of consideration of the phase transition (PT) from the paraelectric (PE) phase to the ordered phase in substances with compositions locates near the triple FE-AFE-PE point of the "composition-temperature" phases diagram. The model of these transitions taking into account the process of local decomposition of the solid solution in the vicinity of the FE-AFE interphase boundary is discussed. Experimental results pointing to the presence of the two-phase FE+AFE states not only at low temperatures, but also at the temperatures essentially higher than the Curie point are given.

The eighth chapter is devoted to discussion of the substances with an insignificant difference in the free energies of the FE and AFE states in the presence of an AC electric field. It is shown that dielectric properties of these materials are defined by the dynamics of the interphase boundaries in the range of low and medium frequencies of the AC. The equation describing the dynamics of the interphase boundary is obtained and solved. It is demonstrated that the relaxation dynamics of domain boundaries is responsible for the dispersion of the dielectric constant. The Vogel-Fülcher relation describing the behavior of the dielectric constant at the phase transition into the PE state near the triple FE-AFE-PE point in the phase diagram is obtained.

The ninth chapter contains the analysis of a variety of experimental results associated with the so-called relaxor state of ferroelectrics. These results are analyzed based on the point of view that takes into account the inhomogeneous state of coexisting phases developed in this review. Special attention is paid to the analysis of the phase diagrams for the mentioned class of materials obtained at extreme values of external thermodynamic parameters such as the pressure and the external electric field.

The tenth chapter focuses on a consideration of properties of the solid solutions that have one component being ferroelectric and the other one being antiferroelectric and that are often referred to as dipole glasses. As a rule, analyzing the behavior of such substances one does not take into account the following factors:
 – stability of the two-phase state due to negative energies of the interphase boundaries,
 – local decomposition of the solid solutions near the said boundaries.
An analysis of the experimental phase diagrams of such materials is presented. It is shown that not all the substances that are referred to as "glasses" are in fact belonging to this class of substances.

The chapter eleven contains consideration of the applied material-science aspects of the FE-AFE phase transitions. The following topics are discussed:
- Materials with the FE-AFE phase transition for the energy accumulation and conversion devices;
- Phase transition via intermediate state and control of piezoelectric parameters;
- Local decomposition of solid solutions, formation of nanostructures and optical materials with negative refractive index;
- Creation of textured nanostructures;
- Effects caused by the AFE nanodomains in the PZT-based coarse-grained ceramics with compositions from the morphotropic boundary region.

## 2. Relative stability of the ferroelectric and antiferroelectric phases

The description of emergence of the FE and AFE states in the frames of the phenomenological approach is beyond questioning at present. Microscopic models (the first principles calculations) also give a good description of behavior of these two classes of substances in a wide range of variation of external factors (temperature, pressure, and electric field).

It is well known that the crystal structure of solids and the possible types of dipole ordering are determined by the chemical bonding. In other words the wave functions of the external electrons of atoms and the overlapping of these wave functions are responsible for the bonding type. Basically, calculations based on this approach reliably predict the type of dipole ordering, which is observed experimentally. Likewise the types of the crystal lattice distortions following from results of these calculations correspond to the experimentally observed. It would seem clear to all. Nevertheless, the mechanism of the phase transition between the FE and AFE states at the relatively small changes in the external thermodynamic parameters (temperature, field intensity or pressure) remains unclear. A phrase of the type that "there exists a delicate balance of interactions the violations of which causes the phase transition" would be not entirely relevant here (by the way the phrase is absolutely true). Any phase transition is the result of violation of some delicate balance of interactions. But it does not follow even from the first-principles calculations when such a violation must occur and when must not. The most remarkable is the fact that these phase transitions are accompanied by a significant hysteresis, but this circumstance is not even discussed during the analysis of the results of calculations. However, the nature of this hysteresis is obvious in the frames of the phenomenological description; namely, a potential barrier in the dependency of the free energy on the order parameters of these phases is the reason for it. Nonetheless the experimental values of the thermal hysteresis are different from the calculated ones when such calculations are carried out (it has to be noted that such calculations are extremely rare the literature).

Experimental results on the relative stability and the crystal structure of the AFE and FE phases, as well as the results of calculations of the dipole structure in the frames of the models providing sufficient proximity to the real structure allowed revealing of the main factors determining the presence (or the absence) of a phase transition. The main conclusion of all calculations reduces to the fact that the difference in energies of these two types of dipole ordering is small. This conclusion is also confirmed by model calculations.

Energy calculation for different types of dipole ordering in the perovskite crystal structure of the ferroelectric barium titanate ($BaTiO_3$ is isostructural to antiferroelectric lead zirconate) has been carried out in [3]. Results of this work are important because they include not only barium titanate, but also all oxide with the perovskite crystal structure. Barium titanate has been taken as a typical representative of this class of compounds. The energy of dipole ordered states was calculated as a function of the polarizability of the oxygen ions. The Slater model was used as an initial calculation model in [4]. This model assumes a simple temperature dependence of the ionic polarizability of titanium. This temperature dependence of the polarizability of titanium ions is a consequence of the anharmonicity of the potential energy of titanium. The non-linearity of the electronic polarizability of oxygen was taken into account during calculations of the energy of different types of dipolar ordering. The model provided a good agreement between the calculated and experimental values of the spontaneous polarization and the relative shift of the titanium ions and oxygen from their centrosymmetric positions.

This indicates that the results of calculations should be treated with confidence. At present, the calculations carried out in [4] and [3] seem very simplified against the backdrop of the latest models. Nonetheless they give a good description of the "real situation" and are physically illustrative (nothing is hidden behind sets of atomic functions and matrices). The results of [3] and [4] make it easy to establish the connections between the calculated parameters and external factors (an electric field, a pressure, an ion substitution, etc.) leading to a phase transition.

Calculations of the FE ordering and different types of the AFE ordering of the dipole structure were carried in [3]. Two types of the AFE dipole structures (the AP(100)-structure and AP(110)-structure), for which the ordering energy was calculated are presented in Fig. 2.1. The approximation of interacting sublattices had been used in these calculations. The AFE ordering of the whole crystal was considered to be a superposition of the sublattices with the AFE orderings since the perovskite crystal lattice consists of the five ($A$, $B$, $O_I$, $O_{II}$, $O_{III}$) interpenetrating cubic lattices. This circumstance was taken into account during calculations in [3]. Apart from the types of the dipole ordering shown in Fig.2.1 one more dipole ordering type with dipole moments parallel and antiparallel to the [111] crystallographic direction is possible. However, the energy of the this type of ordering is much higher than the energy of the above discussed orderings and, thus, the AP(111)-structure was not considered in [3]. The calculations carried out in [3] were made under assumption that the dipole moments of the ions inside each of the planes under consideration ((100) plane for the AP(100)-structure and (110) plane for the AP(110)-structure) are parallel.

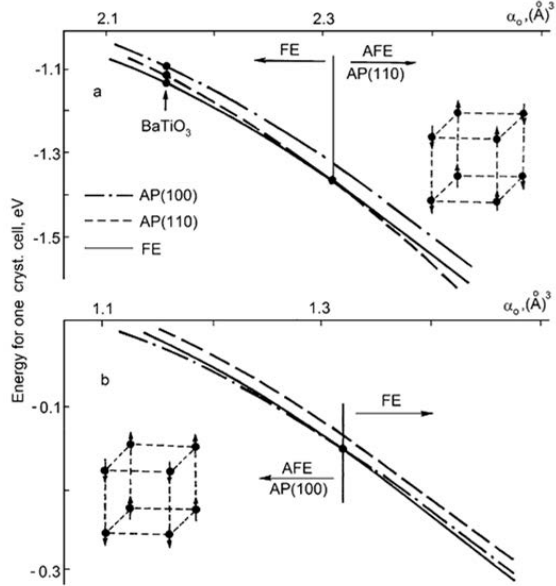

Fig.2.1. Calculated energy for three dipole ordered states in the perovskite $BaTiO_3$-like structure vs. oxygen polarizability $\alpha_O$ for large (*a*) and small (*b*) $\alpha_O$ [3].

Generalizations of results on dependencies of the energies of electrostatic interactions on the polarizability of the oxygen ions for different configurations obtained in [3] are also given in Fig.2.1. The energies of dipole-ordered states for the polarizability values in the range $2.0 < \alpha_O < 2.5$ (Å)$^3$ (see Fig. 2.1a) and for the case of the small values of polarizability $1.1 < \alpha_O < 1.5$ (Å)$^3$ (see Fig.2.1b) are presented in Fig. 2.1 for convenience of analysis of results [3]. The intersection of the curves for AP(100)- and AP(110)- AFE structures happens at the value of polarizability $\alpha_O = 1.8$ (Å)$^3$ and it is not shown in Fig.2.1.

It has to be noted that based on the characteristic value $\alpha_O = 2.16$ (Å)$^3$ in the case of barium titanate the calculation data indicating the ferroelectric ordering are in full agreement with the experimental results. But for the other values of the oxygen polarizability ($\alpha_O < 1.32$ (Å)$^3$ and $\alpha_O > 2.31$ (Å)$^3$) the AFE type of dipole ordering is more stable. In this connection the AFE AP(100)-structure is stable for small values of $\alpha_O$, and the AP(110)-structure is stable for large $\alpha_O$ values. The nonpolar PE disordered dipole structure is energetically stable at $\alpha_O < 1.10$ (Å)$^3$. It should be pointed out that the difference in the energies corresponding to different states of the dipole ordering in the entire range of oxygen polarizability values is very small. This range of values of the oxygen polarizability, $\alpha_O$, is characteristic for the majority of oxides with the perovskite crystal structure.

Calculations carried out in [3] concentrated on the dependencies of the relative stability of the FE and AFE configurations on the polarizability of the oxygen ions only. It is clear that the introduction of other parameters into these calculations gives a possibility to obtain different results. In particular, one can expect that the sequence of stable states will be different and nothing like the FE configuration may occur in the case of barium titanate. For example, a slightly different procedure of calculations of local fields used in [4] lead to the result that the AFE structure *AP*(110) was always the most stable configuration in barium titanate. It was noted in [3] that the regions of stability of the AFE phases may be swapped over if the calculations of internal fields are carried out taking into account that the coefficients of dipole-dipole interactions may be dependent on the polarizabilities of both titanium and oxygen ions. As a result of this changes the *AP*(110) phase becomes to be stable at the oxygen polarizabilities less than 1,7 (Å)$^3$ and the *AP*(100) phase becomes to be stable at the larger values of $\alpha_O$

The main conclusion is that in the case of the perovskite crystal structure the difference in energy of the FE and AFE states is small. This conclusion cannot be considered as a solely consequence of the model used in calculations. In particular, it was already demonstrated in [5, 6] that the parameters of the models in which calculations are based on the lattice dynamics can be chosen in such a way that the same crystal structure may allow both the FE and an AFE orderings. These types of dipole ordering can be obtained by the variation of calculation parameters in rather narrow intervals. We will return to this circumstance in our further considerations

The problem about the relative stability of the FE and AFE states was approached differently in [7]. The lead zirconate (PbZrO$_3$) was considered in this paper. Based on results of numerous experimental studies it was well established that the ground state of this compound is the AFE state. The purpose of calculations was elucidation of the role of the dipole-dipole interaction in the formation of the AFE phase and the reasons for the absence of the AFE phase in barium titanate, lead titanate, barium zirconate, and strontium zirconate. The dipole-dipole interaction was determined by means of the lattice-sum method for three types of dipole configurations (see Fig. 2.2)
- FE configuration,
- the AFE configuration $A_1$,
- the AFE configuration $A_2$.

It has to be noted that the AFE structure $A_2$ corresponds to the real AFE structure of dipole ordering in lead zirconate at room temperature.

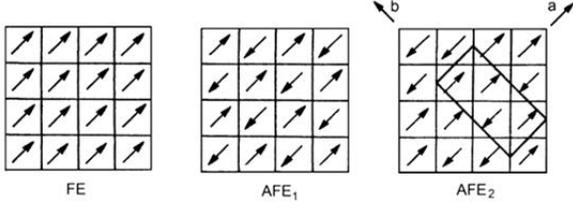

Fig. 2.2. Three possible domain configurations of lead zirconate for calculation of the electrostatic energy in [7].

The calculation of the electrostatic energy of dipole configurations taking into account the ionic and electron components of the polarization was carried out for different values of the effective cation charge of the A-position of the perovskite lattice and the sizes of the elementary cell. The tolerance factor was introduced for more precise estimation of energies of different configurations. This allowed taking into account the mobility of cations in the elementary cell empirically. The energy of the dipole-dipole interaction is:

$$U_{d-d} = -\sum_{ij} D_{ij} x_i x_j \qquad (2.1)$$

In this expression $x_i$ are the displacements of ions from the equilibrium position in the high-temperature phase; $D_{ij}$ are the coefficients dependent on the electron polarizability of ions and the volume of the elementary cell; $i = 1, 2, 3, 4,$ and 5 correspond to the A, B, O$_I$, O$_{II}$, and O$_{III}$ ions of the perovskite elementary cell of ABO$_3$.

The ions A and O$_I$ are mobile at small values of the tolerance factor. In this case only $x_1 \neq 0$ and $x_3 \neq 0$, and all other $x_i$ are zero in (2.1). The electrostatic energy has the following form in this case:

$$U_{d-d} = -D_{11} x_1^2 - 2D_{13} x_1 x_3 - D_{33} x_3^2, \qquad (2.2)$$

The relation $D_{13} = D_{31}$ is taken into account in this expression. Only B-site ions of the perovskite lattice are mobile at larger values of the tolerance factor. In this case $x_2 \neq 0$ only and the rest of the $x_i$ values are zero. The interaction energy for this case has the form:

$$U_{d-d} = -D_{22} \cdot x_2^2. \qquad (2.3)$$

Calculations of the coefficients $D_{11}$, $D_{22}$, $D_{33}$ and $D_{13}$ carried out in [7] were done for different ABO$_3$-type crystals with the perovskite crystal structure. An estimation of the stability of the $A_1$ and $A_2$ AFE phases was done using expressions (2.2) and (2.3) and numerical values of the coefficients $D_{ij}$. It turned out that the FE-type dipole structure possesses the lowest energy in the lead titanate while in the lead zirconate the $A_2$-type of the AFE structure has the lowest energy. The estimation of energies for these compounds was made using formula (2.2). It should be noted here that the stability of the $A_2$-type AFE structure in the lead zirconate is to a large extent provided by the interactions between the Pb and O$_I$ ions. The estimation of the energies of dipole configurations in barium titanate was done using formula (2.3) since barium titanate has larger tolerance factor. The estimation showed that the FE structure possesses the lowest value of energy.

In all cases, the stabilization of a particular type of dipole ordering is provided by geometric factors such as inter-ion spacings (just they determine the value of tolerance factor for each of the compounds under consideration). Inter-ion spacings can be easily changed, for example, by hydrostatic pressure. In this case, the pressure (even very modest from the energy point of view) can lead to a drastic restructuring of the dipole structure by means of the induced transition between the FE and AFE states.

Thus, earlier years calculations demonstrated that the difference in energies of the FE and AFE states in lead zirconate and isostructural compounds is small. The sizes of ions (or the values of the tolerance factor) are the decisive factor in determination of the relative stability of the FE and AFE phases. The followed closely factor affecting the relative stability of phases is the effective charges of the cations in the A-sites of the crystal lattice. These calculations showed how close are the values of energy for the FE and AFE states in the barium titanate and lead zirconate (and correspondingly in solid solutions with compositions close to the lead zirconate). These calculations also demonstrated how difficult is to define the line separating these phases by the results of calculations. The more reliable results of calculations that use contemporary powerful calculation methods were obtained in this century. However these results have lead to the same conclusions.

The most reliable are the results of analysis of the phonon spectrum. Thereunto, there exists a very good agreement between the calculated and experimental phonon spectra.

Calculations of the total energy for the AFE state in $PbZrO_3$ were performed in [8] for the structure of Jona et al. (*Pba*2 space group [9]) and the two structures of Fujishita and coworkers (*Pbam* space group [10, 11], and refined two separate sets of oxygen coordinates). The two structures of Fujishita and co-workers yielded the same energy within the precision of their calculations. This energy is 0.233 eV per formula unit below the cubic perovskite structure but 0.02 eV above the rhombohedral FE structure. This would seem to imply a stability of the FE structure and not the AFE ground state. However, the calculations reveal substantial forces on the O-atoms in both of these structures.

Therefore, the calculations were undertaken to relax the position of the oxygen ions. The calculated energy, with the relaxed O-positions, is 0.274 eV below the cubic perovskite structure, and 0.021 eV below the FE structure on a per formula unit basis.

The calculated phonon dispersion curves along the high symmetry lines of the simple cubic Brillouin zone are shown in Fig.2.3 for $BaTiO_3$ and $PbZrO_3$ [12]. The unstable modes, which determine the nature of the phase transitions, have imaginary frequencies. Their dispersion is shown below the zero-frequency line. The character of these modes has significant implications for the properties of the system.

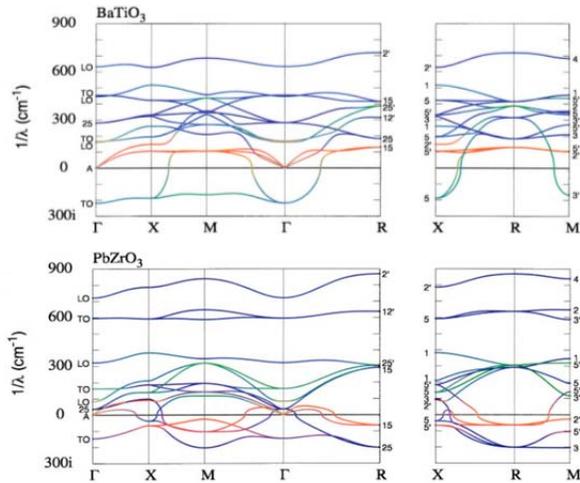

Fig.2.3. Calculated phonon dispersion relations of BaTiO3 and PbZrO3 along various high-symmetry lines in the simple cubic Brillouin zone [12]. A color has been assigned to each point based on the contribution of each kind of atom to the associated dynamical matrix eigenvector (red for the *A* atom, green for the *B* atom, and blue for O atoms). Symmetry labels follow the convention of [13], with the *A* atom at the origin.

In $BaTiO_3$ the most unstable mode is at the Γ-point of the Brillouin zone, and this mode, dominated by the Ti-ion displacement against the oxygens is the one that freezes in to give the FE phase. However, the instability of the crystal lattice is not restricted to the Γ-point. Branches of Ti-dominated unstable modes extend over much of the Brillouin zone. The flat dispersions of the unstable transverse optic mode towards *X* and *M*, combined with their rapid stiffening towards *R* (all these at the edge of the Brillouin zone), confine the instability to three quasi-two-dimensional ''slabs'' of reciprocal space intersecting at the point Γ. This is the fingerprint of a ''chainlike'' unstable localized antiphase distortion for the Ti displacements in real space.

The ground state of PbZrO$_3$ is an antiferroelectric, obtained by freezing in a set of coupled modes, most importantly antiphase modes at $R$ and $\Sigma\left(\frac{1}{4}\frac{1}{4}0\right)$. The unstable branches are dominated by Pb and O displacements, with no significant Zr character. But there is still a polar instability at the Γ-point with the eigenvector that clearly dominate by the displacement of lead against the oxygens while the Zr atom now moves *with* these oxygens.

The authors of [14] studied the relative stability of the FE and AFE phases in lead zirconate at varying the lattice parameters in the vicinity of the experimental value of the lattice constant. Authors compared the orthorhombic AFE state of lead zirconate with the rhombohedral FE one in terms of normal modes of the ideal perovskite pure lead zirconate.

The rhombohedral FE phase was considered as consisting of 10 atoms per unit cell and as the result of the simultaneous freezing in of both the transverse optical polar (TO) $\Gamma_{TO}$ mode polarized along the [111] direction and the $R_{25}$ oxygen octahedron tilting mode with [111] rotation axes. The orthorhombic AFE phase of PbZrO$_3$ has 40 atoms per unit cell and is the result of the 8 frozen (condensed) modes $R_{15}$ (at the center of Brillouin zone), $R_{25}\left(\frac{1}{2}\frac{1}{2}\frac{1}{2}\right)$, $M_5'\left(\frac{1}{2}\frac{1}{2}0\right)$, $X_1\left(00\frac{1}{2}\right)$, $\Sigma_3\left(\frac{1}{4}\frac{1}{4}0\right)$ и $S_3\left(\frac{1}{4}\frac{1}{4}\frac{1}{2}\right)$ in the structure refinement of Teslic and Egami [15]. The relative contributions of each mode to the distortion are 0.301, 0.012, 0.004, 0.002, 0.664 and 0.017, respectively. As can be seen the two modes with the $R_{25}$ and $\Sigma_3$ symmetry labels are prevailing ones. It is the $\Sigma_3$ modes, which yield the AFE cation displacement pattern in PbZrO$_3$. Such picture suggests that the FE–AFE phase transition in PbZrO$_3$ and Zr-rich PZT can be described in terms of cooperation and competition among $\Gamma_{TO}$, $R_{25}$ and $\Sigma_3$ (1/4,1/4,0) instabilities of the ideal perovskite structure.

Authors of [14] performed calculations using local density-functional theory with electronic wave functions determined by conjugate-gradients optimization [16]. The ionic potentials were replaced by the pseudopotentials described in [17]. The force constants for the polar $\Gamma_{TO}$ mode and the $R_{25}$ oxygen octahedron tilting mode were calculated in the interval of lattice parameters from 3.96 to 4.28 Å with the 0.08 Å step. Determination of the force constants associated with the $\Sigma_3$ modes required both lower symmetry and more perturbations than for the calculations for Γ and $R$. Therefore the calculations were made for the frequencies of the $\Sigma_3$ mode only for the values of the lattice constant $a$ = 4.10 and 4.12 Å.

The dependencies of the frequencies of modes on the lattice parameter are shown in Fig. 2.4. The labels at the left indicate the irreducible representation for each mode branch. The frequencies of the two lowest $\Sigma_3$ modes are shown as open circles.

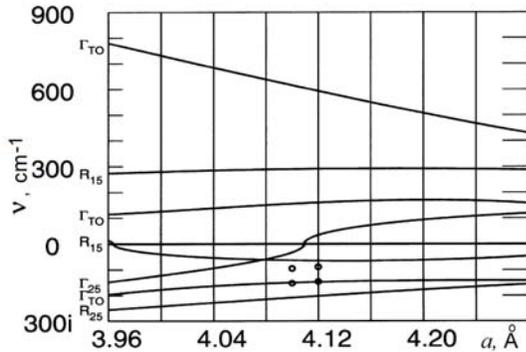

Fig. 2.4. First-principles normal mode frequencies (in cm$^{-1}$) of selected modes in perovskite PbZrO3 as a function of lattice parameter [14]. The circles indicate $\Sigma_3$ modes. As $a$ goes from 4.12 to 4.10A°, the ferroelectric $\Gamma_{TO}$ instability goes from 147i to 150i cm$^{-1}$, while the AFE ($\Sigma_3$) instabilities go from 145i to 151i cm$^{-1}$ and from 88i to 96i cm$^{-1}$.

The number of instabilities at a given wave vector (corresponding to a point in the first Brillouin zone) can change as the lattice constant (or pressure) is changed. For example, at $a$ < 4.11 Å the $\Gamma_{25}$ oxygen "buckling" mode becomes unstable. At $a$ < 3.96 Å the lowest $R_{15}$ mode becomes stable. Although these changes probably play a little role in the FE–AFE transition in PZT, they may play an important role in some of the high-pressure phase transitions recently reported in pure PbZrO$_3$. Most of the modes shown in Fig. 2.4 become more unstable with decreasing lattice parameter (increasing pressure). The trend is not necessarily monotonic. The lattice parameter dependencies of the lowest $R_{15}$ and $\Gamma_{TO}$ modes reverse at sufficient increase of the parameter. The dependence of the lowest $\Gamma_{TO}$ mode is most remarkable.

The FE–AFE phase transition occurs when one or more competing instabilities are replaced by another. When the lattice parameter reduces from 4.12 to 4.10 Å, the frequency of the mode corresponding to the FE instability goes from 147i to 150i cm$^{-1}$, while the frequency of the lower of the two modes, corresponding to the AFE instability, goes from 145i to 151i cm$^{-1}$. These values show, how small is the energy difference between the FE and AFE states in PbZrO$_3$. Moreover, for a complete explanation of the observed phase transition one will need to take into account other modes involved in the orthorhombic phase distortion, as well as those instabilities of the perovskite PbZrO$_3$ that are *not* frozen in. Anharmonic terms in the expansion of the total energy in powers of the normal mode coordinates involve modes, which are *stable* in the ideal perovskite structure. In other words, one will need to include quantitative information on anharmonic coupling in the expansion of total energy.

Further complications of the analysis of the relative stability of the FE and AFE states appear when one takes into consideration additional factors related to the nonmonotonic behavior of the mode frequencies caused by the change of the crystal lattice parameters (or when the pressure is applied to the crystal). An example of this can be found in [18]. In our opinion the theoretical analysis carried out in [18] shows how energetically close these phases are. First-principles calculations were performed to investigate several low-enthalpy phases of the prototype AFE PbZrO$_3$ under hydrostatic pressure. The dependencies of enthalpy for the orthorhombic *Pbam* and *Pnma* states, the monoclinic *P*2$_1$/*c* and rhombohedral *R*3*c* phases that possess different ionic degrees of freedom (e.g., antipolar motions, oxygen octahedral tilting, and polarization) were obtained.

Calculations revealed that the difference between the enthalpy values for all these phases and that of $Pm\bar{3}m$ becomes more negative as the pressure grows. Such features therefore strongly suggest that some, or all, structural instabilities with respect to the cubic state (e.g., tilting of oxygen octahedra, antipolar displacements, polarization, etc.) existing in these four phases are enhanced during the decrease of the lattice constant (pressure variation). Fig. 2.5 presents the pressure dependencies of enthalpy for these phases with respect to *Pbam*, in order to determine which phase is the most stable one in different pressure ranges.

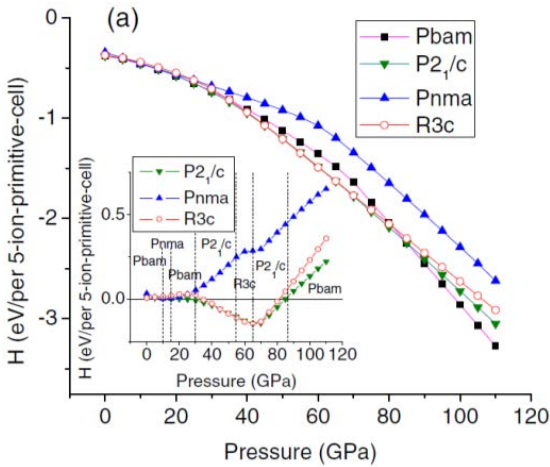

Fig.2.5 Pressure dependency of enthalpy of studied phases of PbZrO$_3$ [18] (per-5-ion-primitive-cell) with respect to the $Pm\bar{3}m$ cubic phase. The inset shows these enthalpies with respect to *Pbam*.

The pressure range from 30 to 85 GPa is the most interesting. Within this interval of pressures the FE state with the R3c symmetry is the most stable state in the AFE PbZrO$_3$. This state is again replaced by the AFE state with the *Pbam* symmetry when the pressure increases.

Thus, the first principle calculations also confirmed a small difference in energies of the FE and the AFE states in both BaTiO$_3$ and PbZrO$_3$.

A vast amount of experimental results of investigations of the relative stability of the FE and AFE states and phase transitions between these phases in PZT solid solutions has been accumulated at present. The stability of the FE and AFE states in 95/5 PZT solid solution (notation 95/5 is used for the Zr/Ti content in the PZT) was experimentally investigated by transmission electron microscopy method in [19]. The PZT solid solution with this particular composition is of interest due to the fact that at the room temperature its composition is located on the border separating these two types of dipole ordering in the "composition-temperature" phase diagram. This PZT solid solution is the isostructural with the barium titanate and lead zirconate by its crystal structure.

The energies of the FE and the AFE states in 95/5 PZT are so close that the domains of these states co-exist within a single crystallite. Fig. 8 in [19] shows a bright-field image obtained at room temperature for PZT 95/5. The coexistence of domains with the FE and AFE structures is clearly seen in this image. The most reliable identification of such coexistence can be achieved by obtaining the selected area electron-diffraction (SARD) patterns that are shown in Fig.2.6.

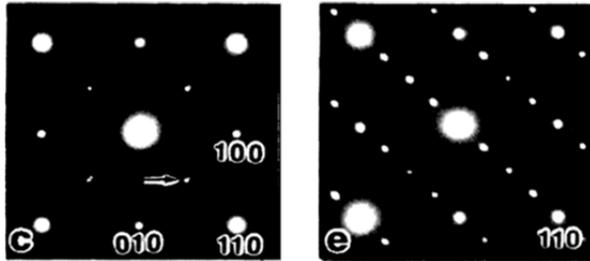

Fig. 2.6. ⟨001⟩ SAED pattern of FE (in the left) and AFE (in the right) phases [19] (Figures c and e from Fig.8 in [19]). The patterns were obtained on a single crystal. The right pattern reveals the existence of ⟨001⟩ superlattice reflections, which are isomorphous with that of AFE phase of $PbZrO_3$

On the other hand, contemporary calculation methods in solid state physics have achieved such high level that one can calculate the structure of specific substances and can determine material parameters that are not yet known from experiments. Basically all this was demonstrated using the lead titanate and lead zirconate as examples. Naturally, PZT is a more complex system especially when the region of the FE-AFE phase transition is of interest. However, even in this case sufficiently reliable results of the first principles calculations were obtained. Examples of such calculations one can find in [17, 20, 21, 22].

The analysis of the relative stability of the FE and AFE states in 95/5 PZT solid solution was carried out in [22]. It turned out, that it is enough to vary parameters of the model within a rather narrow range (without going beyond the accuracy of the model) in order to obtain the change of the stability of dipole-ordered states and to describe the phase transition between these states. This means that a relatively weak external influence can lead to phase transitions. The fact that this work has demonstrated a small difference in the energies of the FE and AFE states is not the only interesting result of this paper. It also demonstrated that it is impossible to determine the AFE state without taking into account this mall difference in energies of the FE and AFE states and consequently without taking into account the interaction between these states (Fig.2 in [22]).

The first principles calculations carried out in recent years have given practically most of the information about the structure of the substances under consideration. However, all these approaches have one substantial drawback which is particularly important for developers of materials with predetermined physical properties. Despite of a large number of results of diverse set of calculation methods using different types of pseudopotentials and wave functions it is difficult to elicit the physical factors determining the behavior of the studied substances. First and foremost, this applies to the relative stability of different phases, and the boundaries of their existence. Therefore, in conclusion of this brief discussion of results of studies of the relative stability of the FE and AFE states, it is necessary to emphasize some highlights of this problem. First of all, it has to be noted that the truly microscopic model of phase transition between the FE and AFE states is absent up today. This certainly complicates further targeted research and the practical application of these transitions.

The analysis of results of studies of distributions the internal electric fields and their gradients in perovskite compounds with special dielectric properties does not detect significant differences in the nature of the FE and AFE ordering. Calculations of the stability of these structures demonstrate that the unit cell dimensions and the value of the structural t-factor have the strongest influence on the type of the ordering. In particular, the reduction in the cell size and the value of the structural factor stabilize the AFE structure. The values of charges and polarizabilities of ions are the secondary factors from the point of view of their influence.

At the same time, in the case of the phase transition in the concrete compound the model calculations are less reliable. In particular, the calculations for the original (model) structure of barium titanate give the possibility for both the FE and the AFE dipole orderings. Similarly the lead zirconate can be attributed to ferroelectrics. But, this does not show that the physical basis of the model calculations is poor. On the contrary it demonstrates the difficulties in taking into account all the factors. At the same time, the simplification of calculations by neglecting some of the factors leads to deviations from the concrete crystalline structure of the material. It is also useful to emphasize, that the traced relationship patterns in the behavior of the internal fields in ferroelectrics (barium titanate and lead titanate) and antiferroelectrics (lead zirconate) are almost identical under the influence of various external effects.

An estimate of the electrostatic dipole-dipole interaction showed that the stability of the AFE state with respect to the FE one increases with a decreasing electron polarizability of oxygen ions. Estimates of the polarizability of ions led to the conclusion that it should be higher in ferroelectrics [23]. This is consistent with the fact that the sizes of the oxygen ions in the ferroelectric material are greater than in the antiferroelectric one. At the same time the dimensions of the oxygen ions, and consequently, their polarizability decreases in the following cases:

- when the size of the ions in the *A*-sites decreases;
- when the size of ions in the *B*-sites increases, if their radius exceeds the value of $0.414 R_{O^{2-}}$.

Both these factors are equivalent to reduction of the structural *t*-factor. Moreover, the method of determination of the polarizability of the oxygen ion practically reduces its value to the one of the size of "emptiness", occupied by this ion. So the ion polarizability reflects the change in lattice parameters and nothing more. Therefore, the changes in the structure of barium titanate considered at the beginning of this section are actually caused by variations in the lattice parameters entered into the calculations.

Studies devoted to the analysis of external influences on phase transitions should be singled out from the studies examining physical causes of transition between the FE and AFE states. Without dwelling on the nature of dipole instability in perovskite structures and considering only a relative change of the stability of phases relative to one another, it is possible to distinguish the role of the "internal pressure" in the crystal lattice. Though, a physical interpretation still needs to be given to the very concept of the "internal (in other terminology – the chemical) pressure". This concept has a much in common with hydrostatic pressure by the nature of its effects on phase transitions. The factors leading to an increase in the "internal pressure" and, hence, to the improvement of the stability of the AFE phase are the same that contribute to a decrease in the polarizability of the oxygen ions. The same decrease of the polarizability causes a stabilization of the electrostatic energy of the AFE ordering. Thus, one can assume that the decisive cause of the phase transition may be the reduction of the configuration volume of the crystal lattice that is the lowering of the crystal's *t*-factor. External factors that lead to such changes in the crystal (primarily it is hydrostatic pressure) must stabilize the AFE phase.

In this connection the main result of the studies carried out in [24] and shown in Fig. 2.7 should be mentioned. This figure presents the dependence of the structural *t*-factor for the PZT-based solid solutions with compositions Pb(Zr$_{1-x}$Ti$_x$)O$_3$ on the titanium content. The FE phase is stable at larger values of the *t*-factor, whereas at smaller *t*-factor values the AFE phase is stabilized in the solid solution. The value of the *t*-factor, at which the FE and AFE phases change their stabilities, is marked on the dependencies of the *t*-factor on the solid solutions composition. These points correspond to the position of the boundary separating the regions of the FE and AFE states in the diagram of phase states of solid solutions. Diagrams of the phase states, obtained by the same authors a little earlier [25], are also shown in Fig.2.7. Various solid solutions were obtained by doping the PZT with different oxides. As one can see from these *t*(*x*) dependences, the transition from the FE state into the AFE state occurs at practically the same value of the *t*-factor independently on the chemical composition of the solid solution. The value of the *t*-factor is the key parameter for the change of the stable state's type.

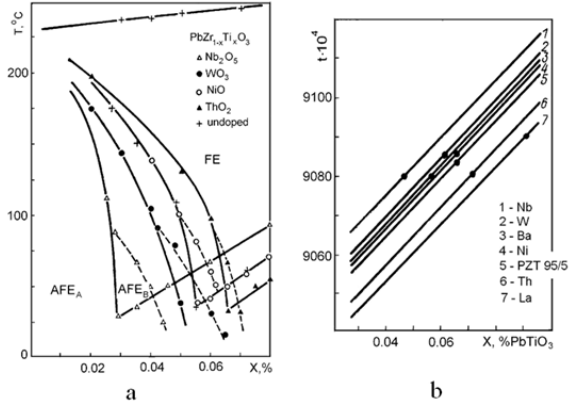

Fig. 2.7 (a) $x-T$ phase diagrams of the Pb(Zr$_{1-x}$Ti$_x$)O$_3$ solid solutions doped by 1 % of different oxides [25]. Points on the phase boundaries between the FE and AFE states were obtained at the room temperature.
(b) Dependencies of the $t$-factor on Ti content for the doped PZT solid solutions [24].
Points of the dependencies of the $t$-factor on Ti-content correspond to the boundary between the FE and AFE state in the "composition-temperature" phase diagram.

The value of an additional "internal (or chemical) pressure" produced by the ion substitution, is determined by the ratio

$$\Delta P = \frac{1}{\kappa}\left(\frac{V_B}{V_{B_0}} - \frac{V}{V_0}\right), \qquad (2.4)$$

where $\kappa$ is the isothermal compressibility; $V_0$ is the volume of the unit cell of an undoped material, obtained by the extrapolation of the dependence $V_0(T)$ from the PE phase to the room temperature; $V$ is the same volume for the doped material; $V_{B_0}$ and $V_B$ are average volumes of the octahedral ion in the undoped and doped substances, respectively. Formula (2.4) is in a good agreement with the experimental data obtained from a study of influence of the hydrostatic pressure on the FE-AFE phase transition in the PZT-based solid solutions [24].

All discussed above demonstrates the fact that if the FE and AEE states were present in the same substance the difference in their energies is small. Further in this paper we will discuss the consequences of such small difference in energies and the physical effects caused by such small energy difference.

## 3. Inhomogeneous state of the coexisting ferroelectric and antiferroelectric phases

Before proceeding to the further discussions of the behavior of the physical system under consideration, the following remarks are to be done.
1. The thermodynamic potential of the system has two minima (a stable and a metastable one) separated by the potential barrier within the region of the phase diagram limited by the lability boundaries of the FE and AFE phases. As a rule, if both the AFE and FE ordered states are realized in a substance (each taking place within its range of thermodynamic parameters), then the difference in energies of these states is small.
2. Under the conditions of real experiment (especially, at the varying temperature) a distribution of local temperatures $T(x, y, z)$ takes place in any actual substance. The probability distribution function $P(x, y, z)$ for $T(x, y, z)$ has a finite nonzero dispersion for measurements carried out in a quasi-static regime. This fact has to be always taken into account (for instance, some phenomena caused by the distribution of local temperatures $T(x, y, z)$ in substances with a single phase transition are considered experimentally in [26] and theoretically in [27]).As a consequence of the above-discussed dispersion of local temperatures a greater part of the sample's volume undergoes the transition into a state with a deeper minimum of the thermodynamic potential in the systems in question in the process of cooling. However, some part of the crystal volume undergoes the phase transition into a state corresponding to a less deep (metastable) minimum of the thermodynamic potential. The volume of the second phase decreases exponentially with the growth of the difference in energy of the phases. If the minima of the FE and AFE phases have close values, then the volumes occupied by these phases are to be approximately equal. As the temperature is lowered the height of the barrier separating these minima increases. As a consequence, the two-phase structure becomes more stable.
### 3.1. Theoretical model

The thermodynamic potential for an infinite sample of a two-phase system is presented as:
$$\varphi = \xi_1\varphi_1 + \xi_2\varphi_2 + W_{int} \tag{3.1}$$
where $\varphi_1$ and $\varphi_2$ are the thermodynamic potentials of the phases; $\xi_1$ and $\xi_2$ are the volume shares occupied by each of the phases; and $W_{int}$ describes the interaction between the domains of these phases.

The energy of the interaction of these phases is represented as follows [28]:
$$W_{int} = \frac{1}{2}\left\{\xi_2\sum_{i,j}\left[E_{\eta_{1,i}}(x_2,y_2,z_2)\eta_{2,j}(x_2,y_2,z_2)\right] + \xi_1\sum_{i,j}\left[E_{\eta_{2,j}}(x_1,y_1,z_1)\eta_{1,i}(x_1,y_1,z_1)\right]\right\} \tag{3.2}$$

Here $\eta_{1,i}$ and $\eta_{2,j}$ are the order parameters of the first and second phases, respectively ($i$ and $j$ are the numbers of the order parameters of these phases); $E_{\eta_{1,i}}(x_2,y_2,z_2)$ are the fields, induced in the domains of the second phase by the nonzero values of the order parameters belonging to the domains of the first phase; $E_{\eta_{2,j}}(x_1,y_1,z_1)$ are the fields induced in the domains of the first phase by the nonzero values of the order parameters of the second phase. Based on the physical nature of the order parameters (or the coexisting phases), it is evident that not all the terms in the expression (3.2) are nonzero. This is a consequence of the symmetry of the order parameters and the fields influencing the order parameters. The only nonzero terms in the expression (3.2) are those that are transformed according to the irreducible representations of the symmetry group containing the identity representation.

The dependencies of the fields $E_{\eta_{1,i}}(x_2,y_2,z_2)$ and $E_{\eta_{2,i}}(x_1,y_1,z_1)$ on the order parameters generating them have a complex form. To determine these dependences one must take into consideration the physical nature of the order parameters, the particular shape of domains, as well as the spatial distribution of order parameters in the domain's volume. The problem of determination of $E_{\eta_{1,i}}(x_2,y_2,z_2)$ and $E_{\eta_{2,i}}(x_1,y_1,z_1)$ must be solved using a self-consistent procedure that takes into account the situation that the fields influence the spatial dependencies of the order parameters defining these fields.

Such a problem does not seem to be solved exactly at present and, therefore, various approximations are to be used to find the solution. We shall restrict ourselves by the expansion of the fields $E_{\eta_{1,i}}(x_2,y_2,z_2)$ and $E_{\eta_{2,i}}(x_1,y_1,z_1)$ into the series of powers of $\eta_{\alpha,i}$:
$$E_{\eta_{1,i}}(x_2,y_2,z_2) = \xi_1 C_{1,i}(x_2,y_2,z_2)\eta_{1,i} + \cdots \tag{3.3a}$$
$$E_{\eta_{2,i}}(x_1,y_1,z_1) = \xi_2 D_{2,i}(x_1,y_1,z_1)\eta_{2,i} + \cdots \tag{3.3b}$$
Note that the considered fields are rather long-range and smoothly vary in space.

In the case when external fields are present, they may be taken into account in (3.1) in a usual way.

Thus, the study of behavior of the system of domains of the interacting coexisting phases is reduced to the investigation of the thermodynamic potential (3.1) with $W_{int}$ in the form (3.2), along with the expressions (3.3) under the condition $\sum_\alpha \xi_\alpha = 1$ ($\alpha = 1, 2$). The next step is a conventional procedure of minimization for the nonequilibrium thermodynamic potential
$$\varphi_\lambda = \varphi - \lambda\left(\sum_\alpha \xi_\beta - 1\right), \tag{3.4}$$
where $\lambda$ is the indeterminate Lagrangian multiplier. This procedure leads to the system of equations that give the equilibrium values of the order parameters:
$$\xi_\alpha(\partial\varphi_\alpha/\partial\eta_{\alpha,i} + E_{\eta_{\alpha',i}}) = 0, \qquad (\xi_\alpha \neq 0), \tag{3.5}$$
$$\varphi_\alpha + \eta_{\alpha,i}E_{\eta_{\alpha',i}} = \lambda = const, \qquad \alpha, \alpha' = 1, 2. \tag{3.6}$$

As seen from Eq. (3.6), the condition for coexistence of the thermodynamically equilibrium structure of domains of the coexisting phases is the equality of their thermodynamic potentials, taking into account the fields (3.3), but not the equality of the "bare" thermodynamic potentials. It also follows from Eq. (3.6) that there is a peculiarity of the considered multiphase structure connected with the spatial dependence of the coefficients in (3.3). This peculiarity is observed inside the region separating the domains of the coexisting phases which is wider than the usual domain boundary. The fields $E_{\eta_{\alpha,i}}$ are spatially dependent, and their intensity decreases while moving inside the domains of the other phase. Therefore, in these boundary regions the thermodynamic potential differs from the thermodynamic potentials characterizing the inner regions of domains. This fact testifies that the phase state inside the region separating the domains will not be similar to that in the inner regions of the adjacent domains. Physical considerations allow us to conclude that this state is transient between the states of the adjacent domains.

For a particular example of the FE and AFE phases, the simplest density of the nonequilibrium thermodynamic potential has the form [28, 29, 30]:

$$\varphi = \frac{\alpha_1}{2}P^2 + \frac{\alpha_2}{4}P^4 + \frac{\beta_1}{2}\eta^2 + \frac{\beta_2}{4}\eta^4 + \cdots + \frac{A}{2}P^2\eta^2. \tag{3.7}$$

Here $\alpha_2$, $\beta_2$ and $A$ are positive values ($A > \sqrt{\alpha_2 \beta_2}$), $\alpha_1$ and $\beta_1$ change their sign at the temperatures $T_{c,f}$ and $T_{c,af}$, respectively:

$$\alpha_1 = \alpha_0(T - T_{c,f}); \qquad \beta_1 = \beta_0(T - T_{c,af}). \tag{3.7a}$$

Considering the thermodynamic potential in the form (3.7), one has to take into account that the antiferroelectric phase transition is a structural transition into the state with the order parameter $\eta$ (see [29], for example), and the latter interacts with the polarization $P$, which is the order parameter of the ferroelectric state [29, 30, 31]. If $T_{c,f} > T_{c,af}$, the expression (3.7) describes the phase transition between the paraelectric and ferroelectric states at varying temperature, if $T_{c,f} < T_{c,af}$, (3.7) describes the paraelectric-antiferroelectric transition.

For definiteness let us consider the case when the ferroelectric phase (phase 1) is more stable in energy than the antiferroelectric one ($T_{c,f} > T_{c,af}$) and investigate the peculiarities of behavior of the system.

To write down the thermodynamic potential for each of the phases let us refer to Fig.3.1 in which the schematic equipotential lines of the nonequilibrium thermodynamic potential (3.7) are presented as the $P$ - $\eta$ plot. This schematic map of equipotential lines corresponds to the temperature interval that includes the local minima, corresponding to possible low-temperatures states of the system. The minimum corresponding to the FE phase has the coordinates ($P_{1,0}$, 0), where $P_{1,0}^2 = -(\alpha_1/\alpha_2)$. The equilibrium value of the energy for this state is $\varphi_{1,0} = -(\alpha_1^2/4\alpha_2)$. The minimum corresponding to the AFE state has the coordinates (0, $\eta_{2,0}$), $\eta_{2,0}^2 = -(\beta_1/\beta_2)$. The equilibrium value of the energy for this state is $\varphi_{2,0} = -(\beta_1^2/4\beta_2)$.

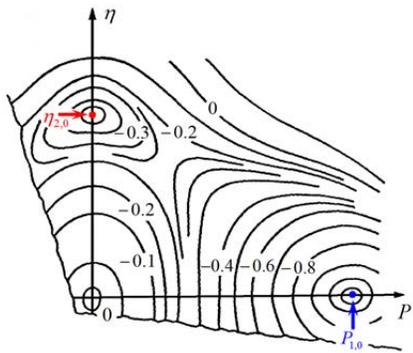

Fig.3.1. A schematic map of equipotential lines for the thermodynamic potential (3.7) in a temperature interval $T < T_{c,af} < T_{c,f}$

Now the nonequilibrium potentials for each of the phases may be written (up to the quadratic terms) as:

$$\varphi_1 = \varphi_{1,0} + U_1\left(P_1 - P_{1,0}\right)^2 + V_1\eta_1^2$$

$$\varphi_2 = \varphi_{2,0} + U_2 (\eta_2 - \eta_{2,0})^2 + V_2 P_2^2 . \tag{3.8b}$$

The coefficients $U_i$ and $V_i$ may be expressed in terms of the coefficients from the expansion (3.7). The most important for our consideration is that they are all positive.

In accordance with expressions (3.2) and (3.3) and taking into account the restrictions imposed by the order parameters symmetry (see the remark accompanying expressions (3.2)), one can present the energy of the interphase interaction as:

$$W_{int} = \xi_2 P_2 E_{P_1} + \xi_1 \eta_1 E_{\eta_2} = \xi_1 \xi_2 C_1 P_{1,0} P_2 + \xi_1 \xi_2 D_2 \eta_{2,0} \eta_1 \tag{3.9}$$

Now, according to (3.1), (3.8) and (3.9), the density of the nonequilibrium thermodynamic potential of the two-phase system can be written as:

$$\varphi = \xi_1 \varphi_{1,0} + U_1 (P_1 - P_{1,0})^2 + V_1 \eta_1^2 + \xi_2 \varphi_{2,0} + U_2 (\eta_2 - \eta_{2,0})^2 + V_2 P_2^2$$

$$+ \xi_1 \xi_2 C_1 P_{1,0} P_2 + \xi_1 \xi_2 D_2 \eta_{2,0} \eta_1, \qquad \xi_1 + \xi_2 = 1. \tag{3.10}$$

A conventional procedure of minimization with respect to $P_1$, $\eta_1$, $P_2$ and $\eta_2$ gives the equilibrium values of these parameters:

$$P_1 = P_{1,0}, \qquad \eta_1 = -\frac{1}{2} V_1 \xi_2 D_2 \eta_{2,0},$$

$$\eta_2 = \eta_{2,0}, \qquad P_2 = -\frac{1}{2} V_2 \xi_1 C_1 P_{1,0}. \tag{3.11}$$

Substitution of the expressions (3.11) into (3.10) gives the density of equilibrium energy for the two-phase system under consideration leads to expression:

$$\tilde{\varphi} = \xi_1 \left( \varphi_{1,0} - \frac{\xi_2^2}{4 V_1} D_2^2 \eta_{2,0}^2 \right) + \xi_2 \left( \varphi_{2,0} - \frac{\xi_1^2}{4 V_2} C_1^2 P_{1,0}^2 \right). \tag{3.12}$$

Comparison of expressions (3.12) and (3.8) shows, that the phase interaction in the form (3.9) lowers the energy of the system. As the final result, this two-phase state may acquire larger energy stability than a single phase state, if the difference of $(\tilde{\varphi} - \varphi_{1,0})$ is negative. The latter condition is fulfilled in the vicinity of the ferroelectric-antiferroelectric stability line. In this case $\varphi_{1,0} \cong \varphi_{2,0}$, $\xi_1 = \xi_2 = 1/2$, and the absolute energy advantage due to the formation of the inhomogeneous two-phase state will be equal to:

$$\Delta W = -\left( \frac{\xi_1 \cdot \xi_2^2}{4 V_1} D_2^2 \eta_{2,0}^2 + \frac{\xi_1^2 \cdot \xi_2}{4 V_2} C_1^2 P_{1,0}^2 \right) \xrightarrow{\xi_1 = \xi_2 = 1/2} \Delta W = -\frac{1}{32} \left( \frac{D_2^2}{V_1} \eta_{2,0}^2 + \frac{C_1^2}{V_2} P_{1,0}^2 \right) \tag{3.13}$$

As seen from (3.3), the fields $E_1$ and $E_2$ and, consequently, the coefficients $C_1$ and $D_2$, are the functions of the coordinates. They decrease as the distance from the interphase boundary increases. Thus, the long-range interaction fields favour the stabilization of the inhomogeneous state to the greatest degree. The stability of the inhomogeneous state is also raised by low values of the parameters $V_1$, $V_2$, $U_1$ and $U_3$.

In the considered case of the coexisting FE and AFE phases the long-range fields have electric (dipole) and elastic nature. Since in each real crystal there exist free charges, crystal lattice defects (e.g. dislocations), the radius of action of these fields is not infinite, and corresponds to the distance equal to several tens of lattice parameters. Therefore the energy advantage of the two-phase state (13) is ensured by the regions separating domains of the phases in the sample's volume.

For a more general case (given by the equation (3.9)) than considered above the expression for the interphase interaction energy has the form:

$$W_{int} = \xi_1 \xi_2 C_1 P_1 P_2 + \xi_1 \xi_2 D_2 \eta_1 \eta_2 . \tag{3.14}$$

This form of the phase interaction energy was considered in [28, 32], and it was found that the energy stabilising the inhomogeneous state (for $\varphi_{1,0} \cong \varphi_{2,0}$, $\xi_1 = \xi_2 = 1/2$) is given by the expression:

$$\Delta W = -\frac{1}{32}\left(\frac{D_2^2}{V_1}\eta_{2,0}^2 + \frac{C_1^2}{V_2}P_{1,0}^2\right) - \frac{1}{512}\left(\frac{C_1^4}{U_1 V_2^2}P_{1,0}^2 + \frac{D_2^4}{U_2 V_1^2}\eta_{2,0}^2\right) \qquad (3.15)$$

As one can see, the inhomogeneous state becomes even more stable when a more general case is considered.

The above cconsideration demonstrates that the interaction between the coexisting FE and AFE phases (the phases described by different order parameters) actually stabilizes the inhomogeneous state of domains of these coexisting phases. The negative value of $\Delta W$ in (3.13) (and (3.15) in a more general case) demonstrates in fact that the considered interphase boundary possesses a negative surface energy. Such a negative energy will lead to the division of the sample volume into unlimited number of very small domains (thus ensuring the largest area of the interphase boundaries and, consequently, the largest energy advantage).

The presence of different phases and the interphase interaction have been confirmed experimentally by an X-ray diffraction and a transmission electron microscopy (TEM). However, it should be emphasized that the notions of phase and phase state are defined and introduced for the case of rather large finite volume of substance (theoretically, for the volume which tends to infinity). This means that division of a finite volume of the substance into domains of the coexisting phases cannot continue indefinitely. This process will have to end at a certain stage. The latter is defined by the least size of the phase domains at which the phases still exist. At further reduction of the domain size the phases disappear and one cannot speak of their existence as well as an interphase interaction and the energy of interphase boundaries.

A simplified consideration of the problem yet reflecting the main physical result is presented in this review. For example, specific effects associated with the conditions of continuity for elastic medium at the interphase boundary have not been taken into account. The interphase domain wall separates the domains with the FE and AFE states. Elementary crystal cells of the FE and AFE phases have different sizes. In connection with the above-said two, distinctive features should be emphasized. First, it has been demonstrated in [33] that at the conjugation of crystal planes of the phases with close crystal structures and with different inter-plain distances (this is the situation when the FE and AFE phases coexist) the intervals between possible dislocations are of order of several tens of *nm*. Second, the TEM studies of the coexisting domains of the FE and AFE phases showed that linear sizes of these domains are also of order of several tens of nanometers [34, 35], and no dislocations at the interphase domain boundaries were found.

This indicates that the crossing of the interphase domain wall (from one phase to the other) is accompanied with continuous conjugation of the crystal planes (free of breaks and dislocations). Such a coherent structure of the interphase domain wall leads to an increase in the elastic energy. This increase is the more essential the larger is the difference in the configuration volumes of the FE and AFE phases.

The above-discussed effect (the increase in the elastic energy) gives a positive contribution into the surface energy density of the boundaries separating ordinary domains in ferroelectrics [36, 37] and weakens the condition of the inhomogeneous state existence. Using an example of the $Pb_{0.90}(Li_{1/2}La_{1/2})_{0.10}(Zr_{1-y}Ti_y)O_3$ system of solid solutions [38], it has been shown that changing the differences in the interplane distances of the FE and AFE phases (by varying the solid solution composition – $y$) one can observe the change from the mechanism providing the elastic strictional blocking in the process of creation of the metastable phase domains to the mechanism that provides the inhomogeneous state of coexisting domains of the FE and AFE phases.

Thus, the analysis of the system described by the thermodynamic potential in Eq. (3.1) with the interphase interaction (3.2) shows that the stable state of this system may be inhomogeneous (see Eq. (3.6) and the commentary to it). It is shown that in the substances characterized by the FE-AFE phase transition and described by the potential (3.7) this inhomogeneous state may be considered as the state of the coexisting domains of the FE and AFE phases with a negative energy of the interphase boundary.

The possibility of existence of the inhomogeneous states with negative interphase boundary energy leads to two significant conclusions. First of all, this is the possibility of a simple, an elegant, and self-consistent physical explanation of different aspects of known phenomena that have not been adequately treated so far. The second conclusion concerns the prediction (and subsequent experimental verification) of earlier unknown phenomena in different physical processes.

## 3.2. Experiments on visualization of the inhomogeneous states

First experiments on confirmation or disproof of the above-presented results were carried out in [34, 35]. A direct observation of coexisting domains of the FE and AFE phases in the samples of the $Pb_{1-3x/2}La_x(Zr_{1-y}Ti_y)O_3$ ( PLZT) system of solid solutions with composition 7/65/35 (the first number is the percentage of the La-content, the second number is the percentage of Zr, and the third number is the percentage of Ti) and in the PLZT samples with composition 8/65/35 was carried out by the TEM in [39]. Compositions of the above-indicated solid solutions are located inside the hysteresis area separating the regions of stable FE and AFE states in their corresponding "composition-temperature" phase diagrams. The last circumstance means that these phases possess practically equal phase stability having close values of their free energies. The AFE state is the main phase state of these solid solution at room temperatures (it possesses a deeper minimum of the free energy) while the FE state is metastable.

A coarse-grained ceramics with 8 to 10 μm sizes of crystallites was used for observations of the domain structure. The spalls of certain crystallites (produced along different crystallographic planes) with a thickness of not more than 0.2 μm were chosen for the observation of domain structure using a JEM-200 transmission electron microscope. These spalls were essentially the lamella-type crystals of PLZT cut out along different crystallographic directions. Such method of sample preparation allowed us to avoid clamping due to the surrounding crystallites which is always present after the sample preparation by thinning of ceramic samples as well as to avoid physical effects caused by this clamping.

TEM images of the structure of the 7/65/35 PLZT for two crystallite spalls are given in Fig. 3.2. The bright-field image for the sample with the (110) spalling plane (the incident electron beam was along the [110] direction) is presented in Fig.3.2a. The two-phase structure in the form of ellipsoidal dark regions in a lighter shade AFE matrix is clearly visible at this orientation of the crystallite. The domain diameters are of 2 − 4 μm, and their density is of the order of $10^{11}$ cm$^{-2}$. The opacity of these regions is caused by the additional scattering of electron beam by the depolarization field of the FE domain. In spite of the difference in the interplane spacings of these two phases the dislocations are not present at the interphase boundaries. Specific manifestations of distortions of the crystal matrix that can be produced by the elastic stresses caused by the metastable phase domains are also absent in this image.

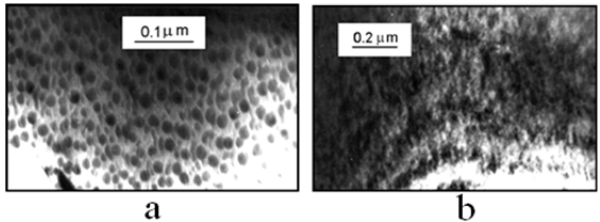

Fig. 3.2. Transmission electron microscopy bright-field images for 7/65/35 PLZT.
*a* – for (110) the foil plane, *b* – for (120) the foil plane [34, 35].

Electron diffraction images of the same crystallite manifest the system of reflections that are unique for the perovskite crystal structure of the PLZT solid solutions without the presence of any reflections corresponding to irrelevant inclusions.

The two-phase domain structure of crystallite oriented in such a way that the incident electrons beam is along the [120] crystal direction is presented in Fig. 3.2b. In this picture the same domain structure is under the oblique incidence of an electron beam which gives the image of the same domain structure form the side. The filamentary structure of domains ingrown through the crystallite thickness is clearly visible. Similar images have been obtained for the other spalls of crystallites in the case when their [110] directions were misaligned with respect to the electron beam.

Analysis of obtained images with different orientations of crystal planes with respect to the incident electron beam showed that the domains of metastable phase have a cylindrical shape with the axes along the [110] direction. These domains are grown through the thickness of the crystallite if their spalling plane was (110). Such shape of the coexisting domains of the metastable FE phase occurs in thin crystals only. In the bulk samples the shape of these domains is ellipsoid of revolution close to a sphere [40, 41, 42].

There exist a number of recently obtained TEM images of the PLZT solid solutions. As a rule these images are obtained on samples with a different prehistory of preparation. Even within one particular publication the authors not always pay close attention to this matter. But the sample prehistory is exactly the feature that determines the phase composition of the solid solutions that fall into the hysteresis region in the phase diagram. With rare exceptions the authors of these publications see the structures similar to the one shown in Fig. 3.2b. As an example we want to mention [43, 44]. These papers contain studies of the structure of annealed (and left to age for some time) samples and the structure of samples subjected to action of an AC electric field exceeding the value of the coercive field in the PLZT solid solutions at the room temperature. Since the authors did not have a concrete goal of searching for the coexisting domains they did not watch closely for the specific orientation of their samples with respect to the electron beam.

In the paper [45] that followed the first publication devoted to the direct identification of the system of coexisting domains the TEM studies were carried out on the samples obtained by thinning of ceramic samples by the mechanical grinding and polishing, the chemical etching, and the ion-beam etching at the final stage. In the process of such preparation of samples for TEM studies the clamping of the crystallite, which was selected for observations, by the neighboring crystallites always takes place. Thus, the imaging of the structure takes place in the presence of the lateral mechanical stress (in the plane of observation). This lateral stress may be compressive and tensile. The image of the domain structure observed in [45] is shown in Fig. 3.3. As one can see the system of cylindrical domains has transformed into the system of stripe domains. Similar transformation of one domain system into another is possible in the system of cylindrical magnetic domains.

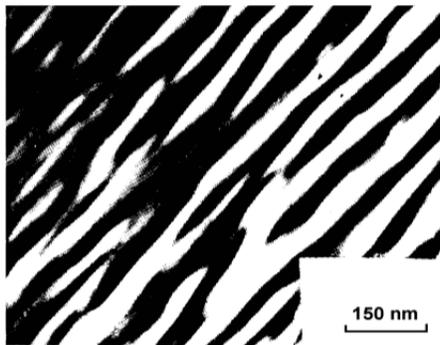

Fig. 3.3. Aligned macrodomain structure in PLZT [45].

In relation to the above said it is interesting to follow up the changes of domain structure existing at the nanoscale during the variation of the relative stability of the FE and AFE phases (for example, using changes in the composition of the solid solution primarily due to the variation of the La-content). Such studies have been carried out by a transmission electron microscopy on the samples of the X/65/35 and X/60/40 series of the PLZT solid solutions [46]. The composition of the second series of solid solutions X/60/40 is located close to the composition of the series X/65/35 in the PLZT phase diagram. The crystal structure of the FE phase (when its stability slightly exceeds the stability of the AFE phase) is a perovskite structure with the tetragonal distortions only in the X/60/40 series of solid solutions whereas the crystal structure of the FE phase in the X/65/35 series manifests the rhombohedral distortions. Increase of the lanthanum concentration in specified solid solutions leads to the transition from the region of the FE states ($X \leq 5$ for the X/65/35 series) into the region of the AFE states ($X \geq 8.5$) through the hysteresis region ($6.0 < X < 8.0$) in the phase diagram.

Bright-field images of the microstructure of the X/65/35 PLZT series with the clearly traceable transition of the domain structure with variation of the lanthanum concentration are given in Fig. 3.4. Superfine domain structure with the nanometer scale exists in solid solutions with $X \geq \sim 8$ (Fig. 3.4a).

Sizes of domains of the AFE phase are of the order of 20 – 30 nm, which is in agreement with results of [34, 35, 38] obtained earlier. Increase of both the share of the AFE phase and the size of its domains takes place when the lanthanum concentration decreases (that is the stability of the AFE phase increases) is clearly visible in the image of the 7/65/36 PLZT (Fig.3.4b). In some cases the FE domains combine into the system of stripe domains in the AFE matrix. Authors of [46] also noted that the recharging of samples taking place during studies in electron microscope did not result in the domain motion. This result confirms the increase of stability of the FE state one more time.

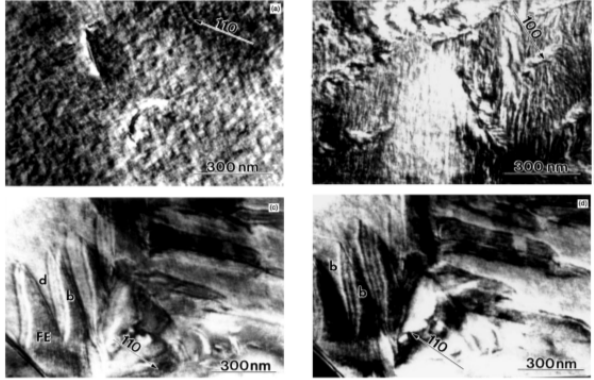

Fig. 3.4. Room temperature transmission electron microscopy images of various PLZT X/65/35 compositions revealing [46]:
(*a*) mottled contrast typical for relaxor ferroelectrics, imaged in dark field (DF), X = 8;
(*b*) domains with striation-like morphology, imaged in bright field (BF), X = 7;
(*c*) ferroelectric domains (labelled FE) and asymmetric δ fringes (labelled b, d), BF, X = 5;
(*d*) the same grain as in (c), imaged in DF, revealing symmetric δ fringes (labelled *b*).

The fully established system of FE domains on micrometer scale is observed in the 5/65/35 PLZT (Fig.3.4*c*, 3.4*d*). If to refer to the phase diagram of the X/65/35 PLZT system of solid solutions it is easy to see that the 5/65/35 PLZT is located in the region of spontaneous ferroelectricity at room temperature and its location is close to the hysteresis region of the FE-AFE phase transition. According to the above-considered model (see section 3.1) the interphase interaction lowers the energy of the two-phase FE-AFE domain with respect to the energy of the homogeneous FE state. This circumstance is manifested in the transmission electron microscopy images.

"Ordinary" ferroelectric domains (such domain is marked as FE in Fig. 3.4c) possess clearly expressed boundaries, the twinning planes are mainly {110}. Domain ordering in these planes is "head-to tail" and they can be considered as 109° domains [47]. Twinning planes {100} separating 71° domains are essentially less common. Authors of [46] noted (even though it is not clearly seen in the images) the presence of spatial regions (with contrast of the type shown in Fig.3.4a) between the "ordinary" FE domains. This contrast is characteristic for the structure of the nanometer scale domains of the coexisting FE and AFE phases. Precisely in these regions of crystallites the mechanical stresses are concentrated. These stresses increase the stability of the AFE state with respect to the FE state and stabilize the nonhomogeneous state of domains of the coexisting phases in the crystal with composition actually corresponding to the region of spontaneous ferroelectricity in the phase diagram.

The so-called δ-fringes (they are marked as *b* and *d* in the Fig. 3.4c and 3.4d) appear in the vicinity of the ("ordinary") domain boundaries inclined with respect to the plane of the foil. This δ-fringe is visible more clearly in the dark-field images. The structure of wedges that limit this δ-fringe represents the line structure of nanoscale domains. The reason for appearance of such structure in the images is some (weak) misorientation of the crystallographic planes in the neighboring nanodomains leading to a misorientation of the scattering vectors of electrons. The most probable reason for appearance of such domain structure is the increase of the stability of the two-phase FE-AFE state and the nucleation of the system of coexisting domains of these phases. Such formations are observed in the regions of concentration of elastic stresses. In these cases the side (in the plane of the image) stresses lead to the transformation of the system of cylindrical domains into the system of stripe domains. In connection with the above-said it ought to be noted that such system of stripe domains was first observed in [48] in 1974 which now seems like a long time ago. Most likely the majority of researches are not aware of this particular study.

Let us now dwell on the X/60/40 system of solid solutions. Transmission electron microscopy images of the samples with different concentrations of lanthanum, X, are shown in Fig.3.5. Dissimilarity of this series of solid solution from the above-considered X/65/35 series is that the boundary of spontaneous ferroelectric state at room temperature is located at slightly higher concentration of lanthanum (at X close but slightly higher than 8) and the crystal structure of this state possesses the tetragonal type of distortions.

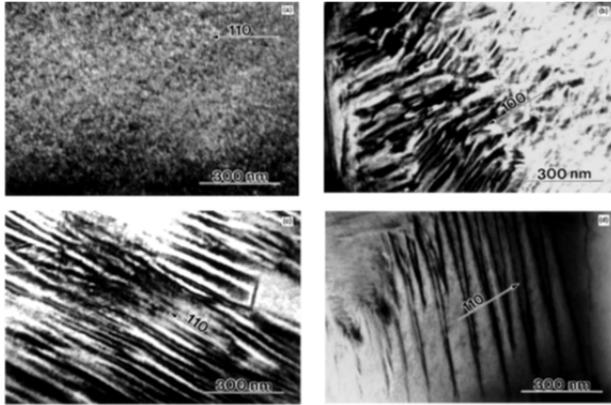

Fig.3.5. Room temperature images for various X/60/40 compositions of the PLZT solid solutions [46] revealing: (*a*) mottled contrast, bright field BF, X = 10; (*b*) coarser domains formed near grain boundaries and mottled contrast at grain centers, BF, X = 9; (*c*) ferroelectric domains with stripe-like morphology, BF, X = 8; (*d*) wedge-shaped domains, BF, X = 8.

In general lines the physical picture observed in this series of solid solutions is similar to the one observed in the X/65/35 PLZT. The microstructures of the 10/60/40 and 9/60/40 are similar and represent the coexisting domains of the FE and AFE phases with sizes of the order of 20–30 nm. Enlargement of the metastable FE domains and their merging is noticeable during the transition from the first system of solid solutions to the second one. One can observe small-scale stipe domains in the vicinity of domain boundaries. The appearance of domains of such form is caused by mechanical stresses. It is quite analogous to the situation taking place in the 8/65/35 and 7/65/35 PLZT solid solutions. The stripe domain structure emerges with a further decrease of lanthanum content in the solid solution (that is at an increase of the stability of the FE state). This stripe domain structure is different from the stripe domain structure of the nanometer scale existed in the 5/65/35 PLZT solid solution (see Fig.3.4c and 3.4d). One can observe the remains of the small scale structure existing in the 10/60/40 PLZT (Fig. 3.5a)

The primary domain structure in the 8/60/40 PLZT contains ordinary $90°$-domains with polarization axes of the neighboring domains directed perpendicularly to one another. These very domains form the domain structure shown in Fig.3.5c according to the authors of [46]. Wedge-shape domains are also present in the 8/60/40 solid solution. The presence of such domains is responsible for appearance of the {100}-type twin planes. As it was noted in [49] such domains appear in the $PbZrO_3$ and also in the PZT solid solutions with small titanium concentrations during the FE-AFE phase transition caused by the change of the temperature. It is again the evidence of the fact that the energy stability of the FE and AFE states differs only slightly.

The above consideration gives the picture of behavior of the microstructure in the PLZT solid solutions at a variation of the relative stability of the FE and AFE phases. This variation has been achieved by changing the lanthanum concentration at the constant ratio of titanium and zirconium concentrations. All presented results are in complete correspondence with the model of the inhomogeneous state of coexisting domains of the FE and AFE phases given in section 3.1.

In conclusion let us discuss the behavior of the same microstructure during the variation of the relative stability of dipole-ordered phases due to the change of the ratio of titanium and zirconium concentrations that takes place along the horizontal cross section of the phase diagram of PLZT with the lanthanum concentration X = 8. Corresponding transmission electron microscopy images [46] are given in Fig.3.6 (the images that were presented earlier are also shown here for the reader's convenience). It has been noted above that at the room temperature the 8/60/40 solid solution is located in the spontaneous ferroelectricity region of the phase diagram. The domain structure of this solid solution corresponds (with the account of all notes and comments given above) to the "ordinary ferroelectric" (Fig.3.6.c). The stability of the AFE state rises with an increase of the concentration of zirconium and the AFE phase becomes more stable than FE in the 8/65/35 solid solution but still close to the region of stability of the FE phase. This means that there are two free energy minima with close depths. Base on the model of interacting phases the consequence of the existence of these minima leads to the presence of nanoscale domains of the coexisting FE and AFE phases. As one can see in Fig.3.6b the domains of the indicated-above phases coexist in this solid solution and the shares of both these phases are approximately equal.

Further stabilization of the AFE phase happens with the further increase in the concentration of zirconium. The free energy of the 8/70/30 PLZT has two minima. The minimum that corresponds to the AFE phase is much deeper than the minimum for the FE phase. As a consequence the share of the AFE phase has to prevail in the solid solution with this composition. This conclusion is amply demonstrated by the image in Fig. 3.6a.

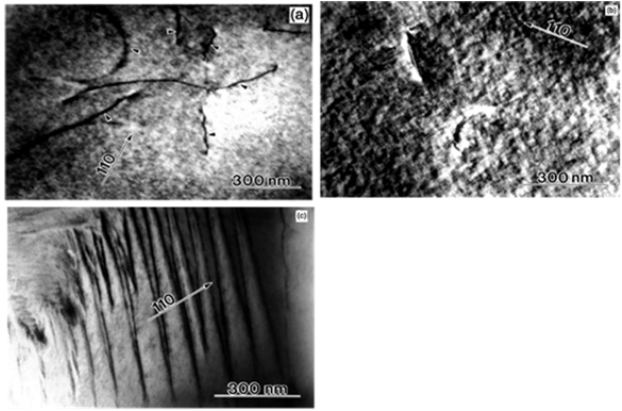

Fig. 3.6. Bright-field images of domain structure in the PLZT solid solutions with lanthanum content X = 8 [46].
Solid solution: a − 8/70/30, b − 8/65/35, c − 8/60/40.

Properties of the PLZT solid solutions with compositions from the region of induced states and peculiarities in their behavior (primarily under the action of an electric field since an external pressure is rarely used due to the cumbersome procedure) are widely discussed in the literature and there is no doubt that the induced phase is a FE phase. All back-and-forth discussions are usually about the nature of the state before the field was applied. Nowadays when the complete set of the phase diagrams for these solid solutions have been obtained (see, for example, [37, 50]) the AFE nature of this initial state is now evident. Peculiarities of this state are determined by the small difference in energies of the two dipole-ordered states. Due to such small difference in energies the metastable FE phase is present in the AFE matrix of crystallites. This two-phase state turns out to be more stable than a single phase one.

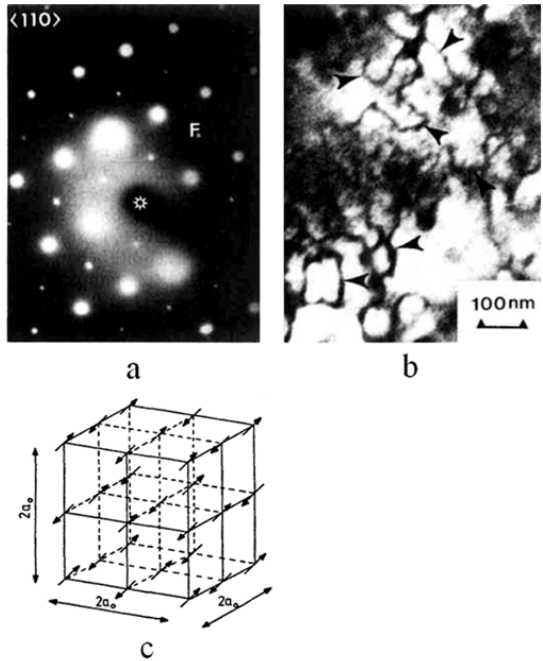

All above-presented results of transmission electron microscopy studies fully corroborate the two-phase character of states in the PLZT solid solutions with compositions form the hysteresis region of the phase diagram. In addition to what was said above we would like to turn to results of [45]. The authors of this article studied the domains of stable phase in solid solutions from the hysteresis region of the phase diagram of PLZT. The system of the $\left(h+\frac{1}{2}, k+\frac{1}{2}, l+\frac{1}{2}\right)$-type reflections was present in electron diffraction images besides of the base Bragg reflections at room temperature. The intensity of reflections was rising slightly with a decrease of temperature. The dark-field image of the part of crystallite obtained using one of these fine structure reflections is given in Fig. 3.7a. The antiphase boundaries (they are marked by arrows) separating the AFE domains (as it will be shown slightly further) are clearly visible.

Fig.3.7. $\langle 110 \rangle$-zone axis diffraction pattern (*a*), dark field

image showing domains associated with *F*-spot in the 8.2/70/30 PLZT, at T = −40 °C (*b*) and anti-parallel ordering in the PLZT antiferroelectric structure (*c*) [45].

Usually the system of fine structure reflections of the $\left(h+\frac{1}{2}, k+\frac{1}{2}, l+\frac{1}{2}\right)$-type corresponds to the one of the following several cases:
– during ordering in the B-sublattice [51]
– during the rotation of octahedrons about the main diagonal of the cubic of perovskite crystal lattice [52]
– in the case of antiparallel dipole ordering (the antiphase displacements of the lead ions along the $M_5'$ mode [53]) in the AFE phase of lead zirconate and in the lead zirconate based solid solutions, for example PZT.

The first of the cases listed above does not suggest the existence of the temperature dependence of the fine structure reflections in the temperature interval close to room temperature. As for the two other cases of appearance of the $\left(h+\frac{1}{2}, k+\frac{1}{2}, l+\frac{1}{2}\right)$-type reflections they accord fully with the AFE state of the solid solutions doped with small amount of lanthanum ions (with the rotations of the oxygen octahedrons about the [111] axis). This situation was analyzed in [45] and the authors suggested the arrangement of the ordered dipoles (given in Fig. 3.7 with antiparallel orientation of dipoles in the lattice sites in the [111] plane of the crystal lattice of PLZT).

The authors of the study [54] entertained slightly different point of view. They related the appearance of the $\left(h+\frac{1}{2}, k+\frac{1}{2}, l+\frac{1}{2}\right)$ reflections with the phase transition into a state analogous to the low-temperature rhombohedral state of the PLZT. While at the same time they noted that the same $\left(h+\frac{1}{2}, k+\frac{1}{2}, l+\frac{1}{2}\right)$ reflections take place in the solid solutions with small concentrations of lanthanum. They stated that the appearance of these reflections in these solid solutions has an unambiguous interpretation as the presence of the AFE state because the softening of the $\Gamma_{25}(R)$-mode plays essential role in the stabilization of the AFE state in the lead zirconate and in the PZT solid solutions.

A different method of studies of the coexisting phases in the X/65/35 series of the PLZT solid solutions was used in [55]. The small scale structure of domains of the coexisting phases was studies by means of a piezoelectric microscope. Hot pressed ceramic samples of the 9.75/65/35 PLZT with the thickness of the order of 0.3 – 0.5 mm polished to the optic quality were used in these experiments. Experimental method permits to register the specific patterns created at the sample surface by the domains in the bulk and, thus, register the "emergence of domains" at the sample surface. Unlike the transmission electron microscopy this method is "less direct," however, it allows extracting effects caused by the application of a DC electric field allowing at the same time to keep track of the material's behavior at the nanoscale level of resolution. To this must be added a much simpler procedure of sample preparation for this studies.

As it follows from the PLZT phase diagrams [34, 35, 37,56] the basic state of the 9.75/65/35 solid solution at room temperature is the AFE state with an admixture of the FE phase as an "impurity". This very structure has been observed by authors of [37] at room temperature. It worth to discuss results obtained in [55] in greater detail because such type of studies has not been adequately covered for the time being (in comparison, for example, with transmission electron microscopy studies). Some important features of the apparatus are the following. The tip with the radius of the tip apex slightly less than 10 nm was used. The tip was retained against a sample's surface by the spring with rigidity of 35 N/m. The visualization of domains was assured by application of 4 V of AC electric potential with a frequency of 5 kHz (far from the resonance frequency of the cantilever).

The topography and piezoelectric response (at two magnifications) from the surface of the ceramic sample of the 9.75/65/35 PLZT is shown in Fig.3.8. As one can see at this magnification the structure of grains is inhomogeneous consisting in polar (piezoelectrically active) and nonpolar domains. One can see in Fig. 3.8 that the polar axes in the nanodomains from neighboring crystallites are oriented in different spatial directions, which are pointed out by the distribution of their response (but this does not mean that they are oriented along different crystallographic axes). Domain sizes are determined in the interval from 30 to 50 nm which is in agreement with the sizes (although slightly bigger) obtained using transmission electron microscopy (Fig.3.2 and Fig.3.3). The dependence of the dipole moment of the polar domain on its area (exposed at the polished sample's surface) is determined as $P(S) = P_0 S^{D/2}$ [55] with D ≈ 1.55. This value is close to 1.50 that has to be observed at completely random distribution of polar directions [57, 58]. Some discrepancy can be due to the depolarization fields generated by polar domains.

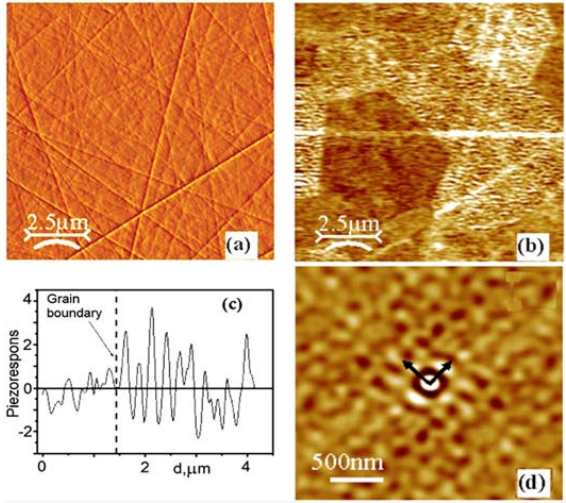

Fig.3.8. Typical topography (a) and piezoresponse (b) and (d) images of the 9.75/65/35 PLZT ceramics. Cross section of piezoelectric response image (c) shows the change of piezoelectric response in adjacent grains [55].

When DC electric field is applied to the nonpolar AFE domain an induction of the FE state takes place (which is demonstrated in Fig. 3.9) and the piezoelectric hysteresis loop is observed when the electric field intensity varies. In the same image one can clearly see the appearance of contrast in the region where the induction of the FE phase took place. The reverse transition takes place when the electric field is removed however this process is the long-term one. The contrast that has been preserved during 7 hours after the field was switched off is visible in Fig. 3.9. The mechanism of such long-term relaxation can be explained by the proximity of the free energies of the FE and AFE phases at room temperature. It will be discussed below in detail.

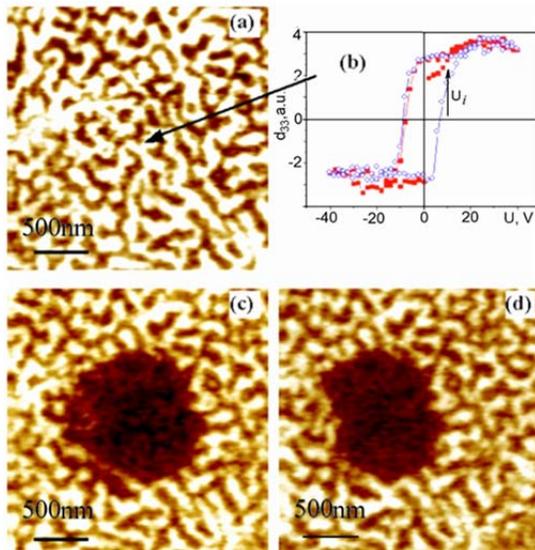

Fig.3.9. Initial piezoelectric response image (*a*), local piezoelectric response hysteresis loop (*b*) (arrow indicates position where the loops was measured), piezoelectric response image just after the application of a bias field (*c*), and piezoelectric response image relaxation during 7 h (*d*) [55].

## 4. Phase diagrams and effects caused by the anharmonicity of the elastic energy

### 4.1. Theoretical analysis

As it follows form expressions (3.13) and (3.15), the smaller are the values of the expansion coefficients $U_1$, $U_2$, $V_1$ and $V_2$, the higher is the contribution of the energy advantage due to the inhomogeneous two-phase state formation into the free energy of the system under consideration. Small values of the coefficients $U_1$, $U_2$, $V_1$ and $V_2$ correspond to gently-sloped potential surfaces for each of the ordered low-symmetry states of the system. Gentle slopes in potential energy lead to the phase transitions (the first order ones) with the extended hysteresis regions separating the FE and AFE states. In what follows, we will consider physical situations and concrete materials with such hysteresis regions. We will also analyze the manifestation of the inhomogeneous state of the coexisting domains of the FE and AFE phases in experiments.

The phase diagrams and phase transitions for the system with the FE-AFE phase transition are described by the model of two interacting order parameters (for the FE and AFE states) based on the nonequilibrium thermodynamic potential of the following form [29, 30, 59]:

$$\Phi = \Phi_P + \Phi_\eta + \Phi_{P\eta} + \Phi_u + \Phi_{Pu} + \Phi_{\eta u}; \tag{4.1}$$

$$\Phi_P = \frac{\alpha_1}{2}P^2 + \frac{\alpha_2}{4}P^4, \qquad \Phi_\eta = \frac{\beta_1}{2}\eta^2 + \frac{\beta_2}{4}\eta^4, \qquad \Phi_{P\eta} = \frac{A}{2}P^2\eta^2.$$

The last two terms have the form:

$$\Phi_{Pu} = Qu_{ll}P^2, \qquad \Phi_{\eta u} = \lambda u_{ll}\eta^2.$$

The elastic energy is usually considered in the harmonic approximation:

$$\Phi_u = \frac{K}{2}u_{ll}^2 + P_{ext}u_{ll}.$$

Here the following notations are used: $P_{ext}$ is the external hydrostatic pressure, $Q$ and $\lambda$ are the striction coefficients, $u_{ll}$ is the sum of diagonal components of the deformation tensor.

Then the conventional procedure consists in the minimization of (4.1) with respect to $u_{ll}$:

$$Ku_{ll} = \sigma; \qquad \sigma = -(QP^2 + \lambda\eta^2 + P_{ext}) \tag{4.2}$$

and the thermodynamic potential assumes the following form:

$$\Phi = \frac{\alpha_1'}{2}P^2 + \frac{\alpha_2'}{4}P^4 + \frac{\beta_1'}{2}\eta^2 + \frac{\beta_2'}{4}\eta^4 + \cdots + \frac{A'}{2}P^2\eta^2. \tag{4.3}$$

The phase diagrams for the potential (4.3) drawn in the $\alpha_1'$, $\beta_1'$-plane and in the "pressure-temperature" plane are presented in Fig.4.1a. These diagrams are in a good agreement with the experimental ones obtained for the PbZrO$_3$ [60] and the PZT with small content of Ti [61]. The peculiarity of both calculated and experimental diagrams is the linear character the boundaries of stability regions (the lability boundaries) for all the phases.

Studies of temperature dependencies of the Mossbauer Effect probability in PbZrO$_3$ and PbTiO$_3$ were carried out in [62]. Based on these dependences the authors estimated the degree of anharmonicity for elastic potentials describing oscillations of the ions occupying B-sites of the perovskite crystal lattice in these compounds. It turned out that in PbZrO$_3$ the elastic potential is practically harmonic whereas in PbTiO$_3$ the deviation from harmonicity is large. The considered phenomenon is the so-called "low temperature" anharmonicity caused by the discrepancy between the size of the ions and the volume they occupy in the crystal lattice.

Similar discrepancy between ion size and the volume occupied by this ion in the crystal lattice is proper to other substances with the perovskite structure when the ions with different radii occupy equivalent lattice sites. Parameters of the crystal lattice in the said substances are determined by the "driving" the lattice towards average radius of ions. However, ions with smaller radius have "freedom" among its crystal surrounding. Naturally, in this case low-temperature anharmonicity should be taken into account.

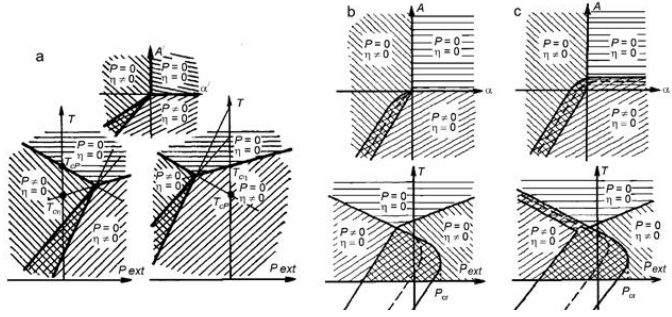

Fig.4.1. Phase diagrams described by the potential (4.3) at different values of expansion coefficients:

a - $A' > 2\sqrt{\alpha'_2 \beta'_2}$; $A' < 0$ - for s$T_{CP} > T_{C\eta}$ and $T_{CP} < T_{C\eta}$ .

b - $A' > 2\sqrt{\alpha'_2 \beta'_2}$; $\alpha'_2 > 0$; $\alpha'_3 \neq 0$;

c - $A' > 2\sqrt{|\alpha'_2| \beta'_2}$; $\alpha'_2 < 0$; $\alpha'_3 \neq 0$.

The diagrams in the expansion coefficients' coordinates of the potential (4.3) are in the upper row. The lower row contains "pressure-temperature" diagrams.

Therefore, the terms of higher orders in deformation are to be taken into account in the expansion (4.1) for the investigation of phase transitions and phase diagrams:

$$\Phi_u = \frac{K}{2} u_{ll}^2 + \frac{G}{3} u_{ll}^3 + \frac{D}{4} u_{ll}^4 + P_{ext} u_{ll}. \qquad (4.4)$$

The equilibrium deformation value in this case is defined by the expression:

$$K u_{ll} + G u_{ll}^2 + D u_{ll}^3 = \sigma; \qquad \sigma = (QP^2 + \lambda \eta^2 + P_{ext}), \qquad (4.5)$$

or, in the approximation quadratic in σ

$$u_{ll} = s_1 \sigma + s_2 \sigma^2 \qquad (4.6)$$

here $s_1$ and $s_2$ are the linear and nonlinear elastic compliance, respectively.

The substitution of expression (4.6) into (4.1) leads to the appearance of terms characterized by the higher orders in $P$ and $\eta$ which renormalize the coefficients in (4.1) as compared to the ones of the initial expression due to the dependence of σ on $P^2$ and $\eta^4$. Thus, taking into account the anharmonicity of the elastic energy leads to a more complex expression for the thermodynamic potential. This in turn must lead to a more sophisticated form of the phase diagrams. For example, Fig.4.1b and 4.1c present the phase diagrams for the substances which undergo the FE-AFE phase transition when only the term of the sixth order in polarization, i.e. $\alpha'_3 P^6$, is taken into account.

Now we discuss the effects that may take place in the case when the degree of *anharmonicity* increases (in this case the character of the phase diagrams in the $\alpha'_1$, $\beta'$-plane transforms from the type Fig.4.1a into the types 4.1b or 4.1c). This situation is shown in Fig.4.2 [63, 64]. Note that all our considerations will use the PZT-based solid solutions as an example, since the sufficient amount of necessary experimental data can be found only for the latter. However, these considerations remain valid for any oxide systems with the perovskite structure. Moreover, we suggest that the effects under discussion may take place in any other types of crystal structures that possess the crystal symmetry allowing the expansion of the thermodynamic potential in the form (4.1).

The upper row of Fig.4.2 presents the qualitative evolution of the "external pressure-temperature" ($P_{ext}$-T) phase diagrams that correspond to the case when the degree of the elastic energy anharmonicity increases. In the PZT-based solid solutions the said value rises with the growth of the concentration of Ti substituting Zr. Therefore the transition from the diagram 4.2a to the diagram 4.2c is realized in the PZT as the AFE state region in the "Ti-composition − temperature" diagram widens towards the region with the higher content of Ti.

Ion substitutions for both *A*- and *B*-sites of the perovskite crystal lattice are equivalent to the action of some "effective" hydrostatic pressure (see (2.4) and [24]). Thus, one can build the "composition-temperature" diagrams, which correspond to the $P_{ext}$-T diagrams shown in the upper row of Fig.4.2. These diagrams (4.2d-4.2f) are presented in the middle row of Fig.4.2.

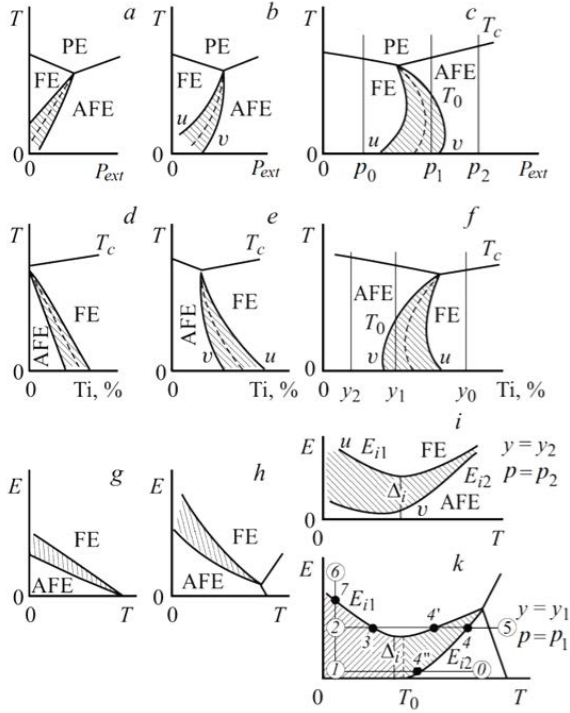

Fig.4.2. Calculated phase diagrams for substances undergoing FE-AFE transformations for different degree of anharmonicity [63, 64].

Here we have taken into account the fact that an increase in the content of Zr in PZT leads to the growth of the "effective" pressure, whereas an increase in the content of Ti lowers it. The "temperature-electric field" ($T$-$E$) phase diagrams corresponding to the "pressure-temperature" ($P_{ext}$-$T$) diagrams of the upper row are shown in the lowest row of Fig.4.2.

The phase diagrams shown in the first column of Fig.4.2 (a, d, g) practically coincide with the known phase diagrams for the $PbZrO_3$ and PZT solid solutions with a small content of Ti (when the degree of anharmonicity is small).

The phase diagrams which may be obtained using the thermodynamic potential (4.1) for the case of a weak elastic energy anharmonicity are given in the second column of Fig.4.2. The difference between these diagrams and those shown in the first column consists only in a slight disturbance of the phase boundary linearity. Phase transitions in the substances where such diagrams are realized will slightly differ from those in the substances characterized by the diagrams from the first column.

The phase diagrams shown in the third column are the most interesting, since in this case the variation of the diagrams assumes a qualitative character. Let us first consider the $P_{ext} - T$ diagram in Fig.4.2c. The most essential (and distinctive) feature of this diagram is the changing sign of the slope of the lability boundaries for the FE and AFE states, when moving along these boundaries. It is just such a character of the phase boundaries that leads to entirely new peculiarities of the FE-AFE phase transformation. The first manifestations of the said peculiarities in experiments have given rise to both surprise and incomprehension. The point is in the following. The diagram in the Fig.4.2c actually represents three different phase diagrams that may belong to three different solid solutions same series of PZT-based compounds having different contents of Ti. The points $P_0$, $P_1$, and $P_2$ are essentially the origins of coordinates for the three solid solutions with the Ti-contents equal to $y_0$, $y_1$, and $y_2$, respectively, shown in Fig.4.2f. In the solid solution with the composition $y_0$, the only FE-PE transition always takes place as the temperature changes, if no external pressure being applied. In the solid solution with the composition $y_2$, the same conditions induce the AFE-PE transition. For the solid solution with the composition $y_1$, the phase transition behavior is quite different. This solid solution is characterized by the trajectory $P_{ext} = P_1 = const$ in Fig.4.2c. This line passes through the hysteresis region as temperature changes; therefore the sequence of phase transitions that takes place during the change of temperature depends on the sample's prehistory. If no electric field is applied then the AFE → PE phase transition arises at the point $T_C$. The lability boundary for the FE phase (described by the curve $T_0(P_{ext})$), does not manifest itself. However, when this phase is induced by applying electric field below the line $T_0(P_{ext})$, it does not disappear after switching of the field. As the sample's temperature rises the following sequence of phase transitions is observed: FE → AFE → PE.

For the solid solutions described by the $P_{ext}$ - $T$ diagrams of the type shown in Fig.4.2c, the "composition-temperature" diagrams (Fig.4.2f) also differ qualitatively from those presented in Fig.4.2d. First of all, in the diagram Fig.4.2f the slope of the line representing the boundary of the region of the FE states is opposite in sign as compared with the one shown in Fig.4.2d. This is connected with the change of the slope of the line representing the boundary of the FE states region in the $P_{ext}$ - $T$ diagram shown in

Fig.4.2c. The second distinctive feature is the appearance of a region of induced states marked by hatching in Fig. 4.2f. For the solid solutions corresponding to this region, both the phase states and the sequence of phase transitions depend on the samples' prehistory. The sequence of phase transitions at the $T_C$ point prior to the action of an electric field is the following: AFE ↔ PE. After the application of the field with an intensity exceeding some threshold critical value at $T < T_0$, the FE → AFE → PE sequence of phase transitions is observed.

Now consider the peculiarities of the $T - E$ diagrams. These diagrams are presented in the lowest row of Fig. 4.2. Fig. 4.2i shows the $T - E$ diagrams for the solid solutions which are characterized by the $P - T$ diagrams of 4.2c type. To understand the behavior of these $T - E$ diagrams one has to return to the $P_{ext} - T$ diagrams of 4.2c type. Consider, for example, the solid solution with the composition $y_1$ which is characterized by the line $P_{ext} = P_1$ in the $P_{ext} - T$ diagram. The application of electric field raises the stability of the FE phase with respect to the AFE one, the lability boundaries for the FE and AFE states (curves $v$ and $u$, respectively) are being shifted to the right (see Fig.4.2c). At the temperatures $T < T_0$ the phase transition into the FE state (i.e. the electric field induces the FE state) will take place at the field intensity $E_{i1}(T)$ which corresponds to the intersection of the phase boundary $u$ shifted by the electric field with the straight line $P_{ext} = P_1$ (see Fig. 4.2c). The boundary $u$ in the $P_{ext}$ -$T$ diagram is substantially nonlinear, therefore the phase boundary $E_{i1}(T)$ in the $T$-$E$ diagram 4.2i is also nonlinear. The form of this boundary is similar to that of the boundary $u$ in the $P_{ext} - T$ diagram 4.2c. Moreover, the presence of the minimum field $\Delta_i$ is a consequence of the fact that the slope of the boundary $u$ changes its sign in the $P_{ext} - T$ diagram. Therefore, the slope of the dependence $E_{i1}(T)$ with respect to the temperature axis changes as well. As the field's intensity lowers, the phase boundaries $u$ and $v$ are shifted to the left (Fig.4.2c); the transition from the FE to the AFE state will take place in the field with the intensity $E_{i2}(T)$, when the phase boundary $v$ intersects the straight line $P_{ext} = P_1$. It is obvious that the reversible transition may occur only at $T > T_0$. Such change in the stability of the AFE and FE states in the $P_{ext} - T$ diagram manifests itself by the presence of the regions of both the irreversible and reversible induction of the FE phase in the $T - E$ diagram 4.2i. The former exists at $T < T_0$ and is preserved after switching off the field; the latter arises at $T > T_0$ and appears when the field is switched off. As it seen from Fig.4.2i, there is the so-called "field gap" $\Delta_i$ in the $T$-$E$ diagram. For $E < \Delta_i$ the phase transition into FE states never takes place at any temperatures.

If the FE phase is induced either in the presence of hydrostatic pressure or in the solid solution with the compositions varying from $y_1$ to $y_2$, then the curves $E_{i1}(T)$ and $E_{i2}(T)$ in the $T - E$ diagram are shifted in the way shown in the upper part of Fig. 4.2i. As it follows from this diagram, there exists such pressure or Ti concentration that starting from them only the reversible FE-AFE transitions in electric field are possible.

Let us now discuss the peculiarities of phase transitions, which are directly defined by the form of the $T - E$ diagrams of the 4.2i type. If the samples are cooled along the thermodynamic path 0-1-2 they are in the AFE state at the point 4. If now the samples are heated along the path 2-5 one shall observe the AFE → FE and the FE → AFE phase transitions at the points 3 and 4, respectively. The reverse movement along the path 5-2 will lead to only the AFE → FE phase transition taking place at the point $4^/$. No phase transitions occur at the points 4 and 3. On cooling along the thermodynamic path 0-1-2-6 the AFE-FE phase transition takes place at the point 7; the reverse movement along the same path results in the appearance of the FE-AFE phase transition at the point $4^{//}$.

The above analysis makes obvious the nature of the distinction between conditions of the "zero field cooling" and "field cooling". It consists in the fact that a substance falls within the hysteresis region in the phase diagram. Being in the said hysteresis region determines the dependence of the sequence of phase transitions upon the prehistory of specimens.

To conclude this section, let us discuss the influence of anharmonicity on the phase states of solid solutions based on the X - $T$ diagram, where X is the content of the substituting ion, which gives rise to the appearance of anharmonicity. As mentioned above, the substituting ions must be the ones that substitute the A-site of the perovskite crystal lattice and have a smaller ionic radius (or the $A^/A^{//}$-complex with a mean ionic radius smaller than that of the A-ion). For the PZT solid solutions, for example, such substitution means the increase of the "effective" hydrostatic pressure with the growth of X (see [24] and Eq.2.4), and the displacement of the boundary separating regions of the FE and AFE states in the "Ti-content-temperature" phase diagram towards the larger content of Ti. At the same time, the form of the said diagram changes from the type Fig. 4.2d to that of the Fig 4.2f with the increase of X. Let us choose a particular Ti content (e.g. $y_1$) such that at $X = 0$ the corresponding PZT solid solution is located far from this boundary (between regions of the FE and AFE states) in the FE region of the "Ti-content-temperature" diagram. Then let us denote the considered solid solutions by the symbol $X/1-y_1/y_1$ ($1-y_1/y_1$ is the symbol for the ordinary PZT). Since at $X = 0$ the solid solution is located in the FE region in the phase diagram and it undergoes the FE-PE phase transition under the change of the temperature. This phase transition corresponds to the point 1 in the diagram in Fig. 4.3. With the growth of X the "Ti content-temperature" diagram changes from the type Fig. 4.2d to the type 6e, and the FE-AFE boundary is displaced to the right. However, the solid solution $X/1-y_1/y_1$ is still within the FE region and undergoes the FE-PE transition, which corresponds to the point 2 in the diagram in Fig. 4.3. The diagram 6e transforms into the diagram 6f at further increase of X. Now (and it is very significant) the FE-AFE boundary approaches to the position of the $X/1-y_1/y_1$ solid solution and the latter is in the hysteresis region of the "Ti content-temperature" diagram. If the sample has not been previously subjected to the action of an electric field, it is in the AFE state at low temperatures, and at changing the temperature it undergoes the AFE-PE phase transition (point 3 in Fig. 4.3). After the action of electric field with intensity $E > E_{i1}$ at $T < T_0$ the FE state is induced in the sample's volume. Under subsequent heating (at $E = 0$) the FE → AFE phase transition will take place at the temperature $T_0$ (the point $3'$ in Fig. 4.3) and then the AFE → PE transition (the point 3). At further increase of X the FE-AFE boundary in "Ti content-temperature" diagram becomes essentially shifted to the right with respect to the point $y_1$. Now the solid solution with composition $X/1-y_1/y_1$ is located within the AFE region of the diagram, and therefore it undergoes the AFE-PE phase transition only (the point 4 in Fig. 4.3).

The same change of states take place at the substitution of the ions in the B-site by the ($B'B''$) complex with a mean ionic radius exceeding that of the B-ion (of course, the radii of $B'$ and $B''$ must be different).

The authors hope that the readers have already understood that in the case of the A-ion the diagram shown in Fig. 4.3 is characteristic of e.g. X/65/35 PZT in which La or Sr are substituted for Pb. In the case of the B-ion the diagram in Fig. 4.3 is characteristic of the lead titanium in which titanium is substituted by the ($Mg_{1/3}Nb_{2/3}$) complex.

All these types of phase diagrams discussed above will be considered for particular substances in the experimental part where the experimental phase diagrams and their evolution at subsequent substitution of the main ions will be analysed.

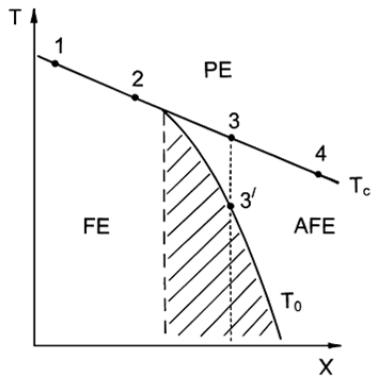

Fig.4.3. Schematic phase diagram for $X/1-y_1/y_1$ series. The dashed region corresponds to hysteresis region in Fig. 4.2.f (or 4.2.c).

**4.2. Peculiarities of experimental "composition-temperature", "temperature-electric field," and "pressure-temperature" phase diagrams**

It was mentioned in Ch.2 (see also [24, 25]) that the substitution of ions with smaller ionic radii for lead in the PZT compounds increases the energy stability of the AFE state with respect to the FE state. As a consequence the boundary separating the regions of the FE and AFE orderings in the "Ti-content - temperature" (Y - T diagram) phase diagram is shifted towards the higher Ti concentrations. The role of anharmonic terms in the crystalline potential grows at the same time. Therefore, the Y-T phase diagram must change from the type presented in Fig.4.2d to that in Fig.4.2f. The Y-*T* diagrams for the PZT-based solid solutions obtained in the process of successive step-by-step substitution of ($Li_{1/2}La_{1/2}$), Sr, and La for Pb are shown in the Fig. 4.4a [32, 63]. The compositions of these solid solutions can be presented by the following formulas $Pb_{1-x}(La_{1/2}Li_{1/2})_x(Zr_{1-y}Ti_y)O_3$ (PLLZT), $(Pb_{1-x}Sr_x)(Zr_{1-y}Ti_y)O_3$ (PSZT), and $(Pb_{1-3x/2}La_x)(Zr_{1-y}Ti_y)O_3$ (PLZT), respectively, As one can see the character of changes in the PZT phase diagram corresponds to that predicted in Ch.4.1. (Fig.4.2)

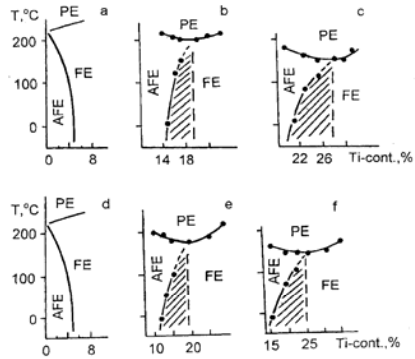

Fig.4.4a. "Composition-temperature" phase diagrams for PZT based solid solutions (X/100-Y/Y) with substitutions of Sr, ($Li_{1/2}La_{1/2}$), and La for lead. All diagrams have been obtained after induction of the FE phase by electric field [32, 63]. Sr (X),%: a − 0; b − 10; c − 20. ($La_{1/2}Li_{1/2}$) (X),%: d − 0; e − 10; f − 15. La (X),%: g − 0; h − 2; i − 4; k − 6; l − 8.

The X-*T* diagrams for PSZT, PLLZT, and PLZT solid solutions presented in Fig 4.1b also correspond to the model diagrams in Fig.4.4.

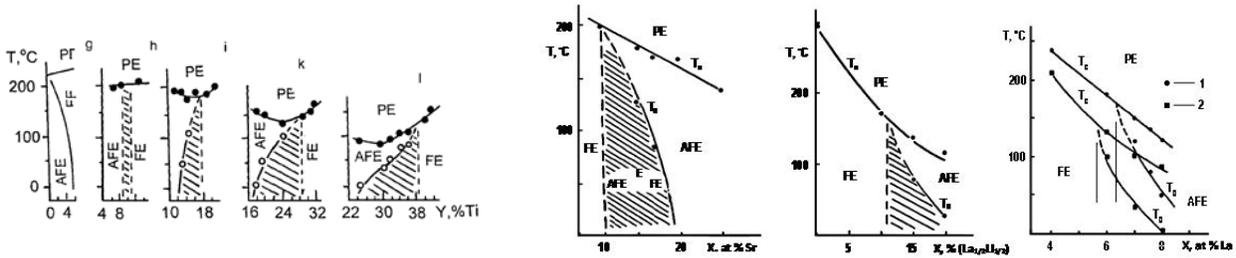

Fig.4.4b. "Composition-temperature" X/80/20 phase diagrams for PZT based solid solutions [63] PSZT, PLLZT, and PLZT (for two Ti-contents, 1 − X/75/25, 2 − X/65/35)

The Y – *T* phase diagrams contain broad intermediate regions (marked by dashes in Fig. 4.4a) between the regions of the FE and AFE states. The same intermediate regions are also present in the X - *T* diagrams (Fig.4.4b). According to X-ray analysis data [65] and TEM investigations [34, 35, 38], these regions are characterized by the coexisting domains of the FE and AFE phases. The phase state of the samples with compositions located within these regions depends on the previous history of samples. Electric field with an intensity exceeding the critical value $E_{i1}$ induces the macroscopic FE state. This state is stable under the heating up to the temperature $T_0(y)$. The dependencies $E_{i1}(T)$, i.e. the phase *T* - *E* diagram, for the 10/85/15 PLLZT and 4/87/13, 4/86/14, 6/82/18, 6/80/20 and 6/65/36 PLZT solid solutions are shown in Fig.4.5 [36, 63, 66] (as it was pointed in Ch.3.2, the notation X/1-Y/Y is used to define the composition of the solid solution obtained from the PZT; here X corresponds to the percentage of the element substituting lead, whereas (1-Y) and Y denote the content of zirconium and titanium, respectively). As seen from the diagrams in Fig 4.5, there exists a minimal critical field $E_{i,min} \cong 13$ kV/cm for the induction of the FE phase. In weaker fields the FE phase cannot be induced, and the macroscopic state of the samples remains unchanged (AFE) at any temperature changes.

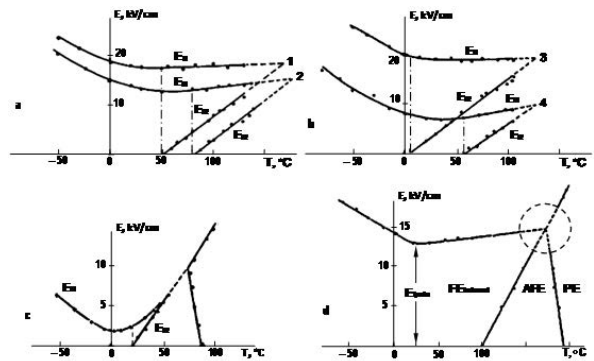

Fig.4.5. T-E phase diagrams for PLZT solid solutions with 4% (a), 6% (b), and 9% (c) content of La and for PLLZT 10/85/15 solid solution (d) [36, 63, 66,].

Content of Zr/Ti: 1 − 87/13; 2 − 86/14; 3 − 82/18; 4 − 80/20; 5 − 65/35.

The value of the minimal critical field, $E_{i,min}$, reduces with an increase of the titanium content in 10/100-Y/Y PLLZT solid solutions. The situation when the value of the minimal critical field becomes equal to zero $E_{i,min}=0$ corresponds to the transition to the region of the spontaneous FE states in the $Y$ - $T$ diagram of PLLZT. Based on the $E$ - $T$ diagrams of the type shown in Fig.4.2 it will be easy to understand the difference in the behavior of the considered solid solutions in the ZFC and FC regimes of investigation. In the first case the solid solutions undergo the PE-AFE phase transition when temperature changes from the values ($T > T_C$) to lower ones. In the second case the changes of the solid solution properties follow the sequence of transitions: PE-AFE-FE at cooling, provided that the electric field intensity satisfies the condition $E > E_{i,min}$. It is quite natural that the results of investigations in these two cases will be absolutely different. Such a problem was discussed in detail in Section 4.1 (Fig.4.2) Experimental Y-$T$ and $T$-$E$ diagrams of the PLLZT, PLZT and PSZT solid solutions correspond to the phase diagrams predicted by the model of the FE-AFE phase transformations (Fig.4.2g-i).

As demonstrated in Section 4.1 (Fig.4.2), the shape of the Y-$T$ phase diagrams typical for the PLZT with the La-content x ≥ 4 and the PLLZT with x = 10% and 15% (Fig.4.4) is the consequence of the specific shape of $P$-$T$ diagrams presented in Fig.4.2c.

Experimental $P−T$ diagrams for the 6/100-Y/Y series of PLZT solid solutions shown in Fig.4.6 were obtained for the first time in [67]. It is easy to see that their character (shown in Fig.4.2c) completely corresponds to the one predicted by the model considered in Section 4.1. Moreover, it should be noted that the experimental $P$-$T$ diagrams for lead zirconate with the region of the FE states correspond to the calculated $P$-$T$ diagrams presented in Fig.4.2a.

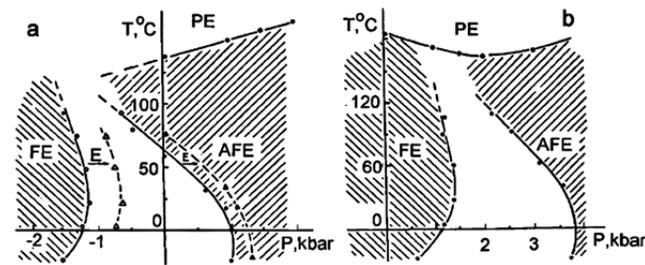

Fig.4.6. P-T phase diagrams for PLZT solid solutions with 6% content of La, Zr/Ti: a - 80/20; b - 65/35 [37, 63, 67]. Arrows show the shift of the phase boundaries by electric field 500 V/cm.

The experimental $P−T$ diagrams define the specific shape of the $T−E$ diagrams (as it was discussed above for diagrams in Fig. 4.2$i$ and 4.2$k$). The model for the FE-AFE phase transformation was also used to predict the evolution of the $T$-$E$ phase diagrams for substances subjected to the action of hydrostatic pressure (Fig.4.2i and the commentary to it). The experimental $T$-$E$ diagrams for the 6/80/20 series of solid solutions (which belongs to the boundary hysteresis region in the Y-$T$ diagram) corresponding to the case when the hydrostatic pressure is applied to a sample are shown in Fig.4.7. They are completely similar to those predicted by the model from Fig.4.2i.

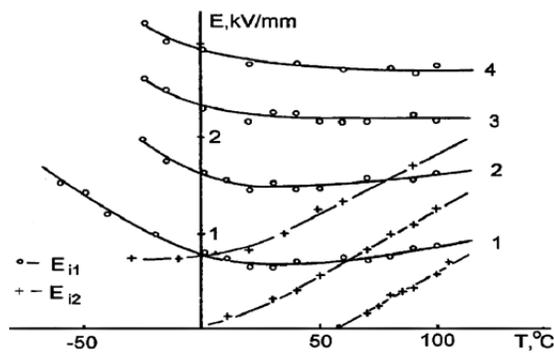

Fig.4.7. $E_{i1}$ and $E_{i2}$ critical fields as functions of temperature for 6/80/20 PLZT solid solution
for different pressures (MPa): 1 - 0.1; 2 - 100; 3 - 200; 4 – 300 [63, 67].

The substance phase state that corresponds to any part of each of the diagrams may be easily determined by comparing the E-$T$ diagrams with the Y-$T$ and $P$-$T$ diagrams. The phases participating in transformations that take place while crossing any phase boundary in these diagrams may be easily found. In particular, the dependence of phase transformations on the history of the sample may be readily interpreted. The authors would like to emphasize that the analysis of the whole set of the phase diagrams testifies that only three states, namely, the FE state, the AFE state, and the PE state participate in the phase transformations (More complete sets of diagrams can be found in [37]). Other states are not present. Nowadays the *T-E* diagrams of the PLZT solid solutions may be found in the literature quite often. In our opinion, it is impossible to build the completely clear picture of properties of the investigated substances on the base of only one type of measurements.

The properties of the PLZT solid solutions are most often discussed in the literature on the base of the ones for the X/65/35 solid solutions. The X-*T* diagram for this series is shown in Fig.4.1b. The X-*T* diagram for X/65/35 solid solutions in which praseodymium is substituted for lead one can find in [37 Ch.3, 63] [37] Ch.3, 63]. The way by which such a diagram may be obtained from the set of diagrams shown in Fig.4.4a of the present paper (the said procedure is realized and shown in Fig 4.4b for X/75/25 and X/65/35 PLZT) is described in Section 4.1 (see Fig.4.3 and comments to it). It is experimental confirmation of the fact that the X-*T* diagram contains the states observed in the Y-*T* diagrams only. It is the most important circumstance for our consideration and the other states are absent and cannot take place at all. The latter circumstance is to be emphasized. In the samples of the X/65/35 series the temperature changes may give rise only to those phase transitions which occur in any of the X/100-Y/Y series sample. This means that in the samples of the X/65/35 series only the transitions between the FE, the AFE and the PE states may take place. Naturally, these phase transitions will have peculiarities defined by a weak difference in energies of the FE and AFE states. Moreover, in the praseodymium series these effects are to be revealed at lower concentrations of the substituting element, since the ionic radius of $Pr^{3+}$ is smaller than that of $La^{3+}$. Such a statement has been completely confirmed by experiments [37 Ch.3, 63] [37] Ch.3, 63] (see Fig.4.8)

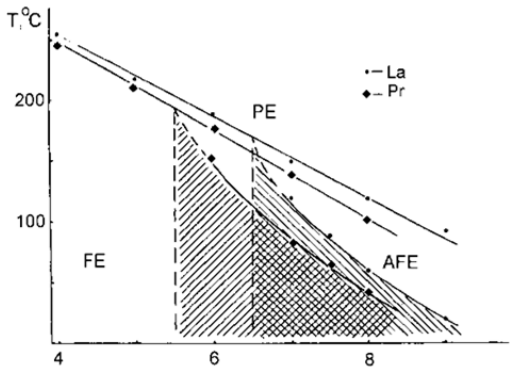

Fig.4.8. Diagram of phase states for the X/65/35 PZT-based solid solutions with substitution of Pr and La for Pb [37] Ch.3, 63].

In this chapter we considered experimental phase diagrams for the PZT based solid solutions only. Experimental phase diagrams for different substances undergoing phase transformations between the FE and AFE states with similar peculiarities of phase transitions (caused by the anharmonicity of crystal lattice elastic potentials) will be demonstrated in Chapter 9 of the present review.

**4.3. Difficulties in the identification of phase states in the lead lanthanum zirconate titanate system of solid solutions**

The PLZT solid solutions are being referred to the class of the so-called relaxor ferroelectrics during recent 20 − 25 years. This period was preceded by 18 year-long studies of physical processes in PLZT compounds (since the synthesis of these materials and the finding out their unique electrooptical properties in 1969). A huge amount of experimental results was accumulated (a detailed review of results of studies of properties of the PLZT solid solutions and phase transitions in this substance as of 1985 one can find in [37]). In particular, the crystalline structure was found to be non-cubic (the degree of distortion being low) at the temperatures below the value $T'_m$ corresponding to the maximum of the $\varepsilon'(T)$ dependence. This structure was characterized as pseudo-cubic. However, later the crystal structure of PLZT was considered cubic at $T < T'_m$ (before the action of electric field). Therefore, we will show that the crystalline structure of the PLZT solid solution system is not cubic at $T < T'_m$, as it must take place for ferroelectrics and antiferroelectrics.

The evolution of the Y – T phase diagram in the process of successive increase of the La concentration in $Pb_{1-3x/2}La_x(Zr_{1-y}Ti_y)O_3$ after the action of a strong electric field and an induction of the FE phase (see Chapter 2) is presented in Fig.4.4a. The said phase diagrams before the action of the field are shown in Fig.4.9.

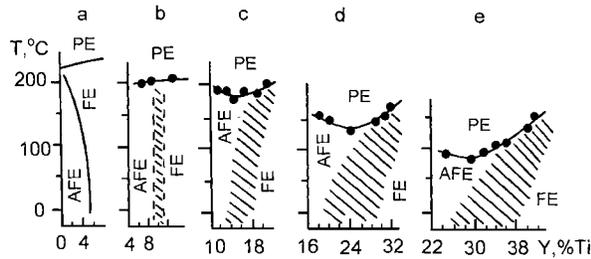

Fig.4.9. "Ti-composition–temperature" phase diagrams for $Pb_{1-3x/2}La_x(Zr_{1-y}Ti_y)O_3$ solid solutions with different La contents before the exposure to the electric field : (a) $x = 0$; (b) $x = 0.02$; (c) $x = 0.04$; (d) $x = 0.06$; (e) $x = 0.08$.

The boundary regions (shown by dashes in the figure) are the hysteresis regions for the FE–AFE transformation. X-ray and TEM studies of the PLZT with compositions from this boundary region of the phase diagram [28, 38, 45 68] have shown that the domains of the FE and AFE phases coexist in the bulk of the sample (the transition electron microscopy images of the coexisting domains of the FE and AFE phases are shown in Fig.3.2 and 3.3). It has to be reminded that clearly manifested boundaries between the single phase FE and AFE states in the diagrams (Fig.4.8) are absent before the exposure of samples to the electric field and the induction of the FE phase.

PLZT samples were obtained by the co-precipitation of components from the mixture of the aqueous solutions of lead and lanthanum nitrates and zirconium and titanium chlorides. After washing and drying the precipitates were calcined at 550 °C and 850 °C. Ceramic samples were sintered at the temperatures 1320–1340 °C in a controlled PbO atmosphere. The grain sizes were from 5 to 7 µm.

As seen in Fig.4.4 and 4.9, the Y – T phase diagrams for all the series of solid solutions with the La content higher than 4% are equivalent from the viewpoint of physics. Therefore, the largest part of our study in the scope of the present part of Chapter 4 was carried out on the PLZT with 6% of La. We investigated the profiles of the X-ray diffraction lines in the solid solutions in question. Our attention was focused on the (200) and (222) X-ray diffraction lines, which are the most characteristic lines for analysis of the crystalline structure of perovskite type compounds. The (200) line is a doublet in the presence of tetragonal distortions of the elementary cell, whereas the (222) line is a singlet in this case. In the presence of rhombohedral distortions the (200) line is a singlet and the (222) line is a doublet. These lines are shown in Fig.4.10. The variation of profiles of these lines with the change of Ti content in solid solutions is clearly seen in this figure.

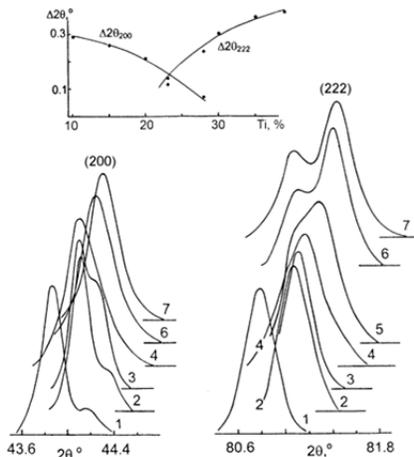

Fig.4.10. Profiles of the (200) (at the left) and (222) (at the right) X-ray lines for the PLZT solid solutions with 6% of La [68]. Dependencies of the positions (the $2\theta_{002}$ and $2\theta_{222}$ angles) of the peaks of components for the (200) and (222) X-ray lines on Ti content in PLZT with 6% of La are presented at the top of the figure.

The complex character of the profiles of X-ray lines undoubtedly proves that the crystal structure is not cubic at the temperatures below the Curie point. The crystal structure in the AFE state is characterized by the tetragonal distortions of the elementary crystal cell, whereas the crystal structure in the FE state undergoes the rhombohedral distortions. In the case of tetragonal distortions, the X-ray patterns contain weak superstructure reflexes that substantiate the AFE ordering in these solid solutions. The splitting of the X-ray lines decreases when the composition reaches the boundary region separating the FE and AFE states in the $Y - T$ diagram. The shape of the X-ray lines is also dependent upon the location of the solid solution's composition in the $Y - T$ phase diagram. In particular, the intensity of components of the composite lines redistributes as the substance reaches the boundary region. Analysis of the intensities of components of the (200) and (222) lines allowed us to obtain the dependence of the fraction of the rhombohedral (FE) phase in the sample volume as a function of the Ti content. This dependence is given in the inset in Fig.4.11. The position of the peak of each component of the X-ray diffraction line (distance between the components of the Bragg 200 and 222 lines) depends on the Ti content in the solid solution. The dependencies of these positions on Ti content for the (200) and (222) X-ray lines are presented in the insert of Fig.4.10.

The dependencies of the crystal cell parameters $a_T$ and $c_T$ for the tetragonal (AFE) phase and crystal cell parameter $a_{Rh}$ and rhombohedral angle $\alpha_{Rh}$ for the rhombohedral (FE) phase on the Ti content are shown in the upper part of the Fig.4.11 to highlight the change of the solid solution crystal structure with the variation of composition.

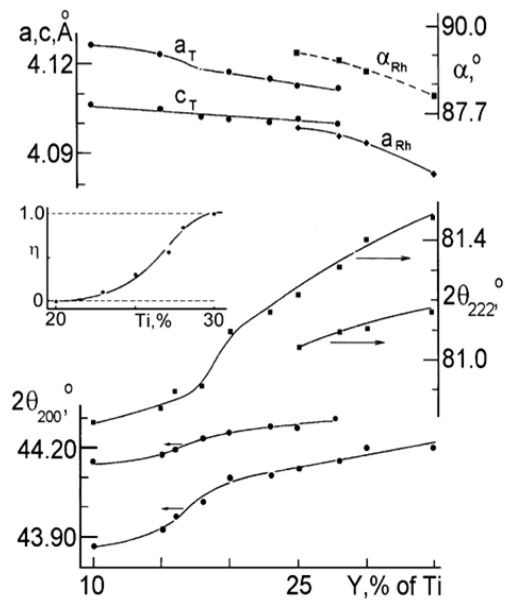

Fig.4.11. Dependencies of the crystal cell parameters on Ti content in PLZT with 6% of La are shown at the top of the figure [68]. The parameters $a_T$ and $c_T$ correspond to the tetragonal (AFE) phase and the crystal cell parameter $a_{Rh}$ and the rhombohedral angle $\alpha_{Rh}$ are for the rhombohedral (FE) phase.
The insert shows the share of FE phase in the sample's volume $\eta(Y)$.

The crystal structure of the solid solutions belonging to the boundary region in the $Y - T$ diagram is two phase. The phases with the tetragonal and rhombohedral distortions of the perovskite elementary cell coexist. The simultaneous finite non-zero splitting of both (200) and (222) X-ray lines (see Fig.4.10 and Fig.4.11) and the widening and asymmetry of the lines (in particular, for the PLZT with $x = 6\%$ and Zr/Ti = 75/25 these lines are quasi-singlet) is the evidence of the two-phase structure of the solid solution with this composition. Thus, the crystal structure of the PLZT with 6% of La is unambiguously non-cubic at the temperatures below the Curie point. These results also demonstrate that the identification of the structure of the PLZT with 6% of La and the concentrations of Ti from the boundary region in the $Y - T$ phase diagram can be hindered by the following circumstance. The domains of the FE and AFE phases coexist in the samples with compositions that belong to the boundary region. So, one can easily make a mistake while identifying the structure of the PLZT from the boundary region, if the examination of the structure of the solid solutions from all regions of the $Y - T$ diagram has not been done. It has to be also noted that the intensity of the X-ray lines superstructure decreases when the composition of the PLZT approaches the boundary region moving from the region of zirconium-rich solid solutions.

The change of the character of the (200) and (222) X-ray lines that reflects a change of the crystalline structure is caused by the interaction between the coexisting FE and AFE phases (considered in Section 4.1). The contribution of this interaction increases as the PLZT composition reaches the value that corresponds to equal free energies of the FE and AFE phases. At the same time, the deviations from properties of the ordinary FE or AFE phases become more pronounced in solid solutions that belong to the interval of compositions where the free energies of the FE and AFE phases are close or equal. These deviations reveal themselves, for example, in measurements of the dielectric or the electro-optical hysteresis loops, in the dispersion of a dielectric permittivity in the vicinity of the point of the PE phase transition, and in the smearing of this phase transition.

In the PLZT with 8% of La the observed picture is even more complicated. An increase of the amount of lanthanum diminishes the energy barrier separating free energy minima corresponding to the FE and AFE states. Therefore, the interaction between these phases manifests itself greatly. The boundary region in the Y – T diagram becomes wider and the degree of crystal lattice distortions decreases, so the X-ray splitting is less pronounced. One more important circumstance has to be noted. The morphotropic phase boundary that is located at the point in the Y – T phase diagram of the $PbZr_{1-y}Ti_yO_3$ that corresponds approximately to the composition Zr/Ti = 53/47 [69] is displaced towards the solid solutions with higher percentages of Zr as the La concentration increases. At 8–9% of La, the $FE_T − FE_{Rh}$ morphotropic boundary is observed in the vicinity of the 65/35 Zr/Ti composition [70]. This composition corresponds to solid solutions, which are called 'relaxor ferroelectrics' and are the object of active discussions in the literature. The substance with this composition contains three phases, and it is manifested in the observed structure of the X-ray lines. Each of these lines is a superposition of the lines of the FE rhombohedral, the FE tetragonal and the AFE tetragonal phases. As far as we know, such fact has not been considered at the identification of the crystalline structure of the PLZT and consequently, at the decomposition of the X-ray diffraction lines into simple components. Here we would like to mention that the transmission electron microscopy investigations [34, 35, 38, 71, 72] of the PLZT show that the size of domains of the coexisting FE and AFE phases is of the order of 20–30 nm.

Complexity of the identification of crystal structure in the PLZT is furthermore complemented by the specifics of the hot pressing method used for the preparation of samples for which the data of crystal lattice investigations are available in the literature. According to [73] the hot pressing prevents the achievement of a high degree of homogeneity due to such factors as violation of the stoichiometry resulting from hot pressing; the "underannealing" effects caused by low temperatures applied in the process of hot pressing; and the presence of residual mechanical stresses arising during hot pressing. Even in the case when the hot-pressed PLZT samples have a high optical quality, some nanometer scale regions containing chemical elements that have not reacted completely are present in the samples' volume. In particular, this was confirmed in [36] by transmission electron microscopy. It was demonstrated in [74, 75, 76] that the hot pressed PLZT samples of high optical quality also contain nanodomains with the composition close to that of pure $PbZr_{1-y}Ti_yO_3$.

The results presented here explicitly demonstrate that the crystal structure of the PLZT series of solid solutions with 6% of La is non-cubic at the temperatures below the temperature of the $\varepsilon(T)$-dependence maximum. One can identify it undoubtedly only if the solid solutions with compositions from different regions of the Y –T phase diagram are investigated simultaneously. There is a considerable probability of error in the structure identification when only one solid solution belonging to the boundary region of the Y –T phase diagram (dashed regions in Fig.4.4 and Fig.4.9) is studied. It is also worth mentioning that the coexistence of the FE and AFE phases or the even more complicated three-phase structure of PLZT with La content 6% could be the reason for relaxor behavior during the phase transition from the paraelectric to the dipole ordered phase [77, 78].

**4.4. Peculiarities of dielectric properties**

Now we shall dwell on another effect caused by the inhomogeneous two-phase state of the samples. As an example, consider the 7.5/100-Y/Y PLZT series. A section of the phase diagram for these substances with the compositions in the vicinity of the boundary between the regions of the FE and AFE orderings is shown in Fig.4.12a.

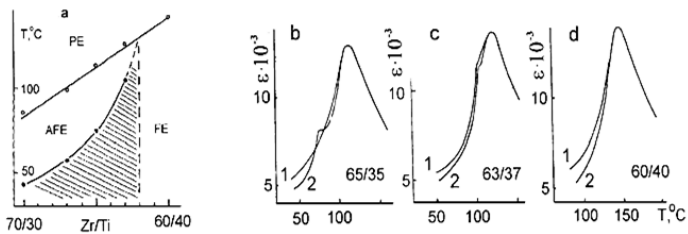

Fig.4.12. The Y-*T* phase diagram (a) and temperature dependencies of the dielectric constant (b, c, and d)
For the 7.5/100-Y/Y series of PLZT solid solutions [50]: 1 − before the application of an electric field; 2 − after the application of an electric field.

It is analogous to the model diagram presented in Fig.4.3 and experimental diagrams in Fig.4.4b and Fig.4.8 for other PZT-based solid solutions. The ε(T) dependencies for this series of solid solutions with compositions located in three different regions of the Y-T diagram are shown in Fig.4.12b-d. These dependencies were measured before (on annealed samples) and after the action of a DC electric field with an intensity of 2.5 kV/mm. For the 7.5/60/40 solid solution the dependencies have no anomalies except a maximum at the point $T_m$. In annealed samples with composition 7.5/65/35 there exists only a maximum at $T = T_m$ associated with the paraelectric phase transition. An additional anomaly is observed at the temperature $T_0$ in polarized samples. It manifests itself only after the action of a high-intensity electric field on samples in which the FE phase has been induced. However, in the 7.5/63/37 solid solution the anomaly at $T_0$ is observed in both annealed and polarized samples.

As seen from the Y-T diagram of the 7.5/100-Y/Y series, the equality of the free energies of the AFE and FE phases takes place in the solid solution containing zirconium and titanium in the ratio 62/38. The composition of the 7.5/63/37 solid solution is located in the vicinity of the boundary (dashed line in Fig.4.12a) therefore the free energies of the FE and AFE states in this solid solution have close values. Moreover, small values of the intensity of the field inducing the FE phase indicate low values of the height of the potential barrier separating the free energy minima corresponding to the FE and AFE states. In view of the above-mentioned, approximately equal quantities of the FE and AFE phases are formed in the volume of the sample in the process of cooling below the temperature $T_m$ (for instance, for the 7.5/65/35 solid solution a larger share of the FE phase is observed only after the action of electric field). Therefore, the manifestation of the FE phase at subsequent heating takes place in all the cases, even if it has not been specially induced by an electric field. The same picture is observed in all PLZT solid solutions that have an approximately equal stability of the FE and AFE states. However, such behavior may be explained only on the base of the corresponding diagrams of phase states.

The described behavior is called spontaneous transition from the relaxor state to the normal FE state (though in fact such a transition is caused by an electric field and manifests itself in experiments performed in the FC regime) in the literature. For a better insight into the phenomenon under consideration, one should refer to the expression which describes the interaction of domains of the coexisting FE and AFE phases (Chapter 3). The condition $T_{c,af} \geq T_{c,f}$ is fulfilled for the 7.5/100-Y/Y PLZT. The spontaneous polarization $P_s$ in the FE domains considerably changes (decreases) under heating in the vicinity of the temperature $T_{c,f}$, whereas the properties (order parameters) in the AFE domains weakly depend on temperature (in the first approximation they may be considered constant). In this case the action exerted by the interphase interaction on the FE subsystem may be described by the effective field $E_s$ such that $W_{int} = -\vec{E}_s \cdot \vec{P}_s$. As discussed in Section 4.1, $T_{c,f}$ is not a true point of the phase transition in this situation, and above it (for $T > T_{c,f}$) $P_s \neq 0$ (though for these temperatures $P_s$ has small values). Whereas, a sharp decrease of spontaneous polarization and the change of the FE ordering happen at the $T_{c,f}$-point itself.

Let us now consider the influence of the coexisting domains of the FE and AFE phases on dielectric properties, using PLLZT solid solutions as example. The phase diagrams for the said solid solutions are shown in Fig.4.4a (after the action of the electric field). The boundary regions (shown by dashes in figures) are the hysteresis regions for the FE–AFE transformation, and for solid solutions from this region of compositions the FE phase is induced in the samples' volume by an electric field. X-ray studies of the PLLZT with compositions from these boundary regions of the phase diagrams [65] have shown that the domains of the FE and AFE phases coexist in the sample volume. It is necessary to note that the legible boundaries between the regions of the single phase (FE and AFE) states and the region of coexisting phases are absent in the phase diagrams before the application of an electric field.

The left curve in the Fig.4.13 shows the dependence $\varepsilon_m$ ($\varepsilon_m$ is the maximum of the ε(T) dependence) on the Ti content in the 10/100-Y/Y PLLZT series [79]. The dependencies of ε(Y) measured at $20^oC$ are presented in the right part of Fig. 4.13. Curve 1 shows the dependence, obtained on the samples that have not been affected by an electric field. In this case the maximum of ε(Y) corresponds to

~ 19 % of Ti content. Curve 2 shows the dependence ε(Y) after the application of an electric field. Now, the maximum of the ε(Y)-curve is located near ~13 % of Ti content.

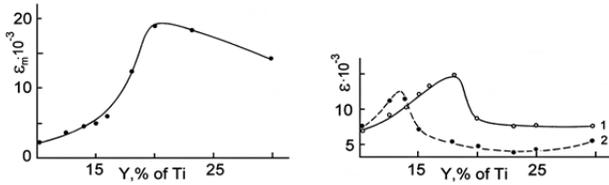

Fig.4.13. Dependence of $\varepsilon_m$ on Y (on the left) and $\varepsilon(20\ ^oC)$ on Y for annealed (1) and polarized (2) samples (on the right) for 10/100−Y/Y PLLZT solid solutions [65, 79].

The discussed dependencies of dielectric constant on the solid solution composition can be consistently explained taking into account the coexistence of domains of the FE and AFE phases in the bulk of solid solutions with compositions falling into the dashed regions in the diagrams of the PLLZT phase states. Low-frequency measurements (at frequencies less than $10^8$-$10^9$ Hz) reveal the essential contribution of domain wall oscillations into the frequency dependence of the dielectric constant of FE. In the substances containing the coexisting domains of the FE and AFE phases the oscillations of the interphase boundaries must have a noticeable influence on the value of ε, because they have high mobility in an electric field. Therefore the behavior of the dielectric constant near the FE-AFE-PE triple point gives essential information about the system of coexisting phases.

The ε(Y) dependence at $20^oC$ obtained on samples which have not been subjected to the action of an electric field is shown in the right part of Fig.4.13 (curve 1). In this case the boundary between the FE and AFE states in the phase diagram is at ~ 19% of Ti (and its location is nearly independent on temperature), and the maximum of ε(Y) is located at the point corresponding to the same composition. After the application of a DC electric field (and induction of the FE state in the dashed region of the Y-T diagram in Fig. 4.4a) the position of the boundary separating the FE and AFE regions is at ~ 13% of Ti concentration. This value corresponds to the position of maximum of the ε(Y) dependence for the polarized samples. This behavior of the ε(Y) dependence is confirmed by the results, presented in the left part of Fig.4.13, where the $\varepsilon_m$(Y) dependence is shown. In this case the maximum of the dielectric constant is also located near the point of the FE-AFE boundary in the Y-T phase diagram.

## 5. The two-phase ferroelectric-antiferroelectric system in an external DC electric field. Intermediate state in ferroelectrics and antiferroelectrics

### 5.1. Model theoretical consideration

Phase transition via the intermediate state (IS) is one of the most interesting phenomena in the physics of magnetism and superconductivity. Such IS represents a thermodynamically stable state of the coexisting phases between which the first order phase transition takes place under the action of the external magnetic field. The presence of such intermediate state was demonstrated by L. Landau for superconductors and by V. Bar'yakhtar and co-workers [80, 81] for magnetic crystals. The existence of the IS in the above-mentioned substances is caused by the action of the demagnetizing field resulting from the finite size of the samples. The IS cannot exist in the infinite superconductor or the infinite magnetic material. Thermodynamic potentials used for the phenomenological description of the phase transitions in magnetic substances on the one hand and the FE and ~~the~~ AFE on the other hand are akin to each other. That is why the suggestions were made that the IS analogous in its nature has to be present in the FE and AFE substances. However, such IS has not been experimentally observed. It was shown in [82] that the IS similar to the ones in magnetic substances and superconductors is fundamentally impossible in both ferroelectrics and antiferroelectrics.

However, it was shown that the IS that represents a thermodynamically stable structure of the coexisting domains of the FE and AFE phases may exist. The nature of this intermediate state is vastly different from the one in magnetic substances and in superconductors. The main difference of the FE and AFE substances from, for example, the magnetic materials consist in the way how an external field is applied to the sample [82] (from the physics point of view this difference is caused by a different form of one of the Maxwell equations applied to magnetic substances and dielectrics).

Let us consider the behavior of a system with the coexisting FE and AFE phases placed into a DC electric field. This system is described by the thermodynamic potential (1.1) and in this case the equation (1.4) is replaced by the expression:

$$\varphi_\lambda = \varphi - \eta_{\alpha,i} E_{\alpha,i;ext} - \lambda \left( \sum_\alpha \xi_\alpha - 1 \right), \tag{5.1}$$

where $E_{\alpha,i;ext}$ is the external field conjugate to the order parameter $\eta_{\alpha,i}$. The conventional procedure of minimization of the nonequilibrium thermodynamic potential leads to the following system of equations for the equilibrium values of the order parameters

$$\xi_\alpha \left( \partial \varphi_\alpha / \partial \eta_{\alpha,i} + E_{\eta_{\alpha',i}} - E_{\alpha,i;ext} \right) = 0, \qquad (\xi_\alpha \neq 0) \tag{5.2}$$

$$\varphi_\alpha + \eta_{\alpha,i} E_{\alpha,i;int} = \lambda = const, \tag{5.3}$$

$$E_{\alpha,i;int} = E_{\alpha,i;ext} - E_{\eta_{\alpha',i}}. \tag{5.4}$$

As seen from eq. (5.3), the condition for the existence of the thermodynamically balanced structure of coexisting domains of the phases is the equality of their thermodynamic potentials that takes into account the external and the internal effective fields.

The fields $E_{\eta_{\alpha,i}}$ are spatially varied. Therefore, the transition from antiferroelectric to ferroelectric phase (AFE → FE) may take place only in certain local regions of the sample (but not in the whole volume of the sample) at the same value of the external field $E_{\alpha,i;ext}$. This means that, within a certain interval of the external electric field intensity, the domains of the phases, participating in the transition, coexist in the sample's volume. By analogy with magnetic materials or superconductors, such a state should be called the intermediate state in the AFE. In view of this, let us obtain a relation between the external field and the field of the phase transition for the case of a thermodynamically balanced structure of the coexisting FE and AFE phases, and define the external field intensity corresponding to the onset and the completion of the phase transition into the IS. The condition of the phase transition is the equality of the internal field within local regions of the sample to the field of the phase transition ($E_{\alpha,i;int} = E_{pt}$). From this condition and from eq. (5.4) we have the following relation:

$$E_{pt} + E_{\eta_{\alpha',i}} = E_{\alpha,i;ext}. \tag{5.5}$$

Now let us express $E_{\eta_{\alpha',i}}$ in terms of the share of the sample's volume which has undergone the transition, and put down the order parameters in accordance with (5.3). Thus, we obtain the dependence of the share of the phase, which has undergone the transition into the FE state, on the external field intensity within the limits of the IS existence.

To write down the thermodynamic potential for each of the phases one has to take into account that the coordinates of the FE and AFE minima now depend on the electric field intensity. The nonequilibrium potentials for each of the phases may be written up to the quadratic terms as (by analogy with eq. (3.8)):

$$\varphi_1 = \varphi_{1,0}^E + U_1 \left( P_1 - P_{1,0}^E \right)^2 + V_1 \left( \eta_1 - \eta_{1,0}^E \right)^2.$$

$$\tag{5.6}$$

$$\varphi_2 = \varphi_{2,o}^E + U_2 \left( \eta_2 - \eta_{2,0}^E \right)^2 + V_2 \left( P_2 - P_{2,0}^E \right)^2$$

The interphase interaction energy has the form (3.14):

$$W_{int} = (\xi_2 P_2)(\xi_1 C_1 P_1) + (\xi_1 \eta_1)(\xi_2 D_2 \eta_2).$$

Let us now refer to equation (5.4). For the considered system with the AFE → FE phase transition induced by electric field, we have the expression for the intrinsic field for the AFE phase domains:

$$E_{2;int} = E_{ext} - \xi_1 C_1 P_1. \tag{5.7a}$$

Since the equilibrium solution for $P_1$ for the thermodynamic potential of the two phase system is $P_1 \cong P_{1,0}^E$ (see expressions (3.11)), we obtain:

$$E_{2;int} \cong E_{ext} - \xi_1 C_1 P_{1,0}^E. \tag{5.7b}$$

For the intermediate state $E_{int} = E_{pt}$. Therefore, by analogy with magnetic materials one can get:

$$\xi_1 \cong \frac{E_{ext} - E_{pt}}{C_1 P_{1,0}^E}. \tag{5.8}$$

Also by analogy with magnetic systems, for the AFE material we have the boundaries of the IS:

$$E_1 \cong E_{pt}; \quad E_2 \cong E_{pt} + C_1 P_{1,0}(E_{pt}). \tag{5.9}$$

These boundaries are determined by the conditions $\xi_1 = 0$ and $\xi_1 = 1$, respectively.

Comparison of expressions (5.8) and (5.9) with the analogous ones for the IS in the magnetic materials [80, 81] shows that the equations describing the IS in the AFE are similar in appearance to those for magnetic substances. However, the fields $E_{\eta_{\alpha,1}}$ and $H_{diam}$ (the latter is the demagnetization field in magnetic materials and it defines the nature and the range of existence of the IS in magnetic materials) differ in their physical nature. This results in some distinctions in the conditions of observation of the IS in magnetic materials and ferroelectrics or antiferroelectrics. First of all, the interval of the existence of the IS in the AFE does not depend on the geometric characteristics of the sample, whereas in magnetic substances this dependence is pronounced. Moreover, in ferroelectrics the IS may also appear in the case when the FE → AFE phase transition is induced by the hydrostatic pressure. In fact, with an increase in the pressure the stability of the FE phase diminishes, but the stability of the AFE state increases. At some critical pressure $P_{pt}$ the thermodynamic potentials of these phases acquire the same value, and this leads to the emergence of the nuclei of a new AFE phase. They induce the fields $E_{\eta_{\alpha,i}}(x, y, z)$, due to which different local regions of the sample are under different conditions. This is just the fact that gives rise to the coexistence of domains of the phases participating in the phase transition.

Thus, in this section it has been shown that the appearance of the IS is possible at the phase transitions induced by an electric field (or the hydrostatic pressure) in antiferroelectrics (or ferroelectrics). Such intermediate state represents the coexistence of phases participating in the phase transition within a rather wide interval of thermodynamic parameters initiating the transition.

## 5.2. Experimental results

### 5.2.1. *Lead lithium lanthanum zirconate titanate solid solutions*

The *E-T* phase diagrams for the 10/100-Y/Y series of PLLZT solid solutions are presented in Fig.5.1 [82]. The Y-*T* phase diagram [65] is also shown in this figure. The measurements performed in a bias DC electric field show that the transition temperature decreases with the increase of the field in the 10/90/10 and 10/87.5/12.5 PLLZT series of solid solutions (the rate of change of the transition temperature $dT/dE = -0.7 \cdot 10^{-3}$ deg·cm·V$^{-1}$). For the 10/75/25 PLLZT solid solution with composition from the FE region of the Y-*T* phase diagram $dT/dE = 3.0 \cdot 10^{-3}$ deg·cm·V$^{-1}$. The $T_c(E)$ dependencies for the solid solutions from both the FE and AFE regions of the Y-*T* phase diagram are described on the base of the ordinary thermodynamic point of view.

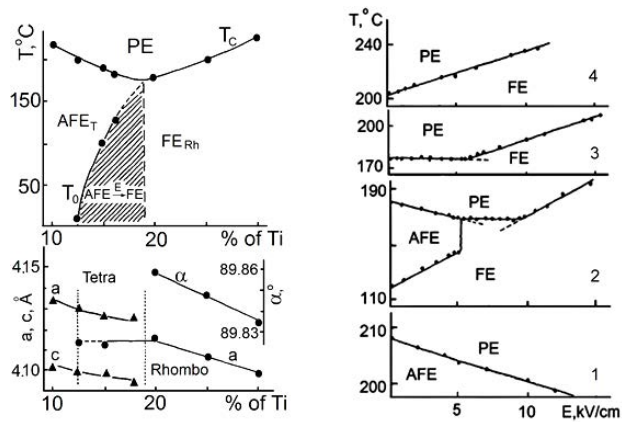

Fig.5.1. The *Y-T* and *E-T* phase diagrams for 10/100-Y/Y series of PLLZT [65, 82].
Ti-content Y,%: 1 – 10; 2 – 16; 3 – 18; 4 – 25.

The solid solutions from the borderland region of the *Y-T* phase diagram possess an unusual dependence of the temperature of the PE transition on the electric field intensity. The *E-T* phase diagrams for the 10/84/16 and 10/82/18 PLLZT solid solutions are shown in Fig.5.1. A salient feature of these diagrams is the presence of a wide interval of the electric field intensity within which the temperature of the PE phase transition is constant. As seen from the *E-T* and *Y-T* diagrams, such a peculiarity manifests itself only in the borderland region of the *Y-T* phase diagram separating the regions of the FE and AFE states. This feature of the *E-T* phase diagrams cannot be described on the base of the generally accepted thermodynamic point of view.

However, all mentioned facts are easily explained based on the model concepts, discussed in Sections 3.1 and 5.1. For the solid solutions from the borderland region in the *Y-T* and in the *E-T* phase diagrams, the FE and AFE states have approximately equal stabilities. Therefore, the domains of these phases coexist in the sample's volume. As seen from expression (5.8), the change in the electric field intensity leads to the redistribution of the shares of the FE and AFE phases in the IS of coexisting phases. The thermodynamic potentials of these phases remain unchanged (constant), until the field intensity lies within the limits of existence of the IS. A consequence of these factors is the independence of the phase transition temperature on the electric field. If a DC field intensity exceeds the upper critical field $E_2$, the $T_c(E)$ dependence becomes a typical linear one. For the 10/82/18 and 10/80/20 PLLZT solid solutions the value of $E_1$ is zero. For the PLLZT solid solution with composition 10/84/16, the low-temperature state is the AFE in the field interval up to 5 kV/cm. At the fields higher than 5 kV/cm there emerges the IS state which is later (at higher fields) suppressed by the FE state.

### 5.2.2. *Lead lanthanum zirconate titanate solid solutions*

The PLZT solid solutions are characterized by somewhat unusual type of the dielectric hysteresis loops, in particular, by the so-called narrow hysteresis loops which result in the appearance of the quadratic hysteresis loops of the electrooptical hysteresis. As shown in Section 3.1, such behavior must be manifested in those substances where the domains of the FE and AFE phases coexist and the AFE→FE phase transition induced by the electric field takes place. This behavior was experimentally investigated in 4 series of the PLZT solid solutions: 6/100-Y/Y, 7.5/100-Y/Y, 8.25/100-Y/Y, and 8.75/100-Y/Y [46]. Since the *Y-T* diagrams of all these series are physically equivalent, below we present only the results for the 8.25/100-Y/Y series. The *Y-T* diagram of the said series is shown in Fig.5.2a. First of all, it should be noted that there exists a field interval within which $T_c$ does not depend on the field intensity (insert in Fig.5.2a) for the solid solutions with compositions located near the FE-AFE boundary. The mentioned effect has been discussed above for the PLLZT solid solutions (sect.5.2.1).

Fig.5.2b presents the dependence of polarization on the electric field intensity (the samples were previously annealed at 600°C). The electric field dependencies of the resonance frequency (strictly speaking, the frequency constant $C_f = rf$, here *f* is the frequency of the first radial resonance, and *r* is the radius of the sample) of the first harmonics of radial oscillations for these samples are shown in Fig.5.2c. The study of the resonance characteristics also gives important information on the elastic properties of these materials; however, for the goals of this review the data presented in Fig.5.2 are sufficient. In contrast to the *P(E)* dependence, the $C_f(E)$ dependencies are nonmonotonic, therefore, they allow to fix the values of the critical field of the PT more precisely.

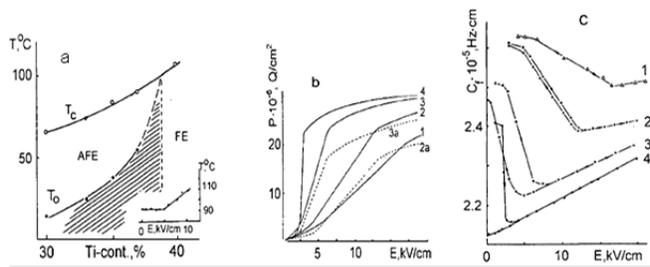

Fig.5.2. Phase Y-T diagram (a), polarization vs electric field (b), and frequency constant vs field (c) for 8.25/100-Y/Y PLZT solid solutions [50, 82].
Content of Zr/Ti: 1 − 72/28; 2 − 70/30; 3 − 67/33; 4 − 65/35.

As one can see from the curves 1 in Fig.5.2b and Fig.5.2c, there are three intervals of the external electric field, where the properties of samples noticeably differ when the field intensity increases. At zero electric field this solid solution is in the AFE state which is preserved within the field interval $0 < E < E_{cr,1} = 5.0$ kV/cm. At the high fields ($E > E_{cr,2} = 12.5$ kV/cm) the sample is in the FE state. When the field intensity is in the interval between 5.0 kV/cm and 12.5 kV/cm the polarization increases linearly. This interval is the region of the IS in the studied solid solutions. An increase of the Zr content in the solid solution leads to the widening of the intermediate state region.

The measurement results discussed above allowed us to build the "composition – electric field" phase diagrams that represent the dependencies $E_{cr,1}(Y)$ and $E_{cr,2}(Y)$. These diagrams containing the intermediate state region are presented in Fig.5.3 for the 8.25/100-Y/Y and 8.75/100-Y/Y PLZT solid solutions. In this figure the symbol IS is used for the intermediate state, the symbol $FE_i$ is used for the induced FE phase, and the symbol $FE_s$ determines the spontaneous FE phase. As one can see, the region of the AFE phase stability is shifted toward the solid solutions with the higher Ti content when the La concentration increases. The intermediate state region is also shifted in the same direction when the La content increases (see Fig. 5.3).

It should be noted that the $P(E)$ dependence is not saturated in the high-intensity fields (up to 20 kV/cm and more). This is connected with the fact that the AFE domains are still preserved in the sample's volume after the FE phase has been induced. Such a situation was observed in experiments on light scattering in the solid solutions with the composition close to that of the 8.25/67/33 PLZT and belonging to the same region of the Y-T diagram [40, 42, 50]. The increase of polarization at $E > 12.5$ kV/cm is connected with the change of the internal state of preserved domains of the AFE phase. The processes caused by the field-induced polarization having the form $P_i = \varepsilon E$ do not seem to be complete, since there is still no saturation in dependencies corresponding to the curves 1a and 2a in Fig.5.2, which present the sample polarization after the subtraction of $P_i$. Completing the discussion of behavior of the 8.25/63/37 PLZT solid solution, we need to note that the processes caused by the electric field are characterized by a weak hysteresis.

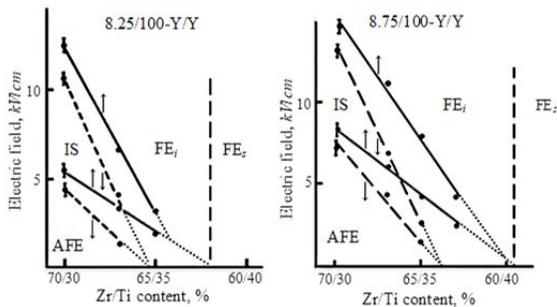

Fig.5.3. "Composition – electric field" phase diagrams for 8.25/100-Y/Y (on the left) and 8.75/100-Y/Y (on the right) PLZT solid solutions.

As one can see from Fig.5.2b and Fig.5.2c, the electric field value at which the induction of the FE phase takes place increases when the solid solution composition moves away from the value corresponding to the equilibrium between the AFE and FE phases (that is, when the concentration of Zr increases). It is not something unusual because the stability of the AFE phase increases with the increase of the Zr content and a stronger field is necessary for the induction of the FE phase.

It follows form the phenomenological consideration [83, 84, 85] that the AFE → FE phase transition has to occur in a sharp abrupt manner. However, a different behavior is observed in our experiments. The transition takes place within a finite interval of electric fields. The width of the transition increases when the solid solution composition moves away from the region of the FE states in the "composition-temperature" phase diagram.

The width of the intermediate state region $E_{end} - E_{onc} = CP_{1,0}(E_{pt})$ (see expression (5.9)) is determined by the value of the induced polarization inside the domains of the FE phase. Since the stability of the AFE state in the solid solutions in question increases with an increase in the content of Zr, the transition field value, $E_{pt}$, also increases. As it follows form the phenomenological theory of the AFE → FE phase transition, the $P_{1,0}(E_{pt})$ value noticeably increases when the induction of the FE state takes place at the higher values of the electric field intensity. It means that the $P_{1,0}(E_{pt})$ value increases with an increase in the content of Zr in the solid solutions under investigation. As a consequence of this increase the interval of the electric field values within which the intermediate state takes place also increases with an increase of the Zr content in the solid solution.

Fig.5.2 shows the change of the resonant frequency with an increasing electric field. It should be particularly emphasized here that this resonance frequency change is not linked to the change of the linear sizes of a resonator in an electric field. The dependencies of the sample deformations on the applied electric field intensity can be found in [86] for the solid solution compositions close to the ones studied here. The relative deformations less than $10^{-3}$ observed in [86] in the fields with intensities of the order of 10 kV/cm cannot be compared with the relative changes of the resonance frequency observed in our experiments. Moreover the relative deformations in the direction along the applied electric field and in the perpendicular direction have different signs. The dependence of the frequency of the longitudinal resonance (it is a thickness resonance for the geometry of our samples) on the applied field intensity at the phase transition via an intermediate state is given in Fig.5.4 for comparison. As one can see, the sign of the effect for both longitudinal and transverse (radial) resonances is the same. These data allow neglecting the changes of the resonator's linear sizes in an applied electric field during the analysis of the resonators behavior in an external electric field.

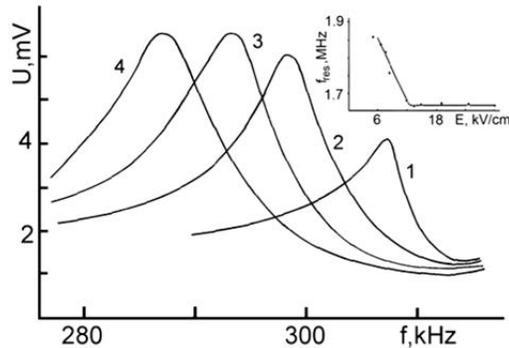

Fig. 5.4. Resonance curves of the 8.25/70/30 PLZT at different intensities of external electric field $E$, kV/cm: 1 – 5.0, 2 – 8.0, 3 – 10.0, 4 – 13.0.
Insert shows the dependence of the resonance frequency of the thickness resonance on the external electric field intensity in the 8.25/70/30 PLZT.

As indicated above, the phase transition via an intermediate state accompanied by the displacement of the interphase boundaries without change of the internal state of domains of both coexisting phases ends when the share of the induced FE phase is close to unity. However, the complete transition to the homogeneous FE state does not take place in the values of the electric field intensities achieved in our measurements. The domains of the AFE phase are still present in the volume of the samples (in the FE matrix). When the intensity of the electric field is higher than the second critical field the further increase of the field intensity leads to a change of the internal state of the AFE domains and the displacement of the interphase boundaries is not a major factor any more. Such process causes another mechanism of the resonance frequency changes. As it is seen in Fig.5.2c, the sign of the change of the resonance frequency becomes opposite to the sign that was observed when the values of the applied field were within the interval $E_{cr,1} < E < E_{cr,2}$. The reason for this change is that the effective rigidity of the system increases (at the expense of the decrease of the contribution caused by the mobile interphase boundaries) when the phase state of the system becomes increasingly homogeneous. The controllability of the resonance frequency becomes higher in the fields $E > E_{cr,2}$. The hysteresis is completely absent and the changes in the piezoelectric parameters are weak.

## 6. Coexistence of phases and ions segregation. Long-time relaxation

### 6.1. Model consideration

The stability problem for the inhomogeneous state of the coexisting FE and AFE phases has been considered in Chapter 3. The negative value of $\Delta W$ in (1.13) or in (3.15) shows that the considered interphase domain wall (IDW) possesses a negative surface energy. It seems that the negative value of this boundary's energy should have stimulated the division of the sample volume into an unlimited number of very small domains. However, it was only a simplified consideration of the problem. In particular, the effects connected with the condition of continuity for elastic medium at the interphase boundary were not taken into account.

The crossing of the IDW (from one phase to the other) goes on simultaneously with the continuous conjugation of the atomic planes (free of breaks and dislocations) (see Chapter 3 of the present paper). This coherent IDW structure leads to an increase of the elastic energy. Such increase is the more essential the larger is the difference in the configuration volumes of the FE and AFE phases. Just this effect defines the positive value of the surface energy density for the boundaries separating ordinary domains in FE [87, 88]. This elastic energy restricts the increase of the area of the interphase boundaries and, consequently, the reduction in size for the domains of the coexisting phases. These stresses weaken the condition of existence of the inhomogeneous state.

In the substances under consideration, namely, the substances where the FE and AFE states are possible, the equivalent crystallographic sites are occupied by ions that are different either in the size or in the value of charge or in both. In a single-phase state (inside the domains of each of the coexisting phases), the ions, forming the crystal lattice, are not subjected to the action of forces in the absence of external factors (more correctly, the resultant force affecting each ion is equal to zero). Opposite situation is observed for the ions located near the "bare" IDW. The balance of forces affecting each of these ions is upset. "Large" ions are pushed out into those domains, which have a larger configuration volume and, consequently, a larger distance between crystal planes. "Small" ions are pushed out into the domains with a smaller configuration volume and a smaller interplanar distance. Such process is accompanied with both a decrease of the elastic energy concentrated along the "bare" IDW and an increase of the energy bound up with the segregation of the substance. The considered process of ion segregation will be completed when the new-formed IDW structure provides the energy minimum. Such a "clothed" IDW will be further called the real IDW or simply the IDW.

Thus, the formation of the heterophase structure of the coexisting domains of the FE and AFE phases is followed by the emergence of the chemical inhomogeneity of the substance. The said process is realized owing the ion diffusion at relatively low temperatures ($T < T_c$). In this case, the diffusion coefficients are too small, and the process is a long-time one. For some PZT-based solid solutions characteristic times of this process exceed 100 h at 20°C. The process of the ion segregation and formation of segregated will be considered in experimental part in greater details.

### 6.2. Experimental results

The above-discussed mesoscopic segregations were observed in [89, 90]. However, at that time there was no complete understanding of the mechanism of their formation even though the dependence of the segregation's parameters on the solid solutions composition (and, consequently, on the location of the solid solution in the phase diagram), as well as the influence of a DC electric field on the formation of segregations were studied [90].

Now we consider the kinetics of the segregation process near the IDW separating domains of the FE and AFE phases in the solid solutions [56, 91, 92,]. The investigation was performed on two series of the PZT-based solid solutions with coexisting domains of the FE and AFE phases. The substitution of the lead ions in the PZT by the isovalent complex $(La_{1/2}Li_{1/2})^{2+}$ (PLLZT) or by the $La^{3+}$ ions (PLZT) allows achieving wider regions of the solid solution compositions where the phase coexistence takes place [34, 37, 38, 50, 65, 66, 68, 77,]. The studies were carried out on the PLZT with 6% content of lanthanum, i.e. on the 6/100-Y/Y PLZT series and on the PLLZT with 15% of $(La_{1/2}Li_{1/2})^{2+}$, i.e. on the 15/100-Y/Y PLLZT series. The phase diagrams for the said solid solutions (before they were subjected to the action of an electric field) are presented in Fig.4.9 for PLZT and in Fig.6.1 for PLLZT.

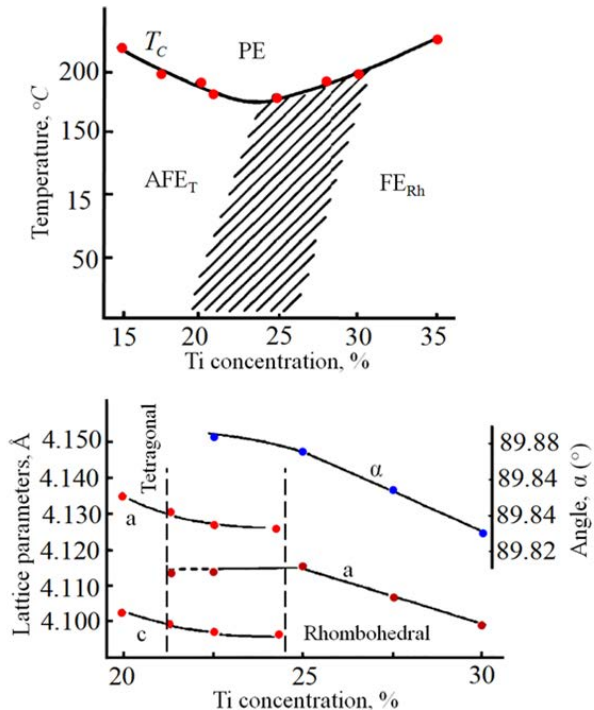

Fig.6.1. The Y-$T$ phase diagram for the 15/100-Y/Y PLLZT series of solid solutions and the dependence of lattice parameters on Ti-content [65, 66].

The decomposition of the solid solution is manifested in the appearance of weak diffusion lines (halos 1 and 2 in Fig.6.2, pattern 2) near the basic diffraction lines of the X-ray diffraction pattern, which are characteristic for the perovskite structure of solid solutions [56, 89, 91]. Therefore, there arises a problem of recording of the X-ray diffraction patterns in a wide range of angles during a short time interval while investigating the kinetics of the long-duration relaxation and the formation of the mesoscopic structure of segregates. For this purpose, the method of registration of scattered X-rays from the sample placed in the Debye X-ray chamber with the subsequent photometry of the X-ray diffraction patterns, is the most suitable (the Debye-Scherrer method). In our studies we used a chamber with a diameter of 57 mm. The registration was carried out by the "plane section" method with the filtered $CuK_\alpha$ radiation. The vanadium oxide selective absorber was used for the X-ray filtration. The layer thickness was chosen experimentally. The registration process lasted for 20 minutes (10 minutes for each of the two positions of the sample plane symmetric about the X-rays direction). The sample rotation speed was 1 r.p.s. The described method was repeatedly used in the earlier investigations while studying the processes of atomic ordering in the oxide ferrite substances [93, 94], as well as in the examination of the clustered structures in the PZT-based solid solutions [56, 89, 91].

X-ray diffraction measurements by the Debye-Scherrer method were carried out according to the following scheme. The samples were annealed during 22 hours at 600°C. After the annealing they were quenched to the room temperature. Then the samples were aged at the room temperature during the period of time $\tau$ and the X-ray diffraction patterns were obtained afterwards.

The solid solutions belonging to the shaded region of the Y-$T$ phase diagram, i.e. the 15/77/23 PLLZT and 6/73/27 PLZT have been selected for the studies of the long-duration relaxation. The domains of the FE phase (with the rhombohedral type of distortions of the elementary perovskite cell) and the AFE phase (with the tetragonal type of distortions of the elementary perovskite cell) coexisted in the bulk of the samples at the room temperature.

As a first step, the X-ray patterns were obtained at 600°C on the annealed PLZT samples These X-ray patterns contained only the strong singlet X-ray lines caused by the coherent scattering from crystal planes of the cubic perovskite lattice. It is important to note that the ferroelectric Curie temperatures of the solid solution under investigation are well below the annealing temperature, but well above the ageing temperature (the room temperature).

After the high-temperature X-ray studies the samples were quenched to the room temperature and then left to age at 22°C during the time interval $\tau$. At the end of each time interval $\tau$ the X-ray patterns were recorded by the Debye-Scherrer method. Right away after quenching ($\tau \approx 0$) the X-ray patterns contained only the strong singlet diffraction lines as in the high-temperature case (Fig.6.2 (pattern 1). The structure of the X-ray patterns became more complicated during ageing. A splitting of the singlet lines took place. Broadened diffuse lines (halos) with a significantly lower scattering intensity appeared in addition to the diffraction lines (Fig.6.2, pattern 2). These new lines were the result of the incoherent scattering from chaotically oriented segregates at the FE–AFE interphase boundaries [58, 60, 61]. We studied the behavior of the halos located in the two angle intervals $\theta = 25°–27°$ (Halo 1) and $\theta = 29°–32°$ (Halo 2). The intensity, location, and shape of the diffuse lines changed with time. The shape and the location of the diffraction Bragg-lines, which characterize the crystal structure of the solid solution under investigation, also changed with time.

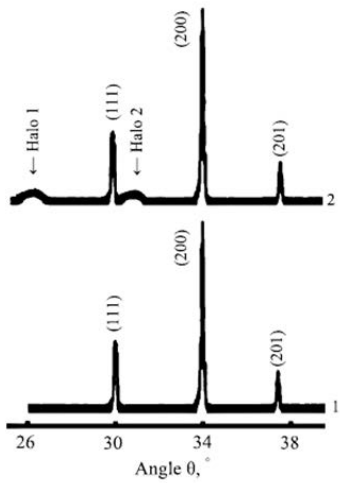

Fig. 6.2. X-ray Debye–Scherrer diffraction patterns of 6/73/27 PLZT solid solution obtained at the room temperature right after the quenching (1) and after ageing during 30 days (2).

The dependencies of the elementary cell volume (the calculations were made based on the position of the (200) X-ray diffraction peak in the pseudo-cubic approximation) on the ageing time for the solid solutions under investigation are shown in Fig.6.3. During the first stage of the aging process (of about 15–20 h for the PLZT and about 3 h for the PLLZT), the volume of the elementary cell of the perovskite crystal structure of the solid solution increases. At longer ageing times, the volume decreases. As one can see, the occurring change of the volume virtually follows an exponential behavior for both the first and the second stage of ageing. There is a peculiarity in behavior of the $V(\tau)$ dependence near $\tau \approx 22$ h which we will discuss later.

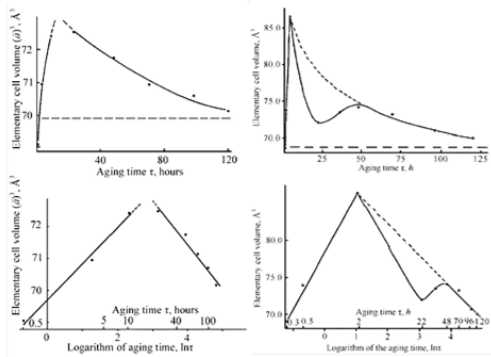

Fig.6.3. Dependence of the elementary cell volume on the ageing time for the 6/73/27 PLZT solid solution (on the left) and for the 15/77/23 PLLZT solid solution (on the right) [56, 92].

The changes of the shape, intensity and position of the diffuse lines with ageing time are presented in Fig.6.4 and Fig.6.5. The following peculiarities of these dependences attract attention. Both the profile and angular position of the diffuse scattering lines change during the aging process.

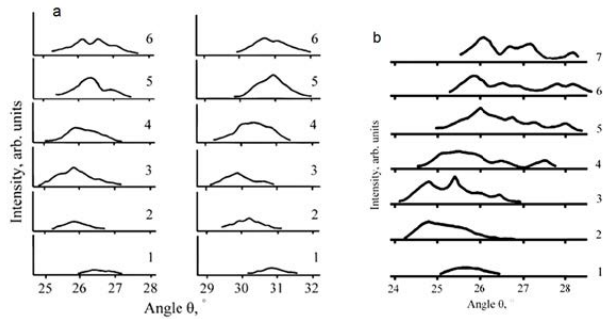

Fig.6.4. Changes of the profile and position of diffuse lines in the process of ageing after the quenching for the 6/73/27 PLZT solid solution and 15/77/23 PLLZT solid solution (b) [56, 92].
Ageing time $\tau$ (hours)
a: 1–0.5, 2–3.5, 3–23, 4–48, 5–72, 6–120.
b: 1–0.3, 2–0.5, 3–3.0, 4–22, 5–48, 6–72, 7–120.

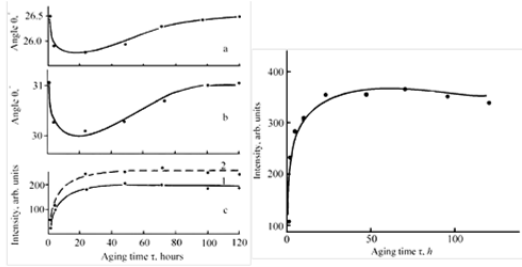

Fig.6.5. The ageing time dependencies of positions of two diffuse lines: halo 1 (a), and halo 2 (b) and their intensity (c) for the 6/73/27 PLZT solid solution (on the left) [56, 92].

It should be mentioned that there is a clear correlation between the change of the positions of the X-ray diffraction Bragg-lines (Fig.6.6) and the change of the location of the diffuse scattering lines. The dependence of the intensity of the diffuse lines on the ageing time reaches saturation in approximately 25 h.

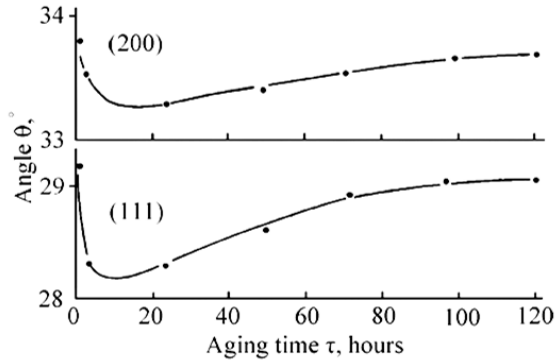

Fig.6.6. Dependencies of positions of the (111) and (200) X-ray diffraction lines for the bulk sample on ageing time for the 6/73/27 PLZT solid solution [56, 92].

Changes in the shape and the position of the X-ray diffraction lines during the samples' ageing process are given in Fig.6.7. We analyzed the behavior of the (111) and the (200) X-ray lines, which are the most typical for the perovskite structure of the solid solutions investigated. The former line is a singlet in the case of the tetragonal lattice distortions, and it is a doublet line in the case of the rhombohedral distortions. On the contrary, the latter line is a doublet in the first case and a singlet in the second case. Analysis of the profiles of the said diffraction lines allows us to deduce a structural relationship between the low-temperature phases during the process of ageing. Immediately after quenching, the phase with the tetragonal type of distortions of the perovskite crystal structure predominates in the bulk of the sample. Then, the phase with the rhombohedral type of distortions grows to dominate, with some ageing. Only with the further ageing the low-temperature phases, namely, the phase with the tetragonal type of lattice distortions and the phase with the rhombohedral type of lattice distortions, coexisting in the bulk of the samples and forming the equilibrium two-phase structure, are established.

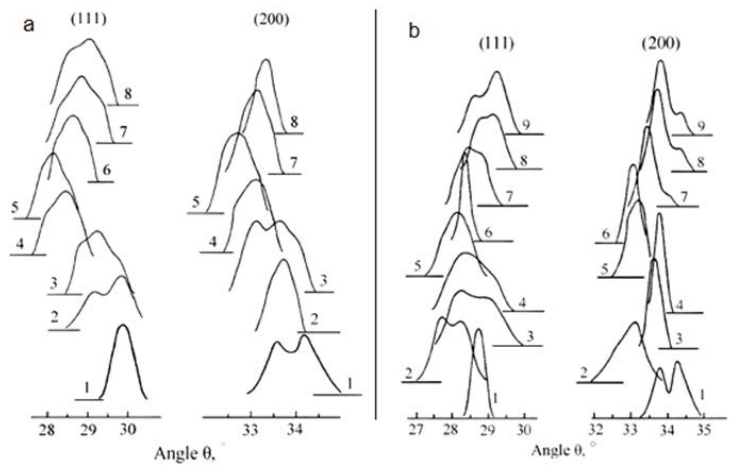

Fig.6.7. (a) - variations in the shape and position of the (111) and (200) X-ray diffraction lines in the process of the samples' ageing after quenching for the 6/73/27 PLZT solid solution [56, 92, 91]. Ageing time $\tau$ (hours): 1–0.25, 2–0.5, 3–3.5, 4–23, 5–48, 6–72, 7–96, 8–120.
(b) - variations in the shape and position of the (111) and (200) X-ray diffraction lines in the process of the samples' ageing after quenching for the 15/77/23 PLLZT solid solution.
Ageing time $\tau$ (hours): 1–0.3, 2–0.5, 3–3.0, 4–22, 5–48, 6–72, 7–120.

The domains of the FE and AFE phases coexist in the sample volume of the investigated solid solutions. It is known, that the characteristic time of the establishment of the equilibrium low-temperature state at the structural phase transitions is of the order of $10^{-4}$–$10^{-6}$ s [95]. The paraelectric-ferroelectric or paraelectric-antiferroelectric phase transitions are also the structural ones. Therefore, all the processes that we investigated in this study (after speed cooling from 600°C) take place in a system which is essentially a two-phase system containing domains of the coexisting FE and AFE phases. Thus, one has to take into account the two-phase nature of the system and the presence of the interphase boundaries while interpreting the results.

The time dependences of the shape and the intensity of the diffuse X-ray lines (halos), as well as the absence of the said lines in the X-ray patterns obtained at 600°C, confirms the connection between the long-duration relaxation and the formation of segregates in the vicinity of the FE-AFE interphase phase boundaries. The establishment of an equilibrium state is a long-time process in the solid solutions in which the state of the coexisting FE and AFE phases is possible. As one can see from the X-ray data, it continues for not less than 120 h. However, taking into account the limited sensitivity of this method one can assert that this process takes even longer.

The segregation process is multistage. It is clearly seen from the results presented in figures 6.2, 6.3, 6.4, 6.5, and 6.6 that there are different relaxation times caused by the different mechanisms. It is necessary to note that along with the segregation there are the mechanisms responsible for establishment of equilibrium values of the structural order parameters. These mechanisms have a characteristic time scale of the order of $10^{-4}$ s (such time intervals are beyond the abilities of our experimental methods). Without elucidation of particular mechanisms responsible for the attainment of the equilibrium state, one can assume that the long time constant of this process is connected with the diffusion processes associated with the local decomposition of the solid solutions along the interphase domain boundaries. The estimation of the size of segregates (using the shape of the diffused X-ray lines) gives values of 8–15 nm [89, 90] (similar approach to estimations of the average size of nanoregions can be found in [96, 97, 98]).

Long-duration relaxation processes are the non-monotonic processes due to the condition of 'strong deviation from equilibrium' in the initial stages following quenching. In the case of 'weak deviation from equilibrium' (at the final stage) the relaxation process is monotonic and is described by an exponential law. Nonetheless, the PLZT and PLLZT solid solutions do differ by the presence of vacancies in the $A$-positions of the crystal lattice. There is a peculiarity of the ageing process for the PLLZT system of solid solutions occurring at the ageing times of 20–30 h not present in the PLZT system. This peculiarity was attributed to accumulation of the elastic stress and the subsequent drop in strain [56, 91]. Otherwise, the compositional (and the structural) relaxation process follows similar patterns for both the PLZT and PLLZT solid solutions systems.

Now let us dwell on the mechanisms defining the kinetics of the processes in question at different periods of time (stages). There are different mechanisms, which contribute to this long-time relaxation. Two of these mechanisms should be pointed out among the others. The contribution of the crystal lattice defects, in particular the oxygen vacancies and the diffusion of the cations in the vicinity of the interphase boundaries, caused by the local mechanical stresses, command the greatest attention. Under the conditions of our experiments the concentration of vacancies in the lead sites remained practically constant because the lead volatilization in the PLZT solid solutions starts only at temperatures $T > 800$°C. The difference in the size (and, consequently, in mobility) and in the charge of the ions, which are located at the equivalent sites of the crystal lattice, should be taken into account. At the same time, a permanent rearrangement of the multiphase domain structure also takes place. This domain structure rearrangement is due to the change of the local composition of the solid solution which, as a consequence, leads to the change of the local phase stability. The complete analysis is still beyond our grasp as the influence of the oxygen sublattice defects on the crystalline structure of these solid solutions is investigated insufficiently at present, but by drawing from the experimental results obtained for related oxide materials with the perovskite or the perovskite-type structures, some additional insights are possible.

The annealing of samples at 600°C leads to the growth of the concentration of oxygen vacancies; the equilibrium concentration of the oxygen vacancies grows rapidly as the temperature rises. Quenching down to room temperature leads to freezing of the nonequilibrium elevated concentration of the vacancies in the bulk. Oxygen vacancies in ionic–covalent compounds, to which the solid solutions with the perovskite structure belong, lead to an increase of the crystal lattice parameters [99, 100, 101, 102]. Alongside the increase in lattice parameter, the vacancies in the perovskite and the perovskite-like compositions favor the increase of the stability of the phases with tetragonal type of crystal lattice distortions, caused by a static $T_2 \times e$ Janh-Teller effect [102, 103] (for example, on the $Ti^{3+}$ ions in the PZT, $BaTiO_3$ et. al.).

During the ageing process the oxygen vacancies move towards the sample surface and leave the sample (actually, the diffusion of oxygen into the bulk of the sample across the surface takes place) and, therefore, the crystal lattice parameters decrease. Since at room temperature, the diffusion coefficient is comparatively low, and the surface maintains a steady state concentration of the vacancies, the said process is the long-time one.

Let us consider the ion diffusion in the vicinity of the domain boundaries separating domains of the FE and AFE phases. Mechanical stresses arise at these interphase boundaries after the quenching and the formation of domains of the coexisting FE and AFE phases inside the sample's volume. These stresses are caused by the difference in the interplanar distances of the neighboring domains. There is an associated increase in the elastic energy. This increase in strain is reduced by the redistribution of the ions in the vicinity of the interphase boundaries and, as a result, by the local decomposition of the solid solution and the formation of segregates. Since mechanical stresses are now the motive force of the ion diffusion, this process must have a higher rate than the process of establishment of the equilibrium concentration of the oxygen vacancies. As is seen from Fig.6.5(c), the ageing time dependence of the intensity of the diffuse X-ray lines, connected with the formation of the segregates, reaches the saturation already in 20–25 h.

The ions that essentially differ in the ionic radii (and the charge) participate strongly in the diffusion processes, which define the formation of the segregates and the local decomposition. These are the lead ions $Pb^{2+}$ and lanthanum ions $La^{3+}$, which occupy the A-sites of the perovskite crystal lattice, and the zirconium $Zr^{4+}$ and titanium $Ti^{4+}$ ions, which occupy the B-sites of the crystal lattice. Obviously, the rates of their diffusion differ as well. This leads to the change of the chemical composition both of the segregates and of the solid solutions inside the domains with time. As the ions with the smaller ionic radii from the domains of one of the coexisting phases reach the interphase boundary, the solid solution inside the domains becomes enriched with the 'larger' ions. Consequently, the position of the solid solution in the $Y$-$T$ phase diagram changes in time, the crystal lattice parameters increase and the type of the crystal lattice distortions changes. This is clearly seen in Fig.6.6. When at the first stage of the local decomposition the "small" lanthanum and titanium ions reach the interphase boundaries, the composition of the solid solution inside the domains corresponds to the PZT with an elevated content of zirconium. Such solid solutions are characterized by the rhombohedral type of the crystal lattice distortions. Therefore, as the intensity of the diffuse lines grows the profile of the X-ray diffraction lines changes. This circumstance points to the fact that the predominating amount of the tetragonal phase is replaced by that of the rhombohedral phase. That is, in the bulk of the sample the share of the phase with the rhombohedral distortions increases.

At this stage of the process the change of the elementary cell volume is defined by the competition of two processes – the reduction of the volume owing to the decrease of the concentration of the oxygen vacancies and the increase of this volume due to the enrichment of the composition inside the domains with the 'larger' ions. The diffusion of oxygen vacancies is the slower process. As a result, during the first stage of ageing the volume increases and the maximum value is achieved in approximately 20 h.

As further ageing takes place, the 'larger' lead and zirconium ions reach the interphase boundaries and the composition inside the domains approaches its nominal formula composition; the solid solution regains its stable location in the composition versus temperature phase diagram (that is, a return to the two-phase region of the state diagram takes place). In the profiles of the X-ray lines this fact manifests itself (after approximately 50–60 h of ageing) by the gradual establishment of the precisely such shape that is characteristic of the domains after the sintering and the ageing on a very long timescale (the line shape that is established after one year).

The changes in the profile of the diffusive scattering also take place during the above-described ageing process. These changes confirm the fact that the concentration of different ions in the segregates varies at the interphase boundaries. The profile of the diffusive lines is defined by the resulting enveloping curve obtained after the summation of the X-ray scattering from the crystal planes in the segregates. Since the local chemical composition of the segregates constantly changes in the process of ageing after quenching, the profile of the diffusive lines changes as well.

As we have indicated in Chapter 4 (Section 4.3), numerous difficulties in the identification of the crystal structure are associated with the presence of the coexisting domains of the FE and AFE phases. Local decomposition of the solid solutions in the vicinity of the boundaries between the coexisting phases and the formation of the mesoscopic structure of segregates at these boundaries in the bulk of the samples poses a particular problem. In the X-ray diffraction patterns, it manifests itself in the appearance of the supplementary diffuse lines that accompany the diffraction lines, which are used for the identification of the PLZT crystal structure. These diffuse lines may often appear as the satellites of the main Bragg peaks. In this case, in the process of the mathematical treatment of the experimental X-ray diffraction patterns according to the existing programs a third phase is brought into consideration. We have faced such a phenomenon while treating the experimental results. This is connected only with the fact that in the development of the corresponding programs the phenomenon of the local decomposition of solid solutions has not been taken into account. As far as we know, the effects associated with the coexistence of domains of the FE and AFE phases and the local decomposition of solid solutions in the vicinity of the interphase boundaries have not been discussed in the literature up to now.

In conclusion, let us consider the peculiarity of the time dependence of the elementary cell volume in the PLLZT solid solutions which manifests itself at the aging time intervals of 22 h. The X-ray patterns obtained for the 10/77/23 PLLZT solid solution after different aging time intervals are presented in Fig 6.8 [56]. A pronounced complex structure of the (200) X-ray line is clearly seen for the aging time interval of 22 h. We estimated the sizes of the regions of the coherent scattering after different aging time intervals, namely, for $\tau = 22$ h and $\tau = 48$ h for comparison. In the first case, the size of the above regions was $3 \cdot 10^{-4}$ cm, and in the second case, it was $1.5 \cdot 10^{-4}$ cm. It has to be noted that the regions of the coherent scattering are inhomogeneous and this circumstance leads to the oblong shape of reflexes in the diffraction pattern obtained after the aging time of 22 h. The ratio of the long, $a$, and the short, $b$, axes of the coherent scattering region is $a/b \approx 2.26$. In the case of $\tau = 48$ h the shape of the coherent scattering regions is close to equiaxial.

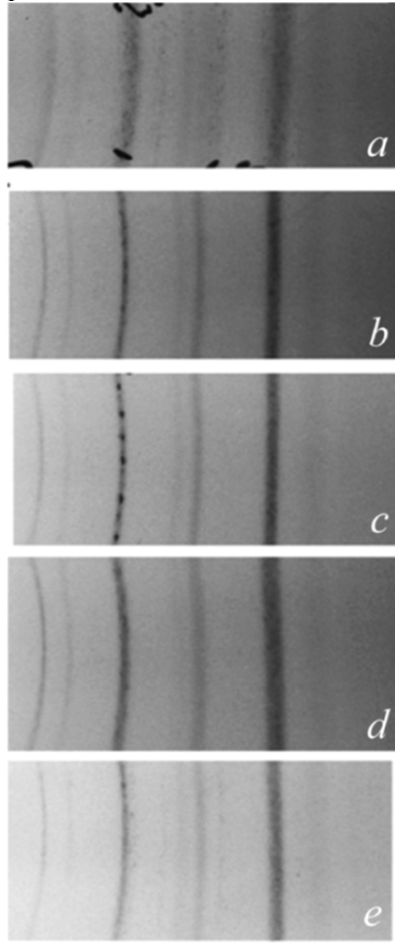

Fig. 6.8. X-ray diffractograms for the 15/77/23 PLLZT solid solution, obtained after different aging time intervals [56].
$\tau$ (hours): a–0.5, b–3.0, c–22.0, d–48, e–96.

The complex structure of the (200) X-ray line is the evidence of the presence of texture in the samples after the aging time interval $\tau = 22$ h. This texture was absent at the shorter aging time intervals. The said texture disappears again at $\tau = 48$ h and the longer intervals of aging time. It can be attributed to the presence of the internal mechanical stress and to the increase of the elastic energy caused by this stress. Such phenomenon is well known and studied in the case of the alloy aging process (accompanied by the process of local decomposition) [104]. The relaxation toward the equilibrium stress-free state (with a smaller value of elastic energy) takes place by means of the ordering in the system of the segregates-precipitates including appearance of the texture in the samples. Hence, we attribute the peculiarity in the V(τ) dependence at τ =22 h to the appearance of the internal mechanical stresses, which are the consequence of the above-considered diffusion processes accompanied by the local changes in the chemical composition of the solid solution. This peculiarity in the V(τ) dependence disappears in the process of further aging because of the relaxation of the mechanical stresses. The absence of such peculiarity in the V(τ) dependence for the PLZT solid solution is explained by the fact that the mechanical stresses relax by means of the redistribution of vacancies in the regions where mechanical stress appears. Let us remind that in the PLZT solid solutions (also investigated in this paper) each two lanthanum ions lead to appearance of one additional vacancy in A-cites of the crystal lattice.

Thus, in spite of the similarity of the local decomposition processes in the PLZT and the PLLZT solid solutions some differences are connected with the essentially different number of vacancies in the crystal lattice in these systems of solid solutions and therefore with the presence of the additional mechanical stress in the PLLZT solid solutions.

## 7. Two-phase nucleation and diffuse phase transitions

### 7.1. Theoretical model consideration

In Chapter 3 we considered stability of the two phase state at the temperatures lower than the Curie point. Some physically correct conclusions regarding the peculiarities in the behavior of the considered system may be also made for the temperatures close to $T_{c,f}$ ($T_{c,f} > T_{c,af}$) and the higher ones. The thermodynamic potential of the system has only one minimum within the temperature interval close to the PE transition. Therefore, the substance described by such a potential is a FE, and the appearance of the AFE phase domains at the temperatures near $T_{c,f}$ is possible only in the form of the fluctuations. The said domains, arising in the volume of the substance, interact with the FE matrix. Such interaction changes the density of the thermodynamic potential in the volume of the sample within which it takes place. For the other parts of the volume the density of the FE state potential remains unchanged.

Thus, for the volume, in which the interphase interaction takes place, the density of the nonequilibrium thermodynamic potential may be written as [28, 32]:

$$\varphi = \frac{\alpha_1}{2} P_1^2 + \frac{\alpha_2}{4} P_1^4 + \frac{\beta_1}{2} \eta_1^2 + \frac{A}{2} P_1^2 \eta_1^2 + \frac{m}{2} P_2^2 + \frac{n}{2} \eta_2^2 + C P_1 P_2 + D \eta_1 \eta_2 \qquad (7.1)$$

In this expression the first four terms correspond to the density of the nonequilibrium thermodynamic potential of the FE phase for $T > T_{c,af}$ ($\beta_1 > 0$, so we assume that $\beta_2 = 0$), the next two terms correspond to the density of the nonequilibrium potential of the fluctuational domains of the AFE phase ($m > 0$, $n > 0$), and the last two terms represent the interphase interaction.

The minimization of (7.1) with respect to $P_2$ and $\eta_2$ yields:

$$P_2 = -(C/m)P_1; \qquad \eta_2 = -(D/n)\eta_1. \qquad (7.2)$$

Substituting the obtained expression into (7.1) we have:

$$\varphi = \frac{1}{2}\left(\alpha_1 - \frac{C^2}{m}\right) P_1^2 + \frac{\alpha_2}{4} P_1^4 + \frac{1}{2}\left(\beta_1 - \frac{D^2}{n}\right) \eta_1^2 + \frac{A}{2} P_1^2 \eta_1^2. \qquad (7.3)$$

As one can see from the last expression, the interphase interaction leads to the renormalization and the increase of the Curie point (see the coefficient at $P_1^2$). Moreover, under the certain conditions this interaction may also stabilize the AFE state in some part of the sample's volume (see the coefficient at $\eta_1^2$). This means that there is a possibility of the existence of the two-phase (FE+AFE) domains at the temperatures exceeding $T_{c,f}$.

It is not difficult to determine the distribution of the Curie temperatures through the crystal's volume (the dependence of $T_C$ on the spatial coordinates). It is enough to use the standard form of the nonequilibrium thermodynamic potential (3.7) for the system in which the minima of thermodynamic potentials for two dipole-ordered states have close values of their depths and take into account that (3.7a)

$$\alpha_1 = \alpha_0(T - T_{c,f}); \qquad \beta_1 = \beta_0(T - T_{c,af}).$$

As it was shown previously in Ch.3 (eq. (3.13)) the energy change in the region of space where the interphase interaction takes place is given by the expression (3.13):

$$\Delta W = -\left(\frac{\xi_1^2 \cdot \xi_2^2}{4V_1} D_2^2 \eta_{2,0}^2 + \frac{\xi_1^2 \cdot \xi_2^2}{4V_2} C_1^2 P_{1,0}^2\right) \xrightarrow{\xi_1 = \xi_2 = 1/2} \Delta W = -\frac{1}{32}\left(\frac{D_2^2}{V_1} \eta_{2,0}^2 + \frac{C_1^2}{V_2} P_{1,0}^2\right),$$

which has to be added to (3.7). In this expression the order parameters $\eta_{2,0}$ and $P_{1,0}$ correspond to the part of the crystal which is not affected by the interphase interaction ($\eta$ and $P$ – in terms of eq.(3.7)), that is these are parameters determined by the thermodynamic potential (3.7). That is why the coefficients of this expansion are renormalized as follows:

$$\alpha_1 = \alpha_0\left[T - \left(T_{c,f} + \frac{\xi_1^2 \cdot \xi_2^2}{4}\frac{C_1^2(x,y,z)}{V_2}\right)\right] \xrightarrow{\xi_1 = \xi_2 = 1/2} \alpha_1 = \alpha_0\left[T - \left(T_{c,f} + \frac{1}{32}\frac{C_1^2(x,y,z)}{V_2}\right)\right]$$

$$\beta_1 = \beta_0\left[T - \left(T_{c,af} + \frac{\xi_1^2 \cdot \xi_2^2}{4}\frac{D_2^2(x,y,z)}{V_1}\right)\right] \xrightarrow{\xi_1 = \xi_2 = 1/2} \beta_1 = \beta_0\left[T - \left(T_{c,af} + \frac{1}{32}\frac{D_2^2(x,y,z)}{V_1}\right)\right] \quad .(7.4)$$

As one can see the local temperatures of the phase transition into the PE state increase and the spatial dependence of the Curie temperatures appears as a consequence of the special dependence of the coefficients $C_1(x, y, z)$ and $D_2(x, y, z)$. Let us remind that the coefficients $V_1$ and $V_2$ in equation (7.4) are determined by the expansion coefficients of the thermodynamic potential (3.7) and are positive (see Chapter 3). That is why the spatial distribution of the Curie temperatures is determined not only by the interphase interaction but also by the "initial" properties of the system itself. In particular, as it was mentioned in preceding chapters, the decrease of the coefficients $V_1$ and $V_2$ takes place in the systems in which the elastic potential anharmonicity manifests itself and the flattening of the minima of the thermodynamic potential (3.7) occurs. The widening of the temperature interval of the phase transition into the completely disordered state occurs as a result. It is also obvious that the domains in which the state of the coexisting spatially separated FE and AFE phases is realized (in the first approximation) become stable in the temperature region of the "initial" PE state.

As it follows from (7.3), (7.4), and also form results of the Chapter 3 of this review the local regions of the crystal in the vicinity of the boundaries separating domains with the FE and AFE orderings are characterized by the lower free energy than the main volume. At the temperatures above the Curie temperature such domains appear at first as fluctuations and are stabilized later by the interphase interaction. At the cooling from the high temperatures approaching the Curie temperature from the PE phase the domains with $P \neq 0$ и $\eta \neq 0$ appear at first and then they acquire the features of the two-phase domains in the vicinity of $T_C$. The temperature at which the first nuclei of the dipole-ordered state appear can exceed the Curie temperature by 100 °C and even more.

Our simplified approach does not take into account effects connected with the presence of the elastic stresses caused by the phase coexistence and the interphase boundaries. Therefore some additional remarks have to be made. It is a general knowledge that in the course of the PE-FE phase transition the configuration volume increases, whereas at the transition into the AFE state this parameter decreases. Therefore the existence of the domains of each phase separately at the temperatures above the Curie points is accompanied by the appearance of the substantial elastic stresses, and is not advantageous from the energy point of view.

However, the increase of the energy does not take place for the complex two-phase (FE + AFE) domain. In this case the configuration volume does not vary for the two-phase domain as a whole. The presence of the domains of such structure allows to remove the elastic stresses and to raise the dipole disordering temperature due to the interaction (1.9) as much as possible (i.e. to decrease the free energy to the utmost). The ratio of the phase volumes in the two-phase domain which is present in the PE matrix of the crystal is defined by both the relative stability of the FE and AFE phases (i.e. by the relative difference of their free energies) and the changes in the elastic energy, which may be brought about by each of the phases. A characteristic property of such a domain is that it should easily match the varying external conditions: an electric field and a pressure (including those of the PE matrix). An insignificant polarization of this domain, almost complete absence of elastic stresses, and the absence of distortions in the PE matrix, where the domain arises, provide its existence at the temperatures exceeding both $T_{c,f}$ and $T_{c,af}$.

For simplicity the latter phenomenon may be thought of as a complex two-phase nucleation. It is realized in the form of a spatial domain which lies within the PE matrix of the crystal.

Now let us proceed to consideration of the PE-FE transition itself. The diffuse nature of the PE PT usually is bound up with the "entrapment" of the low-temperature (LT) phase remnants into the high-temperature (HT) region with $T > T_c$. Thus, the ordered LT state is the starting-point for the process in question. On the contrary, in the physics of phase transitions it is the symmetry and the properties of the HT phase define all the possible LT phases and the phase transition under any temperature changes. This idea should be a base while considering the diffuse phase transition.

Typical temperature dependence for the inverse dielectric constant (curve 2) in the vicinity of the point of the FE-PE phase transition is shown in Fig.7.1 [105]. Hear a special attention has to be paid to the value of the Curie-Weiss temperature $T_{cw}$, since it is above the Curie point (to the right of the $\varepsilon(T)$ dependence maximum). Up to now this fact has been neither explained nor even discussed in the literature. However, from the viewpoint of the physics of the phase transitions it is just at this $T_{cw}$ temperature the transition into the ordered state must take place.

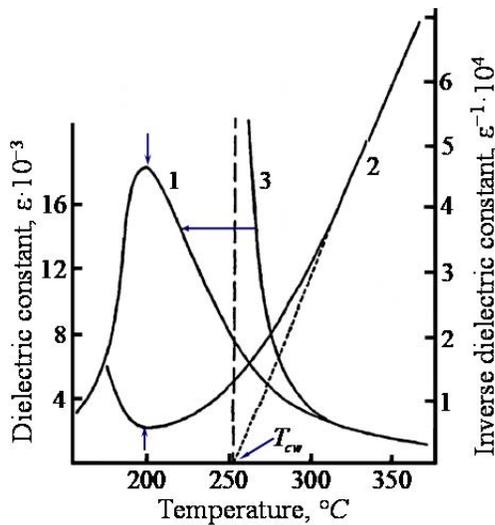

Fig.7.1. Dependencies $\varepsilon(T)$ (curve 1) and $1/\varepsilon(T)$ (curve 2) for a FE with diffuse phase transition. Dependence $\varepsilon(T)$ calculated from Curie-Weiss law (curve 3). Arrow shows the shift of the $\varepsilon(T)$ dependence under effect of ferroelectric-nonactive impurity [105].

The transition from the PE into the FE state is due to the condensation of the polar mode with the wave vector at the center of the Brillouin zone. This mode is responsible for the increase of $\varepsilon$ when the temperature approaches the Curie point. Curve 3 built in agreement with the Curie-Weiss law presents the temperature dependence of $\varepsilon$ which reflects the properties of the HT phase and must describe the behavior of the dielectric constant when approaching the Curie point from above. Nevertheless, in the vicinity of the transition the experimental curves $\varepsilon(T)$ (curve 1) and, consequently, $\varepsilon^{-1}(T)$ (curve 2) are shifted toward the lower temperatures. In other words, at lowering the temperature the PT is blocked, displaced to the lower temperature region. Moreover, due to the diffuseness of the transition the said blocking should differ in intensity in different local regions of the crystal.

In our opinion, the main problem in the diffuse phase transition physics is to clarify the mechanism (or mechanisms) of this "pressing out" of the ordered phase towards the LT region. Our recent results including those presented in this paper allow proposing a mechanism of "pressing out" the paraelectric PT towards the LT region with its simultaneous diffuseness. The decisive factor which defines the behavior of the said PT is the above-considered mechanism of the separation of ions at the interphase FE-AFE boundary in the two-phase domains.

The dynamics of the DPT may be presented as follows. At the first step, in the process of cooling from the high temperatures, the behavior of the substance corresponds to the classical theory of phase transformations. The temperature dependence of the dielectric constant obeys the Curie-Weiss law. Then the above-discussed two-phase domains appear at cooling. The "separation" of the elements forming the substance's crystal lattice at the interphase (FE-AFE) boundary is initiated due to the difference in their ionic radii. There appear the local segregations with the "foreign" composition and the mesoscopic dimensions; in the general case they are not FE-active or AFE-active. It is just at this step the deviation from the Curie-Weiss law manifests itself on the $\varepsilon^{-1}(T)$ and, consequently, $\varepsilon(T)$ dependencies. In the process of cooling the share of such segregations in the volume of the substance increases and the deviation from the Curie-Weiss law becomes more noticeable.

However, the passive dielectric segregations reduce both the share of the FE component in the substance and the value of the dielectric constant. Resent results of the physics of inhomogeneous solids prove that the introduction of the non-magnetic impurities into magnetic materials or the introduction of the passive dielectric impurities into FE and AFE suppresses the formation of the low-temperature ordered phase. The literature on this problem is ample enough, but here we would like to dwell on the well-investigated case of the percolation systems only. It is shown [106] that when approaching the percolation threshold from the ordered phase (this case corresponds to the increase of the content of the passive impurities in the system), the phase transition temperature decreases and tends to zero at $x \to x_c$. At present this problem is investigated in details not only for the case of the point defects in the crystal lattice, but also for the case of the extended (mesoscopic-scale) inhomogeneities [107]. Similar mechanisms work in the systems under discussion. The segregations which are developing in the process of cooling "press out" the LT phase and, consequently, the phase transition, toward the lower temperatures and lead to the diffuse nature of the phase transition. The latter is explained by the fact that in the places of segregation development it is impossible to find out the pronounced boundaries separating the active FE and the non-active parts of the sample. Moreover, there exist local mechanical stresses which are not compensated due to the segregation of the chemical elements.

In this connection it should be noted that for a majority of investigated systems of solid solutions (where the change in the content of one of the components results in the substitution of the FE ordering by the AFE one), the concentration dependencies $T_c(x)$ are characterized by a fall in the vicinity of the phase boundary between the regions of the said states in the $x$-T diagrams. As seen from the results presented here, it is another convincing reason to substantiate our concept of the system's behavior near the triple FE-AFE-PE point.

The distinction from the FE phase transition consists only in the fact that the transition from the PE into the AFE state is caused by the condensation of the non-polar mode with the wave vector at the boundary of the Brillouin zone, whereas the increase of the dielectric constant at $T \to T_c$ results from the softening (but not the condensation) of the polar mode with the wave vector at the center of the Brillouin zone. The letter mode interacts with the soft non-polar mode. Therefore, in the case of the AFE the extrapolation of the straight-line part of the dependence $\varepsilon^{-1}(T)$ does not yield the point of the AFE phase transition (i.e. the point of the loss of the stability for the high-symmetry phase with respect to the relatively small anti-polar displacements of the crystal lattice ions).

Thus, in the proposed model of the diffuse phase transition the basic role belongs to the chemical segregations caused by the difference in the configuration volumes of the FE and AFE phases. Such segregations were revealed experimentally and reported in [89]. At that time the mechanism of their formation was not clear. Detailed studies of the dependence of the properties of segregations on the composition of the solid solution (the position of the solid solution in the phase $x$-T diagram) were carried out in [90]. The influence of the DC electric field on the properties of segregations was also studied in [90]. At present we have experimental results concerning the time development of the process of the chemical segregation at the interphase boundaries.

The diffusion processes bound up with the inhomogeneous deformations of solids have already been considered in the literature. As an example, see the Gorsky effect in [108] (Subsection 11.2).

We would like to note that the considered "stratification" must also take place at the FE (AFE)-PE interphase boundaries. However, in this case the change in the interplanar distances while crossing such a boundary is essentially smaller in comparison with that for the FE-AFE boundary. Therefore, the driving force of the process will be essentially smaller, too. In our opinion, the decomposition of the solid solution near the latter boundary is decisive.

In what was presented above we proposed a mechanism of the diffuse phase transition into the paraelectric state for the substances characterized by a faint difference between the energies of the FE and AFE orderings. Now we shall show that the above-presented ideas are applicable to the understanding of the nature of the diffuse phase transition in a vast class of FE and AFE materials.

The results of experimental investigations of the oxygen-containing octahedral substances obtained in the course of many years show that for all real FE and AFE the difference in energy between different types of dipole ordering is small. From the viewpoint of the free energy this means that for the majority of oxides with the perovskite structure the free energy in the space of the order parameters of the FE and AFE phases has two minima at rather low temperatures. The expansion of the free energy (described in terms of the Landau theory) is to be realized with respect to the powers of two order parameters. However, since the depths of the minima are different, the metastable minimum does not become stable in some substances. Under such conditions only one phase transition (FE-PE or AFE-PE) will be observed in simple experiments, and it is sufficient to take into account only one order parameter in the Landau expansion. But this by no means signifies that the presence of the second minimum cannot be revealed in experiments.

Now let us refer to Fig.7.2. The lower part of this figure contains the generalized phase diagram. Here the temperature is shown on the ordinate and the external parameter (e.g. the hydrostatic pressure P or the chemical composition $x$) is plotted on the abscissa. There is the triple FE-AFE-PE point in this diagram. The low-temperature phases coexist near the FE-AFE boundary. As has been discussed above, the diffuseness of the PE phase transition greatly depends on the type of the LT state (see the upper part of Fig.7.2). In the regions of the FE or AFE states remote from the triple point the diffuseness parameter $\delta$ is small. As the triple point is approached when the external pressure $P$ or the concentration $x$ are changed, the diffuseness of the phase transition increases.

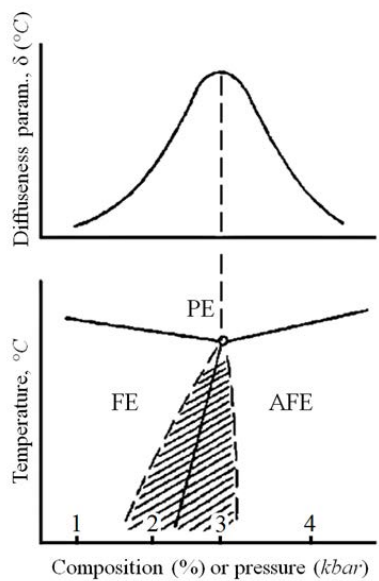

Fig. 7.2. Generalised diagram of phase states (below) and the dependence of diffuseness parameter on external factors (above) near the triple point FE-AFE-PE [105].

The majority of known FE or AFE are located within the interval 1-2 or 3-4 in the generalized diagram of states. (The particular location may be determined only from a complete set of phase diagrams of the substance under consideration, namely, $x$-T, P-T, $x$-E, $x$-P, $E$-P, $E$-T, etc.). If the location of the substance corresponds to the points 1 or 4 (or approaches them), then the free energy is characterized by the presence of only one minimum, and the substance manifests the properties of an ordinary FE or AFE. The experimental data may be easily interpreted. But when the location of the given substance approaches the point 2 or 3 in the diagram of phase states, there appears the second (metastable) free energy maximum, and its depth approaches that of the minimum corresponding to the stable state while reaching the phase boundary. In this case it is difficult to interpret the experimental results. Many physical properties of these compounds are defined by the coexistence of domains with the FE and AFE dipole orderings in the volume of the substance. Though the real FE-AFE transition is not observed in most experiments, such coexistence takes place indeed. What new states have been devised to account for this fact! But the explanation is quite simple: for this purpose one must only have the complete set of the phase diagrams.

In view of the above-said it is clear that the experiments must be performed in a way that allows following the transition from the point 1 to the point 4 in the diagram of states. This can be achieved either by changing the solid solution composition or by applying external factors. In the latter case the experiments using the hydrostatic pressure will have the most unambiguous interpretation since in these experiments the origin of the "unusual" properties is easily revealed. It seems strange that for several decades the experiments aimed at the building the complete sets of phase diagrams have not been carried out while studying many substances.

Thus, the authors adhere to the viewpoint that the developed here ideas about the diffuse phase transition in the vicinity of the triple FE-AFE-PE point are suitable for the description of the paraelectric phase transition in a wide class of FE and AFE. This viewpoint will be consistently confirmed further using a set of experimental phase diagrams as a background.

## 7.2. Experimental results

### 7.2.1. *Diffuse phase transitions near the ferroelectric-antiferroelectric-paraelectric triple point*

Let us now consider the phase transitions into the PE state for the PZT-based solid solutions with compositions located near the boundary between the regions of the FE and AFE orderings in the Y-$T$ phase diagram. The dependencies of the diffuseness parameter of the paraelectric phase transition, $\delta$, on the composition for two series of the PLLZT solid solutions with 10% and 15% of the $(Li_{1/2}La_{1/2})$-complex are shown in Fig.7.3a and 7.3b [65]. The main peculiarity of the presented dependencies is the increase of the diffuseness parameter in the solid solutions belonging to the boundary region separating the FE and the AFE states in the Y-T diagram (see Fig.4.4a, 5.1, 6.1). The displacement of the boundary region towards the higher contents of titanium at increasing the concentration of $(Li_{1/2}La_{1/2})$-complex in the solid solution results in the shift of the maximum of $\delta(Y)$ dependence.

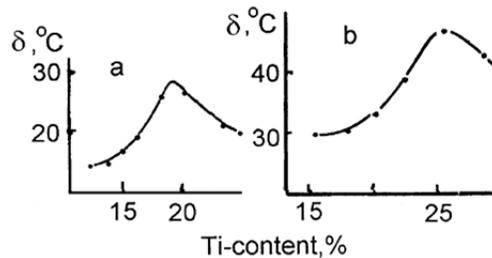

Fig.7.3. Diffuseness parameter $\delta$ vs Ti-content for the 10/100-Y/Y (a) and 15/100-Y/Y (b) PLLZT solid solutions [50, 65, 109].

The dependencies of the diffuseness parameter on the Ti-content in the PLZT solid solutions with sequentially increased contents of lanthanum are presented in Fig.7.4 [109]. Under such a substitution the phase boundary, separating the FE and AFE states in the Y-$T$ diagram (Fig.4.4a and 4.9) is shifted towards the increased values of the Ti content. As seen from Fig.7.4, $\delta(Y)$ dependencies are shifted in the same way, and their maximum coincides with the location of the borderland between the regions of the spontaneous FE and AFE states in the Y-$T$ diagrams. Thus, it is obvious that the maximum diffuseness of the paraelectric phase transition in the PLZT takes place in the vicinity of the equilibrium of the FE and AFE states.

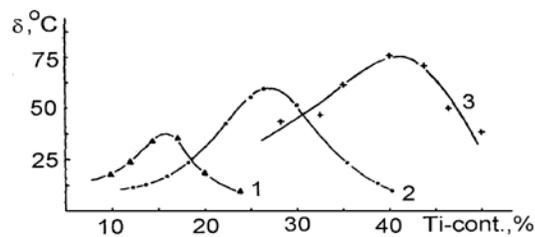

Fig.7.4. Diffuseness parameter $\delta$ vs Ti-content for several PLZT solid solutions [34, 37, 39, 50, 109].
Content of La, %: 1 – 4; 2 – 6; 3 – 8.

Let us consider the influence of the hydrostatic pressure on the diffuse phase transition in the PLZT. Fig.7.5 shows the diffuseness parameter as a function of the hydrostatic pressure for the PLZT solid solutions with compositions 6/65/35 and 6/80/20 ($P – T$ phase diagrams for these solid solutions are presented in Fig.4.6). Under the normal conditions the solid solution with the composition 6/65/35 is in the FE state and undergoes the FE-PE phase transition at a temperature of about 180°C (Fig.4.4a and 4.9). The increase of the pressure leads to transition of the solid solution into the state of the coexisting FE and AFE phases, and then into the single-phase AFE state (Fig.4.6 and 4.7). The solid solution with the composition 6/80/20 is in the state of the coexisting phases under the normal conditions; the increase of the pressure induces the transition into the single phase AFE state. As it may be seen from Fig.7.5, the increase of the pressure up to 3 kbar leads to the growth of the diffuseness parameter in the 6/65/36 and to decrease of this parameter in the 6/80/20 PLZT solid solutions.

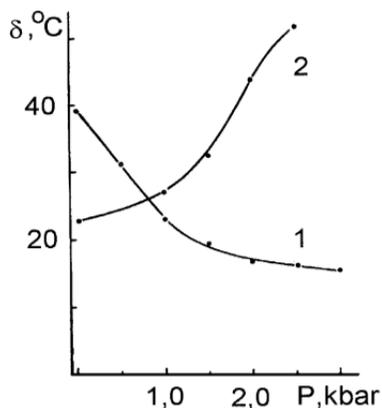

Fig.7.5. Diffuseness parameter δ vs pressure for the 6/100-Y/Y PLZT solid solutions [34, 37, 39, 50].
Content of Zr/Ti: 1 – 6/80/20; 2 – 6/65/35.

Thus, the presented experimental results show that in the vicinity of equilibrium of the FE and AFE phases the transition into the PE state is a diffuse phase transition, and the degree of diffuseness diminishes when moving from the phase equilibrium. We have also stated that such feature does not depend on the origin of the phase equilibrium (a variation of the content of the solid solution components or a variation of the hydrostatic pressure). In the case when the phases are redistributed by the electric field the degree of diffuseness pf the phase transition also changes [109]. In particular, if the field favours the emergence of the state of the coexisting FE and AFE phases, then the diffuseness parameter increases.

To better understand the problem under discussion, let us consider the results of [37, 50, 110] of studies of the diffuse phase transitions in the PZT-based solid solutions with the composition $0.97Pb(Zr_{1-y}Ti_y)O_3+0.03Cd(Ta_{2/5}W_{1/3})O_3$ under a hydrostatic pressure. The addition was introduced with the aim to obtain a nearly vertical FE-AFE phase boundary in the "pressure-temperature" phase diagram (Fig.7.6). As is known, for small values of $y$ the AFE state is stable at the temperatures below the Curie point. The FE state becomes stable at low temperatures when the value of $y$ increases. Due to the choice of the solid solution composition ($y$) the change of the temperature results in the emergence of either the FE-PE or AFE-PE phase transitions at the Curie point (at the atmospheric pressure). We have investigated the solid solutions with $y = 0.02$, 0.03 and 0.04. The phase $P-T$ diagram of the solid solution with $x = 0.02$ is shown in Fig.7.6. As one can see, the pressure $P_0 \cong 3$ kbar corresponds to the equilibrium of the FE and AFE states within a wide temperature range.

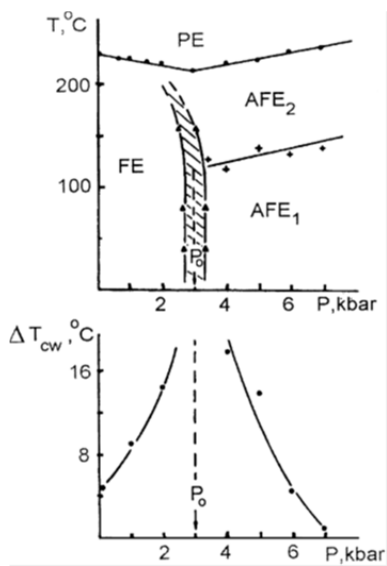

Fig.7.6. . "Pressure-temperature" phase diagram for PZT solid solution (at the top) and the temperature interval of deviation from Curie-Weiss laws pressure (at the bottom) [37, 50, 110].

Experiments showed that the diffuseness of the PE transition depended on the pressure essentially, and the maximum degree of the diffuseness occurred at 3 kbar. For the value of pressure $P >> P_0$ and $P << P_0$ the dependence $\varepsilon^{-1}(T)$ obeys the Curie-Weiss law irrespective of the type of the LT state (FE or AFE). The temperature interval of deviation from this law, $\Delta T_{cw}$, is narrow in the vicinity of $T_c$. This $\Delta T_{cw}$ interval sharply increases when $P \to P_0$ from the sides of both the FE and AFE phases. The dependence $\Delta T_{cw}(P)$ for $y = 0.02$ is also shown in Fig.7.6. While increasing the value of $y$ in the solid solution the boundary between the FE and the AFE states in the $P$-$T$ diagram shifts to the right by ~ 0.9 kbar per 1% of Ti. The anomaly of the $\Delta T_{cw}(P)$ dependence is displaced in the same manner.

The presented results show that the diffuseness of the paraelectric phase transition essentially depends on the proximity to the line of equilibrium of the FE and AFE states in the *P-T* phase diagram. The pressure $P_0$ is the value at which the thermodynamic potentials of the FE and AFE states become equal, and, therefore, the probability of the coexistence of these phases is the highest.

### 7.2.2. *X-ray investigation of the two-phase (ferroelectric + antiferroelectric) nucleation in the paraelectric phase*

We investigated the temperature dependence of shapes of the X-ray diffraction lines at $T > T_c$ (in the PE phase) for the 10/100-Y/Y series of PLLZT solid solutions [39, 50, 65]. For this purpose the (222) and the (400) lines, the most typical for the perovskite crystal structure, were chosen. The former line is a singlet in the case of the tetragonal distortions of the crystal lattice; in the case of the rhombohedral type of lattice distortions it is a doublet. The latter line is a doublet when the distortions are of tetragonal type, and it is a singlet when the distortions are rhombohedral. The analysis of the shape of these X-ray diffraction lines allows concluding, which of the LT phases is extended to the high-temperature region, disposed above the point of the PE phase transition for the main part of the samples' volume.

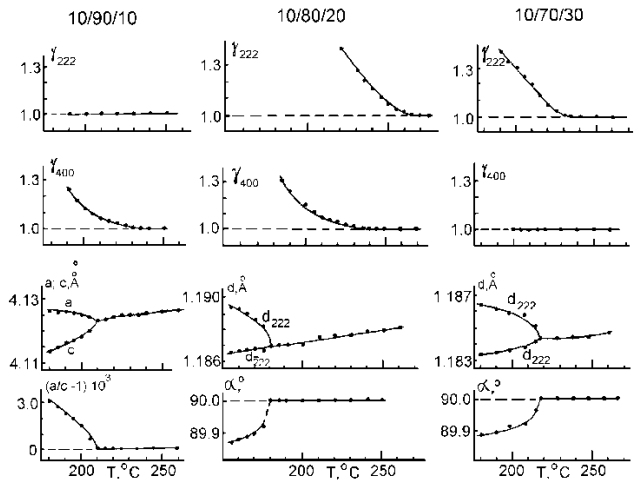

Fig.7.7. Temperature dependences of X-ray parameters for 10/90/10 (to the left), 10/80/20 (in the middle) and 10/70/30 (to the right) PLLZT solid solutions [38, 50, 65].

Fig.7.7 illustrates the temperature dependencies of the asymmetry of the X-ray lines ($\gamma = S_l/S_r$) and of the crystal cell parameters for the PLLZT solid solutions belonging to different regions of the Y-*T* phase diagram. Here $S_l$ and $S_r$ are the areas below the contour of the X-ray diffraction line located from the left and from the right of the vertical line drawn from the vertex of the X-ray diffraction line, $S = S_l + S_r$, where S is the total area of the X-ray diffraction line.

According to the data of the calorimetric, dielectric, and X-ray diffraction measurements, the 10/90/10 PLLZT undergoes the PE phase transition at 209°C. The temperature dependences of the crystal lattice parameters *a*, *c* and (*a/c* − 1) near the point of the PT are shown in the left column in Fig.7.7. The asymmetry of the (400) line is preserved at the temperatures up to 230°C. The (222) X-ray diffraction line is symmetric for the temperatures above 209°C. This fact directly testifies that the diffuseness of the phase transition in this solid solution is connected with the extension of the domains of the tetragonally distorted AFE phase to the high-temperature region above the transition point.

The 10/70/30 PLLZT undergoes the PE phase transition at 218°C. The temperature dependencies of the crystal lattice parameters near the phase transition point are shown in the right column in Fig.7.7. The asymmetry of the (222) line is preserved at the temperatures up to 235°C. The (400) X-ray diffraction line is symmetric within the whole temperature interval above 218°C. This fact shows that the diffuseness of the phase transition in this solid solution is connected with the extension of the domains of the rhombohedrally distorted FE phase to the high-temperature region above the transition point.

For the 10/80/20 PLLZT, located close to the boundary between the regions of the FE and AFE orderings in the Y−*T* diagram the observed picture differs from that described above. In this case both the (222) and the (400) X-ray diffraction lines are asymmetric in the temperature region above the transition point (these data are shown in the middle column in Fig.7.7). The temperature interval, within which the lines' asymmetry is preserved, is much wider than those for the 10/90/10 and 10/70/30 PLLZT. The main part of the samples' volume undergoes the phase transition at 180°C, whereas the complete symmetry of the X-ray diffraction patterns becomes obvious only at the temperatures above 260°C.

Analysis of the profiles of the X-ray diffraction lines, at the temperatures higher than the Curie point revealed that the samples have a complex phase composition. In the first approximation, it may be concluded that the PE matrix of the sample as a whole contains the two-phase regions which, in their turn, consist of the adjoining domains with the FE and AFE orderings.

### 7.2.3. *The nature of the Burns temperature in lead zirconate titanate from the point of view of domains with ferroelectric + antiferroelectric orderings*

The study of phase transitions in the PZT solid solutions shows that a high degree of diffuseness of the PE phase transition is directly related to the coexistence of domains of the FE and AFE phases at low temperatures. X-ray data, discussed above, provide an unequivocal evidence of existence of the two-phase (FE + AFE) domains in the PE matrix of solid solutions at the temperatures above the Curie temperature and directly connect theirs existence to the diffuseness of the phase transition. In our opinion, the first experiments indicating the existence of such domains were published in [111]. We mention these experiments only because the results obtained in this paper seemed to contradict the whole known physics of phase transitions in ferroelectrics and antiferroelectrics. Only much later the results proving the existence of these complex (FE + AFE) domains were obtained and these results put everything in its place – "the physics was saved".

Studies have been carried out both on polycrystalline and single-crystal samples with the $PbZr_{1-x}Ti_xO_3$ ($0.03 \leq x \leq 0.14$) compositions, located near the border separating regions with the FE and AFE types of dipole orderings in the diagram of phase states of PZT. Temperature dependencies $\varepsilon(T)$ demonstrate two anomalies. The first one is located at the temperature $T_C$, corresponding to the transition to the PE state. The second anomaly in the form of a small (barely visible) maximum of the $\varepsilon(T)$ dependence is located at the higher temperature $T_n$. The $T_n$ temperature approaches the $T_C$ temperature when the concentration of lead titanate increases and the value of the corresponding maximum reduces and it becomes indistinguishable on the background of changes in the $\varepsilon(T)$ dependence. However, this second anomaly is clearly manifested in the form of the kink in the $1/\varepsilon(T)$ dependence. The presence of such point of singularity in the $T_n$ dependence in these solid solutions was confirmed by precision X-ray studies of the crystal structure in [112, 113]. Furthermore, it is shown in [113] that difference between $T_n$ and $T_C$ may reach 100 °C for the PZT solid solutions with compositions approaching the lead zirconate.

The dependencies of polarization on the electric field intensity, $P(E)$, for the $PbZr_{0.9}Ti_{0.1}O_3$ solid solution at the temperatures above the Curie point in a temperature range including the point $T_n$ are shown in Fig.7.8. At $T > T_n$ the $P(E)$ dependence is linear, which is typical for conventional dielectrics. In the temperature range $T_C < T < T_n$ the system exhibits pronounced nonlinear properties. It should be noted that the temperature $T_n$ is the threshold above which the non-linearity in the system's behavior disappears.

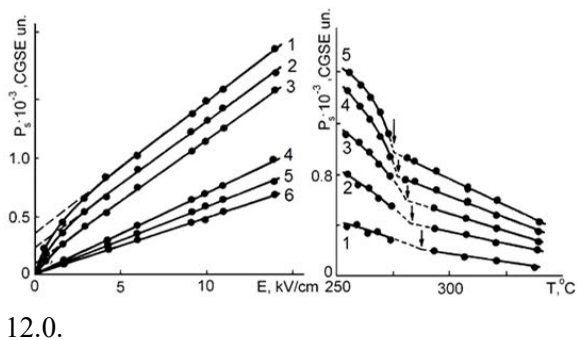

Fig. 7.8. Electric field (a) and temperature (b) dependencies of polarization of the $PbZr_{0.90}Ti_{0.10}O_3$ solid solution in paraelectric phase [111]. Arrows mark the location of the $T_n$ temperature on the curves corresponding to measurements at different intensities of electric field.
a: $T$ °C: 1 – 255; 2 – 260; 3 – 269; 4 – 290; 5 – 310; 6 – 336.
b: $E$, kV/cm: 1 – 2.0; 2 – 6.0; 3 – 7.5; 4 – 10.0; 5 – 12.0.

As one can see in Fig.7.8b, the electric field induces the polar phase in the sample's volume in the temperature interval $T_C < T < T_n$. This polar phase disappears completely when the temperature approaches $T_n$. The induction of the FE phase should be no surprise. All PZT-based solid solutions considered in [111] are in the FE state at temperatures below the Curie point. Their non-linear properties above the Curie point can be attributed to prolongation of the interval of existence of the remnants of the FE phase into the high-temperature region (this is always done in such cases). Quite different phenomenon is unusual, namely, the suppression of the polar state by an electric field. The figure clearly shows that the temperature of the destruction of polar properties, $T_n$, decreases with an increasing field strength. From the point of view of classical and well-established concepts of the physics of phase transitions one faces a paradox here.

This paradox is resolved easily; when one takes into account the phenomenon of the two-phase (FE+AFE) nucleation in the PE state in the temperature interval above the Curie point for the PZT solid solutions with small titanium content. The small difference in the energies of the FE and AFE states is just the case in these solid solutions.

The presence of the polar component of the two-phase (FE+AFE) nucleus also ensures the existence of polar properties of solid solutions in the temperature interval $T_C < T < T_n$ observed in experiments. As far as the reduction of the disordering temperature within the two-phase domain in an electric field is concerned, now everything is in order. Changes of the temperature at which the phase transition in an external electric field $E$ takes place can be determined from the Clapeyron-Clausius relation. This relation applied to the crystals with electric-dipole ordering has the following form [114]:

$$\frac{dT_c}{dE} = -\frac{T_c}{\Delta Q}(\Delta P_s + E \cdot \Delta \varepsilon). \tag{7.5}$$

In this expression $\Delta P_s$ and $\Delta \varepsilon$ are the changes of the spontaneous polarization and the dielectric constant caused by the increase of the temperature up to the disordering temperature; $\Delta Q$ is the latent heat of the transition. In the ferroelectrics the derivative $dT_C/dE$ is positive due to the large negative value of $\Delta P_s$. In the antiferroelectrics $\Delta P_s = 0$, and the derivative $dT_C/dE$ is negative as a result of the positive value of $\Delta \varepsilon$. This means that the external electric field leads to a lowering of the temperature of phase transition between the AFE and PE phases. For example, just this behavior is observed in the lead zirconate. Thus just the AFE component of the two-phase (FE+AFE) domains may only lead to a reduction in the value of $T_n$ when an external DC electric field is applied to the sample.

Thus, the manifestation of both the polarity of the PZT solid solutions at the temperatures above the Curie point and the suppression of polar state (the decrease of $T_n$) by a DC electric field can be observed simultaneously if only the two-phase (FE+AFE) domains are present in the paraelectric matrix of solid solution in the temperature interval $T_C < T < T_n$.

## 8. Relaxation dynamics of interphase (FE-AFE) domain walls

### 8.1. Model consideration

The influence of the DC electric field on a system, in which the free energies of the FE and AFE states slightly differ and the domains of these phases coexist, has been considered in Chapter 4. It has been shown that under the action of the field the state of the substance is changed due to the displacement of the interdomain walls (IDW) separating domains of the AFE and FE phases. The internal state within these domains is being unchanged. Such a motion of the IDW is inertial and is accompanied by the relaxation processes. The relaxation dynamics of the IDW should manifest itself most vividly while investigating the said substances in the AC electric field.

The main properties of the IDW dynamics under the AC electric field action can be obtained analysing the forces acting on the IDW. The basic equation derived from the condition of the balance of forces affecting the IDW, has the form:

$$P_{E(t)} + R(u) + P(u,t) = 0, \tag{8.1}$$

where $u = u(t)$ is the time-dependent displacement of the IDW, $E(t)$ is the electric field intensity; $P_{E(t)}$ is the pressure acting on the IDW and associated with the expansion of the FE domain volume in an external field. The forces which counteract the displacement of the IDW under the effect of the field can be presented as a sum of two terms: here $R(u)$ is the pressure of the forces caused by the interaction of the IDW with immobile defects of the crystal structure, and $P(u,t)$ is the pressure due to the so-called aftereffect forces. The latter forces may be caused by different factors: the interaction of the IDW with the mobile defects of the crystal lattice, as well as the nonzero duration of the thermal processes connected with the phase transitions in those regions of the crystal through which the IDW passes.

Eq. (8.1) describes the equilibrium of the IDW under the action of different forces. While considering this relation as a motion equation (Newton equation), one can easily see that here the terms $m_w \ddot{u}(t)$ ($m_w$ is the effective mass of the IDW) and $\gamma m_w \dot{u}(t)$ are not taken into account. These terms may not be considered in the analysis of motion if the effective mass $m_w$ is small and if the AC electric field frequency is essentially lower than the frequency of the IDW oscillations in the potential well created by the immobile defects of the crystal lattice.

The effective pressure exerted by an external electric field on the IDW can be presented as:

$$P_{E(t)} = E(t) \cdot P_s. \tag{8.2}$$

For simplicity we consider a uniaxial ferroelectric, and the direction of the external field coinciding with the polar axis.

The pressure affecting the IDW due to the force of the static defects can be presented by the first term of the power series expansion in displacements (i.e. by a quasi-elastic force):

$$R(u) = -\alpha u, \quad (\alpha > 0). \tag{8.3}$$

The pressure $P(u,t)$ is defined by the whole of the sample's history. Therefore it must depend on the values of $u$ at any time $t'$ preceding the present measurement time $t$ ($t' < t$). Analytically such dependence can be chosen at different degree of complexity. Here we choose it in the simplest form. Proceeding from the most general consideration, $P(u, t)$ can be expressed as

$$P(u,t) = -w \int_0^t F[u(t), u(t')] g(t-t') dt', \tag{8.4}$$

where $w$ is the width of the IDW, $F[u(t), u(t')] = F[u(t) - u(t')]$, and $g(t-t')$ is the aftereffect function describing the contribution of the relaxation effects existing at the time $t'$ ($0 < t' < t$), to the resultant relaxation at the time moment $t$. For a system with just one relaxation time the function $g(t - t')$ has a simple form:

$$g(t-t') = \frac{1}{\tau} \exp\left(\frac{t-t'}{\tau}\right). \tag{8.5}$$

Further, by analogy with (8.3), for the small displacements of the IDW the function $F[u(t) - u(t')]$ can be presented in a linearized form:

$$F[u(t) - u(t')] = -k[u(t) - u(t')]. \tag{8.6}$$

Thus, the expression which describes the pressure of the aftereffect acquires the form:

$$P(u,t) = kw \int_0^t [u(t) - u(t')] \cdot \exp[-(t-t')/\tau] \frac{dt'}{\tau}. \tag{8.7}$$

The motion equation for the IDW has the following form in the above-mentioned approximation:

$$-\alpha u(t) + kw\int_0^t dt'\left[u(t) - u(t')\right]\exp\left(-(t-t')/\tau\right) + P_s E_0 \exp(i\omega t) = 0$$

or  (8.8)

$$[1+\eta G(t)]u(t) - \eta \exp\left(-\frac{t}{\tau}\right)\int_0^t u(t')\exp(t'/\tau)\frac{dt'}{\tau} = u_0 \exp(i\omega t),$$

where $\eta = kw/\alpha$, $u_0 = P_s E_0/\alpha$ is the IDW displacement under the action of a DC electric field with the intensity $E_0$ and $G(t) = (1-\exp(-t/\tau))$ is the time function. This equation should be solved either by the method of the successive approximations (in the first approximation) or by transforming it into a differential equation which can be easily solved in the stationary case $t \to \infty$ ($G(t) = 1$). Representing the steady-state solution in the form $u(t) = \tilde{u}\exp(i\omega t)$ [115], one can obtain the amplitude $\tilde{u}$ in the form:

$$\tilde{u} = u_0 \frac{(1+i\omega\tau)}{[1+i\omega\tau(1+G(t)\eta)]} \to u_0 \frac{(1+i\omega\tau)}{[1+i\omega\tau(1+\eta)]}. \tag{8.9}$$

Since the IDW displacements under the action of the electric field lead to a change of the dipole moment in the volume of the substance involved by this displacement, it is easy to determine the dielectric susceptibility $\chi$ connected with this process:

$$\chi(t) = P_s S u(t) / E(t), \tag{8.10}$$

where $S$ is the area of the oscillating IDW. Let us put $\chi_0 = P_s^2 S/\alpha$ to be the static susceptibility, then from (8.10) and (8.9) one shall obtain:

$$\chi = \chi' - i\chi'' = \chi_0 \frac{1+\omega^2\tau^2(1+\eta) - i\omega\tau\eta}{1+\omega^2\tau^2(1+\eta)^2}. \tag{8.11}$$

From here it follows that:

$$\chi'(\omega,T) = \chi_0 \frac{1+\omega^2\tau^2(1+\eta)}{1+\omega^2\tau^2(1+\eta)^2} \equiv \chi_0(T)F_1(\omega,T), \tag{8.12a}$$

$$\chi''(\omega,T) = \chi_0 \frac{\omega\tau\eta}{1+\omega^2\tau^2(1+\eta)^2} \equiv \chi_0(T)F_2(\omega,T). \tag{8.12b}$$

As noted earlier, the IDW's relaxation dynamics is influenced by the aftereffect of the first-order PT between the FE and AFE states, happening in that part of the sample where the IDW motion takes place. In this case the said states are separated by a potential barrier, and the thermal activation processes take place. Therefore it is natural to assume that the temperature dependence of the relaxation time has the form:

$$\tau = \tau_0 \exp(\Delta/kT). \tag{8.13}$$

Here $\Delta$ is the activation energy, $\tau_0$ is the reciprocal value of the thermal activation frequency (its order of magnitude equals $10^{-11} - 10^{-13}$ s).

As shown in Chapter 4, the coexistence of the domains of the FE and AFE phases is provided by the equality of their thermodynamic potentials within a wide interval of the field intensity. The change in the intensity of the external field results in the displacement of the interphase boundary, the energy of each of the coexisting phases remains unchanged ($\Delta = 0$). However, the presence of the effect of the "IDW lag", caused by the aftereffect forces (8.4), leads to changes in this simple picture. Now the condition of the field compensation inside the domains is not fulfilled, and $\Delta \neq 0$ (though being a small value). Therefore, under the condition of week fields one may put $\omega\tau \ll 1$.

Under this condition the functions $F_1(\omega,T)$ and $F_2(\omega,T)$ acquire the following forms:

$$F_1(\omega,T) \approx 1-\omega^2\tau^2\eta(1+\eta) \approx 1-2\omega^2\tau_0^2\eta(1+\eta)\Delta/kT, \qquad (8.14a)$$

$$F_2(\omega,T) \approx \eta\omega\tau \approx \eta\omega\tau_0(1+\Delta/kT). \qquad (8.14b)$$

The dependence $\chi_0(T)$, as well as the typical dependencies $F_1(T)$ and $F_2(T)$ for different frequencies near the temperature of the $\chi_0(T)$ maximum are given in Fig.8.1 [116].

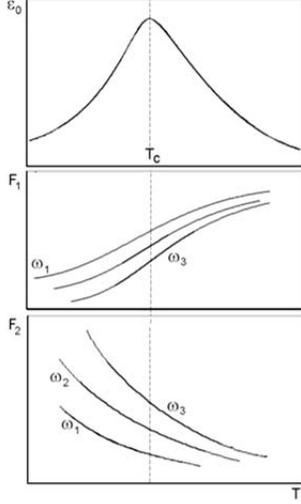

Fig.8.1. Schematic dependences of $F_1$ and $F_2$ on temperature near the maximum of $\varepsilon(T)$ for different frequencies $\omega_1 < \omega_2 < \omega_3$ [116].

Taking into account the fact that the expressions for $\chi'(T)$ and $\chi''(T)$ are the product of $\chi_0(T)$ and the functions $F_1(T)$ and $F_2(T)$, respectively, one can make the following conclusions:

- the $\varepsilon'(T)$ and $\varepsilon''(T)$ dependencies are shifted towards the higher temperatures with the increase of the measuring field frequency;
- the maximum of the real component of the permittivity, $\varepsilon'(T_m') \equiv \varepsilon_m'$, decreases with the increase of the frequency, whereas the maximum of the imaginary component of the permittivity, $\varepsilon''(T_m'') \equiv \varepsilon_m''$, increases; the increase rate of the $\varepsilon_m''$ component is higher in comparison with the decrease rate of the $\varepsilon_m'$ component.

- $T_m'$ is always higher than $T_m''$.

One can obtain expressions describing the shift of the characteristic temperatures $T_m'$ and $T_m''$, caused by the change of the field frequency. In particular, taking into account (8.12) and (8.14) and taking a derivative with respect to (1/T) and equating it to zero, the following expression for $T_m'$ can be obtained:

$$\left(\frac{1}{T_m'}\right) = -\frac{k}{\Delta}\ln\omega - \frac{k}{\Delta}\ln\tau_0 + \frac{k}{\Delta}V(1/T_m'). \qquad (8.15)$$

Here $V(1/T)$ is a slow varying function. The introduction of the effective temperature $T_f$, i.e. the linearization of (8.15) using the Pade approximations method [117], yields another form of this expression:

$$\frac{1}{T_m'-T_f} = -\frac{k}{\Delta}\ln\omega - \frac{k}{\Delta}\ln\tau_0, \qquad (8.16)$$

which can be also written in the form of the well-known Vogel-Fulcher relation:

$$\omega = \left(\frac{1}{\tau_0}\right)\exp\left[-\frac{\Delta}{k(T_m'-T_f)}\right]. \qquad (8.16a)$$

The expression for $T_m''(\omega)$ analogous to (8.16) and (8.16a) are obtained in the same way.

A mathematical expression which relates the location of the $\varepsilon'(T)$ maximum with the electric field amplitude, can be easily derived from (8.16). Consider a DC electric field first. When the electric field is applied to the sample, eq. (8.13) acquires the form:

$$\tau = \tau_0 \exp\left[(\Delta-\alpha P_s E)/kT\right], \qquad (8.17)$$

where the second term in the exponent describes the change of the minimum of the thermodynamic potential of the FE phase, caused by the field. To obtain the dependence $T_m^{/}(E)$ the substitution $\Delta \rightarrow \Delta - A \cdot E$ has to be made in eq. (8.16). It is easily seen that the increase of the field intensity leads to a near linear decrease of $T_m^{/}$.

$$\frac{1}{T_m^{/} - T_f} = -\frac{k}{\Delta - A \cdot E} \ln \omega - \frac{k}{\Delta - A \cdot E} \ln \tau_0. \tag{8.18}$$

Expression (8.18) can be used for consideration of the AC field influence on the location of $T_m^{/}$. The applicability of (8.17) for the description of the IDW motion in the AC electric field in the scope of this model is justified by the fact that the relaxation process of the order parameters to the steady state under the influence of the field and the change of the potential barrier height can be considered simultaneously.

The relaxation dynamics of the IDW is determined by the term $P(u,t)$ in (8.1). Let us remind that this term may be caused by different factors: the interaction of the IDW with the mobile defects of the crystal lattice, as well as the nonzero duration of the thermal processes connected with the phase transitions in those regions of the crystal through which the IDW passes. The characteristic time of the above-mentioned processes is rather long. At the same time the processes of establishment of the order parameter during the structural phase transitions (to which the FE and AFE phase transitions belong) are the fast ones. Characteristic times of these processes are several orders of magnitude shorter. Based on the assumptions used to write eq. (8.1), one can consider the establishment of the potential barrier value, $\Delta$, between two minima corresponding to the FE and AFE phases in eq. (8.13) as instantaneous when the AC electric field is applied to the sample (that is the value $\Delta$ follows the instantaneous value of the field). That is why the dependence $T_m'$ on the AC field amplitude will be the same as the dependence on the DC field intensity, namely, increase of the field amplitude will lead to a near linear decrease of $T_m^{/}$.

We have considered the model of oscillation for the IDW separating the domains of the FE and AFE phases which are in equilibrium. Though the studied version of the model is the most simplified one, the obtained results allow us to predict a wide range of experimental phenomena.

One of the model simplifications consists in the choice of the relaxation function $g(t-t')$ in the form (8.5). This choice corresponds to the case with a single relaxation time. Actually, while considering a real physical system, one should take into account the existence of a spectrum of relaxation times. However, even such slight complication of the model will lead to the certain mathematical (but nonphysical) problems, while investigating eq. (8.8). For instance, solutions of equations analogous to (8.8) were studied in [118, 119, 120] and purely mathematical difficulties were observed. The numerical methods were used to solve these equations, and the physical nature of the obtained solutions became not evident. Moreover, the authors had made simplifying assumptions at the final stage of the solution process. Therefore, we suppose that at the present stage of studies of the relaxation dynamics of the IDW it is sufficient to use the single-time relaxation function $g(t-t')$. The complication of the form of the function $g(t-t')$ may be accomplished based on a subsequent interpretation and comparison of experimental results obtained on different samples (for instance, crystalline, ceramics and film samples) of the same substance.

An important result of the approach presented in the present chapter consists in the fact that the considered model allows us to describe the temperature dependencies of the real and imaginary parts of the permittivity and their evolution under the varying frequency and amplitude of the AC electric field. Equations (8.12) and (8.14) correspond to an isolated IDW. To obtain the dependencies $\varepsilon^{/}(T)$ and $\varepsilon^{//}(T)$ for the whole investigated sample containing a large number of IDWs one has to take into account changes in the quantity and in the area of such interphase boundaries caused by the temperature changes in the sample. At $T<T_0$ the number and dimensions of the IDW weakly depend on temperature; therefore the permittivity of the sample is described by the expressions analogous to those for an isolated boundary. The most essential change in the IDW number and IDW area takes place at the temperatures close to $T_0$ and higher. Moreover, the complex FE+AFE domains exist in the PE matrix of the substance at the temperatures higher than $T_0$ (Chapter 7). That is, at $T>T_0$ the IDWs exist, and theirs oscillations contribute to the permittivity of the substance. However, as the temperature rises, the number and volume of the FE+AFE domains and, consequently, the number and area of IDWs diminish. There exists such a temperature $T_t$ that for all the temperatures $T>T_t$, no FE+AFE domains are present in the PE matrix of the substance, and, consequently, the IDWs are also absent. Thus, there is no contribution of the IDW oscillations into the permittivity. Therefore, at $T>T_t$ the functions $\varepsilon'(T)$ and $\varepsilon''(T)$ do not depend on the AC field frequency and intensity, and $\varepsilon'(T)$ coincides with the function $\varepsilon_0(T)$. Such a situation is shown in Fig.8.2.

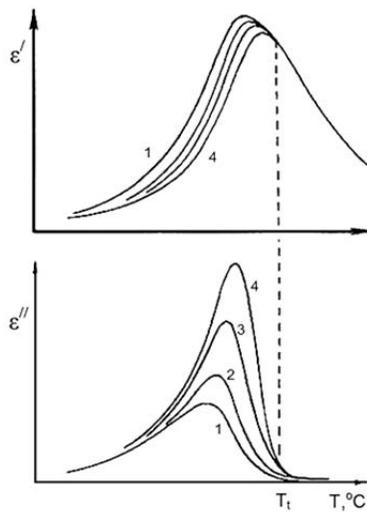

Fig.8.2. Schematic dependences of $\varepsilon'(T)$ and $\varepsilon''(T)$ on temperature Near the phase transition into the PE state for different frequencies $\omega_1<\omega_2<\omega_3<\omega_4$ [116].

## 8.2. Experimental results

Now let us discuss the experimental results on dielectric properties of the substances with a weak difference in the free energies of the FE and AFE states. As an example, we consider the 10/100-Y/Y PLLZT and the 6/100-Y/Y PLZT systems of solid solutions. Phase diagrams of these solid solutions are shown in Fig.4.4e, and Fig.4.4k (see also Fig. 4.9d), respectively.

### 8.2.1. *Lead lithium lanthanum zirconium titanium solid solutions*

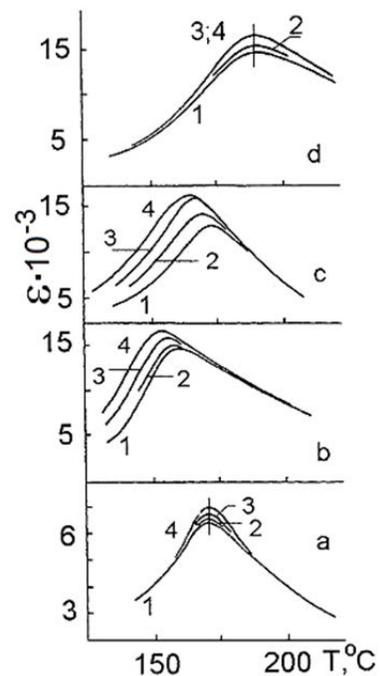

The dielectric response obtained for the PLLZT solid solutions located in the three different regions of the Y-T phase diagram subjected to the action of the AC field with a frequency of 1 kHz and the field amplitude varying between 1 and 150 V/mm is shown in Fig.8.3 [121]. First of all, it should be noted that for the said values of the electric field intensity the dependencies $P(E)$ are linear. The effects bound up either with the reorientation of domains of the FE phase or with the induction of the FE phase manifest themselves in the fields which noticeably exceed the used ones.

Fig.8.3. Temperature dependences of $\varepsilon'$ for 10/100-Y/Y PLLZT [121].
Ti-content Y, %: a – 15; b – 20; c – 23; d – 30;
$E_\approx$, V/mm : 1 – 1; 2 – 50; 3 – 100; 4 – 150.

For the PLLZT with the composition 10/90/10 belonging to the AFE region of the Y-T diagram all $\varepsilon'(T)$ dependencies measured at varying AC field practically fall on the same curve. For the 10/15/85 PLLZT the $\varepsilon'(T)$ dependencies for four values of the AC field are presented in Fig.8.3a. As is seen, the $T'_m$ temperatures coincide, but the $\varepsilon'_m$ values slightly increase with the increase of the field. For the 10/30/70 PLLZT solid solution belonging to the FE region of the Y-T diagram the behavior of $\varepsilon'(T)$ is similar to that for the 10/15/85 PLLZT: $T'_m$ does not depend on the AC field amplitude, and the dielectric constant somewhat increases in the vicinity of $T'_m$ (Fig.8.3d).

A different situation is observed in the solid solutions with compositions located near the boundary between the stability regions of the FE and AFE states in the Y-T diagram that is for the 10/20/80 and 10/23/77 PLLZT.

In the interval of the field amplitude values up to 20 V/mm the observed $\varepsilon'(T)$ behavior is analogous to that described above: $T'_m$ does not depend on the field amplitude and $\varepsilon'(T)$ measured in the fields with different intensities are practically the same. For the AC field amplitudes higher than 20 V/mm the behavior of $\varepsilon'(T)$ changes.

The $\varepsilon'(T)$ dependencies for the 10/20/80 and 10/23/77 PLLZT for the AC field amplitudes higher than 20 V/mm are presented in Fig.8.3b and 8.3c. As is obvious, in these solid solutions both $\varepsilon'_m$ and $T'_m$ noticeably depend on the AC signal intensity. The temperature $T'_m$ diminishes and the value of $\varepsilon'_m$ rises when the AC field amplitude increases. The dielectric constant increases at all temperatures $T < T'_m$ with an increase of the AC field amplitude, whereas at the temperatures $T > T'_m$ the real part of the dielectric constant, $\varepsilon'$, is practically independent on the AC field.

The dispersion of the dielectric constant was investigated within the frequency range from $10^2$ to $10^5$ Hz. For the 10/90/10 and 10/70/30 PLLZT solid solutions which compositions are located within the regions of the uniform AFE and FE states in the Y-T phase diagram, respectively, the real and imaginary parts of the dielectric constant, $\varepsilon'(T)$ and $\varepsilon''(T)$ ($T'_m$ and $T''_m$) do not depend on frequency. However, for the 10/80/20 and 10/77/23 PLLZT the fundamental changes in the behaviour of $\varepsilon'(T)$ have been observed (Fig.8.4) [121, 122]. The $\varepsilon'(T)$ dependencies are being displaced towards the higher temperatures and the value of $\varepsilon'_m$ noticeably decreases with the growth of the field frequency. The $\varepsilon''(T)$ dependencies also essentially change: the $\varepsilon''(T)$ curves are shifted towards the higher temperatures and the value of $\varepsilon''_m$ increases. For example, for the 10/77/23 PLLZT the Vogel-Fulcher relation is fulfilled in the following form: $\omega = (1/\tau_0)\exp\left[-\Delta/k(T_m - T_f)\right]$, with the effective temperature $T_f = (425 \pm 2)$ K (see also [83] for more details).

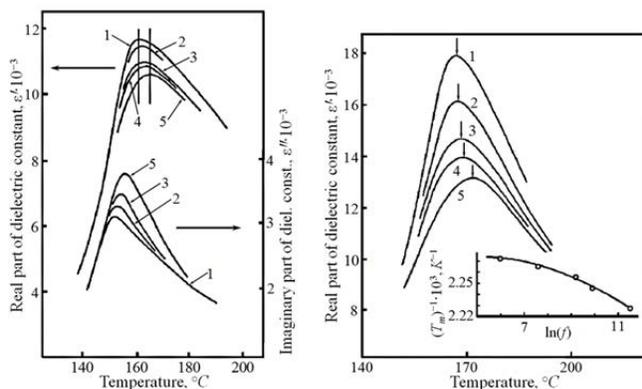

Fig.8.4. Temperature dependencies $\varepsilon'(T)$ and $\varepsilon''(T)$ for the 10/80/20 PLLZT (left) and $\varepsilon'(T)$ for the 10/77/23 PLLZT (right) [121, 122].
Frequency kHZ: 1 – 0.4; 2 – 2.0; 3 – 10.0; 4 – 20.0; 5 – 100.0.

The investigations performed with a DC bias applied to samples, were carried out on the PLLZT solid solution with 15 % of Ti. The E-T phase diagram for the 10/85/15 PLLZT is presented in Fig.8.5a. Only one AFE-PE phase transition takes place on the line $T_C(E)$ for the low-intensity DC bias when the temperature changes (in accordance with Y-T phase diagram). The FE-PE transition is observed for the DC bias values $E > E_{tr} \cong 15$ kV/cm.

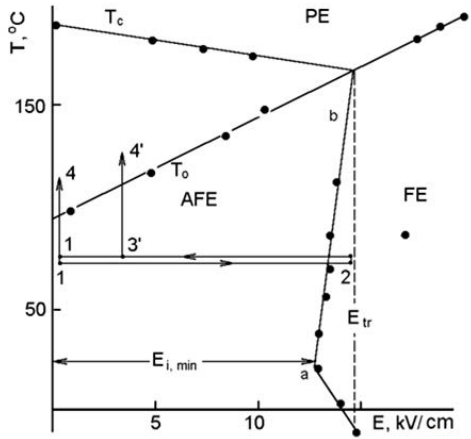

Fig.8.5. Phase $E$-$T$ diagram for the 10/85/15 PLLZT solid solution [121]. Lines with arrows (1 – 2, 2 – 1, 1 – 4, 3' – 4') demonstrate changes of thermodynamic parameters (temperature, electric field intensity) during the determination of the boundaries of the phase states.

The temperature dependencies of the real part of the dielectric constant $\varepsilon'(T)$ for different values of the DC bias and amplitude of the AC field are presented in Fig.8.6 For a zero DC bias (Fig.8.6a) the location of the maximum of the $\varepsilon'(T)$ dependence is independent of the AC field amplitude, and the value of $\varepsilon'_m$ somewhat increases with the rise of the AC field. Similar behavior is observed for a DC bias of intensity $E = 22$ kV/cm (Fig.8.6c). The $\varepsilon'(T)$ dependencies obtained for different AC field amplitudes are practically coincident in the whole temperature range studied.

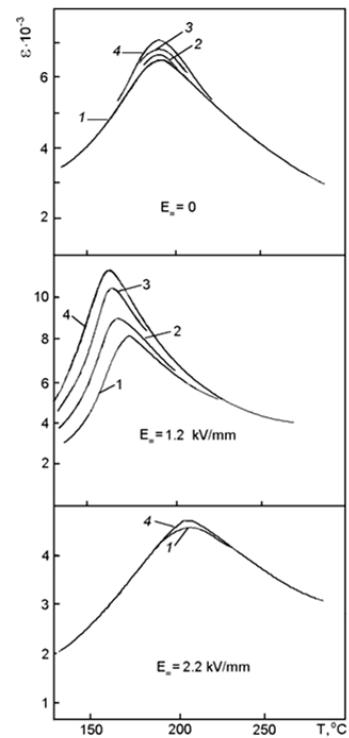

Fig.8.6. Dependencies $\varepsilon'(T)$ for the 10/85/15 PLLZT solid solution at different intensities of the AC and DC field [121, 122].
AC filed amplitude (V/mm): 1 – 1; 2 – 50; 3 – 100; 4 – 150.

A different dielectric behavior is observed for the DC bias $E = 12$ kV/cm (Fig.8.6b (b)). The maximum value of the dielectric constant $\varepsilon'_m$ increases and the temperature $T'_m$ decreases when the AC field amplitude rises. Simultaneously, a large increase of the dielectric constant takes place for the temperatures $T < T_m$.

As seen from Fig.8.3, Fig.8.4 and Fig.8.6, peculiarities in the behaviour of permittivity caused by the increase of the AC field amplitude are observed only in the solid solutions with compositions located near the boundary separating the regions of the FE and AFE states in the Y-$T$ or the E-$T$ phase diagrams. These solid solutions have close values of the free energies for the FE and AFE states, and this leads to the coexistence of domains with of the FE and AFE types of ordering in the volume of the samples. In this case one should take into consideration the contribution of the interphase boundary oscillations. As seen from the present results, the deviations in the behaviour of permittivity from the classic behavior take place only in the case when the contribution of the said boundaries has the form predicted by the model for the FE-AFE phase transformation. At the same time, it is not essential which of the external factors – the change of the DC bias intensity or the change of the solid solution composition – results in the appearance of the equilibrium two-phase state.

The difference in the behavior of dielectric constant under changing amplitude of the measuring AC field should be specially considered for the case of weak and strong fields (with the amplitudes lower and higher than 20 V/mm, respectively). In our opinion, the pinning of IDWs on defects of the crystal structure manifests itself in weak fields (see [121, 122] for more details).

**8.2.2.** *Lead lanthanum zirconate titanate solid solutions*

The dielectric properties of the 6/100-Y/Y PLZT solid solutions and their dependence on the location of the substance in the Y-*T* phase diagram (Fig.4.4k, 4.9d),) are identical to those of the PLLZT solid solutions. For the 6/90/10 and 6/52/48 solid solutions located far from the AFE-FE boundary region of the Y-*T* diagram (and, thus, belonging to the regions of the homogeneous AFE and the homogeneous FE ordering, respectively) the curves ε′(*T*) do not depend on the frequency and the intensity of the AC field. For the 6/72/28 solid solution located near the above-said boundary there is a considerable dispersion of both components of the dielectric constant during the transition into the paraelectric state (Fig.8.7), the Vogel-Fulcher law is fulfilled. Such behavior is called the relaxor behavior in the literature.

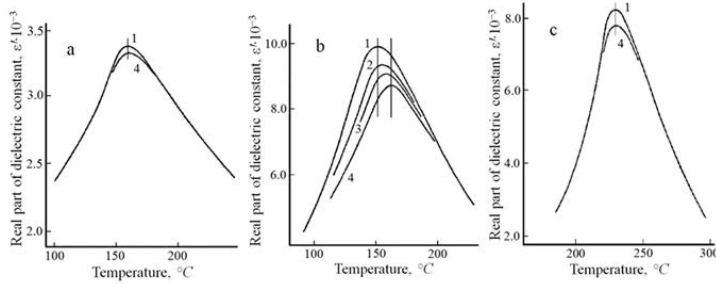

Fig.8.7. Dependence $\varepsilon'(T)$ for the 6/100-Y/Y PLZT solid solutions [121]. Ti-content Y,%: a − 10; b − 28; c − 48. Frequency, kHz: 1 − 1; 2 − 10; 3 − 20; 4 − 100.

The presence of a noticeable dispersion of the permittivity in the vicinity of the $T_m$ temperature and the fulfilment of the Vogel-Fulcher law were already discussed for the 6/65/35 solid solution (located not far from the boundary region in the Y-*T* diagram of the 6/100-Y/Y PLZT system which is characterized by the equal values of the free energy of the FE and AFE states) [43, 44]. However, the authors of the mentioned papers investigated only one solid solution and therefore could not connect the observed dielectric properties and their dispersion with the position of a given solid solution in the phase diagram. Naturally, they could not interpret the observed phenomenon. As seen from the presented experimental results, those properties of the permittivity that serve as a base for attributing the PLZT solid solutions to the class of relaxor FE are caused by the coexistence of the FE and AFE phase domains and by the contribution of the interphase boundaries separating these domains into the value of permittivity (in detail these problems are discussed in [32,], [116] and in Section 8.1 of the present paper). In the solid solutions where the FE-AFE interphase boundaries are absent, no dispersion of the permittivity (or the relaxor behavior) is observed.

## 9. Ferroelectric-antiferroelectric phase transitions and the notion of the "relaxor ferroelectrics"

Lead magnesium niobate $PbMg_{1/3}Nb_{2/3}O_3$ (PMN) is being investigated since the mid-fifties; however, the physical nature of the phenomena observed in this compound is still unclear. We shall not dwell on different models used for the interpretation of behavior of this substance that appeared during such a long period. All these models are based on the idea of the "chemically inhomogeneous substance", the inhomogeneity being predetermined by the substance manufacturing process. The mentioned concept of the "inhomogeneous substance" underlies the model of the relaxor ferroelectrics [123] often used at present while interpreting the properties of the complex-composition oxides with the perovskite crystal structure.

The whole set of phenomena in relaxor FE have not been explained up to now. One of the main results achieved in the frames of the relaxor FE approach is the expression which describes the dispersion of the permittivity in the vicinity of the temperature corresponding to the maximum of the ε(*T*) dependence. As shown in [32], the explanations of physical phenomena observed in such substances as $(Pb,Li_{1/2}La_{1/2})(Zr,Ti)O_3$ (PLLZT), $(Pb,La)(Zr,Ti)O_3$ (PLZT), $Pb(Mg_{1/3}Nb_{2/3})O_3$ (PMN), and $Pb(In_{1/2}Nb_{1/2})O_3$ (PIN) self-consistently follows from the ideas of the FE-AFE transitions. The applicability of the developed model concept to the analysis of behavior of real substances is demonstrated in [50] while analyzing the phase diagrams of the PZT-based solid solutions and the experimental data on the behavior of the said substances under the action of external effects (temperature, hydrostatic pressure, electric field or any combination of these factors). The goal of the present chapter is to demonstrate the suitability of the approach based on idea of the FE-AFE phase transition for the explanation of the whole complex of peculiarities in behavior of the PMN, the PIN, and related substances. The idea to apply the FE-AFE phase transition to the explanation of PMN behavior under external effects was first presented in [124].

### 9.1. Phase states and phase transitions in the lead magnesium niobate

PMN is a typical oxide with the $ABO_3$ perovskite crystal structure. It was demonstrated in Chapter 2 of this review that the energies of the FE and AFE dipole-ordered states have close values in oxides with such structure. In subsequent chapters 3 – 8, we have pointed out the peculiar features in behavior of perovskite materials in this case (with close values of energy of the FE and AFE states) and have also shown that the behavior of such materials can be very different from the one of classical FE and AFE.

We will start this section by showing that the same situation takes place in the PMN that is in this substance the energies of the FE and AFE states differ slightly. Then, based on the available experimental data, we will show that this proximity of values of the energies of the FE and AFE states determine all the features of the PMN behavior under external influences.

Density functional studies using plane wave basis sets and pseudopotentials and all-electron linear augmented plane wave (LAPW) have been carries out in [125] Calculations involving several ordered supercells of the PMN were used to explore the energy landscape to understand the relative energies of possible FE and AFE states for ordered supercells with 1:2 ordering along the [111] and [001] directions. These are idealized supercells, whereas a real PMN is disordered. However, there is the evidence for small ordered regions (for example, see [126, 127]). In the process of calculations the Pb ($5d$, $6s$, and $6p$), Mg ($2s$, $2p$, and $3d$), Nb ($4s$, $4p$, and $4d$) and O ($2s$ and $2p$) were treated as valence. The structures of the ordered supercells used in [125] are shown in Fig.9.1.

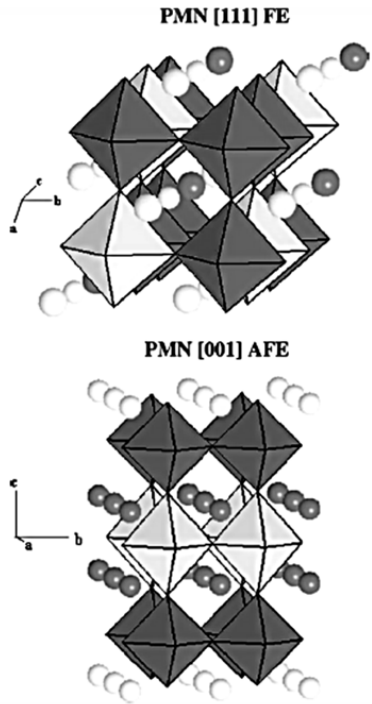

Fig.9.1. The polyhedral representation of supercells for the [111] FE and [001] AFE structures [125]. The coordination polyhedra around the Mg (light) and Nb (dark) are shown. The Pb atoms are shown as spheres (distinct shade and sizes represent different Wyckoff positions).

The space groups $P3m1$, $P\bar{3}m1$, $Cm$, and $P4/mmm$ correspond to the 1:2 chemically ordered supercells of PMN. Results of the energy calculation for these structures are presented in Table 9.1 (Table II in [130]).

| Ordering direction | Space group | $\Delta E$ LAPW | $\Delta E$ ABINIT | a | b | c |
|---|---|---|---|---|---|---|
| [111] | $P3m1$ | 0 | 0 | $(-a',a',0)$ | $(a',0,-a')$ | $(-a',-a',-a')$ |
| [111] | $P\bar{3}m1$ | 12.6 | 11.7 | $(-a',a',0)$ | $(a',0,-a')$ | $(-a',-a',-a')$ |
| [111] | $Cm$ (E) |  | −1.11 | $(-a',a',0)$ | $(a',0,-a')$ | $(-a',-a',-a')$ |
| [111] | $Cm$ (R) |  | −6.67 | $(-a',a',0)$ | $(a',0,-a')$ | $(-a',-a',-a')$ |
| [111] | $C2$ |  | −9.36 |  |  |  |
| [001] | $P4/mmm$ | 22.4 | 19.2 | $(a'',0,0)$ | $(0,a'',0)$ | $(0,0,3.09a'')$ |
| [100] | $Pmm2$ | 20.3 | 18.4 | $(3a''',0,0)$ | $(0,a''',0)$ | $(0,0,1.01a''')$ |

Table 9.1. Calculated energies for different structures of 1:2 ordered PMN structure from [125].

As in may be seen, the nonpolar AFE ($P\bar{3}m1$) structure has a larger energy than the polar FE ($P3m1$) structure with an energy difference of 12 mRyd/15-atom, but lower energy than the $P4/mmm$ and $Pmm2$ polar FE structures. All energy differences, obtained in [125] are small and demonstrate the small difference in energies of the FE and AFE orderings in the PMN structure.

Density functional calculations using the linearized augmented plane-wave method have been performed in [128]. Structural relaxations were performed on 30 atom unit cells with *B*-cations arranged in 1:1 chemical ordering along the [111] direction. The magnitudes of the cation off-centerings with respect to centers of the O-cages are given in Fig.9.2.

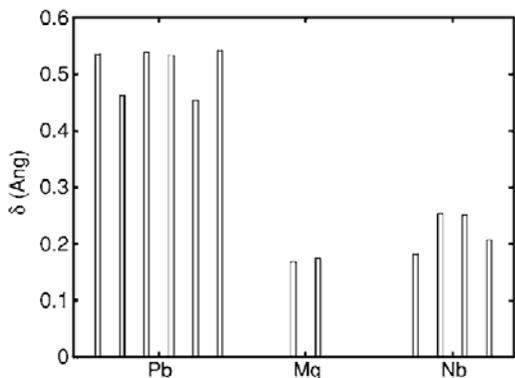

Fig.9.2. Off-centering distance with respect to the octahedral oxygen cages for all Pb, Mg, and Nb in the 30 atom supercell. Note that these are the magnitudes of the off-centerings. The actual displacements are not collinear [128].

As may be seen the largest displacements are for Pb, reflecting the *A*-site driven nature of the PMN. Later, in the end of this Section we will consider results of the neutron diffraction study of PMN, and it will be shown that the softening of optic phonons take place at *M* and *R* points on the boundary of the Brillouin zone in this substance. These phonons are linked with Pb displacements. Just the same phonons define the AFE properties of lead zirconate.

Authors of [129] used a 30-atom supercell in their calculations. The density functional theory calculations were done using the plane wave pseudopotential method. Perdew–Zunger local density approximation was used for the exchange–correlation functional. The aim of studies was (authors focused their attention on the AFE phase in the PMN) to determine stabilizing factors for the AFE ordering. A supercell based on 1:2 ordering model was used for calculations. This 30-atom supercell can be thought of as snapshots of small regions of the real disordered solution. The supercells used in [129] are shown in Fig. 9.3.

Fig. 9.3. Relaxed structures of supercell I (a) and II (b) [129]. They are both projections of supercells at experimental lattice constant on y–z plane. Arrows show the displacement of atoms from perfect perovskite positions. Pb atoms are 1/2 unit cell above the plane, and apical O atoms are omitted.

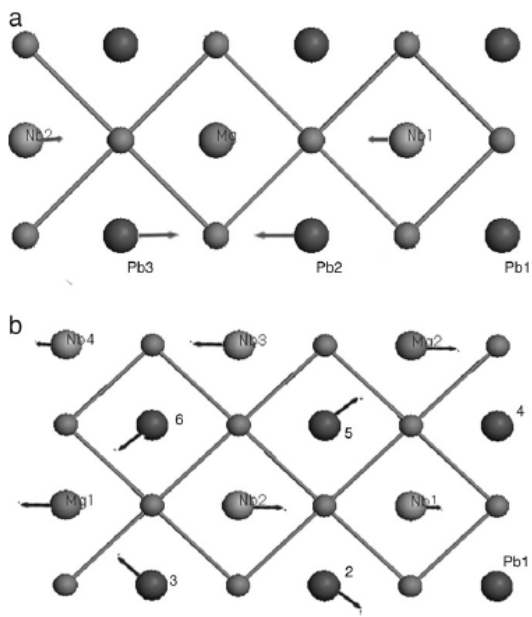

The pair distribution functions obtained from the relaxed structure of PMN is shown in Fig. 9.4. The pair distribution function of the supercell I fits better with the experimental one than that of the supercell II, because the compositional long-range order occurs in the real PMN and supercell I is ordered, while supercell II is disordered. As can be seen the results of calculations are in good agreement with neutron-scattering experiments [130, 131], this indicates that used supercells are sufficient for capturing the local structure of PMN.

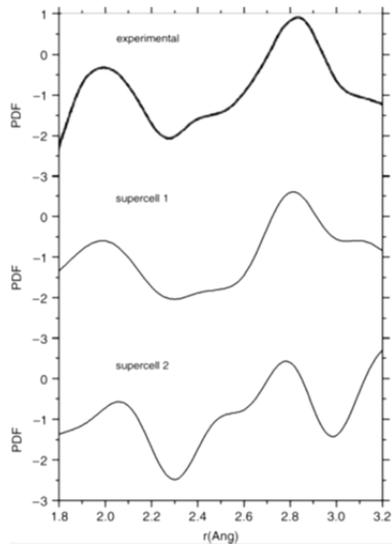

Fig.9.4. Comparison of PDFs obtained from the relaxed structures of two supercells [129] and from the experimental results [130, 131]. The PDF provides information about the interatomic distance in PMN. The first peak at about 2.0 Å is due to the B-site–O nearest neighbors, and the second peak at about 2.8 Å is due to Pb–O and O–O pairs.

The relaxed structures are shown in Fig.9.3. Arrows indicate the displacement directions of cations. The two relaxed supercells both are AFE, with anti-parallel polarization along [001]. The results of [129] are in good agreement with previous works. The two Nb atoms in supercell I are equivalent due to the symmetry along $z$-direction and the same takes place for $Pb_2$, $Pb_3$. The pairs of atoms $Pb_1$ and $Pb_4$, $Pb_2$ and $Pb_5$, $Pb_3$ and $Pb_6$, $Nb_1$ and $Nb_3$, $Nb_2$ and $Nb_4$, $Mg_1$ and $Mg_2$ in supercell II are also equivalent. The total energies of the supercells I and II are equal to 1762.7446 Ry and 1762.7588 Ry, respectively. As it was noted by authors of [134], the small energy difference between them indicates that these two local structures all possibly exist in real material. In order to avoid 'local minimal state' the authors of [129] also performed geometry optimization from a FE phase (mutually displacing $B$-site ions some 0.04 Å from their ideal positions along (001)). After such optimization of geometry the AFE phase was still realized. Authors of [134] have also examined the reason why the ions displaced in a manner of antiferroelectric phase as shown in Fig. 9.3.

Thus, results of the above-cited studies suggest that the PMN perovskite structure is characterized by a weak difference in the energy between possible FE and AFE states. We do not discuss the reasons for this since it was done in the articles considered above.

In short, it all boils down to the following. It was established by the first-principle calculations of the local PMN structure that both in ordered and disordered supercells the AFE ordering of ions may be present. The interactions between Pb–O and Pb–B-site could be the determining factor for an existence of such a phase. First, Pb and Nb ions tend to move towards the oxygen while the Mg ion tends to move away from the oxygen. Second, the repulsion between Pb–Nb is more intensive than that between Pb–Mg. The two interactions lead to anti-parallel displacement of Pb's in the AFE phase.

Further, the approach used in discussions of properties of the PLZT and PLLZT solid solutions (see chapters 3 – 8 of the present review) is rather hard for consideration of peculiarities in the behavior of PMN, inasmuch as for this substance the number of different types of phase diagrams available in the literature is very limited. Taking into account that the PMN is being studied since middle fifties such a situation is quite surprising.

We have managed to find only two types of the phase diagrams: the Y-T diagram and the T-E one for the $PbTiO_3$-PMN (PT-PMN) system of solid solutions. Therefore, in the course of our discussion they will be often compared with diagrams for other substances for which the nature and the properties of the phase transitions have been established reliably. First of all we shall analyse the phase diagram of the lead titanate in which titanium ions are being sequentially substituted by the ionic complex $(Mg_{1/3}Nb_{2/3})$.

However, before considering the influence of the PMN on the stability of phase states in lead titanate let us consider the changes in the phase states of other substances with perovskite structure taking place at the formation of the lead titanate based solid solutions. For lead-containing oxides with perovskite structure the substitution of the A-site ion by those with the smaller ionic radii increases the stability of the AFE state with respect to that of the FE state. The same picture is observed when the $B$-site ion is substituted by the ions with greater radii [24, 132]. The most well-known is the phase diagram for the solid solutions obtained by the substitution of titanium by zirconium, i.e. the diagram of the PZT solid solutions. Below the Curie point, when the content of zirconium reaches 95%, the FE ordering is superseded by the AFE one. The same picture is observed at the substitution of titanium by hafnium.

The "composition – temperature" phase diagrams (both the Y-T and the X-T) for different compounds are presented in Fig.9.5. The diagram in Fig. 9.5a describes the case when the titanium ion is substituted not by an individual ion but by the ($Zr_{0.6}Sn_{0.4}$) or the ($Zr_{0.4}Sn_{0.6}$) complex [37, 50, 63]. As one can see, in this case the described above behavior also manifests itself: the FE state is substituted by the AFE one. The only difference from the situation observed in the PZT system of solid solutions consists in the fact that now a more noticeable contribution to the system's free energy is due to the effects connected with the anharmonicity of the elastic crystal potential (this is caused by the presence of $Sr^{4+}$, $Sn^{4+}$ and $Ti^{4+}$ ions with different ionic radii in the equivalent crystal sites). The diagrams have a broad region (dashed region in the Fig.9.5a) which actually is the hysteresis region of the FE-AFE transition. Analogous phase diagrams have been obtained at the substitution of the complexes ($Lu_{0.5}Nb_{0.5}$) or ($Yb_{0.5}Nb_{0.5}$) for titanium [133, 134, 135]. They are presented in Fig.9.5b and Fig.95c.

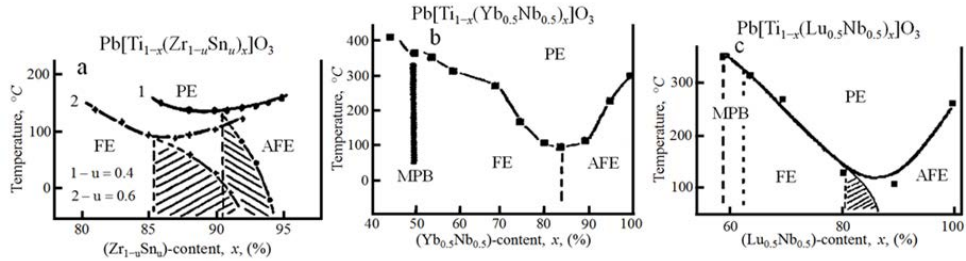

fig.9.5. Phase diagrams "composition-temperature" for the solid solutions on the base of the lead titanate with *B*-site substitution: (a) – $PbTiO_3$ – $Pb(Zr_{1-u}Sn_u)O_3$ [63], (b) – $PbTiO_3$ – $Pb(Yb_{0.5}Nb_{0.5})O_3$ [133], (c) – $PbTiO_3$ – $Pb(Lu_{0.5}Nb_{0.5})O_3$ [134, 135].

The following examples are connected with the situation when the ion substitutions are made in the *A*-site. The $X-T$ diagrams for the PZT with the 80/20 composition when the lead ion is substituted by strontium $Sr^{2+}$ [37, 63] or $(La_{1/2}Li_{1/2})^{2+}$ complex are shown in Fig.4.4. The 80/20 PZT is in a FE state at $T < T_c$. The low-temperature AFE state becomes stable with the increase of the content of strontium or ($La_{1/2}Li_{1/2}$) complex. As in the previous case, there appears an intermediate hysteresis region.

The $X-T$ diagrams obtained at the substitution of lanthanum for lead have already been discussed in Chapter 4. They are totally identical to the ones for strontium $Sr^{2+}$ [37, 63] or $(La_{1/2}Li_{1/2})^{2+}$ complex substitutions (Fig.4.4).

Now let us pass to the consideration of the PT-PMN solid solution system. From the viewpoint of physics and taking into account the above-considered examples it seems natural to expect that the substitution of ($Mg_{1/3}Nb_{2/3}$) complex for titanium should lead to the change from the low-temperature FE ordering to the AFE ordering, and the Y-T diagram of the PT-PMN system should be analogous to those discussed above in this section and presented in Fig.9.5 and Fig.4.4b. The hysteresis region should also manifest itself. This is confirmed by experiments. The Y-T diagram of the PT-PMN system constructed using the results of the papers [136, 137, 138, 139, 140, 141, 142] is shown in Fig.9.6. It is analogous to other diagrams Fig.9.5 and Fig.4.4b and is analogous to the model diagram in Fig.4.3 (see also [20, 22]) for systems with the FE-AFE phase transition). It needs to be noted that we do not consider here the part of the phase diagram of the PT-PMN system corresponding to the morphotropic region. Therefore, one may state that the PNM belongs to the AFE region of the Y-T phase diagram at the room temperature and at the temperatures below 220 *K* it belongs to the hysteresis region of the FE-AFE states. If electric field has not been applied, the macroscopic state of the samples is the AFE for the temperatures $T < T_c$. After the action of the field with an intensity higher than the critical value, the FE state is induced in the sample's volume at the temperatures below $T_0 \approx 220$ K. Subsequent heating leads at first to the FE→AFE phase transition at the temperature $T_0$ and then to the AFE→PE transition. It should be noted that the FE and AFE states in the PMN are not entirely identical to the classic ones realized in the systems in which the interaction of domains of these phases is negligibly small. For the PZT-based solid solution the said problem has been discussed above in the present paper; for the PMN it will be considered below.
.

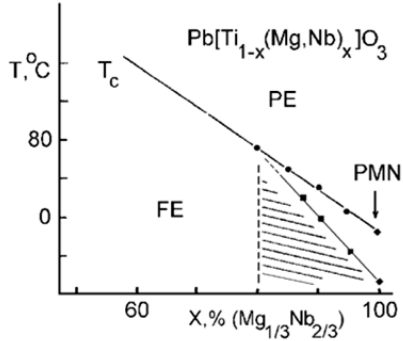

Fig.9.6. Phase diagram "composition-temperature" for solid solutions on the base of the lead titanate with substitution of titanium by complex ($Mg_{1/3}Nb_{2/3}$) [136-142].

The understanding of the origin of phase states and of their FE-AFE nature in the PMN allows to predict a number of its basic properties and to search for their experimental verification. The analysis of the Y-T diagram (Fig.9.6) of the PT-PMN solid solutions allows drawing conclusion that the PMN should have the P-E and the T-E diagrams analogous to the ones presented in Fig.4.2c and 4.2i, respectively. The T-E diagrams for this substance (obtained from the literature data) completely confirm this statement.

The T-E phase diagrams for different PMN samples are presented in Fig. 9.7. The diagram in Fig. 9.7a is taken from [143] for ceramic samples (this is a generalized diagram based on the experimental data [144, 145, 146]). The diagrams in Fig. 9.7b and in Fig. 9.7c are for the crystals and different orientations of the applied electric field [147]. These diagrams are completely identical to the model diagram Fig. 4.2i for the substances undergoing the FE-AFE PT.

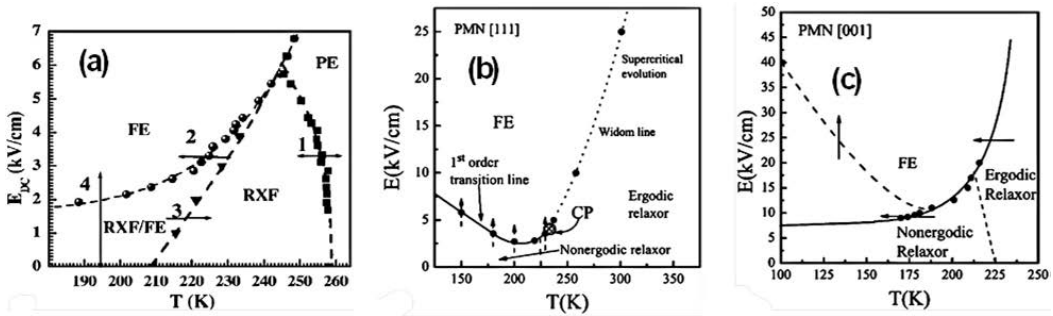

Fig. 9.7. "Temperature-electric field" phase diagrams for PMN.
(a) The standard empirical history-dependent phase diagram presented in [144- 146]. "FE" stands for a ferroelectric state, "PE" for a paraelectric state with nanodomains, and "RXF" stands for a glassy relaxor state in terms of [143.
(b) Phase diagram of the PMN crystal for the bias field in the [111]-direction [147]. (c) Phase diagram of the PMN crystal for the bias field in the [001]-direction [147]. Here, arrows show the path of the approach to the FE state for different transition lines.

The minimal critical field $E_{i,min} \equiv \Delta_i$ exists in T-E diagrams of the PMN. Moreover, the phase diagrams and phase states in PMN satisfy the requirements characteristic to the cases of the FC and ZFC states which follow from the model T-E diagram shown in Fig.4.2i (see the commentary to Fig.4.2i in the present paper). The difference of the FC and ZFC behavior in the PMN is caused by the existence of a minimal field for induction of the FE phase. Such situation has been discussed in Chapter 4 and has been confirmed experimentally for the PZT-based solid solutions.

As we discussed in Section 4.1 the application of a hydrostatic pressure leads to an increase of $E_{i,min} \equiv \Delta_i$ (for example, for the PZT based solutions see Fig.4.6 and Fig.4.7). On the other hand an increase in the lead titanate content in $(1-x)Pb(Mg_{1/3}Nb_{2/3})O_3$–$xPbTiO_3$ leads to an increase of the stability of the FE phase and to a decrease of $E_{i,min} \equiv \Delta_i$. The FE state appears spontaneously with a decrease of the temperature below the Curie point when the equality $E_{i,min} = 0$ is achieved. Based on the "composition-temperature" phase diagram for the $(1-x)Pb(Mg_{1/3}Nb_{2/3})O_3$-$xPbTiO_3$ solid solutions presented in Fig. 9.6 the vanishing of $E_{i,min}$ in the "temperature-electric field" diagrams has to take place in the vicinity of $x \approx 0.20$. This conclusion is confirmed by the experimental results presented in Fig. 9.8 [148].

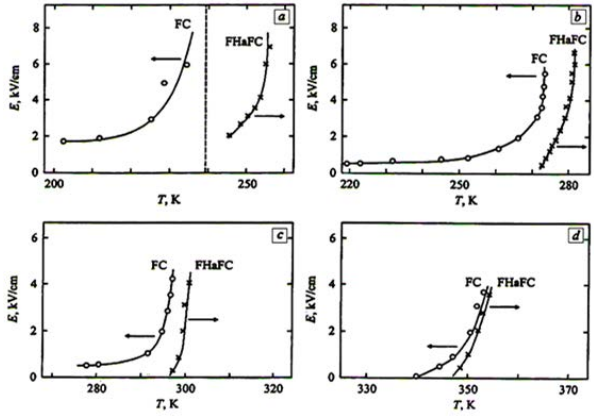

Fig.9.8. The *T-E* phase diagrams for single crystals with compositions $(1-x)Pb(Mg_{1/3}Nb_{2/3})O_3 – xPbTiO_3$. DC electric field was applied along [001] direction [148]. Arrows show the directions of change of external parameters (the temperature or DC field). Content of $PbTiO_3$, x: a – 0.06, b – 0.10, c – 0.13, d – 0.20.

As one can see, the results obtained on the single crystal samples show that the intensity of electric field necessary for the induction of the FE phase decreases with the increase of the $PbTiO_3$ content in the solid solution. The value of parameter $E_{i,min} \equiv \Delta_i$ also decreases and at the compositions in the vicinity of 20% of $PbTiO_3$ this value is equal to zero. Such situation completely matches the phase diagram presented in Fig.9.8d.

Thus, the phase diagrams for the PMN-PT system of solid solutions available in the literature can be completely described in the frames of the view point of the FE-AFE transitions developed in Section 4.1 and coincide with the experimental data for other systems of solid solutions for which the presence of the FE-AFE transitions has been uniquely established. It has to be especially emphasized that the difference between the FC and ZFC behavior in the PMN is also a consequence of the FE-AFE transition in an external electric field (see Fig.4.2i and comments to it).

Now let us formulate other statements about the behavior of the PMN from the point of view of the generalized FE-AFE PT model considered in Chapter 3, and then verify their validity for explanation of experimental results. Such statements are as follows.

First, the AFE state realized in the PMN at cooling differs from the one observed e.g. in the "pure" PZT. This distinction consists in the fact that in the PMN the mentioned state is not the one-phase state: it is characterized by the presence of domains of the FE phase. By analogy with the PLZT or the PLLZT, the size of the metastable phase domains must be of the order of 100 Å or smaller.

Second, since the equivalent crystal lattice sites (*B*-sites) are occupied by the ions with different sizes, there should be observed the "stratification" (local decomposition) of the substance in the vicinity of the interphase FE-AFE domain boundary and the formation of small domains with compositions rich in either Nb or Mg.

Third, the PMN will preserve its inhomogeneous structure at the temperatures essentially higher than $T_m$ (but lower than 350°C). The said inhomogeneous structure has to lead to the diffuse nature of the PE phase transition.

All these conclusions concerning the structural inhomogeneity of the PMN which follow from the model of the FE-AFE transformations have been completely confirmed by experiments. But the described inhomogeneities, though experimentally revealed long ago, still have no clear interpretation.

Before continuing with the further discussion of the PMN properties, one peculiarity presented in Fig.9.9 and 9.10 needs to be noted and connected with the so-called Burns-temperature, $T_B$. The hatched area in Fig.9.9 that presents the temperature interval in the vicinity of $T_B$ marks the interval of temperatures within which one experiences difficulties in precision measurements (with accuracy of up to 1 K) during the determination of this temperature. In the literature, any deviation in the material's behavior in the PE phase (at the temperatures $T_m < T < T_B$) form the behavior typical for classical FE was associated with the heterogeneity of chemical composition acquired in the process of samples' preparation (both single crystals and ceramic samples). Such association was an unbreakable rule followed by almost all researchers. Along with this all experimental data existing in the literature indicate that the behavior of the PMN at high temperatures ($T > T_B$) is typical for a homogeneous substance.

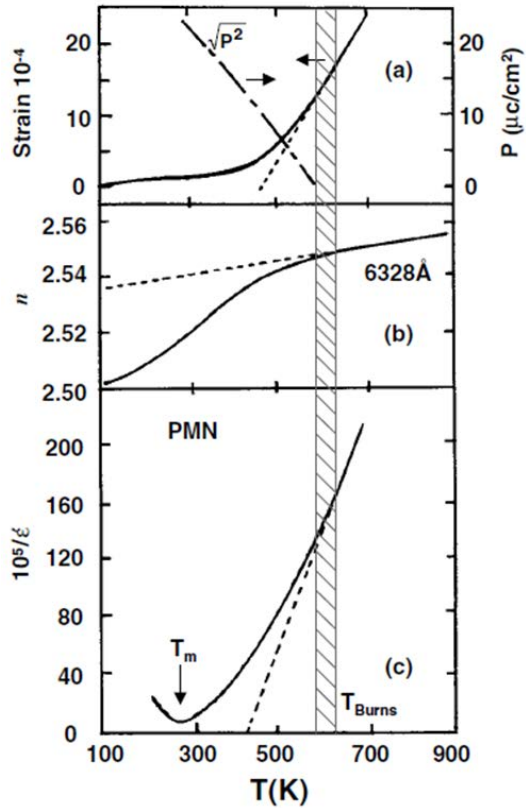

Fig.9.9. Temperature dependences of the linear thermal expansion (a), the refractive index $n$ (b), and the reciprocal dielectric constant, $1/\varepsilon$ (c) for PNM [149]. $T_m$ is the temperature of the peak in dependence of the dielectric constant on temperature, $\varepsilon(T)$.

Infrared investigations revealed that dielectric constant in perovskite relaxors at the temperatures $T > T_B$ is essentially the ionic feature (see, for example, [150]). The dielectric constant temperature dependence is determined by the classical ferroelectric-type phonon softening. Therefore, microscopically, the dielectric permittivity in the PE phase of relaxors (at $T > T_B$) follows the behavior characteristic for normal classical displacive ferroelectrics.

At the same time, the behavior of the physical characteristics at the temperatures $T < T_B$ is inherent to a substance with non-uniform distribution of chemical elements in the crystal lattice sites. In the case of PMN these non-uniformly distributed ions are $Mg^{2+}$ and $Nb^{5+}$ located in the $B$-sites of the perovskite crystal structure of PMN. Fig.9.10 shows the temperature dependence of the dielectric constant over a wide temperature range. As can be seen, at $T_m < T < T_B$ the behavior of $\varepsilon(T)$ can be approximated by the quadratic dependence $\varepsilon(T) \sim T^2$ with the high degree of accuracy [151, 152]. This behavior of $\varepsilon(T)$ is characteristic for ferroelectrics with diffuse phase transition caused by the chemical heterogeneity of a substance [153].

At the temperatures $T > T_B$ the dependence $\varepsilon(T)$ is linear which in turn is a characteristic feature of the homogeneous ferroelectrics with the location of $Mg^{2+}$ ions and $Nb^{5+}$ in their "legitimate" sites identified by the PMN chemical formula.

All these results have been confirmed by all publications related to the analysis of the temperature behavior of the dielectric constant in PMN, and point to a spontaneous breaking of the homogeneity of PMN samples during the cooling process and its recovery during the heating.

Fig.9.10. Fitting to the Curie-Weiss (dashed line) and the quadratic law (solid line) of the dielectric permittivity of $Pb(Mg_{1/3}Nb_{2/3})O_3$ single crystal. Crosses are the experimental data measured at 100 kHz [151, 152].

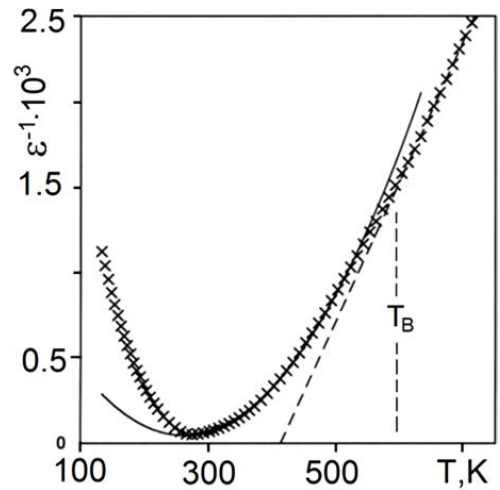

Thus, it is apparent that, if the chemical heterogeneity of ferroelectrics and antiferroelectrics is considered as a reason for "non-classical" behavior of the PMN and related substances (named "relaxor ferroelectrics") one must be sure to provide an explanation of the fact that this heterogeneity appears spontaneously in the cooling process at a relatively low temperature $T_B$ and disappears during the heating process at the same low temperature. To the best of the knowledge of the authors of this review, it is not done to date. It will be demonstrated below that the nature of the spontaneous breaking of the homogeneity of ferroelectrics and antiferroelectrics in the cooling process and restoring it (again spontaneously) during the heating process can be explained taking into account a relatively weak difference in energies of the FE and AFE states in complex oxides with the perovskite crystalline structure.

The main feature of all substances with the small difference in energies of the AE and AFE states is the coexistence of these phases at low and at high temperatures. In the latter case, complex (FE+AFE) domains of the said phases exist in the paraelectric matrix of a substance. The feasibility of such inhomogeneous states was theoretically justified in Section 3.1 of this review and the evidence of their existence that has been proven experimentally in the PZT based solid solutions is discussed in Chapters 6 and 7.

The phase diagrams in Figs.9.5 − 9.8 demonstrate their similarly to the phase diagrams of the PZT-based solid solutions with a small energy difference between the FE and AFE states, and the AFE state may be realized in the PMN. Also similarly to the PZT-based compounds an electric field applied to PMN causes the phase transition from the AFE state into the FE one. Later in this chapter we will present an analysis of the data available in the literature that will demonstrate that the co-existence of the dipole-ordered FE and AFE states is possible in the PMN and related substances. It will become clear that such coexistence of the FE and AFE states in the PMN is similar to the one observed in PZT-based solid solutions. Just this coexistence of phases defines all peculiarities of properties of the PMN as well as peculiarities of its behavior.

Let us address studies of the PMN's behavior in the presence of the hydrostatic pressure. It is known that under the hydrostatic pressure the stability of the AFE state relative to the FE one increases (a comprehensive analysis of this problem one can find in [37]). In particular, in the presence of complex (FE+AFE) domains of these phases in the sample, an increase in pressure should lead to the increase of the share of the AFE phase in the sample's volume and, finally, to a single-phase AFE state. The data available in the literature completely confirm this conclusion.

Experiments on the X-ray diffuse scattering (XDS) in the PMN under the high pressure were performed at room temperature in the pressure range from 0.1 MPa to 10 GPa at the European Synchrotron Radiation Facility on the ID30 high-pressure beam line [154]. The Bragg reflections, the diffuse scattering, and the weak superstructure reflections were observed at low pressure. The XDS along with weak superstructure reflections corroborate the evidence for the local deviations from the average (cubic) structure in the PMN. As rule, both the structural deviations ($BO_6$ octahedra tilts or/and cation displacements) and the local substitution-type (chemical) disorder have to be considered as the origin of the XDS. As was shown in [154], the pressure dependent changes of the XDS (Fig.9.11) cannot be explained by a chemical disorder since it is essentially pressure independent.

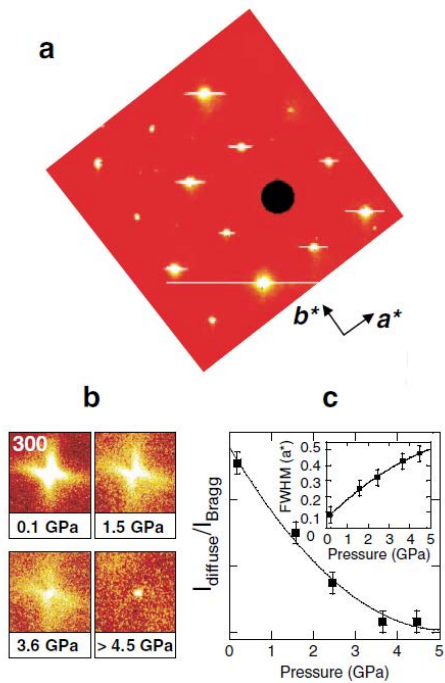

Fig. 9.11 (a) X-ray scattering imaging pattern for $PbMg_{1/3}Nb_{2/3}O_3$ at a hydrostatic pressure of 6.5 GPa.
(b) Illustration of the pressure dependence of the diffuse scattering around the 300 reflection. (c) Evolution of the intensity and FWHM (inset) of the diffuse scattering around the 300 reflection [154].

A transition to a long-range phase occurs at the high pressure according to results of the Raman spectroscopy studies [155] revealing a non-cubic long-range structure that appears with the increase of the pressure. The authors connect the low pressure XDS with the two-phase structure of the substance. The pressure dependent evolution of $(135)_{1/2}$ superstructure and the (400) and the (220) Bragg reflections are illustrated in Fig.9.12 in which we combined figures 4 and 5 form the article [154] to make it more vivid.

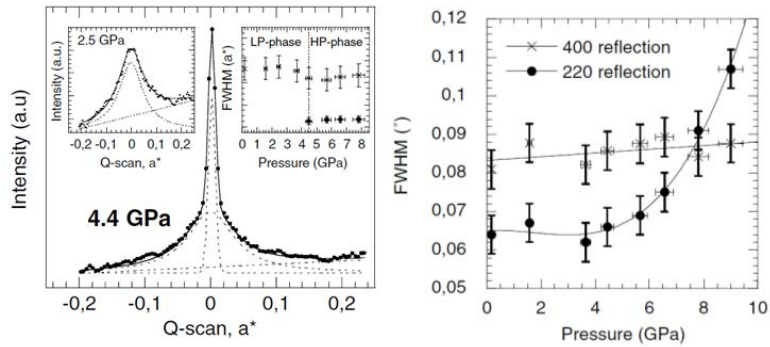

Fig.9.12. On the left: The pressure dependence of a representative 1/2(135) superstructure reflection. The main graph illustrates the superimposition of diffuse scattering with a ''celldoubling'' Bragg reflection at 4.4 GPa. The left inset illustrates the diffuse (chemical) scattering as typically observed in the low-pressure (LP) phase ($p<4:5$ GPa). The right inset presents the evolution of the full width at half maximum of the diffuse and sharp features as a function of pressure [154].
On the right: Pressure-dependent evolution of the full width at half maximum of the 400 and 220 reflections [160].

The superstructure reflections appear to be composed of two components for the pressures above 4.0 GPa: one diffuse reflection (already present at ambient conditions) superimposed with a new sharp Bragg reflection (it is only one component of the said diffusive line for $P < 4.0$ GPa). As shown in the inset in Fig. 9.12 (on the left), the full width at half maximum of the diffuse reflection is rather pressure independent, which adds a further support to its chemical origin. On the other hand, the pressure-induced sharp reflection is related to a doubling of the crystal cell, which might originate in the antiphase tilts of octahedra and/or the antiphase cation displacements. Such a superimposition of these two components related to a unit cell doubling is reminiscent of the temperature-dependent investigation of $1/2(hkl)$ reflections in [156], where a new sharp component has been observed at low temperatures. However, the fact that the Raman spectra at high pressure [155] and the low-temperature ones are different, suggests that the two phases are not identical. The details of this new structure were also observed by the Raman spectroscopy [155] and it was shown that the high-pressure phase of the PMN is characterized by the nonpolar $B$-sites, i.e., the Nb and the Mg cations are not displaced. Combining the complementary information, the authors of [154] proposed that the high-pressure phase of the PMN has antiparallel displacements of the Pb cations.

  The PMN and the 5% La-doped PMN (PLMN) ceramic samples were studied by the selected-area electron diffraction (SAED) [157], the high-resolution electron microscopy (HREM), and the computer simulations. Exceptional super lattice diffraction spots (this is the name given by the authors of [157]) at the positions $k/2\langle 110\rangle$ and $h/3\langle 210\rangle + m\langle 001\rangle$ (where $k$, $h$ and $m$ are integers) with diffusive scattering were observed at room temperature. The SAED patterns of three different zone axes are shown in Fig.9.13 (taken from [157]). The diffraction sports from the disordered PMN matrix are indexed in this figure with the simple cubic indices. The letter "O" denotes the $1/2\langle 110\rangle$ superlattice diffraction spots from the 1:1 chemical ordering, which were well studied by many researches. The exceptional superlattice spots are indicated by letter "N".

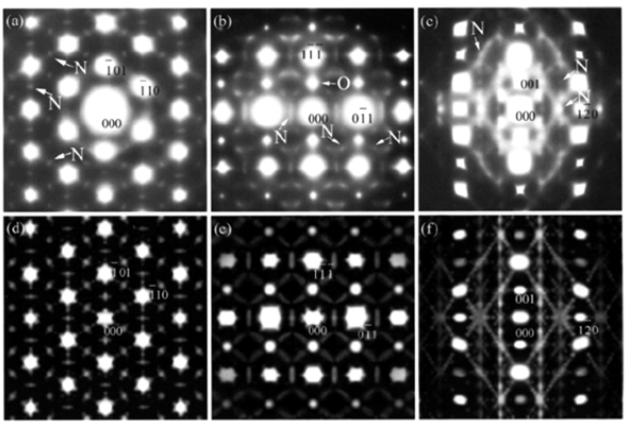

Fig.9.13. Experimental and simulated SAED patterns, indexed with simple cubic indices. (a) experimental [111] zone axis; (b) experimental [211] zone axis; (c) experiment [210] zone axis; (d) simulated [111] zone axis; (e) simulated [211] zone axis; (f) simulated [210] zone axis [157].

According to [157] the main features of these patterns can be summarized as follows:

-new spots at positions $k/2\langle 110\rangle$,

- new spots at positions $h/3\langle 210\rangle + m\langle 001\rangle$ with diffuse elongation along $\langle 001\rangle$,

- intensities fluctuation on "N" spots along $\langle 210\rangle$ and $\langle 001\rangle$, respectively,

- diffuse fringes along $\langle 213\rangle$ in $\langle 210\rangle$ zone-axis pattern.

The intensities of the "N" spots are very weak; therefore a long exposure time and a special contrast technique were used, i.e. developing the strong and the weak patters with different time for registration of the strong "1:1" and the weak "N" spots. Authors of the said paper showed that the positions of "N" spots cannot be caused by the multiple scattering from the strong diffracted beams (the latter can lead to extra diffraction spots in the SAED patterns). Analyzing different structural factors that can lead to the effects observed in this study, the authors of [163] came up with the possible pattern of structural deformations that can lead to the appearance of the local AFE ordering in the PMN shown in Fig. 9.14 [157].

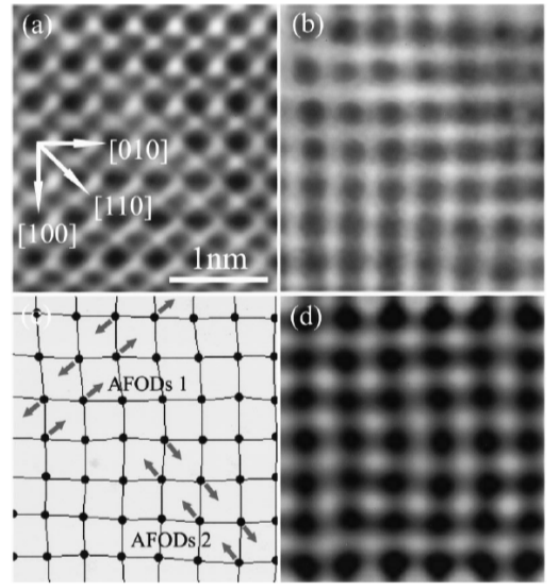

Fig. 9.14. (a) HREM of PMN along [001] direction, Scherzer defocus 561 nm; (b) the fast Fourier transform (FFT) image of (a), the $(100), (010), (\bar{1}00), (0\bar{1}0)$ and transmitted reflections are used; (c) a schematic diagram of the $Pb^{2+}$ displacements in (b); (d) the computer simulated image of (b); AFOD stands for the antiferroelectric ordered dipoles [157].

The above-given experimental data uniquely indicate the composite structure that is present at the temperatures $T < T_{m(c)}$. This composite structure constitutes the AFE matrix with inclusions that are domains of the FE phase. Peculiarities of behavior of this structure under the external influences are determined by the structure of these domains of the coexisting phases and their interactions. Effects caused by such internal structure of the PMN samples will be considered below.

As it was demonstrated earlier the relaxor properties (and diffuse phase transitions) of the PLLZT and PLZT solid solutions are caused by the small difference in energies of the FE and AFE states and the coexistence of domains of the FE and AFE phases in the volume of samples (see Chapters 3, 6 and 7). Similarly to the PZT-based compounds the relaxor properties of the PMN and related materials are also defined by the same mechanism. The corresponding data can be found in the literature.

The $(1-y)PbTiO_3 - yPb(Ni_{1/3}Nb_{2/3})O_3$ and $(1-y)PbTiO_3 - yPb(Mg_{1/3}Nb_{2/3})O_3$ systems are identical. Both these systems have the identical Y-T phase diagrams (Fig.9.6). In both cases the solid solution with $y = 0.4$ is in the FE state at the temperatures $T < T_c$. The paraelectric phase transition in both solid solutions is sharp [158, 159]. According to the above-developed concept, the increase of stability of the AFE state should lead to a diffuse phase transition and to the manifestation of relaxor properties. Such increase in stability of the AFE state was achieved by the substitution of Zr ions for Ti ones [159]. This predicted change in behavior of the $Pb[Ti_{0.4}(Ni_{1/3}Nb_{2/3})_{0.6}]O_3$ solid solution becomes apparent in experiment (see Fig.9.15 combined using the figures from [159]).

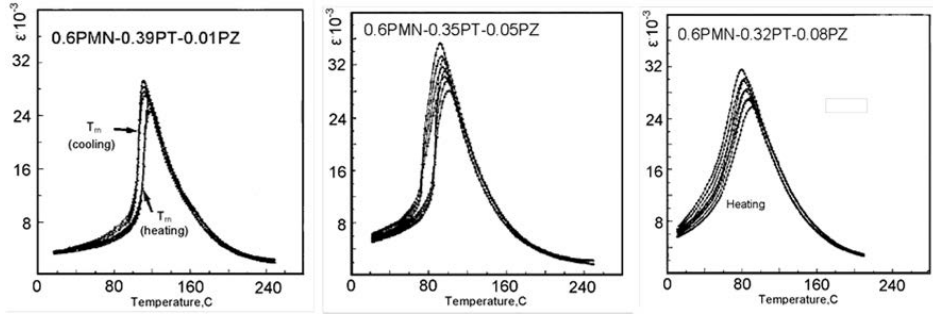

Fig.9.15. Temperature dependences of relative dielectric permittivity for 0.6PNN−(0.4−$x$)PT−$x$PZ solid solutions [159]. The measurement frequencies from the top curve to the bottom one are 0.1 kHz, 1 kHz, 10 kHz, and 100 kHz.

The discussed phenomenon corresponds completely to the model phase diagram in Fig.3.3. In this case the ion substitution takes place in the *B*-sites of the perovskite crystal lattice. The inhomogeneous structure inside the domains of the coexisting phases (that have close values of their thermodynamic potentials) in the PMN has to be preserved up to a temperature of approximately 350 $^{\circ}$C. The diffuse phase transition from the PE phase is the consequence of the presence of this structure. Any influences leading to the change of the relative stability of the FE and AFE phases have to lead to the change of the shares of these phases in the sample's volume and as a consequence it will lead to the noticeable change in the diffuseness of the phase transition. The pressure is the main among these factors.

However, we must emphasize that due to a lack of experimental results on the influence of the hydrostatic pressure on properties of the PMN and the PMN–PT systems it is difficult to compare approach developed in present review with the experimental data for these compounds.

On the other hand, the Pb(Zn$_{1/3}$Nb$_{2/3}$)O$_3$-PbTiO$_3$ (PZN–PT) system is the physical "twin" for the PMN-PT system. Both systems are characterized by the identical "composition-temperature" (x-*T*) phase diagrams. Results of studies of the thermal phase transformations in the PZN–PT system of solid solutions under the action of the hydrostatic pressure and the electric field are available and are published in [160]. These results are in perfect agreement with the ideas developed in the present review. In particular, the 0.905PZN-0.095PT solid solution undergoes the FE-PE phase transition. This transition is weakly diffuse and the so-called "relaxor properties" are practically absent (SEE Fig.9.16 that represents a combination of figures 1, 4 and 9 from [160]).

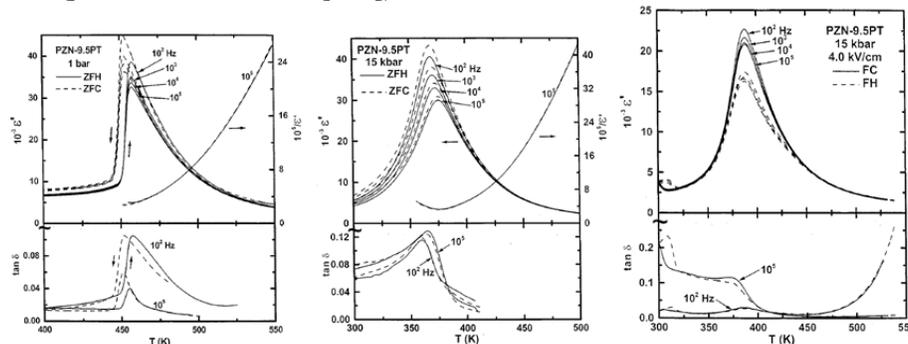

Fig.9.16. Temperature dependencies of the dielectric constant ($\varepsilon^{/}$) and dielectric loss (tan$\delta$) for PZN-9.5 PT from [160]. The experimental conditions (DC bias and pressure) are explained in the upper left corner of each figure. Results presented in the first and second figures (from the left) were obtained without DC bias; the dependencies shown in the rightmost figure were obtained in the presence of a DC bias field of 4.0 kV/cm.

Hydrostatic pressure increases the stability of the AFE phase and leads to the displacement of the position of the above-said solid solution in the x–$T$ phase diagram to the region of the AFE states. This means that the pressure increase results in the rise of the AFE phase stability and coexistence of the FE and AFE phase domains in the bulk of the sample for the temperatures below and above the Curie point. Because of this reason the paraelectric phase transition becomes diffuse and the relaxor properties are observed (Fig.9.16).

Then the electric field is applied to the sample under the pressure. This action causes the return of the above-said solid solution into the FE region of the x–$T$ phase diagram. As a result the stability of the FE phase rises and the stability of the AFE phase decreases. So, the sample returns in the state which existed before the application of the hydrostatic pressure (the initial uniform state). After the said procedure the phase transition into the PE state is not the diffuse one anymore and the relaxor properties disappear (Fig.9.16).

Now we would like to present and discuss the results of investigations of the diffuse phase transition and relaxor properties in the PMN-PbZrO$_3$ system of solid solutions [161]. In the PMN structure the substitution of the complex $(Mg_{1/3}Nb_{2/3})^{4+}$ for $Zr^{4+}$ ions leads to the increase of stability of the FE state. This means that the introduction of Zr ions into the crystal lattice has to lead to a "transition" from the state of coexisting FE and AFE phases to the single-phase state. This transition has to result in reduction of the degree of diffuseness of the phase transition. Temperature dependencies of the dielectric constant for the (1-x)PMN-xPbZrO$_3$ system of solid solutions are presented in Fig.9.17 [161].

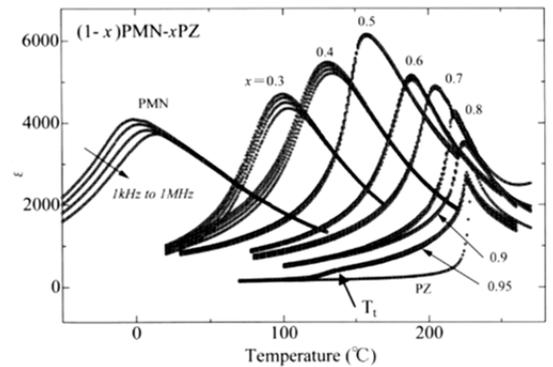

Fig.9.17. Temperature dependences of the dielectric constant in the (1-$x$)PMN-$x$PZ ceramics at 1 k, 10 k, 100 k and 1 MHz [161].

Modification of these dependencies of the dielectric constant due to an increase in the lead zirconate content (it can be clearly seen in this figure) completely confirms the conclusions of the model of the coexisting FE and AFE phases in application to the lead manganese niobate. Let us note that these conclusions were discussed in details in Chapter 7 of the present paper.

In addition to the above-given results let us note a well-known experimental result that the substitution of the $(Mg_{1/3}Nb_{2/3})^{4+}$ complex by $Ti^{4+}$ ions leads to the increase of stability of the FE phase in the lead manganese niobate (see diagrams in Fig. 9.6 and 3.3) and the transition of the substance toward the region of the single-phase FE state in the diagram of phase states (Fig. 9.3) As this takes place, the degree of diffuseness of the phase transition decreases and the relaxation properties of the (1-x)PMN-xPbTiO$_3$ solid solutions get suppressed.

Discussions of the idea that the introduction of an additional third ion into the *B*-sites of the crystal lattice leads to an increase in the degree of lattice disorder and, consequently, to an enhancement of relaxor behavior (increase in the diffuseness of phase transition) are quite frequent in the literature. The above presented examples of substitutions of the $(Mg_{1/3}Nb_{2/3})^{4+}$ complex for $Zr^{4+}$- and $Ti^{4+}$-ions show that introduction of additional ions in the crystal lattice leads to the disappearance of relaxor properties (decrease of the phase transition diffuseness) but not vice versa. In both cases introduction of ions leads to displacement of the solid solution (displacement of its composition) in the diagram of phase states of the system PbZrO$_3$ − Pb(Mg$_{1/2}$Nb$_{1/2}$)O$_3$ − PbTiO$_3$ toward the region of the single-phase low-temperature state, namely, toward the region of the AFE state in the case of substitution of Zr ions and toward the region of the FE state for substitution of Ti ions.

All experimental data on studies of the diffuse PE phase transition in relaxors show that the high-temperature state in the temperature interval $T_m < T < T_d \approx$ (300-350) ºC is not homogeneous. However, the main result is that the degree of the phase transition diffuseness and along with it the degree of inhomogeneity depend on the location of the substance's composition in the diagram of phase states.

Everything is identical to the situation that was observed in the PLLZT solid solutions and was analyzed in Chapter7. That is why one can make a conclusion that the composite FE+AFE domains exist in the PE matrix of the PMN at the temperatures higher than $T_m$ (this excess can be as much as several hundred degrees).

Let us now consider the available literature data that confirm the existence of such two-phase FE+AFE domains in the PE matrix of the substance at the temperatures above $T_{m(c)}$. The results of investigations of the PMN and PMN-based solid solutions up to the temperatures 600-700 $^{\circ}$C are available.

There exist a large number of publications containing a substantial amount of data on the diffuse X-ray scattering and the diffuse neutron scattering as well as on investigations of the Raman spectra and the infrared light scattering along with the data of acoustic studies by means of the Brillouin light scattering. Without going into details of all these studies we want to emphasize that all the authors pointed out the appearance (at the decrease of temperature) of peculiarities in the PMN behavior at the temperatures below $T_d \approx 350$ $^{\circ}$C (see Fig. 9.18 collected from [162, 163, 164, 165, 166]).
In spite of a large number of experimental studies the nature of the phase state below this temperature is absolutely unclear.

Amongst all observed peculiarities of different physical characteristics we will dwell on the peculiarities of the temperature dependence of the optic and acoustic oscillations frequencies. Both the optic and acoustic modes undergo softening below the temperature $T_d$ and have the pronounced minima (Fig. 9.18b). Practically all the researchers are agreed now (it started after the articles [74 - 76] were published) that the temperature $T_d$ determines the beginning of the process of formation of the nuclei of a new phase (with nanoscale sizes) at cooling. In order to explain the difference between $T_d$ and the temperature that corresponds to the minima of the frequency of the oscillation modes the opinion has been expressed that these nuclei have a dynamic nature at the beginning and then become thermodynamically stable. The softening of the transverse optic (TO) mode is attributed to the polar character of these nuclei. This explanation is clear and straightforward. However, in regard to the softening of the acoustic modes the majority of researchers admit almost complete absence of understanding of the physical mechanism of this phenomenon (we are not discussing here different attempts to give the explanation because they are far from consistent explanation of this phenomenon).

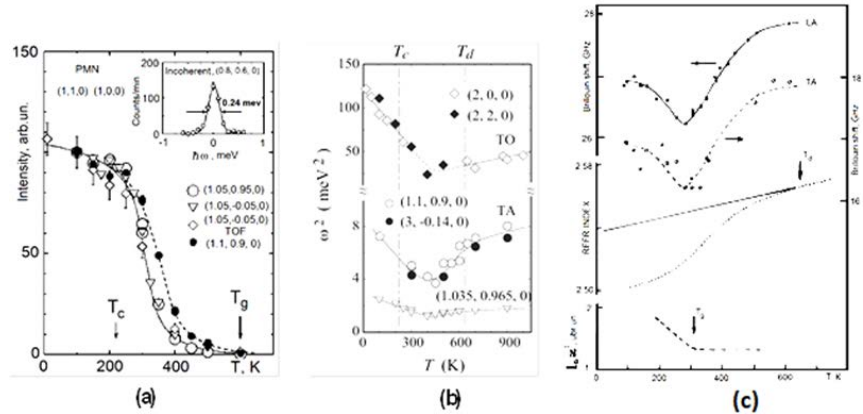

Fig. 9.18. (a) Temperature dependence of the elastic diffuse scattering intensity measured at $\zeta = 0.05$ and 0.1. Data are normalized at 100 K [162]. Open diamonds show the results from [163].
(b) The square of the phonon energy is plotted versus temperature for the zone-center TO modes [164] measured at (200) (open Diamonds) and (220) (solid diamonds) taken from [165] and [166], as well as for three TA modes measured at (1.035, 0.965, 0) (open triangles), (1.1, 0.9, 0) (open circles), and (3,−0.14, 0) (solid circles). Linear behavior is consistent with that expected for a conventional ferroelectric soft mode

(c) The temperature dependence of Brillouin shift for *LA* and *TA* acoustic phonons (upper part) [167], the refraction index [74 - 76] (in the middle) and characteristic value $J\chi^2$ from the neutron scattering [168] in PMN crystal

As we have noted above the available data point to the composite structure of nuclei of the new phase in the PE matrix of the substance. These nuclei are the two-phase FE+AFE domains. In Chapter 7 the experimental and theoretical justification of the appearance of these domains in the PE matrix of the substances possessing the triple FE-AFE-PE point at the temperatures significantly higher than $T_c$ was discussed in details. The literature data given above in this chapter confirm the same character of the high-temperature nuclei in the PMN. Based on such approach it is easy to explain the softening of the acoustic modes. This softening is connected with the AFE component of the two-phase high-temperature ($T_c < T < T_d$) domains. As we said above the relation between the shares of the FE and AFE phases inside the two-phase nucleus is varied under the external influences including heating and cooling. Such behavior of the shares of phases has to lead to the changes in intensities of the lines corresponding to the optic and acoustic modes. The interaction of phases that is the interaction of order parameters inside the domains described by the term $P^2\eta^2$ and the interdomain interaction described by expressions (3.2) and (3.9) (see Chapter 3) has to lead to the changes in the linewidth (damping of oscillations).

This approach is not something new and it was successfully applied for the explanation of the peculiarities in behavior of substances possessing a small difference in the free energies of the FE and AFE states. This subject was discussed in detail for different situations in [6] where the literature references can be found. Here we want to point out the following. Studies of the Mössbauer spectra in lead zirconate (PbZrO$_3$) [169] revealed that the peculiarities in behavior of the PbZrO$_3$ at the temperatures above $T_c$ are similar to the ones of the PMN at the temperatures above $T_c$. The phase transition takes place at the temperature $T_c$ = 236 ºC, however, the minimal value of intensity of the Mössbauer line (decrease of the probability of the Mössbauer effect) is noted at the temperature significantly higher than $T_c$. The temperature dependence of intensity of the Mössbauer effect in the classic FE of the BaTiO$_3$-type is quite different. The decrease in the intensity of the line reaches 40% whereas at the PE phase transition in the lead titanate (PbTiO$_3$) this decrease reaches only 10%. Naturally, that the authors relate the dielectric anomaly in the PbZrO$_3$ with the displacement of TO mode with $\vec{k} = 0$. The longitudinal acoustic (LA) mode is responsible for the AFE transition. Lead ions in the PbZrO$_3$ are displaced in the direction [110] which is the result of the lattice instability with respect to the LA mode at the boundary of the Brillouin zone in the [110] direction.

The possibility of existence and interaction of the FE and AFE oscillation modes in the lead magnesium niobate (Pb(Mg$_{1/3}$Nb$_{2/3}$)O$_3$) should not cause any surprise. In complex oxides with perovskite structure the free energies of the FE and AFE states are close. The typical result confirming this statement is the conclusions derived from the calculations of the stability of different dipole-ordered states in the classic ferroelectric BaTiO$_3$ (see Chapter 2 of the present review). It was shown that a small variation of parameters is enough for both the FE and AFE ordering to be realized in the barium titanate. On the other hand it was shown (see Chapter 2) that two types of ordering can be present in the lead zirconate the AFE ordering (as it is in fact realized) and the FE ordering.

It has to be noted in connection with the above-said that in the course of discussions of experimental results researches were reasonably close to accept the existence of the two-phase FE+AFE domains at the high temperatures and as a consequence the interaction between the FE-active and the AFE-active oscillation modes. For example, studies of the local lattice dynamics using the dynamic pair-density function determined by the pulsed inelastic neutron scattering in the PMN [170] demonstrated that the dynamic local polarization sets in around the so-called Burns temperature $T_d$ through the interaction between the off-centered Pb ions and the soft phonons, and the slowing down of the local polarization with decreasing temperature produces the polar nanoregions and the relaxor behavior below the room temperature.

The major increase of the intensity of diffuse scattering that takes place more than 100 ºC below $T_d$ (see Fig. 9.18a) during cooling of samples can be easily explained by the presence of the two-phase nuclei in the PE matrix of the PMN at the temperatures $T > T_c$. First two-phase (FE+AFE) nuclei appearing at the temperature $T_d$ have their own refractive index different from the one of the PE matrix. The deviation of dependence of the refractive index on temperature from the linear one in the PMN crystals is manifested exactly at the same temperature. The difference in the crystal lattice parameters for the FE and AFE phases constituting the two-phase nucleus is not very pronounced at this stage. However, this difference increases with the decrease of the temperature. That is why the local decomposition (the separation of ions $Mg^{2+}$ and $Nb^{5+}$ having different ionic sizes) does not happen at this initial stage (see Chapter 7 for details). The difference in crystal lattice parameters of the FE and AFE phases constituting a nucleus increases with the further decrease of temperature and at the same time the elastic energy in the vicinity of the boundary between these phases increases. This increase in elastic energy becomes sufficient to start the local decomposition process in the vicinity of the boundary between the FE and AFE phases. This process takes place at the temperatures considerably lower than $T_d$. It is precisely the beginning of the local decomposition is accompanied by the increase of the diffuse X-ray or the diffuse neutron scattering.

The detailed analysis of all peculiarities of behavior of the lead magnesium niobate and similar substances is not the main goal of this review. That is why we will not dwell in detail on such phenomenon as the "waterfall" phenomenon in relaxors. The review of recent studies of this phenomenon one can find, for example in [171]. In our opinion based the above-presented ideas, this phenomenon is connected with the effects of the local decomposition process in the temperature interval between $T_d$ and $T_c$. The "softening" of the crystal lattice and the damping of the oscillation modes take place along with the intense decomposition. The disruption of the translational symmetry occurs simultaneously not only in the vicinity of the FE-AFE interphase boundaries but in a whole volume of the crystal. The process of the ion redistribution among the lattice cites involves the whole crystal volume. It was discussed in Chapter 6 (see Fig. 6.3 and Fig. 6.7). The intensity of the mentioned process is the maximal at the maximum intensity of the local decomposition process. Outside the temperature interval within which the intense decomposition takes place the "waterfall" phenomenon is absent.

There exist a lot of literature data on the shape of the lines of diffuse scattering and their differences during scanning in the reciprocal space in the vicinity of different Bragg peaks. The relationship between the shares of the FE and AFE phases inside the two-phase nucleus and the orientations of the polar axis (for the FE part of the two-phase domain) and the antipolar axis (for the AFE part of the domain) are given by the condition that this nucleus leads to the lowest possible value of the elastic energy increase. Such advancement of the physical process has to lead to a very interesting effect. Let us remind that the FE and AFE components of the two-phase domain are characterized by the different signs of changes in the configuration volume in comparison with the configuration volume in the PE state. The diffuse scattering in the (1-x)PMN-xPbTiO$_3$ solid solutions at the temperatures above the Curie point is observed both for the small values of $x$ and for the big ones. In the first case the low-temperature FE phase of the solid solutions is characterized by the rhombohedral type of distortions of the crystal lattice and in the second case these distortions are tetragonal. The nucleation of the two-phase domains is possible in both cases (see Fig. 7.2 and explanations in the text of Chapter7). However, the morphology of these domains will be different because they differ in the orientations of the polar axes of the FE component of the domain in the first and second cases. This has to lead to a different shape of profiles of the diffuse lines during the scanning in the vicinity of the same Bragg peaks. The difference in the shape of the profiles obtained during scanning in the same substance but in the vicinity of the different Bragg peaks can be also explained by the difference in the morphology of the two-phase domains. These effects are well studied experimentally.

Several different explanations of the unusual development of ferroelectric-like properties of the PMN have been suggested, namely, a model based on compositional fluctuations, a model of superparaelectric behavior, and a model of glass-like behavior. As a rule these models do not explain an entire set of properties and one has to use modifications of these models for such explanations. All models as well as their modifications are based on the chemical inhomogeneity (compositional fluctuations) of the PMN which is created as a result of the sample preparation. That is why a method of fabrication of the single crystalline and the polycrystalline samples has to influence the results significantly. We want to point some technological factors, which can influence the results of experimental investigations, and subsequently the interpretation of these results. Herewith the greatest influence on the experimental results one may expect during the high-temperature studies.

The review of experimental results on inelastic neutron scattering and the model calculations for the PMN one can find in [171] where the results obtained by different authors are considered and the comparison of measurements on different samples are made. It is natural that due to the various kinds of inherent disorder (chemical, strain, polarization) in the relaxor materials, they are not in a uniquely defined equilibrium state and, thus, their properties could depend on the specific crystal growth method and the sample history in a rather complicated way. For example, comparison of the PMN samples grown by different methods [172] allowed to show that some of them do not undergo the expected phase transition under electric field.

The damping of modes of the inelastic neutron scattering spectrum in the PMN could be influenced by the geometric aspects of homogeneity regions, namely, by the sample's state. Such influence has been examined in [171]. In this paper, similar spectra obtained on different PMN samples in different experiments that were performed in the same way, namely, for the same orientation $\vec{q} \parallel [110]$ of the phonon wave vector for the (even/even/even)-type Brillouin zone are collected and directly compared. The results of such comparison for transverse modes are presented in Fig. 9.19. Clearly, all spectra show one mode below 5 MeV and another one above 10 MeV, in agreement with the expectations for the rough positions of the TA and the lowest frequency TO modes for this phonon wave vector. Samples used in these measurements had different origin, for example, the sample in [166] was grown by the modified Bridgman method, the sample in [173] by the top-seeded solution method from the PbO flux, and the sample in [174] by the Czochralski method. Despite the different origin of these samples, the differences in their spectra at this wave vector ($\vec{q} = 0.2\vec{c}^*$) are rather minor. At the same time, a closer look suggests that the upper peak in the spectra corresponding to the 'no-waterfall' case [175] is indeed somewhat less damped, more intense and has a noticeably lower frequency.

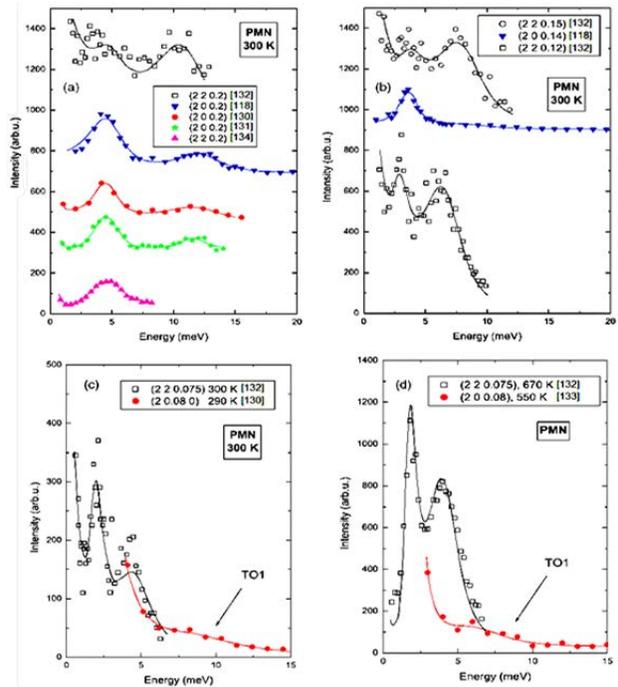

Fig.9.19. Spectra of inelastic neutron scattering with $\vec{q} \parallel [100]$ in (even/even/even)-type Brillouin zones in PMN obtained by different authors [171]. (a) spectra taken at $q = 0.2$ r.l.u.; (b) spectra taken at $q = 0.12-0.15$ r.l.u.; (c, d) spectra taken at $q = 0.075-0.08$ r.l.u. Spectra (a, b, c) are taken at room-temperature, (d) are taken at higher temperatures around 600 K. The intensity scale for the experimental data and original fitting curves in (a) were adapted so that the TA mode intensity is similar in all scans.

This difference becomes more apparent at the lower wave vectors. At the wave vectors $\vec{q} = (0.12 - 0.15)\vec{c}^*$ the position of the second peak in the 'no-waterfall' experiment strongly shifts downwards in a similar way as the acoustic mode does, so that it keeps (due to the Bose-Einstein thermal factor) practically the same intensity as the acoustic one, while in the other experiments the upper mode does not show such strong dispersion (Fig.9.19d). The spectra taken at even smaller wave vectors ($\vec{q} = (0.075 - 0.08)\vec{c}^*$, Fig.9.19c) clearly show that the position of the upper peak in these two experiments differs by more than 3 MeV. Similar discrepancy is also observed at the higher temperatures (Fig.9.19d), because the TO mode frequency in the experiment [175] is almost temperature independent. Such a huge variation in the TO mode frequency could be comparable to the changes associated with some structural phase transition or the anomalous crystal structure of the specimens.

Many researchers refer the presence of structural inhomogeneities in the PMN to technological factors. Hence the following conclusion may be drawn: the inhomogeneities formed in the process of the crystal growth (or the process of the ceramic samples manufacturing), i.e. at the temperatures of 1000 °C and higher, must be preserved at all low temperatures. However, the model of the FE-AFE states for the PMN yields quite different statement: since the structural and the chemical inhomogeneities are caused by the coexistence of domains of the FE and AFE phases, they will be preserved until such coexistence takes place and will disappear when the coexistence of phases is absent. For the PMN the highest temperature at which such inhomogeneities are observed should correspond to 300-350°C (and this is fully analogous to the situation observed in the PLLZT solid solutions by the X-ray method (discussed here in Chapter 7). In the course of thermocycling carried out at the temperatures close to the above-mentioned one such inhomogeneity must be formed at cooling and be completely broken down at heating (thus making the structure homogeneous). Naturally, this process may be somewhat protracted, since at the said temperatures the diffusion of elements is rather slow. Thus, we have formulated the first basic test for the applicability of the model of the FE-AFE transformation to the PMN. Contemporary methods of transmission electron microscopy allow direct verification of the presence or absence of domains of the coexisting FE and AFE phases in the high-temperature region as well as the verification of the presence of the local segregations of ions of different sizes in the vicinity of the boundaries separating domains of the above-mentioned phases.

Now let us point to the second experiment for checking the validity of the developed ideas for the explanation of the PMN properties. As mentioned above, the PMN must be characterized by the P-T diagrams similar to the ones presented in Fig.4.2c and in its turn it unambiguously calls for the E-T diagrams and their evolution given in Fig.4.2i. This experiment may be easily realized.

Here the discussion of the problem of phase transformations in the PMN may be finished. Based on available experimental data we claim that all the peculiarities in the behavior of the PMN are caused by the FE-AFE-PE transformations taking place under the action of external factors. Other states do not participate in phase transitions since they are absent in this substance. However, each of the mentioned states has peculiarities connected with the interaction of the coexisting domains of the FE and AFE phases.

From the view point of the crystal lattice dynamics, an appearance of the AFE state is connected with the lattice oscillations with the wave vector lying on the Brillouin zone boundary. In comparison to the zone center, a little experimental attention has been paid to studies of effects caused by the wave vectors at the zone boundary of the PMN based substances. Usually the main attention has been paid to the effects taking place when the wave vectors were close to the zone center. Interesting temperature-dependent superlattice peaks have been observed in the PMN using TEM, neutron scattering, and X-ray diffraction techniques at the reciprocal-lattice vectors $\left(\frac{1}{2} \frac{1}{2} 0\right)$ and $\left(\frac{1}{2} \frac{1}{2} \frac{1}{2}\right)$ ($M$ and $R$ points of the Brillouin zone, respectively), but authors of these investigations have mainly discussed the rotational modes of the oxygen octahedron.

Concluding this section we point to a series of even "more convincing" experimental results that require to pay a greater attention to AFE "arguments" during the discussion of properties of the PMN and PMN-based solid solutions. Firstly, based on the chronology of their appearance the publications [156, 176, 177] should be noted.

Synchrotron X-ray scattering studies of AFE nanoregions produced by antiparallel short-range correlated $\langle 110 \rangle$ $Pb^{2+}$-displacements were systematically performed on $Pb(Mg_{1/3}Nb_{2/3})O_3$ single crystals in these publications. Separation of $1/2(hk0)$ superlattice reflections from underlying diffuse scattering background allowed to study them as separate entities. The temperature dependence of the 1:1 chemical short-range order peak, e.g. $1/2(115)$ was studied also. The said nanoregions with correlated $Pb^{2+}$-displacements were shown to be different from the chemical nanodomains and the polar FE nanoregions. Instead, they are formed by antiparallel short-range correlated $Pb^{2+}$ displacements that lead to unit cell doubling in $\langle 110 \rangle$ directions. Based on structure factor calculations, authors consider that their results imply that both the FE and AFE fluctuations are present in the crystal structure and their interaction specifies the PMN properties.

A large amount of data obtained during studies of the Brillouin zone phonons in PMN is presented in [178]. This paper reports lattice-dynamical measurements (made using different neutron inelastic-scattering methods) of $Pb(Mg_{1/3}Nb_{2/3})O_3$ at momentum transfers near the edge of the Brillouin zone.

At first, we will consider the elastic scattering measured near all three high-symmetry points on the zone boundary of the Brillouin zone − $M$, $X$, and $R$. The temperature dependence of $M$, $R$ and $X$ zone-boundary elastic scattering is illustrated in Fig.9.20. The intensities measured at both the $M$ and $R$ points show the strong temperature dependence. As it can be seen, the scattering starts to appear at the temperatures close to the Burns temperature; however, the most significant increase in the intensity begins near 400 K. The most probable nature of this behavior we have discussed above.

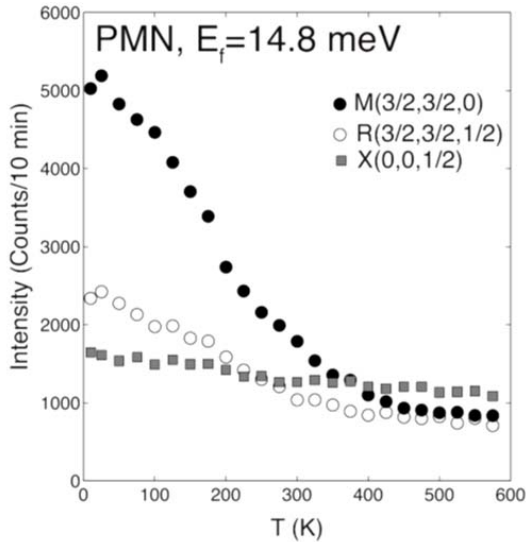

Fig.9.20. The temperature dependence of the elastic scattering measured at $Q_M = \{½, ½, 0\}$, $Q_R = \{½, ½, ½\}$, and $Q_X = \{0, 0, ½\}$ [178].

The enhancement of the elastic scattering as a function of temperature at the $M$ and $R$ points is suggestive of incipient structural instabilities; however, it is also relatively smooth and is not associated with a sharp transition to a long-range structurally distorted phase (that is, with structural phase transition in the crystal as a whole)

As the scattering is diffuse, these data infer the onset and subsequent growth of short-range, antiferrodistortive correlations below 400 K. This finding is significant because 400 K is the same temperature at which the zone-center TO mode reaches a minimum in energy [165] and at which the strong, zone-center diffuse scattering, which is associated with the formation of static polar nanoregions, first appears [162].

Inelastic scattering shows that the dispersion curves reveal several possible incipient soft-mode positions − these being where the optic modes reach a minimum energy as a function of reduced wave vector. Aside from the soft TO mode at the Γ-point, optic mode minima are seen at the $M$ and $R$ points; these minima correspond to the locations where strong, temperature-dependent elastic scattering was observed.

Time-of-flight neutron spectra illustrate three key features regarding the nature of the lattice dynamics near the zone boundaries.

First, a soft mode is evident at the $M$ point in the PMN given the substantial increase in spectral weight that occurs at low energies upon cooling from 600 to 300 K.

Second, another soft phonon is observed at the $R$ point.

It has been shown that these soft modes cannot be the tilt instabilities in cubic perovskites, which are also associated with phonon instabilities at these same points along the edge of the Brillouin zone.

A third feature, which is common to both the *M* and *R* points, is that the strong softening of these zone-boundary modes has an unusual dispersion that is localized in momentum but broad in energy, which is what gives rise to the appearance of a column of scattering in **Q**-*E* space. Unusual "columns" of phonon scattering that are localized in momentum, but extended in energy, are seen at $\mathbf{Q}_R = \{½\, ½\, ½\}$ and $\mathbf{Q}_M = \{½\, ½\, 0\}$ high-symmetry points along the zone edge. These columns soften near 400 K which is similar to the onset temperature of the zone-center diffuse scattering, indicating a competition between *ferroelectric and antiferroelectric* distortions.

This feature is anomalous amongst the ferroelectrics and perovskites, where soft modes are usually described by a well-defined dispersion that is characterized by a single, temperature-dependent value of energy for a given momentum transfer, as seen near the Γ-point in several studies of PbTiO3, PMN, and PMN-60PT. Since the soft modes observed at both the *M* and *R* points exhibit a continuum of energies at these specific momentum transfers, they cannot be understood by such a well-defined function.

Authors of [178] proposed a model for the atomic displacements associated with phonon modes which is based on a combination of structure factors and group theoretical analysis. This analysis suggests that the scattering is not from rotational modes of oxygen octahedral, but from zone-boundary optic modes that are associated with the displacements of $Pb^{2+}$ and $O^{2-}$ ions. Thus, the neutron scattering results point to existence of the antiferroelectric distortions competing with the ferroelectric distortions characterized by strong, zone-center diffuse scattering and a soft optic mode in the relaxor PMN.

## 9.2. The ferroelectric-antiferroelectric nature of the ordering phase transitions and the physical behavior in Pb(B′$_{1/2}$B″$_{1/2}$)O$_3$ compounds

Here we consider lead indium niobate $Pb(In_{0.5}Nb_{0.5})O_3$ as a first example of such compounds. On the one hand, the consideration of peculiarities of the phase transitions in the lead indium niobate is impeded, since up to now very few investigations of this substance have been performed using the pressure and the high-intensity electric field. Moreover, the number of different types of phase diagrams to be examined while discussing the available results of investigations is also very limited.

On the other hand, such a consideration is somewhat promoted by the fact that, as established reliably and unambiguously, the low-temperature state in the lead indium niobate (and in related substances such as $Pb(B'_{1/2}B''_{1/2})O_3$) is either the FE or AFE. For the lead indium niobate the type of the dipole ordering in the low-temperature phase depends on the presence or absence of the spatial ordering in the arrangement of indium and niobium ions (for the said ions the existence of a particular type of compositional ordering is defined by the conditions under which the heat treatment of the substance has been carried out).

Fig.9.21 presents the phase diagram of the lead indium niobate [179]. As one can see, the spatially-ordered arrangement (order parameter s ~ 1) and the spatially-disordered arrangement (order parameter s ~ 0) of *B*-ions correspond to the AFE and FE states, respectively. In the completely ordered crystals the transition into the PE state is well-defined and is not diffused. There are practically no distinctions in the ε(*T*) dependencies measured at different frequencies of the AC electric field. In the crystals with the minimal degree of the composition ordering the paraelectric PT is diffused, the temperature dependencies of the permittivity manifest the relaxor character completely.

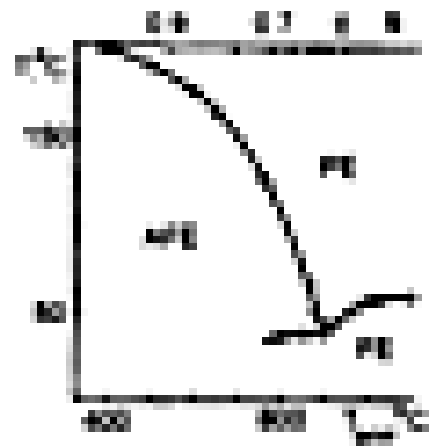

Fig.9.21. Phase diagram of lead indium-niobate [179].

The above mentioned permittivity behavior is usually interpreted based on different versions of the composition fluctuations model (the said model is generally used while attempting to explain the nature of the relaxor ferroelectric state). However, if the composition fluctuation approach was applied to the description of the FE-PE or the AFE-PE transformations in the lead indium niobate, this approach has more drawbacks than advantages. Some results of experimental studies of the PT may be explained in the frames of a theory of the two-component ordered alloys of the (B′B″) type (see [37], section 6.5). However, in this case the displacements of ions associated with the changes in the degree of compositional order must be realized at a distance not exceeding the lattice constant. Naturally, under such conditions the regions with predominate content of certain ions ($In^{3+}$ or $Nb^{5+}$) cannot be formed.

The model of composite fluctuations is not suitable for one more reason. The discrepancy of the composition fluctuations model consists in the following. A strong diffuse character of the paraelectric phase transition and the relaxor behavior are observed after the high-temperature ($T > 1000°C$) annealing, whereas a well-defined sharp phase transition (without diffuse character) occurs after the annealing at the temperatures from 400 to 600°C. The annealing carried out at such low temperatures will hardly increase the degree of the crystal homogeneity, in view of the fact that the said compositional ordering cannot be achieved in the process of the high-temperature annealing.

In our opinion, a high degree of diffuseness of the PE phase transition in the lead indium niobate is connected with a small difference in the free energies of the FE and AFE states which must inevitably lead to the coexistence of these phases even in the absence of the composition fluctuations. Changes in the degree of the composition ordering during the heat treatment process result in the displacement of the energy equilibrium between the FE and AFE states and the rise of the stability of one of the phases. The latter factor leads to the essential change of the degree of diffuseness of the phase transition. The phase diagram of the lead indium niobate (Fig.9.21) clearly shows that a high degree of diffuseness is present only in the substances displaced by the heat treatment towards the boundary separating the regions of the FE and AFE orderings.

Now let us experimentally verify our model of the diffuse phase transition, based on the FE-AFE phase transition, for the lead indium niobate and for a general case of the $Pb(B'_{1/2}B''_{1/2})O_3$-type compounds. Unfortunately, up to now nobody has attributed the diffuseness of the transition (and the nature of the so-called relaxor state) to the location of the substance in the diagram of phase states. Therefore, the required experimental data to be taken from the literature are very few and obtained for different substances. Though it is difficult to compare such results we shall try to perform this comparison using the scheme tested above on the PZT-based and the PMN-based solid solutions.

In the $(1-x)Pb(Lu_{1/2}Nb_{1/2})O_3-xPbTiO_3$ system of solid solutions (the phase diagram is presented in Fig.9.5) $Pb(Lu_{1/2}Nb_{1/2})O_3$ is in the AFE state at the temperatures below the Curie point [180]. When the content of the lead titanate in the solid solution is increased the low-temperature AFE ordering is substituted by the FE ordering. The boundary concentration region separating the regions the FE and AFE states in the Y-T phase diagram is located in the vicinity of 20 mol% of the $PbTiO_3$. The results of X-ray and dielectric investigations of this system of solid solutions presented in [135] corroborate the model proposed in Chapter 7, which relates the degree of diffuseness (and the relaxor behavior) to the relative stability of the FE and AFE phases. In $Pb(Lu_{1/2}Nb_{1/2})O_3$ the PE phase transition is clearly not a defuse one – for $T > T_C$ the Curie-Weiss law is satisfied. An increase in the lead titanate content leads to the diffuseness of the phase transition and the relaxor behavior. In the solid solution with 20% of $PbTiO_3$ such effects are maximally vivid, whereas in samples with the 10% of lead titanate they manifest themselves weaker. With further increase of the lead titanate content the PE phase transition becomes not a diffuse one again (Fig.9.22) [135]). According to the X-ray analysis data, the degree of composition ordering of Lu and Nb ions constantly diminishes as the $PbTiO_3$ concentration in the solid solution increases – any special features of compositional ordering as function of $x$ in the vicinity of $x = 0.20$ are absent.

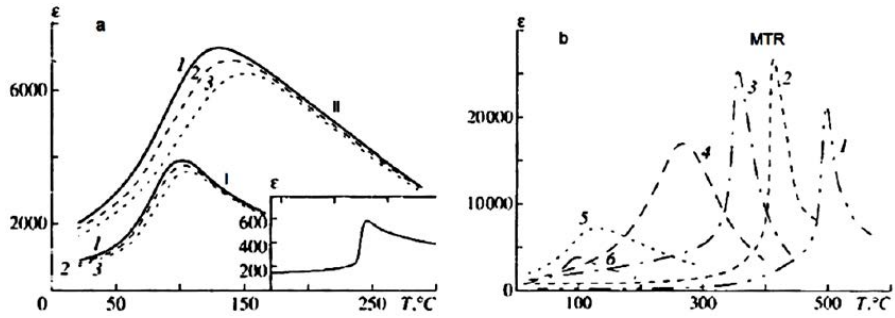

Fig.9.22. (a) Dielectric constant ε vs. temperature for $(1-x)Pb(Lu_{1/2}Nb_{1/2})O_3-xPbTiO_3$ solid solutions measured at different frequencies $f$ [135] for small values of $x$.
Content of $PbTiO_3$, $x$: I – 0.10, II – 0.20;
$f$, kHz: 1 – 1.0, 2 – 10.0, 3 – 100.0.
The insert shows ε(T)-dependence for $Pb(Lu_{1/2}Nb_{1/2})O_3$.
(b) Dielectric constant ε vs. temperature for $(1-x)Pb(Lu_{1/2}Nb_{1/2})O_3-xPbTiO_3$ solid solutions for large values of $x$ [135].
Content of $PbTiO_3$, $x$: 1 – 0.80, 2 – 0.50, 3 – 0.41, 4 – 0.30, 5 – 0.20, 6 – 0.10.

Let us consider another example. The boundary between the FE and AFE phases in the $x$-$T$ phase diagram of the $(1-x)Pb(Yb_{1/2}Nb_{1/2})O_3-xPbTiO_3$ system is located near $x = 0.2 - 0.3$ [133] (see Fig.9.5). Solid solutions with composition located far from $x = 0.2$ in the $x$-$T$ diagram are ordinary ferroelectrics ($x \geq 0.5$) or antiferroelectrics ($x < 0.1$). Phase transitions in these solid solutions are not diffusive. The solid solutions with $x = 0.2 - 0.3$ manifest relaxor properties as it is shown in Fig.9.23 133, [181].
Presented-above results for the $(1-x)Pb(Yb_{1/2}Nb_{1/2})O_3-xPbTiO_3$ system are conflicting with developed in literature concepts that the introduction of additional third ion into the $B$-sites of the crystal lattice leads to an increase in the degree of the lattice disorder and, consequently, to an increase in the diffuseness of the phase transition. As can be seen, the significant increase of the degree of diffuseness in the $Pb(Yb_{1/2}Nb_{1/2})O_3-PbTiO_3$ solid solutions occurs near the boundary of the regions with the FE and AFE orderings in the diagram of the phase states. The degree of diffuseness decreases when the position of the solid solution (its composition) is shifted from the said boundary.

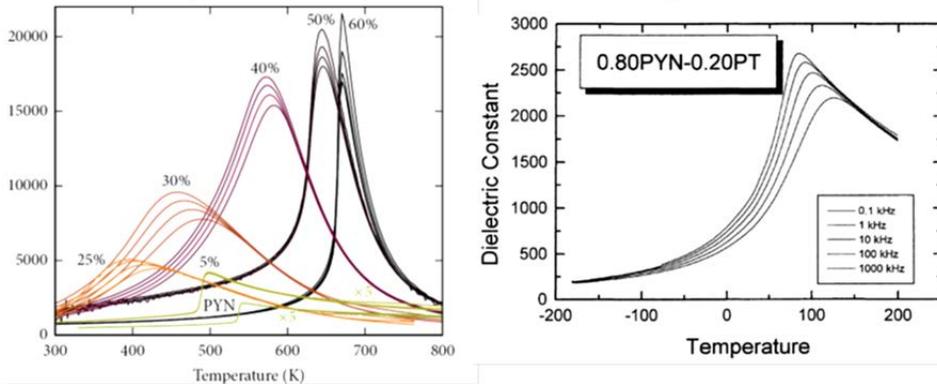

Fig.9.23. On the left: the real part of the dielectric constant ($\varepsilon'$) measured at 1kHz on PYN-PT ceramics exhibiting antiferroelectric (pure PYN and PYN-5% PT), relaxor (for 25, 30, PbTiO3 content), or ferroelectric (50 and 60% PbTiO3 content) behaviors [181].
On the right: the temperature dependence of the dielectric constant at different frequencies for x = 0.2 [133].

The $Pb_{1-x}Ba_x(Yb_{1/2}Ta_{1/2})O_3$ solid solutions based on the cognate compound $Pb(Yb_{1/2}Ta_{1/2})O_3$ were investigated in [182]. Phase transition from the FE state into the PE one in the last substance is not diffusive. Substitution of Ba for Pb leads to a diffuseness of the phase transition. The maximum degree of diffuseness takes place for x ≈ 0.10 – 0.15. An increase in barium content above x ≈ 0.15 leads to a reduction of the degree of diffuseness of the phase transition.

Similar results were obtained for the $(1-x)PbZrO_3-xPb(Yb_{1/2}Nb_{1/2})O_3$ system of solid solutions at the substitution of $(Yb_{1/2}Nb_{1/2})$ complex for Zr in [183]. This system of solid solutions is more complex from the point of view of physics than the one discussed early. The end components of this system are both antiferroelectric. The FE phase appears in these solid solutions with an increase of the $(Yb_{1/2}Nb_{1/2})$-content in the interval of compositions x ~ 0.15 – 0.17. The degree of the PE phase transition diffuseness increases just at this $(Yb_{1/2}Nb_{1/2})$-content as it is clearly seen in Fig.9.24 [183]. With a further increase in the content of $Pb(Yb_{1/2}Nb_{1/2})O_3$ the degree of diffuseness of the phase transition initially increases, reaching its maximum value in the region of x ≈ 0.5 [184], and decreases after that. The transition is sharp and relaxation properties are not manifested in solid solutions with a large content of $Pb(Yb_{1/2}Nb_{1/2})O_3$ and in the $Pb(Yb_{1/2}Nb_{1/2})O_3$ itself. This means that the diffuseness of the PE transition (and relaxor properties) increases in the solid solutions with equal stability of the FE and AFE states from the point of view of their energy and manifests in solid solutions located near the phase FE – AFE boundary in the $x-T$ diagram of phase states.

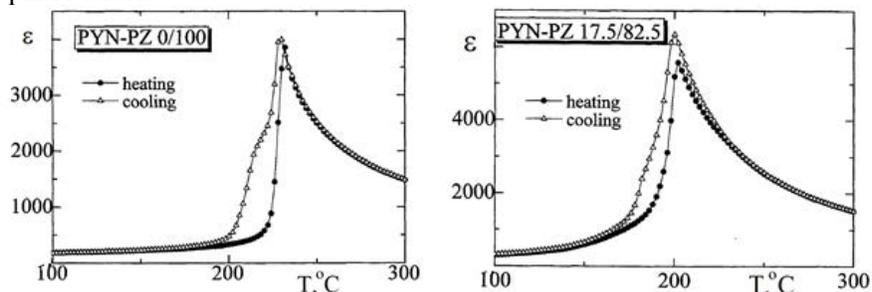

Fig.9.24. Dielectric constants as a function of temperature for PYN-PZ 0/100 and 17.5/82.5 [183].

Now let us consider $Pb(In_{1/2}Nb_{1/2})O_3$-based solid solutions. We mentioned above the ideas that were developing in the literature stating that the introduction of additional third ion into *B*-sites of the lattice of substances discussed here supposed to lead to an increase in the degree of the lattice disorder and, consequently, to an increase in the phase transition diffuseness. As we have demonstrated during the discussion of phase transitions in the $(1-x)Pb(Yb_{1/2}Nb_{1/2})O_3-xPbTiO_3$ system of solid solutions available experimental data do no agree with such statement. Another example of disagreement of the mentioned concept one can find in [185]. The authors of this paper studied the $(1-x)Pb(In_{1/2}Nb_{1/2})O_3 - xPbZrO_3$ system of solid solutions. They also noted that that for the concentrations of $PbZrO_3$ x < 0.3 a diffuse phase transition was observed, but the maximum of dielectric constant increased, and the degree of the phase transition diffuseness decreased with an increase of *x*. Fig.9.25 shows one of the results of [196], namely, the temperature dependencies of dielectric constant for 10 and 50% of $PbZrO_3$. At one can see, the degree of diffuseness for compound with 50% of Zr substitution is much smaller than that for 10% of Zr substitution.

Fig.9.25. Temperature dependence of the real part e; of the complex relative permittivity of solid-solution (l-*x*)PIN-*x*PZ at *x* = 0.10 (in the left) and *x* = 0.5 (in the right) for different frequencies [185].

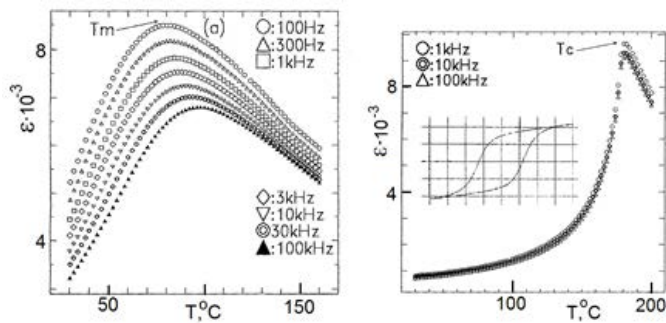

The results of studies of dielectric properties and phase transitions in the $xPb(In_{1/2}Nb_{1/2})O_3-(1-x)PbTiO_3$ solid solutions one can find in [186]. Experimental results presented in this paper totally reject the statement that the introduction of additional third ion into *B*-sites of the crystal lattice of substances in question leads to an increase in the degree of disorder in the lattice and, consequently, to an increase in the diffuseness of the phase transition. Fig.9.26 shows the behavior of dielectric constant in the vicinity of the PE phase transition. It is clearly seen that the introduction of Ti-ions in the crystal lattice leads to a decrease in diffuseness but not on the contrary.

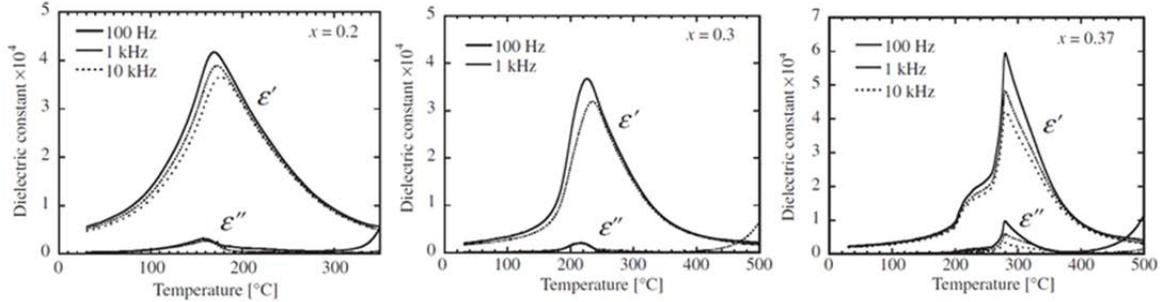

Fig.9.26. Temperature dependence of dielectric constant in the (111) direction in PIN–*x*PT single crystals [186].

Such behavior was also observed in the (1-x)PMN-xPbZrO$_3$ solid solutions [161] and the results are shown in Fig.9.17.

In the above-discussed example the introduction of titanium ions into the Pb(In$_{1/2}$Nb$_{1/2}$)O$_3$ lattice shifts the location of the solid solutions in the x-T phase diagram towards the region of the spontaneous FE phase (away from the FE-AFE phase boundary), thus resulting in the suppression of the relaxor properties and in promoting the manifestation of the ordinary FE properties. Moreover, an electric field raises the stability of the FE phase and must leads to a decrease of diffuseness of the PE phase transition for solid solutions studied in [186] and experiments have confirmed this behavior.

The properties of the Pb(In$_{1/2}$Nb$_{1/2}$)O$_3$ - Pb(Yb$_{1/2}$Nb$_{1/2}$)O$_3$ system of solid solutions were considered in [187]. In this case the second component is antiferroelectric, and an increase in its content leads to the displacement of the solid solution in the phase diagram from the FE-AFE phase boundary towards the region of the spontaneous AFE state (note that Pb(Yb$_{1/2}$Nb$_{1/2}$)O$_3$ is in the AFE state at temperatures lower than ~ 302 °C). As one can see from Figs. 2, 3, and 4 in [187], the relaxor properties are being suppressed in the process of moving away from the FE-AFE phase boundary in the diagram of phase states. The behavior of the PE phase transition under the action of hydrostatic pressure will be considered latter.

Thus, the results of last cited three papers [185 − 187] demonstrate that an increase in stability of both FE and AFE phases in Pb(In$_{1/2}$Nb$_{1/2}$)O$_3$-based solid solutions due to the second component (PbTiO$_3$, PbZrO$_3$, and Pb(Yb$_{1/2}$Nb$_{1/2}$)O$_3$, respectively) leads to a decrease in a degree of diffuseness of the phase transition. Such situation is caused only by the fact that the position of solid solutions in the diagram of phase states shifts away from the state of equilibrium of both dipole ordered phases – away from the boundary between the regions of the FE and AFE states.

It should be emphasized that in all considered examples the relaxor properties are suppressed despite the presence of an additional ion (Zr, Ti or Yb) in the B-sites of the crystal lattice. In the approaches based on the relaxor ferroelectric model [123] it is assumed that the introduction of an additional ion into a crystal lattice increases the compositional disorder and strengthens the relaxor properties. As one can see, such interpretation contradicts the presented experimental results.

Thus, the considered data show that the relaxor character of behavior of the $Pb(B'_{1/2}B''_{1/2})O_3$-type compounds is defined by the location of a given solid solution composition in the diagram of phase states. The degree of manifestation of the said behavior depends on the position of a given compound with respect to the boundary separating the regions of the FE and AFE states in the phase diagram. Therefore, the relative stability of the FE and AFE states of the substance in question may be considered as a decisive factor for its relaxor behavior. If the stability of these states is the same (this signifies the coexistence of the FE and AFE domains in the sample's volume), the relaxor properties manifest themselves completely.

The confirmation of this statement will be given below using the consideration of experiments performed under hydrostatic pressure which effectively changes relative stability of the FE and AFE states (the change of the relative stability of the FE and AFE states under external actions is discussed in [37, 50, 63] in detail).

Now let us consider few results obtained in experiments using the hydrostatic pressure. An opinion that hydrostatic pressure facilitates the transition into relaxor state became established in the physics of ferroelectrics after papers [188] (see also references in [189, 190] were published. Using an example of the $Pb(B'_{1/2}B''_{1/2})O_3$-type compounds we will show that this is not always true. Furthermore, in these compounds the so-called relaxor properties are the consequence of the realization of the AFE and FE states.

If the composition of a solid solution is located near the FE-AFE boundary in the diagram of phase states, but on the side of the AFE states, a hydrostatic pressure increases the stability of the AFE state in this solid solution and its position moves away from the said boundary. In this case relaxor properties at the PE phase transition become weaker. If a the composition of a solid solution is located on the side of the FE states near the FE-AFE boundary a hydrostatic pressure leads to a decrease in stability of the FE state in this solid solution (its position moves toward the said boundary) and relaxor properties should be more pronounced. For the PZT-based solid solutions the validity of such statement was demonstrated in Chapter 7 of this review. Now we will show that in $Pb(B'_{1/2}B''_{1/2})O_3$-type oxides this effect takes place also.

The behavior of the $(1-x)Pb(In_{1/2}Nb_{1/2})O_3-xPb(Yb_{1/2}Nb_{1/2})O_3$ and $(1-x)Pb(In_{1/2}Nb_{1/2})O_3-xPbZrO_3$ solid solutions under the action of a hydrostatic pressure has been investigated in [187, 191]. The main result is presented in Fig.9.27. This figure shows the $\varepsilon(T)$ dependencies obtained at different pressures for solid solutions with compositions form both sides of the AFE-FE boundary. In the case of the $(1-x)Pb(In_{1/2}Nb_{1/2})O_3-xPbZrO_3$ solid solution (with 34% of $PbZrO_3$) an increase in pressure leads to the enhancement of relaxor behavior, whereas in the $(1-x)Pb(In_{1/2}Nb_{1/2})O_3-xPb(Yb_{1/2}Nb_{1/2})O_3$ (with 16% of $Pb(Yb_{1/2}Nb_{1/2})O_3$) an increase in pressure leads to the suppression of a relaxor behavior during the PE phase transition. If ones tries to consider the behavior of both these solid solutions under the action of the pressure simultaneously, the described-above results are completely contrary to the provisions of [188 - 190], because in the first case the pressure leads to an enhancement of relaxor properties but in the second case induces their suppression. However, based on the concept of behavior of the substance with the equal stability of the FE and AFE phases, presented in [28, 32, 105] above-discussed results are absolutely clear. The $0.66Pb(In_{1/2}Nb_{1/2})O_3-0.34Pb(Yb_{1/2}Nb_{1/2})O_3$ solid solution is in the FE state at the temperatures below the Curie point (for example, see Fig.9.27). The pressure applied to this substance increases the stability of the AFE state (that is the pressure shifts the position of this solid solution in the $x$-$T$ phase diagram toward the FE-AFE boundary). The share of the metastable AFE phase in the volume of a sample increases creating conditions for an increase of the contribution of oscillations of the interphase domain boundaries to the dielectric constant and as a result the relaxor properties are enhanced (see Chapter 8 of the present review).

In the second case, the position of the $0.84Pb(In_{1/2}Nb_{1/2})O_3-0.16Pb(Yb_{1/2}Nb_{1/2})O_3$ solid solution is in the AFE region of the $x$-$T$ phase diagram (or near the FE-AFE boundary on the AFE side) without pressure. The pressure applied to this solid solution increases the stability of the AFE state (shifts the position of the solid solution in the $x$-$T$ phase diagram away from the FE-AFE boundary) as a result the share of unstable domains with the FE state decreases which leads to a smaller contribution of oscillations of the interphase domain boundaries to the dielectric constant. Relaxor properties of this solid solution are suppressed at a pressure of 0.6 GPa − position of $T_m$ is only independent on the field frequency. Moreover in this case as one can see from the right part of the Fig.9.27 the pressure separates the FE and AFE states and displaces the FE states to the side of lower temperatures.

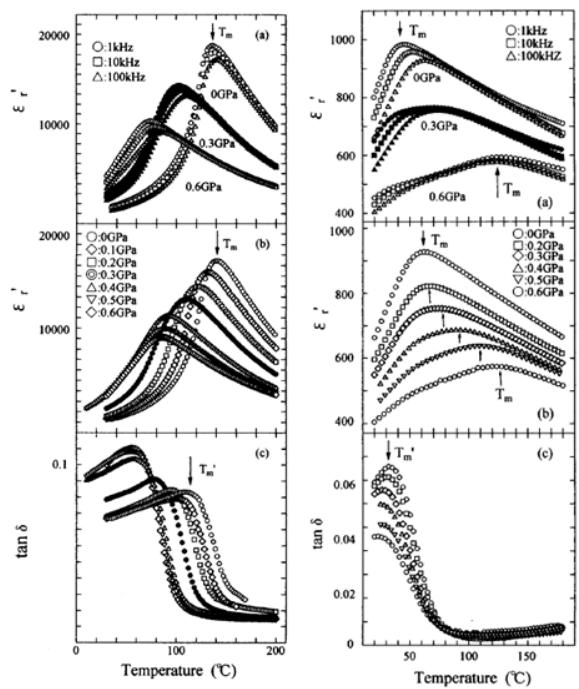

Fig.9.27. Om the left: The temperature dependence of (a) $\varepsilon'_r$ at different frequency and pressure, (b) $\varepsilon'_r$, and (c) tanδ at 100 kHz at various pressures for Zr-PIN at $x = 0.34$ [191].

On the right: The temperature dependence of (a) $\varepsilon'_r$ at different frequency and pressure, (b) $\varepsilon'_r$, and (c) tanδ at 100 kHz at various pressures for Yb-PIN ($x = 0.16$) [191].

It was reported in [192, 193] that the pressure-induced structural phase transition from the FE relaxor state to the AFE state takes place at the pressure $P = 0.4$ Gpa (see. Fig.9.28) in the disordered (with the degree of order $s \sim 0.4$) lead indium niobate (the phase diagram for $Pb(In_{1/2}Nb_{1/2})O_3$ is given in Fig.9.20).

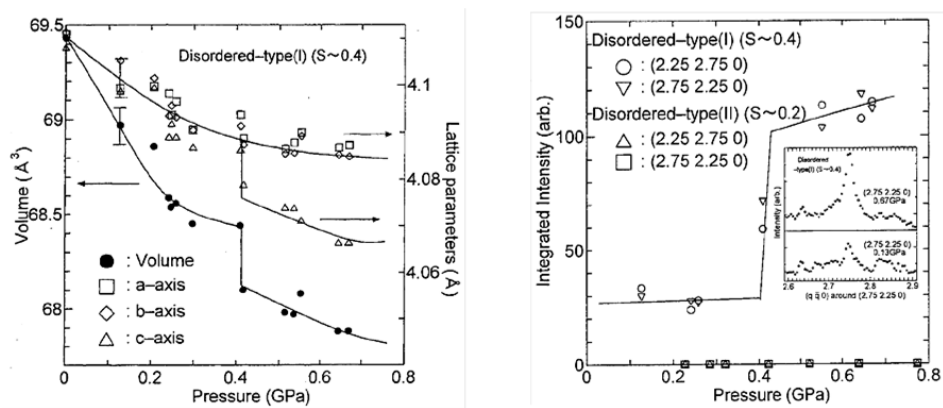

Fig.9.28. The pressure dependencies of the crystal sell parameters for $Pb(In_{1/2}Nb_{1/2})O_3$ disorder-type crystals at room temperature (on the left), and of the integrated intensities of (h/4 k/4 0) reflections, associated with the shift of lead cations in antiparallel (on the right) (Figures 3 and 5 of [192]).

Pressure induced changes of this crystal dielectrics properties were studied in [193]. As one can see from Fig.9.29, the relaxor properties are preserved up to a pressure of 0.4 GPa until the position of the crystal in the phase diagram is located in the vicinity of position where the energy equilibrium for the FE and AFE phases takes place.

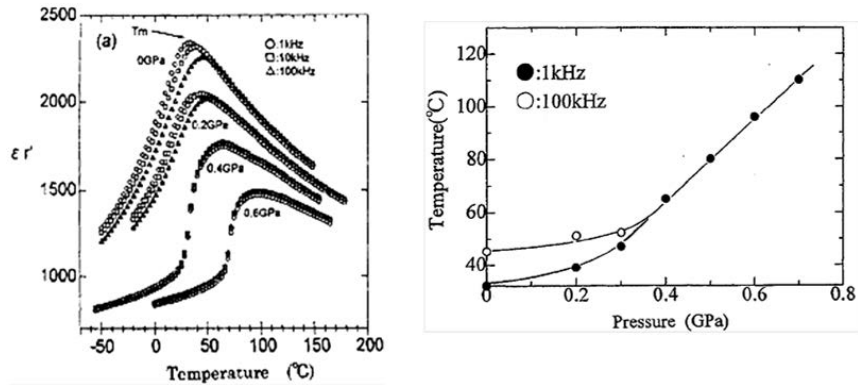

Fig.9.29. The temperature dependencies of dielectric constant at different frequencies and pressure for disordered crystal Pb(In$_{1/2}$Nb$_{1/2}$)O$_3$ [193] (on the left). The pressure dependence of $T_m$ indicating $\varepsilon'_m$ for 1 and 100 kHz (on the right).

In conclusion of this section we consider the (1-x)Pb(Fe$_{1/2}$Nb$_{1/2}$)O$_3$ – xPb(Mg$_{1/2}$W$_{1/2}$)O$_3$ system of solid solutions. Depending upon the composition this system can be in the FE state (at the rich content of Pb(Fe$_{1/2}$Nb$_{1/2}$)O$_3$) or in the AFE state (at the rich content of Pb(Mg$_{1/2}$W$_{1/2}$)O$_3$) at the low temperatures [194, 195]. The boundary between the regions of the FE and AFE states in the x-T phase diagram is located in the vicinity of a 55% concentration of the Pb(Mg$_{1/2}$W$_{1/2}$)O$_3$ (see Fig.10.12).

Fig.9.30. Composition-temperature phase diagram of the (1-x)Pb(Fe$_{1/2}$Nb$_{1/2}$)O$_3$ – xPb(Mg$_{1/2}$W$_{1/2}$)O$_3$ system of solid solutions [194, 195]

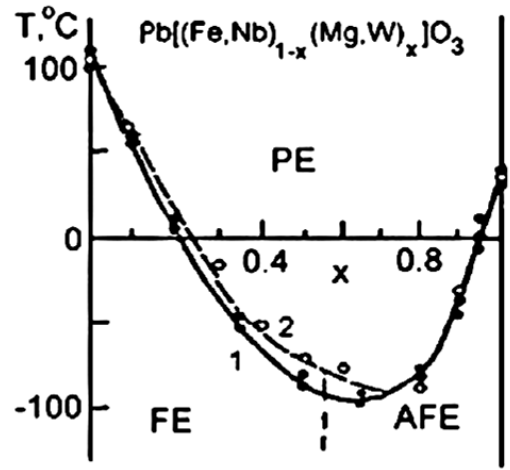

The maximum diffuseness of the paraelectric phase transition (and maximal manifestation of relaxor properties) corresponds to this region of concentrations. In solid solutions with compositions displaced from the said borderland in the x-T diagram the PE transition is not diffuse.

To complete the discussion of the topic in question let us dwell on properties of the Pb(Co$_{1/2}$W$_{1/2}$)O$_3$. In this compound the AFE state is realized at first (at the temperatures close to 298° C) at lowering the temperature form the PE phase and then the FE state is observed (at $T \sim$ 235°). The AFE state is somewhat complicated by incommensurate modulation of the structure, but this is insignificant for our analysis. The behavior of this compound under the hydrostatic pressure is investigated in [196, 197]. The "pressure-temperature' (P−T) phase diagram of Pb(Co$_{1/2}$W$_{1/2}$)O$_3$ obtained in these papers is given in Fig.9.31.

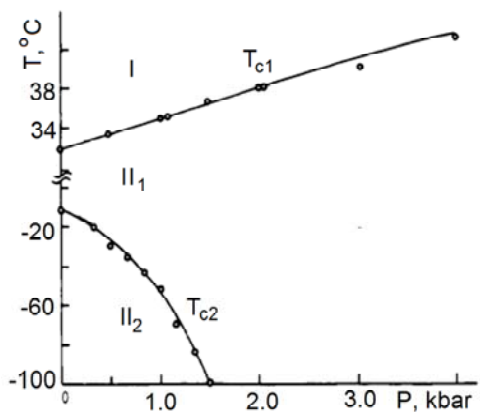

Fig. 9.31. Pressure-temperature phase diagram of Pb(Co$_{1/2}$W$_{1/2}$)O$_3$ [196].

It is similar to that of the 6/100-Y/Y PLZT series (Chapter 4) (naturally, one must take into account the shift of the diagram along the axis of pressures when Y is changed in PLZT). At a pressures of the order of 1.25−1.6 kbar the sequence of transitions and their dependence of the sample's history in the Pb(Co$_{1/2}$W$_{1/2}$)O$_3$ are analogous to those in the (6-9)/65/35 PLZT solid solutions which are called relaxors. Fortunately, the P-T phase diagram for the lead cobalt tungstate is available, and on its base the phase states can be identified at any variation of external parameters.

Finalizing our discussion we have to outline that the considerable attention has been paid to the spontaneous transition from the relaxor state to the FE state in discussions of properties of the $Pb(B'_{1/2}B''_{1/2})O_3$-type compounds in the literature. Such transition for the PLZT system of solid solutions has been considered in detail in Chapters 3 – 8 of the present review. In the compounds discussed in this section the nature of this transition is the same: a weak difference between the thermodynamic minima corresponding to the FE and AFE states.

All above-presented results corroborate the proposed model based on the coexistence of domains of the FE and AFE phases.

## 10. Coexistence of the ferroelectric and antiferroelectric states and "dipole-glass" behavior

### 10.1. Model concepts

The concept of "dipole glass" was introduced by analogy with the "spin glass" concept that was actively studies in physics of magnetism in the mid-1970s [198, 199, 200]. The "dipole glass" concept was used in discussions of experimental investigations of the "relaxor ferroelectrics".

The substances that are usually attributed to "dipole glasses" possess the following main properties:

1) An essential dispersion of the dielectric permittivity in the region of diffuse maximum of the $\varepsilon'(T)$ and $\varepsilon''(T)$ dependences and the fulfilment of the Vogel-Fulcher law:

$$\omega = \left(\frac{1}{\tau_0}\right)\exp\left[-\frac{\Delta}{k(T_m - T_f)}\right],$$

where $T_m$ is the temperature corresponding to the maximum of the $\varepsilon'(T)$ and $\varepsilon''(T)$ dependencies $(T'_m$ or $T''_m)$, $T_f$ is an effective temperature different for $(T'_m$ and $T''_m)$;

2) An increase in the frequency of an AC measuring field causes a decrease of the maximum value of the real component of permittivity, whereas the maximum value of the imaginary component increases, the temperature of the maximum of the $\varepsilon'(T)$ dependence being always higher than that of the $\varepsilon''(T)$ maximum;

3) An increase in the measuring field amplitude $E_0$ leads to a linear decrease of $T'_m$ and to an increase of $\varepsilon'(T'_m)$; the slope of the $T'_m(E_0)$ dependence decreases with increasing AC field frequency;

4) Hysteresis loops of these substances have a specific shape: narrow dielectric hysteresis loops with a small residual polarization and narrow quadratic loops of electro-optic hysteresis;

5) The presence of effects pointing to the existence of micro/nano domains of the polar phase at temperatures essentially higher than $T'_m$ and $T''_m$;

6) The dependence of properties of these substances on the sample's history;
7) The long-duration relaxation;
8) The high degree of diffuseness of the paraelectric phase transition.

The above-listed properties may be somewhat varied, or may manifest themselves not in the complete set. As a rule, the fulfilment of the first, sixth and seventh condition is dominating.

As was shown in the Chapters 3-9 of the present paper, the above set of properties is typical for the substances in which the FE and/or AFE types of dipole ordering may be realised, and the difference in free energies of these types of ordering is small (i.e. the FE-AFE phase transition may take place under the action of such external factors as temperature, field or pressure).

In this section, we will focus our main attention on the process of the long-time relaxation of properties and physical characteristics of these substances after their state of thermodynamic equilibrium was disturbed by external influences. This long-time relaxation along with the pronounced frequency dependence of parameters (for example, dependence of dielectric or magnetic characteristics on the frequency of measuring field) is considered as a characteristic feature of the spin or dipole glass. The substances manifesting such properties are referred to as "glasses" practically immediately.

We will demonstrate that the specified above peculiarities in physical behavior are manifested in the substances in which the FE and AFE phases are realized and the difference in their free energies is small. After that, we will discuss the possible phase diagrams of substances that can be misguidedly attributed to the class of "glassy materials". By no means have we wanted to cast doubt on the existence of "dipole glasses" in the nature. Our main purpose is only to call attention to the fact that one has to be very careful during the interpretation of experimental results in substances in which the inhomogeneous states with negative energies of interphase boundaries can take place. In fact, the peculiarities in behavior of such systems can be misleading during the interpretation. More over the existence of such states (stable in the absence of external influences) was considered impossible in a wide class of substances until recently.

It is shown in Chapters 6, 7 that the formation of the heterophase structure of coexisting domains of the FE and AFE phases is accompanied by the emergence of the chemical inhomogeneity of the substance. Here we want particularly emphasize that just the phase inhomogeneity (the coexistence of the FE and AFE phases) leads to chemical segregations, but not on the contrary.

The process of chemical segregation in the vicinity of the FE-AFE phase boundary can take place only at those temperatures when the dipole ordered FE and AFE states exist and coexist. Driving forces of chemical segregation are absent at the high temperatures (when the uniform PE state is present in the substance) and the annealing of the substance leads to the chemical homogeneity. Thus, the considered process of chemical segregation in the vicinity of the FE-AFE phase boundary is reversible in the course of temperature cycling during cooling and heating.

If the inter-domain wall (IDW) is displaced under the action of electric field (or appears after cooling from the PE state), then the process of chemical segregation will occur in the vicinity of the position of this new IDW. "Old" chemical segregations will be cleaned out. However, not only the IDW itself controls the process of chemical segregation in the vicinity of its location. There exists an influence opposite to the chemical segregation on the IDW during its motion (see Eq.8.1). The said segregation becomes a mobile defect of the crystal lattice. The interaction between the segregation and the IDW should be the most pronounced when the formation of such segregation is accompanied by a violation of local electroneutrality.

The mentioned interrelation between the IDW and the formation of the mobile defects of the crystal lattice causes peculiarities of the IDW dynamics. The rate of ionic diffusion is low at the temperatures, at which the FE and AFE states are realized. In Chapter 8, we have discussed the relaxation dynamic of the IDW and have obtained Eq.8.8 describing its motion. This equation contains the time-function $G(t) = 1 - \exp(-t/\tau)$. In Chapter 7 we adopted condition $G = 1$ (for $t \to \infty$). In view of the fact that the above mentioned process of chemical segregation is a long-duration one, the value of $G = 1$ in Eq. (8.8) has to be substituted by the time function $G(t)$, since the condition $t \to \infty$ is not fulfilled in real experiments. In this case more complicated expressions for $\varepsilon'$ and $\varepsilon''$ are obtained instead (8.12) [201]:

$$\varepsilon'(\omega,T) = \varepsilon_0 \frac{1+\omega^2\tau^2[1+\eta G(t)]}{1+\omega^2\tau^2[1+\eta G(t)]^2}, \qquad (10.1a)$$

$$\varepsilon''(\omega,T) = \varepsilon_0 \frac{\omega\tau\eta G(t)}{1+\omega^2\tau^2[1+\eta G(t)]^2}. \tag{10.1b}$$

Under experimental conditions (when the period of oscillations of the measuring field is much shorter than characteristic time of diffusion processes) $\omega\tau \to \infty$ we have:

$$\frac{\varepsilon'}{\varepsilon_0} = \frac{1}{1+\eta G(t)}. \tag{10.2}$$

As one can see from this expression, the dielectric constant reaches its equilibrium value during a long period of time after the action of any factor leading to the shift of the IDWs (or after their appearance during cooling from the high temperatures) because for the temperatures $T < T_c$ coefficients of diffusion are small. For a particular case of $\eta = 1$ Fig.10.1 shows the time dependence of $(\varepsilon'/\varepsilon_0)$ for some values of $\tau$ [201].

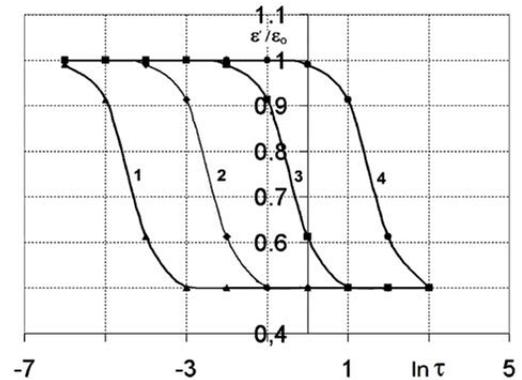

Fig.10.1. Time dependence of the function $\varepsilon'/\varepsilon_0 = [1+(1-\exp(-t/\tau))]^{-1}$ [201] for different values of $\tau$: 1 – 0.0001; 2 – 0.01; 3 – 1.0; 4 – 100.0.

Let us remind (see eq. 8.13) that the relaxation time $\tau$ is determined by the activation energy $\Delta$ for the processes considered here. The lengthy process of change of the chemical composition near the "bare" IDW takes place during the formation of segregates. It is obvious that the change of the activation energy ($\Delta = \Delta(t)$) takes place along with the process of segregates formation. Therefore, the characteristic relaxation time also depends on time ($\tau = \tau(t)$). A small number of experiments on the establishment of dependency $\tau(t)$ has been done up to now. We will use the results of investigation of the long-time relaxation given in Chapter 6 (see Fig. 6.5 and 6.6). The process of formation of segregates along the IDWs takes place during more than 30 hours.

Time dependencies of the intensities of diffuse X-ray lines for the 15/77/23 PLLZT and 6/73/27 PLZT solid solutions corresponding to the time interval of formation of segregates are presented in Fig. 10.2 (the logarithmic scales are chosen along both axes) [201].

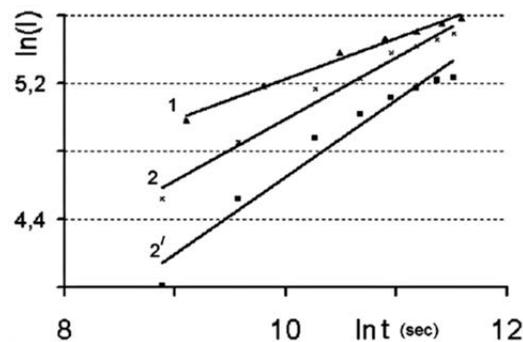

Fig.10.2. Time dependences of integral intensities of diffusive X-ray lines for 15/77/23 PLLZT (1) and 6/73/27 PLZT (2 and 2′) solid solutions [201].

As one can see in this figure, the linear dependence is observed with a high level of accuracy. It clearly demonstrates that the time dependence of the relaxation time has a power character $\tau(t) = At^n$ with $n < 1$. Hence, the time dependence of dielectric permittivity in the process of aging (in the course of formation of segregates in the process of their growth) is given by the formula

$$\frac{\varepsilon'}{\varepsilon_0} = \frac{1}{1+\eta[1-\exp(-t^{1-n}/A)]} \quad (n<1). \tag{10.3.}$$

The dependencies described by the formula (10.3) are presented in Fig. 10.3 [201].

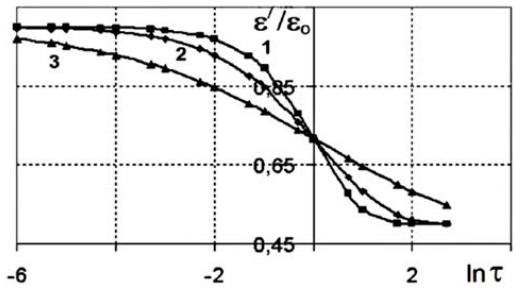

Fig.10.3. Time dependence of function

$$\varepsilon'/\varepsilon_0 = \left[1 + \left(1 - \exp(-t^{(1-n)}/\tau)\right)\right]^{-1}$$ for different $n$: 1 – 0.4, 2 – 0.6, 3 – 0.8 [201].

Relaxation curves typical for spin glass systems are presented in Fig. 10.4a and 10.4b. These curves describe the aging experiments measured after the switching of the external field and after heating up (the data is taken from the review [202]).

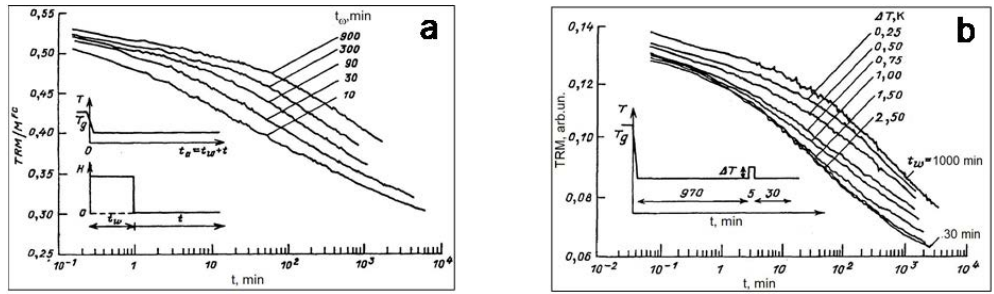

Fig.10.4. Magnetic relaxation in aging experiments [202]:
(a) – after magnetic field was switched off;
(b) –experiments with heating cycle.

As one can see, the dependence (10.3) completely describes the time behavior of the systems, which are referred to as glasses in the process of aging. It has to be stressed that the local decomposition of the solid solution caused by local mechanical stresses in the vicinity of IDWs was the physical factor determining the shape of dependency (10.3).

## 10.2. Experimental results for perovskite oxides

All results of experimental studies presented in this section were obtained on substances possessing the phase diagrams containing the regions of the coexisting domains of the FE and AFE phases. These phase diagrams are given in previous sections of this paper and will be used without references in what follows unless otherwise specified.

The aging of the PLZT ceramic samples with composition 10.5/65/35 after quenching from high temperatures was investigated in [203]. We selected this study as an example due to the following reason. It was unambiguously shown that in the samples with similar composition and the same location in the phase diagram (see Chapter 6 of present paper) the local decomposition of the solid solution and the formation of segregates near the IDWs separating domains of the FE and AFE phases take place after quenching from high temperatures. As one can see in Fig.10.5 the real aging process may well be described by the formula (10.3). Moreover, we have to ascertain that the expression (10.3) can describe even more protracted relaxation processes.

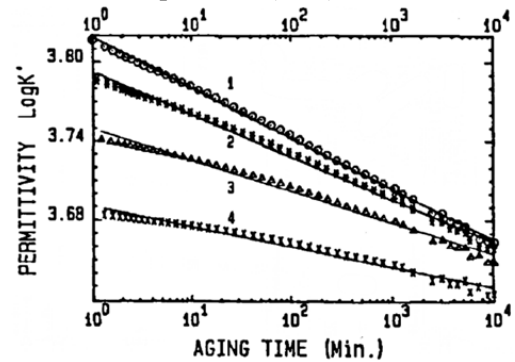

Fig.10.5. The permittivity as function of aging time for quenched (400°C) 10.5/65/35 PLZT samples [203].

Phenomena of the so-called thermal dielectric, field, or mechanical memory that were experimentally observed in "relaxor ferroelectrics" of both the PLZT and PMN types are even more interesting from the point of view of physics. All above-mentioned effects have common physics basis, namely, the local decomposition of solid solutions in the vicinity of IDWs. Therefore, we will discuss in detail only the phenomenon of thermal dielectric memory. This effect implies that when the samples are cooled to some particular temperature $T = T_{age}$ after annealing at high temperatures and are subjected to aging (during the time interval of 20 hours and longer) at this temperature a specific characteristic feature is observed on the $\varepsilon(T)$ dependencies at following cycling. This feature consist in a "drop-shaped" behavior of the $\varepsilon(T)$ dependence in the vicinity of the aging temperature $T_{age}$. Typical experimental results are presented in Fig. 10.6 [203, 204].

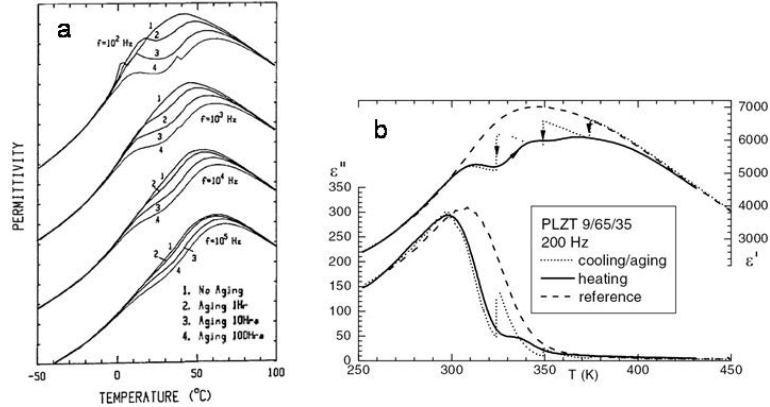

Fig. 10.6. (a) – Temperature dependences of dielectric constant for different aging time τ at ~ 23 $^{o}$C for 10.5/65/35 PLZT samples annealed at 400 $^{o}$C [203];
(b) – Dielectric permittivity of 9/65/35 PLZT samples at 200 Hz during multiple aging stages (24 h each) and subsequent heating curves with memory, compared with the reference curve measured on continuous cooling [204].

All above-presented results can have simple unambiguous explanation if one takes into account the inhomogeneous structure of coexisting domains of the FE and AFE phases present in the sample. Domain structure is formed in the bulk of the sample after the high-temperature annealing. The shares of each phase (and consequently the sizes of domains) are determined by the relation between the free energies of these phases. Diffusive local decomposition of the solid solution constantly takes place in the vicinity of the interdomain boundaries. However, this is the long time process and it almost does not manifest itself in material parameters. This influence is extremely small due to the fact that the characteristic time of measurement (during the temperature measurements), which is of the order of several degrees Celsius per minute, is much shorter than the characteristic relaxation times of the diffusion processes (which are of the order of tens of hours). The same is true for the external field or mechanical stress measurements. During aging at the temperature $T_{age}$, which lasts for long time (tens of hours), the local decomposition of the solid solution takes place and the structure of the formed segregates repeats the structure of the interphase boundaries. The longer is the aging time the more pronounced is the spatial structure (the distinctive network) of segregates and the larger volume of the sample is occupied by segregates. The dielectric permittivity is reduced as a result (Fig.10.6). The formed spatial structure of segregates is conserved during the temperature cycling (with the several degrees per minute rate of the temperature change) following the aging.

The relation between the shares of each phase and the sizes of domains of these phases are constantly changing in the process of temperature cycling. The sizes and the structure of domains coincide with the preserved spatial structure of segregates when the temperature $T_{age}$ is achieved. At this moment, the effective pining of interphase boundaries takes place and the contribution of the oscillations of interdomain boundaries, which determine the main contribution to the dielectric permittivity decreases (see Chapter 8). This pining of the interdomain boundaries leads to the appearance of the "drop-shaped" feature on the dependence $\varepsilon(T)$ (see Fig 10.6).

The nature of the effect of dielectric memory at the temperatures both above $T_m$ and below $T_m$ is the same as it clearly seen from the results of [204] (see also [205]). However, the authors of numerous papers who were trying to connect this phenomenon with relaxor ferroelectrics were forced to come up with different mechanisms for these two cases. One does not need to do that using the model suggested here. As shown in Chapters 6, 7 and 8 coexisting domains of the FE and AFE phases exist both above $T_m$ and below $T_m$ (let us remind that the X-ray studies carried out above $T_m$ demonstrated the coexisting domains of the FE and AFE phases constitute the two-phase FE+AFE domains). Thus, the IDW structure that determines the effect of the dielectric memory exists both below $T_m$ and above $T_m$.

The shares of the AFE and FE phases and, thus, the spatial structure of IDWs can be also changed by means of other thermodynamic parameters such as external field and mechanical stress. The long time aging of the samples at some nonzero values of the above-mentioned parameters will lead to formation of the new spatial structure of segregates and the effects of the filed on mechanical dielectric memory will be manifested during the cyclic changes of these parameters after the aging [206, 207, 208]. These effects are presented in Fig.10.7 and Fig.10.8.

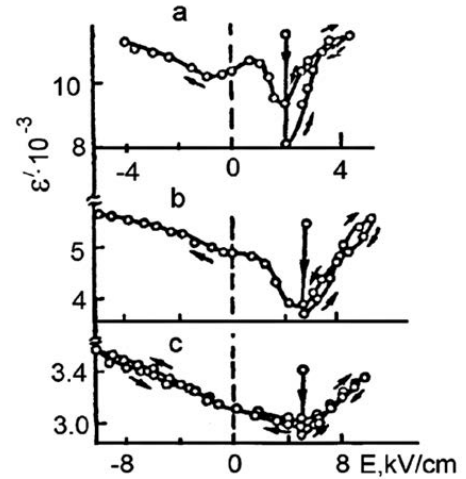

Fig.10.7. Dependencies of dielectric constant on a DC electric field in the PLZT solid solutions after aging during 20 h [206]. The solid solutions compositions: a – 8/65/35, $T_{age}$ = 57 °C; b – 11/65/35, $T_{age}$ = 22 °C; c – 13.5/65/35, $T_{age}$ = 22 °C.

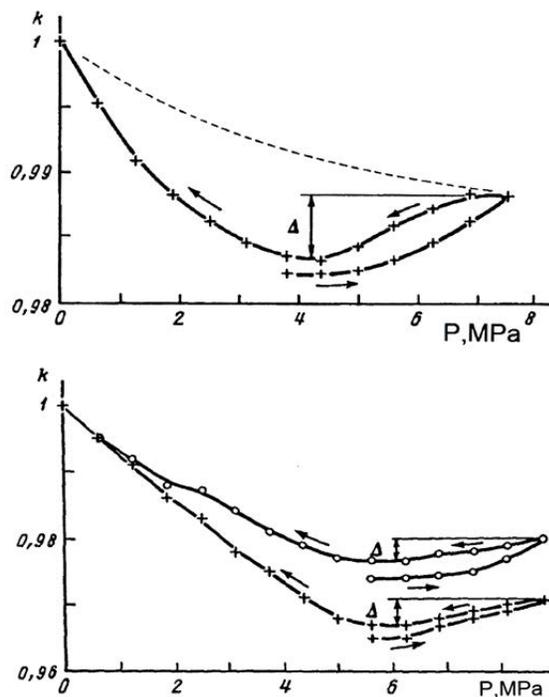

Fig.10.8. Dependencies of the normalized dielectric constant on uniaxial stress after aging during 20 h for the 8/65/35 PLZT (on the top) [207]. Dashed line shows the dependence, obtained without aging (added by authors of the present paper). Dependencies of the normalized dielectric constant on uniaxial stress after aging during 48 h (1) and 120 h (2) in DC electric field $E$ = 1850 V/cm, ($T_{age}$ = 27 °C, $P_{age}$ = 3.8 MPa) for 8/65/35 PLZT (on the bottom).

The effects discussed above are as a rule attributed to the so-called dipole-glass-like behavior in relaxor systems. However, based on this dipole-glass approach it is impossible to describe and to explain all manifestation of other phenomena (see previous chapters). We demonstrated that rather simple explanation could be given if one takes into account the coexistence of the AFE and FE phases.

We have considered above the effect of dielectric memory using the PLZT family of solid solutions as example. Similar effect is also characteristic of the PMN-based solid solutions. The result obtained for single crystals with composition $(1-x)Pb[(Mg_{1/3}Nb_{2/3})O_3 - xPbTiO_3$ with x = 0.10 and 0.12 one can find in [205] (the phase diagram for the PMN−PT system is presented in Fig. 9.2 and is analogous to the experimental PLZT phase diagram presented in Fig.4.4b and the model diagram in Fig.4.3). The above-mentioned results for PMN−PT are analogous to the results for the PLZT given above. Some examples for different aging temperatures (both above and below $T_m$) are shown in Fig.10.9. The physical mechanism of the effect is the same as in the PLZT and it was discussed above.

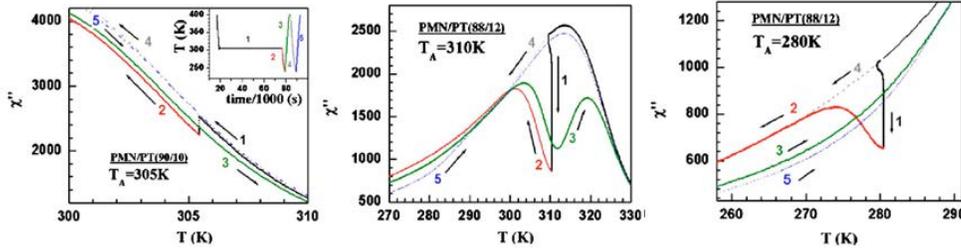

Fig.10.9. Dielectric memory effect in $(1-x)Pb[(Mg_{1/3}Nb_{2/3})O_3 - xPbTiO_3$ solid solutions [205]. In a typical aging memory experiment (see inset for temperature history profile) the sample is cooled to $T_A$ (curve 1). After aging at $T_A$, the sample is cooled to a lower excursion temperature $T_{EX}$ (curve 2) and then immediately reheated past $T_A$ (curve 3). For comparison, reference starting curves (4 and 5) are also taken at the same sweep rates without stopping at $T_A$.
Compositions of solid solution: a – 90/10, b – 88/12, c – 88/12;
$T_A$, K: a – 305, b – 310, c – 280.

## 10.3. Peculiarities of the "composition-temperature" diagrams of the order-disorder KDP-type systems

All up to date concepts about the behavior of the KDP-type systems are based on microscopic calculations that are based on the Ising Model taking into account proton tunneling. A great number of results that give rather good description of the above systems have been obtained using this model. One can say nothing negative about this model. However, the Model is the model that contains only those assumptions that were introduced by the original author (or the authors who were introducing subsequent additions and corrections to the original model). The above-said is primarily connected with phase transformation in these systems. In particular, similar models are not practically working in the case of the ferroelectrics and antiferroelectrics with the perovskite crystal structure (although attempts of application of such models have been made).

On the other hand, the phenomenological approach is based on the information about the structure of the substance under consideration and on the thermodynamics. The information about the crystal structure can be known or not known (so some assumptions can be made but need to be justified). It is true that some simplifications are made in the process of construction of expansions for the thermodynamic potentials but in every case, these simplifications are discussed along with the limits of application of results obtained with these simplifications.

It is difficult to determine the mechanisms of the phase transition in the frames of the phenomenological approach; however, one can obtain additional data about the behavior of the phases in the process of the transition. In this section of our review, we will consider experimental data available in the literature in the frames of the phenomenological approach. As it will become clear below moving along this path it is possible to find explanations for results that "were not so very well explained" in the frames of the Ising-like models.

There are solid solutions among the KDP-type systems in which one of the components is a ferroelectrics and the other is an antiferroelectrics. The solid solutions with an equal stability of these phases exist amongst the solid solutions form this series. The coexistence of different dipole-ordered phases in the volume of the samples is possible form the energy point of view. The question about the existence (or absence of such inhomogeneous state) is as a rule solved in the frames of the Landau phenomenological theory of phase transitions. The physics of domain structures (in ferroelectric, magnetic and other systems) is the most well-known example in which the structure of domains and their behavior under external influences are found by means of phenomenological approach. Other approaches simply do not exist.

The behavior of the KDP-type systems in which the energy difference between the ferroelectric and antiferroelectric states is small and the coexistence of domains of these phases is possible is the main object of our interest in this chapter. That is why the analysis of behavior of such systems is carried out in the frames of phenomenological approach leaving aside the discussions about the nature of these ordered states. As it was said above, there is no discrepancy in these matters.

The H-bonded KDP-type family contains both ferroelectrics ($KH_2PO_4$ − KDP, $RbH_2PO_4$ – RDP are examples) and antiferroelectrics ($AsH_2PO_4$ – ADP is an example). The crystal structure of the PE phase of compounds belonging to this family is a body-centered tetragonal (bct) structure with 2 formula units per lattice site. Each P centered tetrahedral group $PO_4$ is H-bonded to four $PO_4$ groups in such a way that, viewed along the tetragonal $z$-axis, its upper oxygen ions are linked to the bottom oxygen ions of two $PO_4$ groups, while its bottom O-ions connect to the upper O-ions of the other two $PO_4$ groups, the H-bonds are nearly perpendicular to the $z$-axis. Along this axis the $(PO_4)^{+3}$ ions alternate with $(NH_4)^+$ ions. In the high temperature PE phase the mean position of each acid H is centered between the oxygens. At the temperatures below $T_C$ the upper and bottom O-ions of each $PO^4$-group are approached by the hydrogen ions, thus becoming polarized perpendicular to the $z$-axis, and the crystal structure becomes orthorhombic. The $PO_4$ dipole moments are aligned in chains, neighboring chains being polarized antiparallel to each other. The structure of this antiferroelectric (AFE) phase is depictured in Fig.10a [209] as seen along the $z$-axis. The $ab$ plane projection of the structure of $NH_4H_2PO_4$ (on the left) and $KH_2PO_4$ (on the right) is shown in Fig.10c. The arrows point along the dipole moment of the $H_2PO_4$-unit.

Three possible long-range orders are generated as described above; two of them correspond to the macroscopically polarized FE phases and the third being the AFE phase of ADP [210]. A fourth long-range ordered structure is achieved when both upper and bottom oxygens of every $PO_4$-group are approached by hydrogens, thus leading to a $z$-polarized FE phase, which is the case of KDP (see Fig.10b). Along standing question was about the reason for a difference between the low-temperature phases of KDP and ADP [211]. It has been shown (see [212]) through first-principles calculations for the ADP that the total energy of a $z$-polarized FE structure is only a few milli-electron-volts higher than that of the AFE, and the formation of N–H–O bridges is crucial in stabilizing this compound. This is consistent with the suggestion, based on results on electron spin probe measurements [213, 214], that the FE and AFE regions coexist in the PE phase as temperature approaches $T_C$ in several ADP-type antiferroelectric compounds.

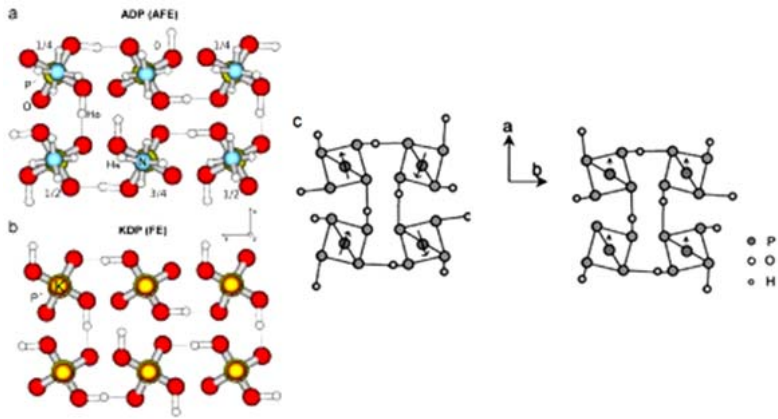

Fig. 10.10. The AFE structure of ADP (a) and FE structure of KDP (b) viewed along the $z$ axis. Acid and ammonia hydrogens of ADP are denoted $H_O$ and $H_N$ respectively. Numbers attached to each $PO_4$ group in (a) indicate its $z$-coordinate in $c$-lattice parameter units. Dashed lines indicate the H-bonds between $PO_4$ groups [209]. The $ab$ plane projection (c) of the structure of $NH_4H_2PO^4$ (on the left) and $KH_2PO_4$ (on the right). The arrows point along the dipole moment of the $H_2PO_4$-unit [88, 215].

The relative energy stability of various FE and AFE structured clusters involving up to 6 formula units embedded in the PE matrix of ADP (for temperatures above $T_C$) was studied using the density functional theory method in [209]. The energies of the FE and AFE clusters were computed as a function of the cluster size. It was shown that FE clusters can be stable although the low-temperature state of the ADP is an antiferroelectric. For clusters with small and large size the energy difference is negligibly small value. The FE and AFE clusters can swap their values of energy under different calculation conditions. Such situation is shown in Fig.10.11 [212].

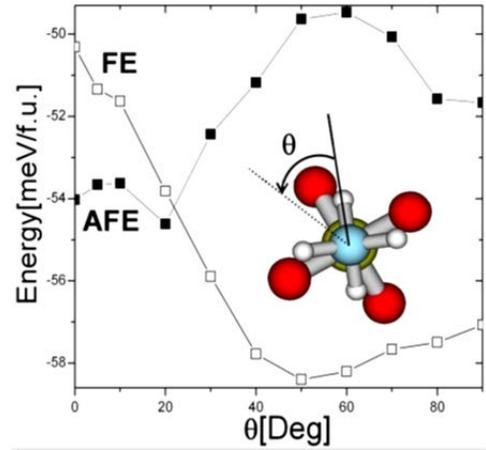

Fig.10.11. *Ab initio* energy of the AFE (solid squares) and the FE (open squares) phases relative to the PE phase, as a function of the rotation angle θ of the $NH_4$- ion [212]. θ = 0 is the ammonium angular position corresponding to the minimum-energy relaxed states in $NH_4H_2PO_4$.

The distortions of the $NH_4$- group turn out to be essential in stabilizing the AFE configuration against a *c*-polarized FE phase, as well as the other FE states polarized perpendicular to the *c*-axis. The energies of the said FE states are only a few milli-electron-volts above the energy of the AFE phase. This explains the observation of the FE-AFE phase coexistence near the AFE transition in antiferroelectric KDP-like structures.

Fig.10.11 is presented here to draw ones attention to the fact that for discussed structures the difference in energies of the FE and AFE states is really small, and the account of additional factors can change their relative stability.

All this clearly demonstrates that the difference in energies of the FE and AFE states in KDP-like structures is very small. As a consequence, all regularities of behavior that have been discussed during the analysis of the perovskite ferroelectrics and antiferroelectrics must be fulfilled in the KDP-like systems.

At first, we will demonstrate that one can give a consistent explanation of the effects that were attempted to be attributed to the dipole-glass state in the so-called relaxor ferroelectrics simply by taking into account the coexistence of the FE and AFE phases. In what follows, we will try to show how this approach can be slightly extended.

Let us consider experimental "composition – temperature" phase diagrams of the solid solutions in which one of the components is a ferroelectric and the other is an antiferroelectric. The phase diagrams of the solid solutions for which the FE and AFE nature of the low-temperature state has been unambiguously identified are shown in Fig. 10.12 [133, 135, 194, 195].

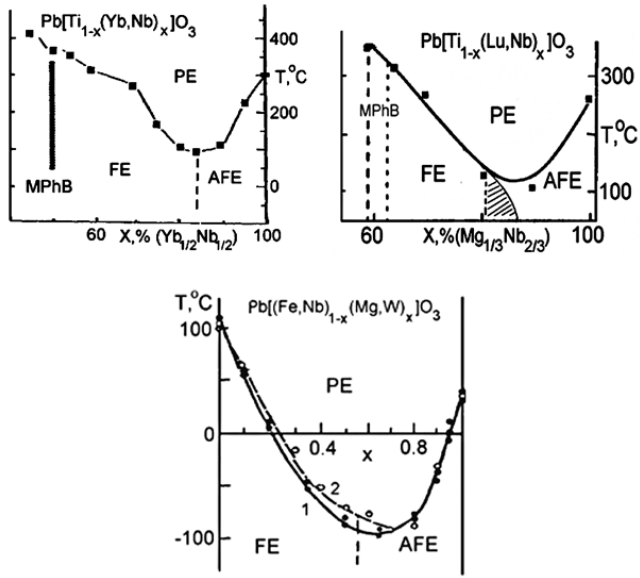

Fig.10.12. Phase diagrams of perovskite solid solutions which have one component being ferroelectric and the other one being antiferroelectric. These phase diagrams have been given in the text above (upper diagrams in Fig. 9.5 and the bottom one in Fig. 9.30) and here we give them for convenience of reading.

The presence of a sharp drop in the $T_c(x)$ dependencies observed in the solid solutions located near the boundary separating the regions with the FE and AFE orderings is a common feature of all presented phase diagrams. Such drop in the value of $T_c$ may reach 200°C and more. It should be also noted, that for all solid solutions with the phase diagrams shown in Fig.10.12 the diffuseness of the PE phase transition increases noticeably as the solid solution composition approaches the FE-AFE phase boundary in the phase diagram. As shown in Chapter 7, such behavior is caused by the existence of the two-phase (FE+AFE) domains in the PE matrix of the substance at $T > T_c$ and the presence of ion segregations in the vicinity of the interdomain boundaries.

In all above-considered cases of oxides with perovskite structure (Fig.10.12) the solid solution components corresponding to the pure FE and AFE states have high values of the Curie point. Due to this, the minimal values in the $T_c(x)$ dependencies do not cryogenic temperatures. The model phase diagram in the case when the FE and AFE components of the solid solution have low values of the Curie temperature is schematically shown in Fig.10.13a. The dependence of the segregation temperature $T_{seg}(x)$ on the solid solution composition is denoted by dotted line. $T_{seg}$ is the temperature at which the two-phase FE+AFE domains start to appear in the PE matrix of the solid solution when temperature of the sample is decreasing from high temperatures. The bell-like dependence of the diffuseness parameter $\delta(x)$ (see Ch.7) is typical for all discussed solid solutions and promotes such behavior of the $T_{seg}(x)$ dependence.

To conclude the consideration of the diagram shown at the top of Fig.10.13, we would like to discuss possible nature of the $T_{quant}(x)$ line in the phase diagram. In our opinion, this feature is clear. To begin with, consider the $(1-x)BaTiO_3 – xSrTiO_3$ system of solid solutions (a lot of attention has been paid lately to $SrTiO_3$-based solid solutions as an example of quantum ferroelectrics). The Curie point of these solid solutions lowers with an increase of the strontium titanate content.

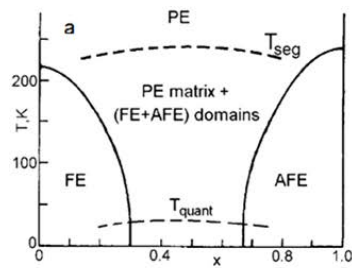

Fig.10.13. Phase diagrams of solid solutions with order-disorder phase transitions. From top to bottom: model diagram;
a – model diagram,
b – $K_{1-x}(NH_4)_xH_2PO_4$ [218];
c – $Rb_{1-x}(NH_4)_xD_2AsO_4$ [219] (the line $T_{seg}$ is added by the authors of the present paper based on [216, 217]);
d – $Rb_{1-x}(NH_4)_xH_2PO_4$ [220].

However, the PE phase transition typical for ordinary FE does not take place at changing temperature in these solid solutions with large $x$-values. When the $T_c(x)$ dependence reaches the region of about (20-30) K further decrease of the Curie point is not observed, and the character of the $\varepsilon(E)$ dependence changes. Such behavior is usually explained by the quantum effects contribution, which determines the nature of the $T_{quant}(x)$ line in the phase diagram presented at the top of the Fig.10.13.

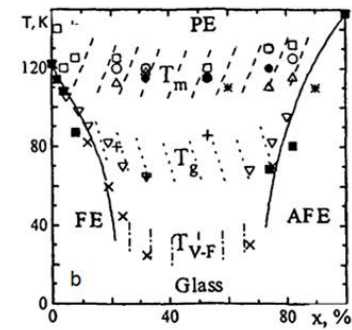

The majority of studies of the dipole-glass state are performed on solid solutions of the following series $K_{1-x}(NH_4)_xH_2PO_4$, $Rb_{1-x}(NH_4)_xH_2PO_4$, $Rb_{1-x}(ND_4)_xD_2AsO_4$. All these compounds fall into the class of order-disorder ferroelectrics. Let us remind that the first component of these solid solutions is a ferroelectric while the second is an antiferroelectric and the difference in energies of the FE and AFE states is small. Experimental phase diagrams of $K_{1-x}(NH_4)_xH_2PO_4$, $Rb_{1-x}(NH_4)_xH_2PO_4$ and $Rb_{1-x}(ND_4)_xD_2AsO_4$ solid solutions are presented in Fig.10.13. These diagrams were constructed using the results of measurements by different methods and presented in [218, 219, 220] (see also a review [221] ). One can clearly see that these experimental diagrams coincide practically completely with the model one shown in the top of Fig.10.13. However, the latter is typical for those substances in which the phases with the FE and AFE orderings are realized and domains of these phases coexist, while the experimental diagrams belong to the substances classified nowadays as "dipole glasses".

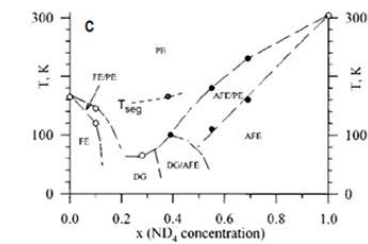

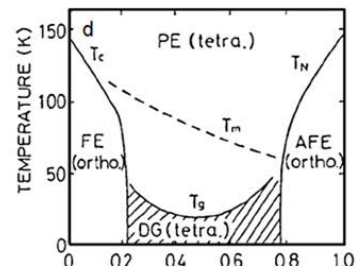

For this reason, it seems expedient to analyze some experimental results of investigations of phase transitions in the $K_{1-x}(NH_4)_xH_2PO_4$ and $Rb_{1-x}(NH_4)_xH_2PO_4$ systems of solid solutions (and other related substances). We are going to consider the results that could not find their explanation in the scope of traditional approach but can be explained from the viewpoint of the effects discussed in the present research. We would like to emphasize that in the majority of cases such experimental results are consistent with the recently developed ideas about the FE-AFE phase transformations with taking into account the interactions between the coexisting phases (interaction between the domains of these phases). Moreover, there exist results the interpretation of which from another viewpoint seems to be artificial.

In this respect, we would like to note the experimental studies of the diffuseness of the PE transition in the $K_{1-x}(NH_4)_xH_2PO_4$ ($0 \leq x \leq 0.24$) solid solutions as an example [222]. These results are presented in Fig.10.14.

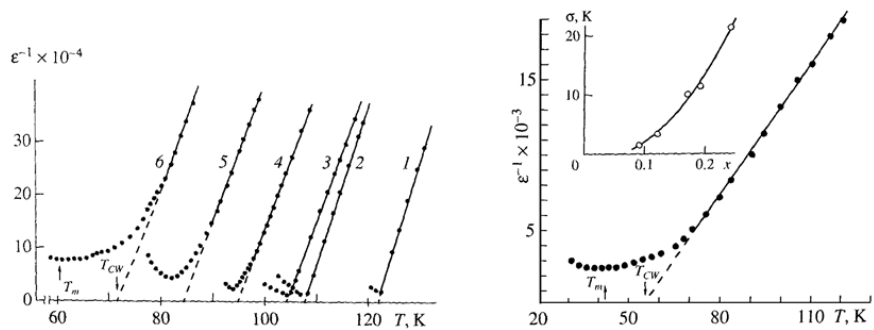

Fig.10.14. On the left: $\varepsilon^{-1}(T)$ dependencies obtained on the crystals of the $K_{1-x}(NH_4)_xH_2PO_4$ series. x: 1 – 0; 2 – 0.04; 3 – 0.05; 4 – 0.09; 5 – 0.12; 6 – 0.19 [222].
On the right: $\varepsilon^{-1}(T)$ dependence for the $K_{0.76}(NH_4)_{0.24}H_2PO_4$ crystal. Dependence of the diffuseness parameter on the composition is given in the insert [222].

The $\varepsilon^{-1}(T)$ dependencies for the solid solutions with compositions $0 \leq x \leq 0.24$ are presented here. It needs to be reminded that the phase boundary between the "dipole-glass" state and the AFE state in the phase diagram is located at $x = 0.20$ (see Fig.10.13). As one can see the diffuseness parameter $\sigma(x)$ increases with an increase of $x$ and does not have any peculiarities at $x = 0.20$. It is an interesting detail that the temperatures at which the deviation from the Curie-Weiss law starts ($T_{seg}(x)$ in our notations) are very close to the line $T_m(x)$ in the diagram for $K_{1-x}(NH_4)_xH_2PO_4$ in Fig. 10.13. Both these experimental results are in complete agreement with the idea about the coexistence of domains of the AFE and FE phases that have been presented above and have been discussed in the course of this article. We would like to call ones attention to the fact that the character of the $\varepsilon(T)$ dependencies does not change at all when the composition of the solid solution crosses the phase boundary $x = 0.20$ in the phase diagram of the $K_{1-x}(NH_4)_xH_2PO_4$ solid solutions. It is apparent that the mechanism of diffuseness of the phase transition for the solid solutions with $x < 0.20$ and $x > 0.20$ is the same.

The results of studies of diffuseness of the phase transition in the presence of a DC electric field applied to samples are more interesting. The diffuseness parameter reduces as the field intensity increases (Fig.10.15) in the $K_{0.76}(NH_4)_{0.24}PO_4$ solid solutions [222].

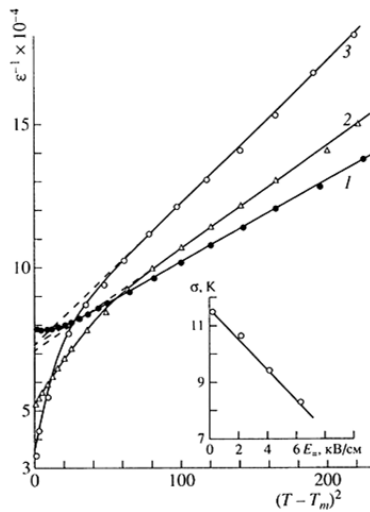

Fig.10.15. Temperature dependencies $\varepsilon^{-1}(T)$ for the solid solution $K_{1-x}(NH_4)_xH_2PO_4$ with $x = 0.19$ in external electric field $E$. Dependence of the diffuseness parameter on the electric field $E$ is given in the insert [222].
$E$, kV/mm: 1 – 0; 2 – 4; 3 – 6.

An electric field leads to an increase in the degree of diffuseness of the ordinary FE-PE phase transformations. However, in the substances with a small difference in energies of the FE and AFE orderings that possess the coexisting domains of the FE and AFE phases in the bulk of the samples, the behavior of the diffuseness parameter presented in Fig.10.15 is normal. The stability of the FE state increases compared to the stability of the AFE state when the field intensity increases. This is equivalent to the shift of the position of the state of the sample away from the FE-AFE-PE triple point in the diagram of phase states (from the point 2 to the point 1 in Fig.7.2). As seen from the upper part of Fig.7.2, the diffuseness parameter of the phase transition decreases. This fact was revealed in experiments on $K_{0.76}(NH_4)_{0.24}PO_4$.

The described above experiments can be unambiguously explained if one takes into account the coexistence of domains of the AFE and FE phases in the solid solutions that have one of the components being a ferroelectric and the other one being an antiferroelectrics. Both an increase in the electric field intensity and a decrease of $x$ lead to the same effect, namely, the rise of the stability of the FE phase relatively to the AFE phase. This in turn leads to a reduction of the share of the AFE phase in the sample's volume and as a consequence to the reduction of the diffuseness of the phase transition (see Chapter 7 for details).

Presented results point to the essential influence of the coexistence of phases on the kinetics of the phase transition in order-disorder type substances in which one component is a ferroelectric and the other one is an antiferroelectric. The most complete presentation of results of investigation of phase coexistence in the systems under discussion one can find in [216, 217, 219, 223] where the $Rb_{1-x}(ND_4)xD_2AsO_4$ solid solutions were considered as an example (phase diagram is presented in Fig.10.13). The authors of above mentioned publications have clearly demonstrated the existence of broad regions in the vicinity of the FE-PE and the AFE-PE phase transition lines. These regions are characterized by the coexistence of domains of ordered phases in the paraelectric matrix of the samples, although, in our opinion this is not the main circumstance. Before further discussions of this topic, we believe that it necessary to point out the resemblance of all phase diagrams depicted in the Fig. 10.13. This indicates that the phenomena under discussion are common to the whole class of the analyses substances.

There is a further point to be made here. As it was noted in [209], the first publications that proved the existence of domains of the energetically less stable phase (these are the domains of the FE phase in the case of antiferroelectric $NH_4H_2AsO_4$ and $ND_4D_2AsO_4$) in the PE matrix of the KDP-type substances were [213, 214]. It was shown in [212], that the FE states in KDP lie only a few milli-electron-volts above the AFE phase, which explains the observation of the FE-AFE phase coexistence near the AFE transition (in the PE matrix).

The temperature dependencies of the LA[100] Brillouin backscattering phonon spectra [216] and the Raman vibration modes [217] have been studied in the mixed FE-AFE system of $Rb_{1-x}(ND_4)_xD_2AsO_4$ solid solutions with compositions corresponding to the FE side of the phase diagram. The same investigations for this system of solid solutions with compositions corresponding to the AFE side of diagram were carried out in [219]. Authors of these publications observed the shift of the frequency of the Brillouin phonon spectrum line and the line's half-width at half maximum in the solid solutions with $x =$ 0.10 near a temperature of 160 $K$. They attributed observed anomalies of the above mentioned properties to the onset of the short-range AFE order caused by the freezing-in of $(ND)_4$ reorientations and suggested an increase of the competition between the AFE and FE phases. According to opinion of the authors of [216, 217], one can expect that such competition between the FE and AFE orderings that can suppress a long-range-order FE transition is responsible for both the phase coexistence and the presence of a broad damping peak in the Brillouin backscattering spectrum centered at $T \sim 146$ K (Fig.10.16 in the left). The rapid growth of the FE ordering near 130 K is responsible for the Landau-Khalatnikov-type maximum.

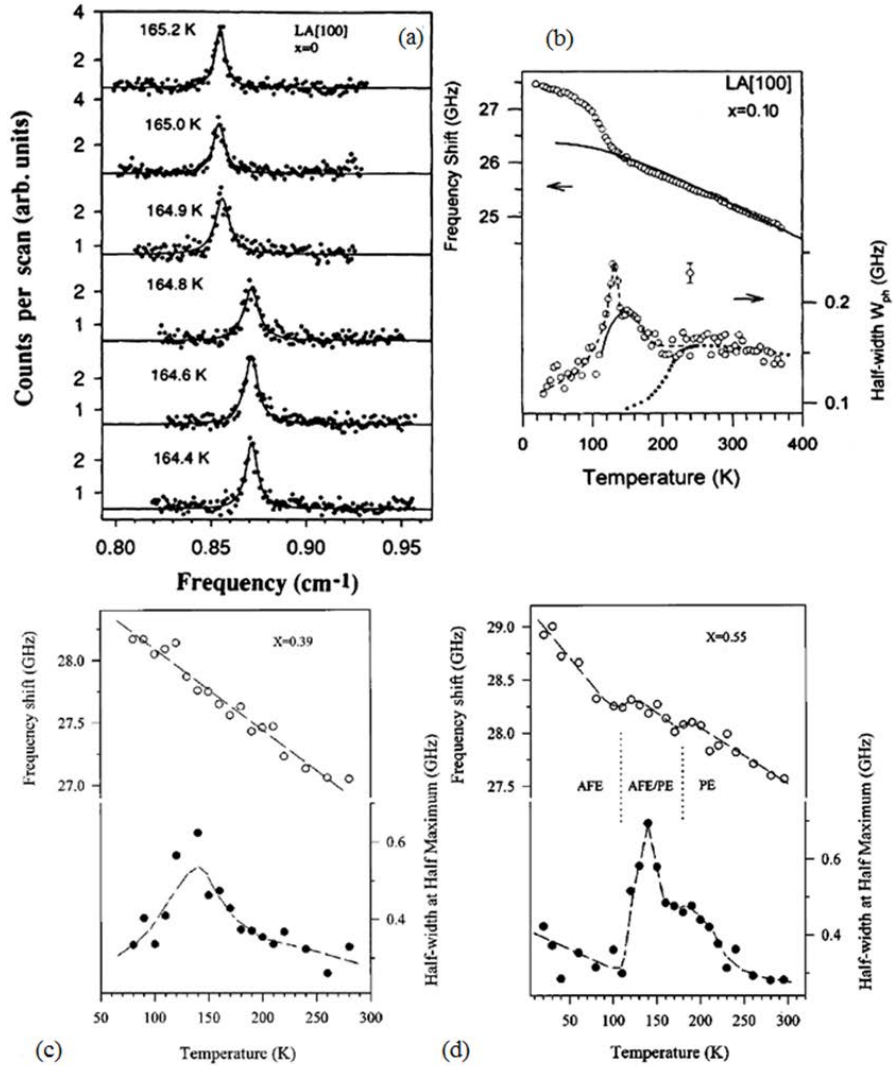

Fig.10.16. Upper row: (a) Anti-Stockes component of LA [100] Brillouin frequency shift for temperatures around the maximum value of half-width (for x = 0.10 as example) in $Rb_{1-x}(ND_4)_xD_2As_O_4$ mixed crystals; (b) frequency shift and half-width vs temperature of the LA[100] phonons for x = 0.10; the solid line for frequency shift is the Debye anharmonic calculation; the solid line and solid circles for half-width are the estimates of fluctuations and pure lattice anharmonic contributions, respectively; the dashed line is a guide to the eye [216].
Lower row: Brillouin shift (open circle) and half width (solid circle) vs. temperature of the LA[100] phonons for (c) $x = 0.39$ and (d) $x = 0.55$; the dashed lines are guides to the eye and the dotted lines are the estimates of various phase boundaries [219].

Analogous results were also obtained for solid solutions with compositions from the AFE region in the phase diagram for the $Rb_{1-x}(ND_4)_xD_2AsO_4$ system. Temperature dependencies of the same parameters for the solid solution with x = 0.55 are given in Fig.10.16 on the right for illustration purpose [219]. As can be seen one may safely suggest that in both cases the inhomogeneous states of the complex FE+AFE domains exist in the paraelectric matrix of the $Rb_{1-x}(ND_4)_xD_2AsO_4$ system of solid solutions with compositions both from the FE and AFE parts of the phase diagram. These inhomogeneous states exist in wide intervals of thermodynamic parameters (such as the temperature and composition). It does look like nobody took into account the possibility of these states until the present time. The behavior of the diffuseness parameters of the phase transitions discussed above for the $K_{1-x}(NH_4)_xH_2PO_4$ system of solid solutions have the same nature.

Let us now discuss the situation that takes place in the $Rb_{1-x}(ND_4)_xD_2AsO_4$ series of solid solutions with compositions that are characterized by approximately equal stability of the FE and AFE states. Temperature dependencies of parameters of the LA[100] Brillouin backscattering for the solid solution with x = 0.39 are presented in the lower left part in Fig.10.16. The temperature dependence of the half width of the anti-Stokes line of the Brillouin phonon spectra manifests a wide and intensive maximum in the vicinity of $T = 140\ K$. As one can see from the phase diagram of $Rb_{1-x}(ND_4)_xD_2AsO_4$ (in Fig.10.11) the line on which the point with this temperature has to be located is absent. It seems likely that the nature of this anomaly is not clear to the authors of mentioned studies. The satisfactory explanation for it can be found if to use the concept about the coexisting domains of the FE and AFE phases both above and below $T_C$.

In the substances, with a small difference in energies of the FE and AFE phases the domains of the FE and AFE phases coexist in the sample. A number of factors determine the dimensions of these domains. The most decisive factor, in our opinion, is the difference in the interplane distances of the crystal lattices in these phases. As noted above, the IDW separating the domains of the FE and AFE phases is coherent. The conjugation of the plains along these boundaries is accompanied with elastic stresses that increase the total energy of the system. Therefore, the smaller is the difference in the crystal lattice parameters of the coexisting FE and AFE phases, the larger is allowed area of the IDW and the smaller are dimensions of the FE and AFE phase domains.

Hence, it is extremely difficult to identify these states by means of the X-ray method or different spectroscopy methods. We as well as other authors encounter this difficulty during investigations of the FE-AFE solid solutions with perovskite type of the crystal structure. Form the physics standpoint such situation may also take place in the system of solid solutions under discussion. In this case the anomaly at a temperature of $140\ K$ in $Rb_{0.61}(ND_4)_{0.39}D_2AsO_4$ (see Fig.10.16) may be related to the formation of the complex FE+AFE domains in the paraelectric matrix of the substance (see Ch.7). The temperature dependence of the Brillouin shift in this case is shown in Fig. 10.16.

In conclusion of this chapter we want to stress once again that we have not dealt with the mechanisms of the ordered state (FE or AFE) in the KDP-type substances. These problems are solved at the level of microscopic models reliably enough. We have only considered the physical effects that can be the result of small difference in energies of the mentioned states. At the model level, this problem is not yet resolved as of today. Chances are that it has no solutions at this level. The phenomenological approach discussed in the present review has allowed us to understand some experimentally observed effects that previously have not found their explanation.

## 11. Applied material-science aspects of ferroelectric-antiferroelectric phase transitions

Ferroelectric materials have found their application almost immediately after they had been discovered. Initially these applications were mainly linked to military equipment. Then more and more fields of civil use have been found. Nowadays ferroelectrics have taken their place in many fields of science and technology. A wide variety of these new applications was primarily provided by piezoelectric ferroelectric ceramics. Moreover, in many cases, ferroelectrics have no competitors, and it is not clear if such competitors would appear in the near future.

The main physical effect that caused widespread use of ferroelectrics is the piezoelectric effect. This effect is absent in antiferroelectric substances. Therefore, the applications of antiferroelectrics were delayed by several decades. New antiferroelectric ceramic materials that have been recently developed promoted applications of antiferroelectrics in modern technology.

The publication [224] is probably the first paper devoted to consideration of some aspects of practical application of antiferroelectrics. It also contained consideration of the physical effects (known at that time) based on which one can expect their applications. This work mainly discussed nonlinear properties that are manifested in such materials during the induced phase transition from the AFE phase to the FE phase under the action of an electric field. This is illustrated in Fig. 11.1 (the data for an ordinary ferroelectric are also given for comparison). A ferroelectric phase is induced in an antiferroelectric at an increase of the intensity of applied electric field and the change of the dielectric constant takes place. The change of the dielectric constant can occur in a threshold manner, which allows this phenomenon to be used in various kinds of sensors. Antiferroelectrics with the so-called "narrow hysteresis loops" have been developed recently. In these substances, the changes in both the polarization and dielectric constant occur smoothly and with a small hysteresis, which make these materials attractive for applications in field-tunable devices.

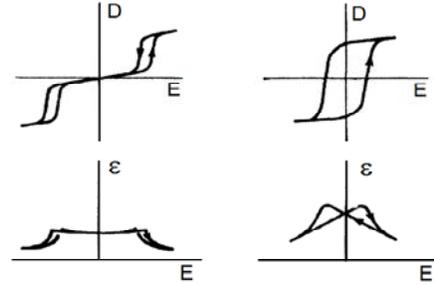

Fig.11.1. Hysteresis loops of electrical induction and dielectric constant for an antiferroelectric (in the left) and a normal ferroelectric (in the right) [224].

However, first working devices with antiferroelectric ceramics as a working element were devices for the storage of electric energy. The subsequent release of the stored energy took place under mechanical load (hydrostatic or uniaxial pressure) in the process of the phase transition from the FE state into the AFE state [225]. The first application of such phenomenon was made in the military areas.

## 11.1. Materials with ferroelectric-antiferroelectric phase transition for energy accumulation and conversion devices

The basic operation principle of devices based on a phenomenon of ferroelectric energy accumulation-conversion consists in an accumulation of the electric energy in the process of polarization of material followed by a release of energy for the load resistance during the depolarization. Application of an electric field polarizes the working element and it accumulates the electric energy with the density $P_r^2/2\bar{\varepsilon}$, where $P_r$ is the remanent polarization, $\bar{\varepsilon}$ is the average value of dielectric constant in the process of polarization [226,227]. FE materials with compositions located near the boundary between regions of the FE and AFE states in the diagram of phase states are used for this purpose.

Such accumulators differ by the principle of realization of depolarization process, i.e. by a way in which the accumulated energy is released. The simplest way is the phase transition into a non-polar (PE or AFE) state. These transitions may be realized due to the effect of temperature or mechanical stresses. The working principles and examples of practical applications of the converters, in which the driving force of depolarization process is temperature, are presented in [227].

The most advanced are investigations and developments of the devices in which the accumulated energy is released due to the action of a pulsed mechanical load (impact). Polarized FE materials are used to obtain electric pulses with a power up to 1000 kW and higher. At pulsed mechanical loads, the polarized materials demonstrate both piezoelectric and ferroelectric properties. However, the electric energies released due to the piezoelectric effect may be of the order of $2 \cdot 10^4$ J/m$^3$, whereas the energies that can be released in the process of depolarization of the FE are of $10^6$ J/m$^3$ and more [228, 229]. The dependencies of the voltage on the ceramic working element electrodes and active load current are shown in Fig. 11.2. Using these dependencies one can estimate the electric parameters which may be obtained when using FE energy converters [229].

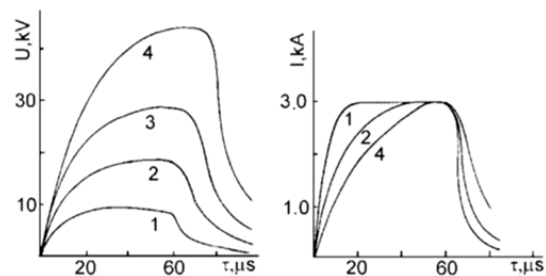

Fig.11.2. Time dependences of voltage on ceramic working element electrodes and active load current [229]. $R_{\text{load}}$, :1 – 1.02, 2 – 5.96, 3 – 8.90, 4 – 10.00.

Lead zirconate-titanate solid solutions are conventionally used as a material for the working elements of converters. The phase diagram "composition – temperature" of these solid solutions is presented in Fig.11.3 [69]. At the beginning, the solid solutions with compositions located in the region of the morphotropic phase boundary (with approximately equal molar shares of lead titanate and lead zirconate) were used. These substances are characterized by the phase transition into the PE state at mechanical loading. In this case, the value of the energy released for load resistance is limited due to depolarization caused by the instability of domain structure, which occurs long before the phase transformation itself. This leads to the essential decrease of the effective value of $P_r$ and the premature leakage of electric charge [226, 230, 231, 232]. Such materials have not found practical application.

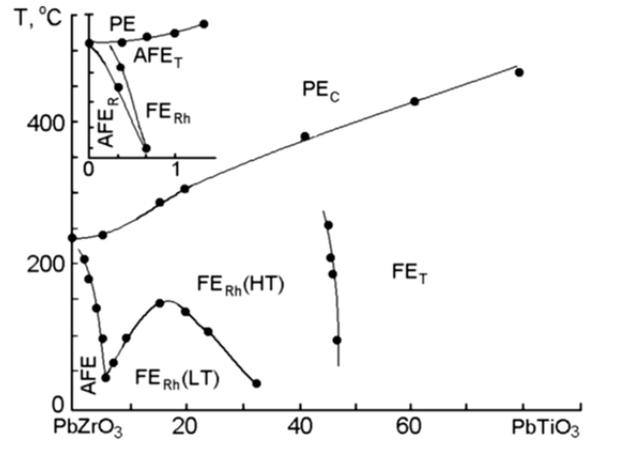

Fig.11.3. Lead zirconate-titanate phase diagram after [69].

The converters in which the energy accumulated in the FE state is released during the phase transition into the AFE state are utilized in practice. Materials with such type of phase transformation are most suitable for the above-mentioned kind of devices due to the following factors. In these materials, the accumulated energy has maximum value. In particular, for the Pb(Zr$_{0.965}$Ti$_{0.035}$)O$_3$ + 1 mol%Nb$_2$O$_5$  $P_r = 38$ μC/cm$^2$ и $\varepsilon = 250$ and the accumulated energy density is 3.3 J/cm$^3$, whereas for the Pb(Zr$_{0.525}$Ti$_{0.475}$)O$_3$ (the solid solution with composition from the morphotropic region) $P_r = 35$ μC/cm$^2$, $\varepsilon = 1250$ and the accumulated energy density is only 0.7 J/cm$^3$. Accordingly, in the former case the released energy densities are higher: 1.8 J/cm$^3$ for Pb(Zr$_{0.965}$Ti$_{0.035}$)O$_3$ + 1 mole % Nb$_2$O$_5$ and 1.4 J/cm$^3$ for Pb(Zr$_{0.965}$Ti$_{0.035}$)O$_3$ + 1 mole % WO$_3$ solid solutions (these data correspond to quasi-static regime of the working element loading) [226]. Such energies are two or three times as high as those obtained while using the solid solutions belonging to the morphotropic region of the phase diagram [69]. In both morphotropic and FE-AFE materials under shock-loading condition of experiment the said energy values are 2 to 2.3 times lower.

However, even if the accumulator-converter working elements are based on materials with the FE-AFE transition, only a certain part of the accumulated energy can be effectively used. In the case of shock-loading regime (used in real converters), this magnitude is considerably lower; it is defined by the electrical strength of the material. The most detailed consideration of the problems connected with electrical breakdown of the working elements one can find in [233]. The resistance of the working element at the wave front undergoes substantial changes when a shock wave is passing through it. For instance, at a shock amplitude of the order of $2 \cdot 10^3$ MPa the estimated resistivity of the 95/5 PZT ceramics is $10^2$ Ω·cm and the resistivity of the barium titanate ceramics is $10^4$ Ω·cm, whereas in the unloaded state their resistivity fall in the range of $10^{10} - 10^{12}$ Ω·cm. Such essential increase in the material conductivity is connected with leakage currents caused by the charge released at the shock wave front.

The dependencies of the electrode voltage of the working elements that were manufactured using the 95/5 PZT ceramics on the amplitude of shock pulse are shown in Fig.11.4. The working elements are polarized up to 10, 20 and 30 μC/cm$^2$. The value of remnant polarization defines the value of the accumulated electrical energy. As seen from the figure, the electrical breakdown occurs practically at the same electrode voltage ~ 5 kV/mm independently from the initial value of polarization. The breakdown does not allow the release of the total value of energy accumulated in the process of polarization.

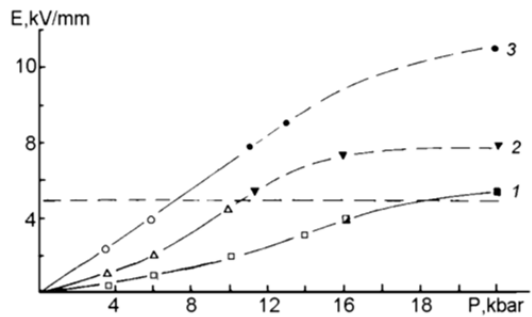

Fig.11.4. Dependences of the electrode voltage of the 95/5 PZT ceramic working element on the shock- pulse amplitude [233]. Black points correspond to electric breakdown of working elements.
$P_r$, μC/cm$^2$: 1 – 10, 2 – 20, 3 − 30.

The performed analysis shows that materials that can be used as working elements in the energy accumulation and conversion devices must satisfy the following criteria:
− the presence of the FE-AFE phase transition in such materials under the action of mechanical stresses;
− the phase transition must occur at the pressures lower than those which correspond to the process of domain disorientation;
− the materials must possess high electrical strength;
− they must be easily manufactured and have a low cost.

Among the existing materials, the PZT-based ceramics meet the above-said requirements. It should be noted that it is not necessary to achieve extreme values of remnant polarization and permittivity. The efficiency of the working elements is most adequately defined by the electric strength of material. We have found a way to increase the electric breakdown voltage in these materials bringing it up to 10 − 12 kV/mm. Using this approach, we have managed to develop a series of ceramic materials with the electric energy release on the active loading under shock-loaded compression conditions up to 3.5 − 4.0 J/cm$^3$.

## 11.2. Phase transition via intermediate state and control of piezoelectric parameters

The majority of physical characteristics of a substance undergo significant changes during phase transformations. This circumstance may be used for development of materials with operational characteristics that can be controlled by variation of external thermodynamic parameters that give rise to this transition. However, the main difficulty in using phase transitions is in the fact that, as a rule, they occur quickly and within a narrow range of variation of external parameters. The phase transition via an intermediate state considered in Chapter 5 of the present review is stretched over a wide interval of external parameters causing this transition. During the phase transition via an intermediate state, the share of the FE phase gradually increases in the bulk of the substance, whereas the share of the other phase diminishes. The AFE state is piezoelectrically inactive while the FE state is piezoelectrically active. Because of this, one may expect a gradual change of piezoelectric properties of the substance during the transition from the AFE into the FE state induced by the electric field when this transition is realized via intermediate state.

The phase transition vi an intermediate state in lead zirconate-titanate based solid solutions and the behavior of piezoelectric properties of these solid solutions in the process of transition have been investigated in [32, 82, 234].

The share of the FE phase increases linearly when an external DC field increases within the interval of field intensities corresponding to the intermediate state (in compliance with (5.8)). The increase of the share of the polar phase leads to changes in piezoelectric characteristics of samples. The experimental dependencies of the piezoresonance parameters on an electric field are presented in Fig.11.5 for the 8.25/100-Y/Y series of PLZT solid solutions. The dependence of the resonance curves on the intensity of an external DC electric field for the 8.25/70/30 series of PLZT solid solution was shown in Fig.5.4.

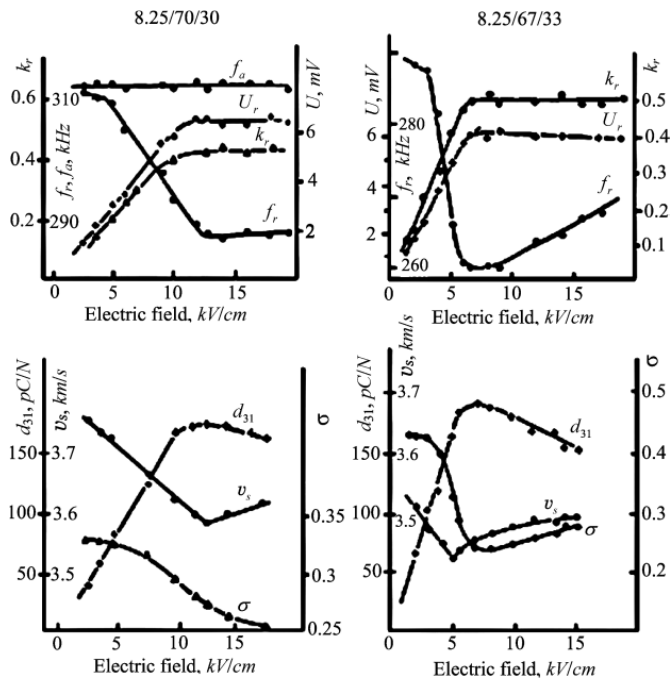

Fig.11.5. Influence of an electric field on the parameters of the 8.25/100-Y/Y series of PLZT solid solution (radial resonance) [82, 234].
From the left: 25/70/30 PLZT, from the right: 8.25/67/33 PLZT.

The above-presented results show that the AFE $\rightleftarrows$ FE phase transition via the intermediate state leads to a gradual change of the piezoactivity of material. Furthermore, this transition is almost without hysteresis. This circumstance allows using this phase transition for the effective control of piezoelectric parameters of material by the external electric field. Significant change of the resonance frequency, which exceeds similar characteristics of known materials by several orders of magnitude, has engaged out attention (Fig.11.6).

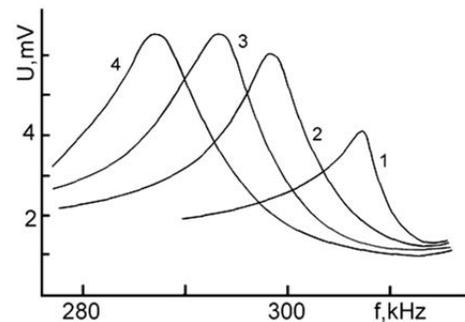

Fig. 11.6. Resonance curves of the 8.25/70/30 PLZT at different intensities of external electric field [82, 234] [82, 234].
$E$, kV/cm : 1 – 5.0,  2 – 8.0,  3 – 10.0,  4 – 13.0.

## 11.3. Local decomposition of solid solutions, nanostructures and optical materials with negative refractive index

The first report on the possibility of existence of materials with negative refractive index for electromagnetic radiation and the analysis of the core set of their physical properties appeared in [235].

Materials with negative refractive index have been hypothetical for a long period until the first experimental confirmations of hypothesis expressed in [235] appeared [236, 237]. Since that, time researches have taken an active interest in materials of this kind and the number of publications on this topic constantly increases. Comprehensive reviews of different physical properties as well as applications of negative refractive index materials have touched practically all aspects of this interesting part of material science existing up to date [238, 239, 240, 241].

First materials with the negative refractive index only in high frequency region of electromagnetic radiation spectrum represented a composite construction of complex-shape elements (for example, a periodic array of split-ring resonators with wires placed uniformly between the split rings and so on) [236, 237. 242]. It is practically impossible to manufacture materials for the optical range of spectrum with similar design.

The metamaterials with the negative refractive index in the near infrared and optical ranges of spectrum were the metal-dielectric composites [243, 244, 245]. The method of manufacturing of such composites described in [243 − 245] is quite laborious and is difficult to reproduce.

The easily realizable method for creation of dielectric-metal composite structures with the periodic arrangement of metallic inclusions was suggested in [246]. Dimensions of the said inclusions are from 8 to 12 nm. The distances between the periodically arranged inclusions are from 0.2 to 2.0 μm. The above-mentioned method is based on controlled use of the process of the local decomposition of sold solutions with the structure of coexisting domains of the FE and AFE phases. The decomposition of the sold solution takes place along the boundaries separating domains of the FE and AFE phases (see Chapters 3, 6 and 7 of the present review). The segregates formed along these boundaries during the local decomposition of the solid solution may have metallic type of conductivity.

Conductivity of the majority of oxides with the perovskite crystal structure (the PZT-based solid solutions are among them) can be varied by means of ion substitutions in lattice sites in a very wide range from pure dielectric state (with resistivities of the order of $10^{14} - 10^{16}$ Ω·cm) to the conducting state (with resistivities of the order of 100 mΩ·cm).

The chemical composition of segregates precipitated along the interphase boundaries is slightly different from the maternal composition of the base solid solution. Selecting a specific chemical composition of the maternal solid solution and controlling the process of local decomposition of this solid solution one can achieve the result that segregates may possess a diverse set of physical properties (they can be magnetic, dielectric, or conducting). We performed the decomposition of the PZT-based solid solution in such a way that the segregates in the vicinity of interphase boundaries had high (near-metallic) conductivity [246]. PZT-based solid solutions are the well-studied substances in which both the FE and AFE orderings take place. The characteristic feature of these substances is a small difference in free energies of the FE and AFE states in the solid solution at specific concentrations of components ($PbZrO_3$ and $PbTiO_3$). At the same time, the domains of the above-mentioned phases coexist in the volume of solid solutions (Chapters 3, 6, 7).

As was shown in previous chapters one can change the relative stability of phases and along with it the volume share of phases in the sample by changing the position of solid solution in the diagram of phase states (in the PZT-based solid solutions it is achieved by changing the content of Zr/Ti). The sizes of domains of the metastable phase and their density (the period of the domain structure) are changed when the share of this phase in the solid solution varies. The parameters of the domain structure can be also changed by variation of an external electric field intensity as well as mechanical stresses (pressure, uniaxial or biaxial compressions or tensile stress), because these external influences change the relative stability of the FE and AFE phases. Under the lateral strains the cylindrical domain structure in thin sample is transformed into the stripe-domain one. The mechanism of the local decomposition of the solid solution in the vicinity of interphase boundaries was considered above in the Chapter 6. It is connected with a redistribution of the "small" and "large" ions in domains with the "small" and "large" interplane distances. In the vicinity of the IPB the "large" ions are driven out into the domains with the larger configuration volume and correspondingly with the larger distances between atomic planes. At the same time, the "small" ions are driven out into the domains with smaller distances between atomic planes. The *A*- and *B*-positions of the perovskite crystal lattice of the solid solutions are occupied by ions with different ionic charges ($Pb^{2+}$, $La^{3+}$, $Li^+$ in *A*-sites and $Zr^{4+}$, $Ti^{4+}$, $Nb^{5+}$, $Mg^{2+}$ in *B*-sites). Because of this the local decomposition of the solid solution along the interphase boundaries can be accompanied by the local disturbance of electro-neutrality. Therefore, the segregates that are formed on the interphase domain walls acquire the metallic conductivity whereas the rest of the sample stays in a dielectric state forming the dielectric matrix.

The wavelength dependencies of transparency for one of the discussed materials with composite structure of conducting segregates are presented in Fig.11.7 as an example.

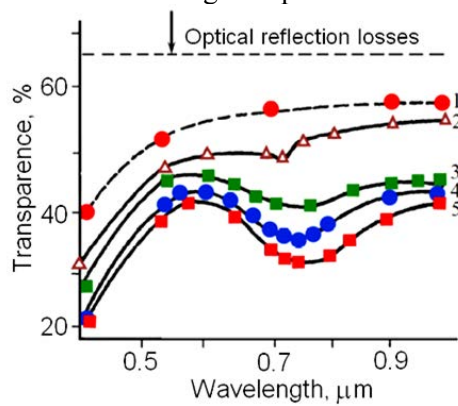

Fig.11.7. Dependence of the light transmission coefficient on the wavelength for the transparent composite material (the dielectric matrix containing conductive metallic segregates) [246]. The increasing numbers near the curves correspond to the materials with increasing values of the conductivity of segregates.

The dips in the curves correspond to the presence of the negative refraction regime. The modulation depth can be controlled by changing the conductivity of segregates. It is also possible to select the optic wavelength range by changing the position of the solid solution in the "composition-temperature" phase diagram of the substance. One can also control the wavelength range changing the position of the solid solution in the "electric field-temperature" phase diagram by varying the potential difference between the element's electrodes.

However, the most interesting results can be obtained using thin film structures. The change of the region of the negative refraction in this case can be achieved both by the application of the electric field to the film substrate (if the ferroelectric crystal is chosen as a substrate material) and by the flexural strains of the substrate. The modulation of the transmitted light has been observed in both these situations.

Thus, the controlled decomposition of solid solutions can be used for manufacturing of composite substances (dielectric mediums containing periodic arrangement of metallic inclusions) with the negative refractive in the optic range of spectrum

## 11.4. Creation of textured nanostructures

The microstructure of the solid solutions can be controlled due to the fact that the inhomogeneous state of coexisting domains and the structure of segregates along the interphase boundaries appear in the process of the phase transition. In turn, the phase transition can be easily controlled (an example of the smooth phase transition is the transition via the intermediate state considered in Chapter 5). After the phase transition the structure of segregates is preserved. Thereby, the textured structures may be created in the materials under consideration when external influences are applied to a material during the phase transition and these structures will remain stable after switching-off external influences.

It has been shown in Chapter 7 that in the PZT-based solid solutions with compositions near the triple point in the phase diagram the two-phase (AFE+FE) domains appear in the PE phase in the process of cooling the sample at the temperatures still above $T_C$.

The presence of such two-phase domains allows one to use the piezoelectric resonance method both for investigations of phase transition and for the study of creation of textured structures. It was demonstrated in [247, 248]. The $Pb_{0.90}(Li_{1/2}La_{1/2})_{0.10}(Zr_{1-y}Ti_y)O_3$ system of solid solutions was studied in these papers. The phase diagram of these solid solutions is shown in Fig.5.1. The said method cannot be applied to ordinary ferroelectrics because in these substances the FE component is absent at the temperatures $T > T_C$ (of course, when an external electric field is also absent).

Two different regimes of the thermoelectric treatment were used to prepare samples for measurements in the above-mentioned studies. In the first case, the samples were annealed at 500 °C for 1 hour and after that, they were cooled at the rate of 4 degree/min. An electric field (800 V/mm) was applied to the samples at 300 °C (which is above the Curie point) and the cooling continued in the presence of the field until the temperature reached $T_C + 20$ °C, at this temperature the electric field was switched of and the samples were cooled down to room temperature. Measurements of the piezoresonance were performed at room temperature.

In the second case, the thermoelectric treatment was slightly different. After the annealing and cooling the electric field was applied to the samples at $T_1 = T_C + 20$ °C (the condition that $T_1 > T_C$ was observed all the time). The samples were kept at this temperature in the presence of the applied electric field for 0.5 hour, after that the field was switched off and the samples were cooled down to room temperature and the resonance measurements were carried out. Recall again, all thermal treatments were carried out when samples were in the PE state.

Dependencies of the piezoelectric coefficient $d_{33}$ on the Ti-content for the 10/100-Y/Y system of PLLZT solid solutions subjected to the above-described regimes of the thermal treatment are shown in Fig .11.8 (curves 2 and 3 refer to the first and second regime, respectively). Both dependencies have well pronounced maxima located near Y = 20 (which corresponds to 20% Ti). Solid solutions do not manifest piezoelectric properties when the Ti-content moves out from this value.

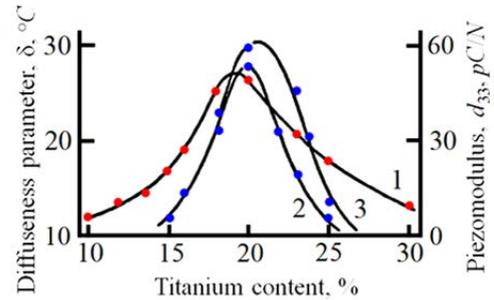

Fig.11.8. Dependencies of diffuseness parameter δ (curve 1 – this curve is presented for comparison) and piezoelectric modules (curves 2 and 3) on composition for the 10/100-Y/Y series of PLLZT solid solutions measured after different regimes of thermal treatment in paraelectric phase [247, 248].

Analogous results have been obtained for the PLZT system of solid solutions with the AFE-FE boundary in the diagram of phase states.

As is seen from above, the piezoelectric activity is manifested only in the solid solutions that have an approximately equal stability of the FE and AFE states. The two-phase domains that appear in the PE matrix of the sample in the process of cooling down from high temperatures have random distributions of spontaneous deformations and, correspondingly, random distributions of the axes of spontaneous polarization in the absence of an electric field. The distribution of the planes of the interphase boundaries is also random. The process of diffuse segregation along the boundaries leads to the fixation of these random spatial distributions in the course of further cooling. The growth of the volume occupied by two-phase domains also takes place during the temperature decrease. At the temperatures close to $T_C$ the whole volume of the sample undergoes transition into an inhomogeneous state of the coexisting FE and AFE phases with the directions of spontaneous polarization pinned by segregates. If one would try to polarize such a sample, the degree of polarization is going to be quite small and the level of piezoelectric properties will be minor.

In the case when the cooling of the sample from high temperatures is carried out in the presence of the electric field, the distribution of the axes of spontaneous polarization (along with the distribution of directions of interphase boundaries) of nucleated two-phase domains will be given by the direction of the field. In this case, texture is created in the samples after cooling. This texture is stabilized by the segregates on the interphase boundaries at room temperature (or temperatures close to room temperature). The samples are macroscopically polarized and it makes it easy to excite the piezoelectric resonance in these samples. The piezoelectric modules are nonzero even in the case when additional polarization (after the thermal treatment) of the samples has not been carried out.

If the solid solution composition is located far from the FE-AFE boundary in phase diagram (correspondingly, far from the triple point), the two-phase domains do not appear in the sample's volume at $T > T_C$, and after cooling to the room temperature the sample will not manifest any piezoactivity.

In the case when the additional polarization of the solid solution was created by application of an electric field in the process of the prior thermoelectric treatment, the resulting material will possess a set of properties that cannot be achieved by any other treatments.

11.5. Effects caused by antiferroelectric nanodomains in lead zirconate titanate based coarse-grained ceramics with compositions from the morphotropic boundary region

PZT-based solid solutions currently have the widest range of applications as piezoelectric ceramic materials. They represent not less (or may be even more) than 95% of such materials manufactured in the world. PZT-based solid solutions have no competition due to the simplicity of the technological process of their manufacturing. Another important reason for their popularity is that there is no replacement for these materials at present despite all the efforts and costs that were invested in development of lead-free compounds. That is why investigations of PZT-based solid solutions are still a topical issue and are carried our intensively and will be continued in the future. The main goal of contemporary studies of these compounds is the improvement of their material parameters and operational characteristics. This is the reason why we want to pay a greater attention to this topic in our review.

The PZT-based solid solutions used for manufacturing of the working elements for devices have compositions located within the so-called morphotropic boundary (MPB) region in the "composition-temperature diagram of phase states of PZT [73] (see Fig. 11.3). For a long time (over 50 years), it has been considered that domains with the rhombohedral and tetragonal types of crystal lattice distortions exist within this region and these domains possess the FE type of electric-dipole ordering.

The monoclinic phase in PZT solid solutions with compositions belonging to the MPB region was first reported in [249, 250, 251]. Shortly afterwards a large number of publications emerged either reporting the confirmation of existence of the monoclinic phase in PZT or trying to give a theoretical explanation of observed results using a phenomenological approach or first principle calculations. All results of research done on this topic are presented and analyzed in a series of reviews [252, 253, 254, 255]. In these reviews, one can find all references on original studies devoted to investigations of the MPB region in PZT.

Main results of the crystal structure studies of the said solid solutions are based on the analysis of the profiles of lines obtained by means of the X-ray or neutron scattering. The analysis itself has been based on the models taking into account concrete individual crystal structures (these are the tetragonal, the rhombohedral, and three monoclinic structures in the case of the PZT) or their coexistence in the bulk of the sample under investigation. The structure that has been accepted as a "true" structure is the one for which the calculated profiles of the lines were the closest to the experimentally obtained profiles.

As it follows from the above-mentioned procedure, only effects and phenomena that have been well known previously, as well as the types of the crystal structures can be obtained by means of this type of analysis of the X-ray and neutron diffraction measurements data. Any phenomenon unfamiliar to the individuals analyzing data was not taken into account and did not play any role in the final solution of the problem.

In the case of the PZT and PZT-based solid solutions, such unaccounted phenomenon is the local decomposition of the solid solution in the vicinity of the coherent interphase boundaries (CIPB). The boundaries between the domains of the tetragonal and rhombohedral phases coexisting within one single-crystalline grain are exactly such CIPB boundaries. These two coexisting phases are the phases with different inter-plane distances and different parameters of their crystal lattices. The local decomposition of solid solution takes place in the immediate vicinity of such CIPBs to lower the elastic energy. Analogous phenomenon in the case of the PZT-based solid solutions with coexisting domains of the FE and AFE phases was considered in Chapter 6. The ions with larger ionic radii are expelled into the domain with larger parameters of the crystal lattice and the ions with smaller ionic radii are moved into the domain with smaller crystal lattice parameters (the ions in equivalent crystallographic positions are referred to here). Due to this decomposition, the chemical composition of solid solution in the vicinity of these interphase boundaries differs from the chemical composition inside the domains of both phases. As a result the segregates with the crystal lattice close to the parent solid solution composition but still different are formed. Such local decomposition of the solid solution in the vicinity of coherent interphase boundaries in the PZT-based solid solutions has been discussed in details in Chapters 6 and 7.

X-ray and neutron diffraction investigations of the crystal structure of PZT solid solutions from the MPB region of compositions represent the larger group of published studies. These studies are based on the analysis of the profiles of lines of the diffraction pattern due to the elastic scattering of incident radiation. The local decomposition of solid solutions and the formation of segregates are manifested in the appearance of additional lines in the X-ray and neutron diffraction patterns. These new lines are usually wider and have lower intensity than the Bragg lines of the main perovskite crystal lattice. These wider and weaker lines are located at the angles close to the corresponding angles of the Bragg lines. The volume occupied by the new phases is small. The crystallographic planes of these regions of new phases are randomly located, and as a result, the corresponding X-ray lines have an appearance of the diffuse lines. This means that the profiles of the X-ray or neutron diffraction lines appear somewhat changed which can be attributed to the appearance of new phases. As far as we know, the local decomposition of solid solution has never been taken into account during the analysis of the crystal structure in the MPB region.

Spectroscopy methods are used in another group of studies of the crystal structure of the PZT-based solid solutions. The Raman light scattering has been used for the most part. Profiles of the spectral lines in the infrared absorption spectrum (and the lines in the reflection spectrum) are used to evaluate possible structural peculiarities of the PZT solid solutions. The situation in this approach is analogous to the situation that took place in the treatment of the data of X-ray and neutron experiments.

New weak intensity lines that appear as a result of the local decomposition of solid solution in the vicinity of the interphase boundaries modify the profiles of lines observed in X-ray and neutron scattering experiments. This modification of line profiles can lead to the ambiguity in the identification of the crystal structure of solid solutions if the above phenomenon has not been taken into account.

All said above is also relevant to the analysis of the electron diffraction results. The presence of segregates may lead to the additional splitting of main electron diffraction reflexes as well as to the appearance of additional reflexes.

The ions of zirconium and titanium are shifted inside the domains with different crystal lattice parameters in the process of local decomposition. As a result local zirconium-enriched regions appear. The antiferroelectric state can be realized in these local regions. It is difficult to elicit directly such-like segregates with the AFE state in the PZT solid solutions at present. It has been done for the PZT-based solid solutions with compositions from the ferroelectric-antiferroelectric morphotropic region of the "composition-temperature" phase diagram where the sizes of segregates were of the order of several tens of nanometer (see Chapters 3, 6, 7 of the present review). In addition to that, several recently appeared publications pointed out the possibility of existence of the AFE ordering in the PZT solid solutions with compositions from the MPB region (the region between the tetragonal and rhombohedral phases) of the phase diagram. We will give the results of these publications in the course of discussion of our results.

The presence of segregates leads to a series of specific effects characteristic to the boundaries between the coexisting domains with the AFE and FE types of ordering (see Chapters 6, 7 and 8).

The search for such effects and their investigations in the PZT-based solid solutions with the compositions from the MPB region of the "composition-temperature" phase diagram has been undertaken in [256]. In no case, we try to assert the absence of the monoclinic phase in the PZT solid solutions. We only want to emphasize that there are physical effects that have not been taken into account during the interpretation of experimental results and during the identification of the crystal structure of solid solutions.

We have developed a procedure for manufacturing of the PZT ceramics (with different ion substitutions) with compositions located near the boundary between the tetragonal and rhombohedral phases in the "composition-temperature" phase diagram with a grain size exceeding 20 μm [257]. Manufactured ceramic samples are characterized by a high degree of the composition homogeneity in the bulk of the substance. This has been confirmed, first of all, by the grain size and, secondly, by the extremely small width of the MPB region (less than 0.5%) in the investigated seven-component PZT-based solid solutions with compositions $(Pb_{0.9}Ba_{0.05}Sr_{0.05})_{0.985}(Li^{+}_{0.5}La^{3+}_{0.5})_{0.015}(Zr_{1-x}Ti_{x})O_3$.

According to the data available in the literature, the grain size of the PZT ceramic samples usually used in experiments is of the order of several micrometers. That is why the rhombohedral and tetragonal phases are in different grains and up to a recent time it has been very difficult (or even completely impossible) to investigate effects caused by the interphase domain boundaries. In this case, the grain boundaries play the role of the interphase boundaries. It is common knowledge that the inter-grain boundaries accumulate a lot of impurities as well as lattice imperfections due to impurities. It is understood that there is no gradual conjugation of crystal planes in this case. The sizes of crystal grains in our samples have been of the order of 30 to 40 microns. The domains of two phases coexisting within one crystallite are separated by the coherent interphase boundaries of the atomic scale and character. There were no discontinuities of crystal plains (it was confirmed by TEM studies) and the boundaries had a coherent character without dislocations and the concentration of the elastic stress. A high degree of the homogeneity of samples has allowed us to study the kinetics of development of segregates in the vicinity of CIPBs along with the physical effects caused by these CIPBs.

In our opinion, it is important to note one more peculiarity of the absolute majority of publications on the monoclinic phase in PZT-based solid solutions with compositions from the MPB region of the phase diagram. Because of the wide region of coexistence of phases, the exact phase boundaries between various phases cannot be located using diffraction data. The wide region of phase coexistence is a convincing characteristic feature of the inhomogeneity of samples. The last circumstance is an additional factor contributing to the uncertainty in the determination of the crystal structure of solid solutions.

### 11.5.1. *Peculiarities of properties of coarse-grained ceramics*

The "temperature-composition" phase diagram for the solid solutions with compositions given by the formula $[(Pb_{0.9}Ba_{0.05}Sr_{0.05})]_{0.985}(Li^{+}_{0.5}La^{3+}_{0.5})_{0.015}(Zr_{1-x}Ti_x)O_3$ is shown in Fig.11.9 [257].

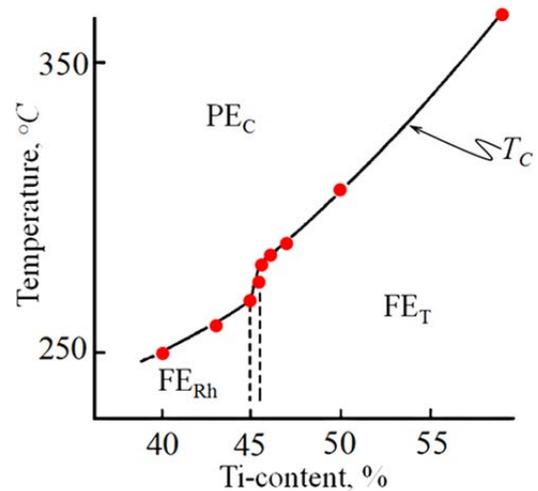

Fig.11.9. The "composition-temperature" phase diagram of the $[(Pb_{0.9}Ba_{0.05}Sr_{0.05})]_{0.985}(Li^{+}_{0.5}La^{3+}_{0.5})_{0.015}(Zr_{1-x}Ti_x)O_3$ system of solid solutions with compositions in the vicinity of the MPB boundary [256, 257].

Profiles of the (200) and (222) X-ray lines for the above solid solutions with compositions from the MPB region of the phase diagram are presented in Fig. 11.10. The coexisting domains with different types of the crystal lattice distortions are present in the bulk of these samples. Availability of such coarse-grained ceramics (let us recall, that the size of ceramic grains have been in the range of 30 − 40 μm) made it possible to investigate specific effects caused by the coexistence of domains with the FE and AFE orderings within a single crystalline grains.

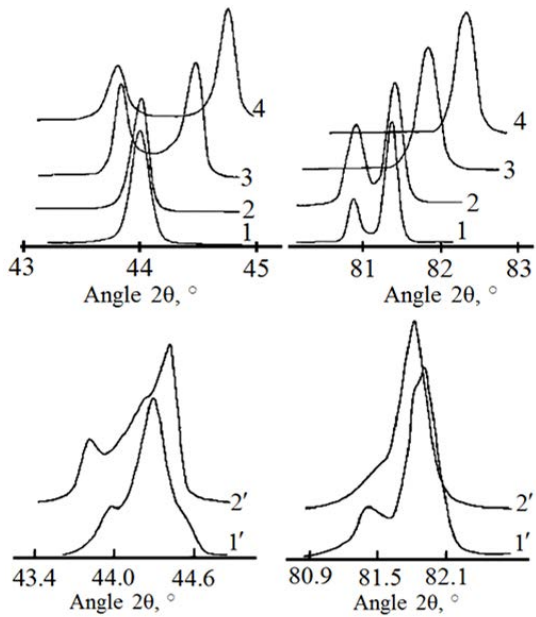

Fig.11.10. Profiles of the (200) and (222) X-ray diffraction lines for the $[(Pb_{0.9}Ba_{0.05}Sr_{0.05})]_{0.985}(Li^{+}_{0.5}La^{3+}_{0.5})_{0.015}(Zr_{1-x}Ti_x)O_3$ system of solid solutions with different content of titanium. Titanium content, x, %; 1 − 0.35; 2 − 0.45; 3 − 0.46; 4 − 0.47; 1′ − 0.4550; 2′ − 0.4575 [256, 257].

The concentration dependencies of the main dielectric and piezoelectric parameters of the $(Pb_{0.9}Ba_{0.05}Sr_{0.05})_{0.985}(Li_{0.5}La_{0.5})_{0.015}(Zr_{1-x}Ti_x)O_3$ system of solid solutions are depicted in Fig.11.11. As it seen from these dependencies, all parameters have a pronounced maximum located in the MPB region of Zr/Ti compositions. On the one hand, this fact confirms the conclusions about the nature of behavior of these parameters in the vicinity of the MPB made earlier. However, a special attention has to be paid to the following fact. The MPB region of solid solutions under consideration is very narrow (not exceeding 0.5%), but the interval within which the parameters change is wide. In particular, for the coefficient of the electromechanical coupling, $K_r$, dielectric permittivity, $\varepsilon$, and polarization this interval is of the order of 10%. At the same time, the interval of high values of the piezoelectric modulus $d_{33}$ is narrow.

Fig.11.11. Dependencies of main physical parameters of the $(Pb_{0.9}Ba_{0.05}Sr_{0.05})_{0.985}(Li_{0.5}La_{0.5})_{0.015}(Zr_{1-x}Ti_x)O_3$ system of solid solutions on titanium content [256, 257].

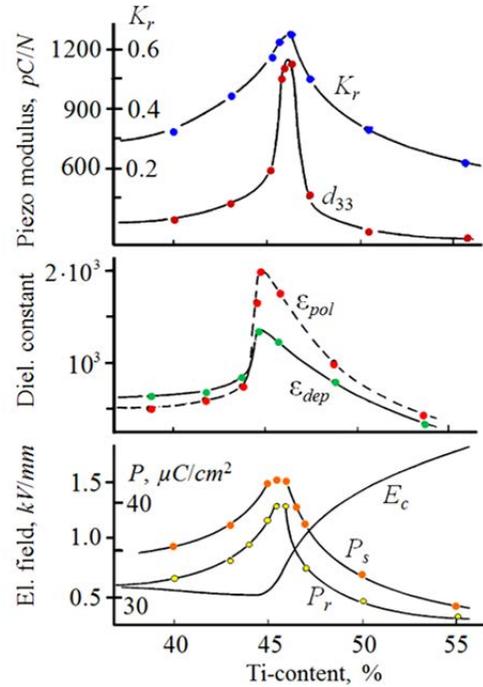

It should be noted that the values of the piezoelectric modulus $d_{33}$ are extraordinarily high for our PZT-based ceramics (more than 1100 pC/N) with compositions belonging to the MPB region of the phase diagram. For comparison we have to remind that for well-known piezoelectric ceramic materials, the value of piezoelectric coefficient $d_{33}$ does not exceed 600 pC/N.

As a rule, extreme values of parameters of the PZT solid solutions are explained in the literature by the coexistence of the tetragonal and rhombohedral FE phases. It is generally considered that the polarization, dielectric, and piezoelectric parameters of solid solutions from the MPB region are defined by the so-called orientation polarization.

However, we have obtained the PZT-based solid solutions with the width of the MPB region less than 0.5%, but the interval of increase of the main physical parameters exceeding 10%. It is clear that in the present case the above-mentioned mechanism cannot explain such behavior. Taking into account the new (monoclinic) phase discussed recently in the literature does not help either since the concentration interval of its existence is narrow (between the regions of tetragonal and rhombohedral phases). Such situation requires searching for new explanations of properties of those solid solutions, which are located within the limits of the MPB region.

**11.5.2.** *Diffuseness of the phase transition from paraelectric to ordered state.*

The degree of diffusion of the paraelectric phase transition for the substances, in which domains of the FE and AFE phases coexist, depends on the location of this substance on the diagram of phase states. The results of investigation of the PE phase transition for a number of PZT-based solid solutions, in which the presence of the coexisting phases has been confirmed by the methods of the transmission electron microscopy and X-ray analysis of the crystal structure, are presented in Chapter 7. The diffuseness of the phase transition is also defined by the location of the composition of considered solid solution in the phase diagram. A considerable increase of the phase transition diffuseness is typical for those solid solutions, which are located in the vicinity of the boundary between the FE and AFE states in the phase diagram. Such behavior of the phase transition from the paraelectric phase into ordered state is characteristic of the substances, in which the difference in energies of the FE and AFE states is small and the domains of these phases coexist.

The dependence of the temperature interval, $\Delta T_{cw}$ within which the deviation from the Curie-Weiss law takes place on the Ti-content for the $(Pb_{0.9}Ba_{0.05}Sr_{0.05})_{0.985}(Li_{0.5}La_{0.5})_{0.015}(Zr_{1-x}Ti_x)O_3$ system of solid solutions is given in Fig.11.12. This temperature interval characterizes the diffuseness of the paraelectric phase transition. As one can see the diffuseness of the paraelectric phase transition in this system of solid solutions also depends on the solid solution composition. The maximum diffuseness of the phase transition is observed in the solid solutions with Zr/Ti compositions belonging (according to the X-ray data) to the MPB region of the diagram of phase states.

Fig.11.12. The temperature interval $\Delta T_{cw}$ within which the deviation from the Curie-Weiss law takes place (diffuseness parameter of the PE phase transition) vs. Ti-content in the system of solid solutions with compositions $(Pb_{0.9}Ba_{0.05}Sr_{0.05})_{0.985}(Li_{0.5}La_{0.5})_{0.015}(Zr_{1-x}Ti_x)O_3$ [256, 257].

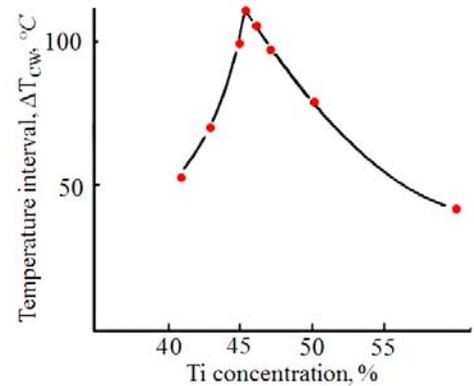

Such behavior of the diffuseness parameter of the paraelectric phase transition suggests a possible presence of domains with the AFE ordering in solid solutions with compositions falling within the MPB region.

### 11.5.3. *Local decomposition of solid solutions in the paraelectric phase*

Consider now another effect manifested on coherent interphase boundaries between the domains of the FE and AFE states in the substances with a small energy differences between these states.

The coexisting domains of the tetragonal and rhombohedral FE phases have different interplanar distances. The transmission electron microscopy was used to demonstrate the coherent character of the CIPBs in the PZT-based solid solutions in Chapter 3. No concentration of deformations was revealed in the vicinity of CIPBs. This allowed us to make a conclusion about the inner structure of the said boundaries. The transition through CIPB (the given CIPB may be called seed or 'bare') is accompanied by a continuous conjugation of the crystal planes. Such a coherent behavior must be accompanied by an increase in the elastic energy of the lattice along these boundaries.

In the said substances, equivalent sites of the crystal lattice are occupied by ions, which have different sizes and/or different charges. As we repeatedly discussed in this review, the 'larger' ions are pushed out into the domains with a higher configuration volume, and consequently, with larger inter-planar distances, but the 'smaller' ions are pushed out into the domains with lesser inter-planar distances. In the considered solid solutions the A-sites of the perovskite crystal lattice are occupied by ions with different ionic charges ($Pb^{2+}$, $Ba^{2+}$, $Sr^{2+}$, $La^{3+}$, $Li^+$), therefore, the local decomposition of the solid solution along CIPB may be accompanied by the local disturbance of electronegativity. Thus, the formation of the heterophase structure is accompanied by the disturbance of samples' chemical homogeneity. At the high temperatures, when the phases do not coexist, the samples remain homogeneous.

The local decomposition of the PZT-based solid solution in the vicinity of CIPBs leads to the appearance of the two types of local domains. One type is the local domain enriched with the lead zirconate and the other type is the local domain enriched with the lead titanate. The latter type of titanium-enriched domains as well as the whole matrix must exist in the FE state at the temperature below the Curie point. This is the reason why it is difficult to reveal their manifestation against the general background. However, the lead zirconate enriched domains may be in the AFE state, which differs markedly from the FE state. Thus, within the solid solution from the MPB region of compositions there exist specific FE+AFE domains formed as a pair of these phases as a result of the local decomposition of the solid solution. This system of domains of coexisting phases manifests a number of specific effects, which are not characteristic of other dipole ordered systems. This circumstance gives an opportunity to observe these characteristics experimentally.

As was shown in Chapter 7, the complex (FE+AFE) domain with inhomogeneous structure can exist at the temperatures above the temperature of the paraelectric phase transition when the transition takes place inside a single-phase domain. The FE and AFE phases constituting the complex domain are separated by the interphase boundary. Appearance of such complex domains happens under definite conditions in the solid solutions with particular compositions that we have learned to proportion. The temperature interval above the Curie point where these domains may appear can reach and even exceed 100 °C. The discussed existence of complex domains above the Curie temperature has been first revealed by X-ray method. This implies that the composite domains with the FE-AFE interphase boundary arise in the paraelectric state of the substance's matrix at cooling down from the high temperatures. As it was described above the local decomposition of the solid solution takes place in the vicinity of these boundaries and the segregates appear (in a general case, these segregates are ferroelectrically active).

One can easily control the microstructure of solid solutions since the inhomogeneous structure of the FE-AFE domains and segregates in the vicinity of the interphase boundaries appear in the process of the phase transition which itself can be easily controlled by an electric field or pressure. Considered physical process allows creation of textured domain structures already in the paraelectric phase with the help of an external electric field applied to a sample at the temperatures far above the Curie point. Analogous process has been considered and discussed above in Section 11.4 for the PZT-based solid solutions from the FE-AFE region of the diagram of phase states.

In the case when the solid solution composition is located in the section of the phase diagram far from the MPB region (and the decomposition of this solid solution does not take place) it can remain macroscopically depolarized at room temperature even after special thermoelectric treatments carried out at the temperatures well above the point of transition into dipole-ordered state. Such treatments can be the cooling of samples in the presence of an electric field or the isothermal annealing in the presence of a field. Piezoelectric coefficients of the solid solutions subjected to these treatments are equal to zero. Such behavior is characteristic for all ordinary ferroelectrics.

For substances with roughly equal stability of the FE and AFE states the two-phase domains appear at $T > T_C$ in the paraelectric matrix of samples during cooling down from the high temperatures. If at $T > T_C$ DC electric field acts on the sample, the sample is piezoelectrically active after cooling to room temperature. When the solid solution composition is located in the section of the phase diagram where the decomposition of this solid solution does not take place, it remains macroscopically depolarized at room temperature after any thermoelectric treatments with or without the involvement of electric field at $T > T_C$. All above-said is demonstrated in Fig.11.8. The above-discussed behavior is characteristic for the substance with coexisting domains of the FE and AFE phases.

Let us now discuss the PZT-solid solutions with compositions belonging to the MPB region of the phase diagram. The results obtained as a result of special thermoelectric treatments with application of an electric field (at $T > T_C$) to the system with the $(Pb_{0.9}Ba_{0.05}Sr_{0.05})_{0.985}(Li_{0.5}La_{0.5})_{0.015}(Zr_{1-x}Ti_x)O_3$ compositions are presented in Fig.11.13. In this case, the thermoelectric treatments have been performed at the temperatures above the Curie point and an electric field E = 120 V/mm has been applied to samples at the temperatures $T_C(x) + 50$ °C. The samples were kept in the field during 20 min (curve 1) and 40 min (curve 2). After that, the field was switched-off, samples were cooled down to the room temperature without electric field, and then the piezoelectric coefficient $d_{33}$ was measured. Obtained results are analogous to the results obtained on the samples with the coexisting FE and AFE phases. The solid solutions become piezoelectrically active only if their Zr/Ti content corresponds to the MPB region of the phase diagram. This is one more experiment indicative of the existence of domains of the AFE phase in the solid solutions with compositions from the MPB region.

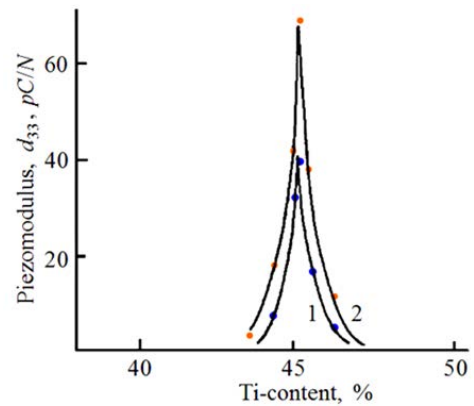

Fig.11.13. Piezoelectric modulus vs content of titanium in the $(Pb_{0.9}Ba_{0.05}Sr_{0.05})_{0.985}(Li_{0.5}La_{0.5})_{0.015}(Zr_{1-x}Ti_x)O_3$ system of solid solutions subjected to the thermoelectric treatment during time $\tau$ in the paraelectric phase at the temperature $T_C + 50$ °C in a DC electric field of 120 V/mm [256]. The field was switched off after the thermal treatment and the samples were cooled down to room temperature without the field.
Treatment time, $\tau$, min: 1 – 20; 2 – 40.

**11.5.4**. *Paraelectric phase transitions in an external electric field*

Let us now dwell on another series of experiments demonstrating peculiarities in characteristics of the substances with the coexisting domains of the FE and AFE phases in the bulk of the samples. The point of the paraelectric phase transition in these materials shifts under the heating in the presence of an external electric field. The transition point in ferroelectrics is shifted toward the higher temperatures whereas in antiferroelectrics the transition point is shifted toward the lower temperatures. Such behavior is caused by a decrease in the energy of ferroelectrics and by an increase in the energy of antiferroelectrics in an applied electric field. The dipole ordering temperature (the transition point) does not depend on the field intensity in the substances with coexisting domains of the FE and AFE phases. The redistribution of the shares of these phases (in favor of the FE phase) takes place in an applied field and the total energy of the system remains constant (see Chapter 5 of the present review). The so-called intermediate state is realized in an applied field when the domains of the FE and AFE phases coexist in the sample's volume. An applied electric field shifts the FE-AFE interphase boundaries in such a way that the share of the FE phase rises, but at the same time the internal state inside each domain remains unchanged. When the substance is in the intermediate state the energy of domain of each phase is independent on the electric field and the total energy of the whole sample is independent on the field. Therefore, in this case the temperature of the phase transition into the paraelectric state is constant in a variable field.

The dependencies of the Curie point on an electric field for the $Pb_{0.90}(Li_{0.5}La_{0.5})_{0.10}(Zr_{1-x}Ti_x)O_3$ series of solid solutions located at different distance from the region separating the FE and AFE states in the phase diagram are presented in Fig.5.1 along with the "composition-temperature" phase diagram for this system of solid solutions. As one can see, the temperature of the paraelectric phase transition does not depend on the electric field when the solid solution composition permits the realization of the intermediate state of the coexisting FE and AFE domains in an electric field.

Analogous results for the $(Pb_{0.9}Ba_{0.05}Sr_{0.05})_{0.985}(Li_{0.5}La_{0.5})_{0.015}(Zr_{1-x}Ti_x)O_3$ system of solid solutions are presented in Fig.11.14. In the case of the above compounds the peculiarities typical for substances with the coexisting domains of the FE and AFE phase, are present only in the solid solutions belonging to the MPB region of compositions.

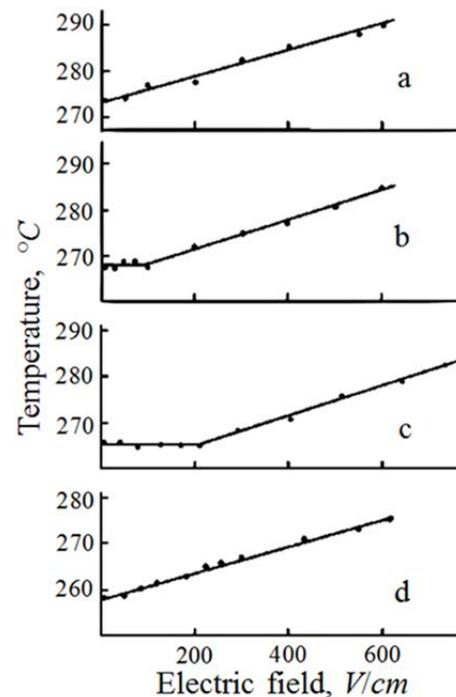

Fig.11.14. Dependencies of the Curie temperature on an external electric field intensity in the $(Pb_{0.9}Ba_{0.05}Sr_{0.05})_{0.985}(Li_{0.5}La_{0.5})_{0.015}(Zr_{1-x}Ti_x)O_3$ system of solid solutions [256].
The content of titanium, x, %: a – 47.0; b – 45.750; c – 45.50, d – 43.0.

This experiment also indicates the presence of domains of the AFE phase in the PZT-based solid solutions with compositions belonging to the MPB region (see diagram in Fig. 11.9).

**11.5.5. Concluding remarks**

Thus, we have demonstrated that the solid solutions with compositions from the MPB region have the same peculiarities in the behavior of their physical characteristics as the solid solutions located in the vicinity of the FE-AFE phase boundary in the "composition-temperature" phase diagram. The MPB region is usually located far from this boundary and the peculiarities in behavior of such solid solutions are not considered to be connected with the possible presence of domains with the AFE state. Moreover, the energies of such states are widely spaced, which allows not considering them in the solid solutions belonging to the MPB region. However, one's understanding of phenomena will be changed substantially if one takes into account the local decomposition of solid solutions in the vicinity of the tetragonal-rhombohedral interphase boundary.

In the frame of such approach, one has to take into account that the inhomogeneity in distribution of zirconium and titanium ions in the vicinity of the interphase boundary can lead to the appearance of nanodomains greatly enriched with zirconium. The AFE type of dipole ordering can exist inside such Zr-enriched nanodomains. The coexistence of such AFE domains with the domains with the FE ordering defines the peculiarities in behavior of the investigated solid solutions. These nanodomains with the AFE ordering can be identified by the transmission electron microscopy.

Results presented here find unique and consistent explanation in the framework of the concept about the coexistence of domains of the dipole ordered FE and AFE phases and about physical processes occurring at the interphase boundaries between these domains. The AFE state alone is unfavorable in the solid solutions belonging to the MPB region since the MPB region and the region of the AFE states are located quite far from one another in the "composition-temperature" diagram of phase states in the PZT. The appearance of the AFE state is only possible because of additional factors, in particular, as a result of the local decomposition of solid solutions and zirconium enrichment of local regions in the vicinity of interphase boundaries (together with local mechanical stresses or without such stresses).

Let us consider data available in the literature that additionally argues for the point of view described this review. Since suggested here approach to the interpretation of results of investigations of the PZT-based solid solutions from the MPB region of the phase diagram is not considered as generally accepted due to its relative novelty there are not enough experimental data in the literature that can directly confirm or disprove its validity.

The majority of contemporary considerations of the transition from the rhombohedral phase to the tetragonal phase (in other words the analysis of the conjugation of phases) and the relaxation of possible stresses appearing on the coherent boundaries between the coexisting domains of these phases in the sample treat this transition as taking place by a rotation of the polarization vector. From the crystallographic point of view, such rotation must lead to the appearance of the monoclinic phases [258, 259].

Our approach suggests taking into consideration a new phase for explanation of structural peculiarities of the substances belonging to the MPB regions in the PZT-based solid solutions. Essentially, such approach is not new. Similar assumptions have been made earlier. An exception is only the circumstance that nobody suggested an appearance of the AFE phase domains in these solid solutions.

Studies [260, 261] were among the first that used the transmission electron microscopy in a selected-area electron diffraction (SAED) mode. The SAED pattern observed in these papers showed the doublet and triplet splitting (for high-index planes) of the diffraction pattern. Authors attributed the triplet spitting to the twin-related tetragonal (T), rhombohedral (Rh) adjacent FE domains. The interpretation of the diffraction pattern was done on the base of succession of domains of the $T_1-Rh-T_2-Rh-T_1$ phases. That is the authors of [260, 261] consider that SAED points to a presence of one more phase with tetragonal distortions in solid solutions belonging to the MPB region in the PZT and this phase is the border phase between the domains with the rhombohedral and tetragonal crystal structures. It is important to highlight in this regard that in the PZT-based solid solutions with the large concentration of zirconium one of the AFE phases has exactly the tetragonal type of lattice distortions. Investigations of the PZT solid solutions belonging to the MPB region by synchrotron radiation were carried out in [262, 263, 264]. Authors of these articles observed lines, which in no way were related to the monoclinic phases. Using transmission electron microscopy, they directly observed the nanodomains with sizes of the order of 10 − 15 nm. As one can conclude from the photographs given in these papers, the location of these nanodomains corresponds to the interphase boundaries. These boundaries are quite wide due to the changes in the distribution of ions in equivalent crystallographic positions in the vicinity of coherent interphase boundaries. Authors of [263] traced the behavior of these nanodomains in an applied electric field. The applied field transformed these nanodomains into the domains of the rhombohedral FE phase. That is the field caused the induced phase transition into the FE state. Since the crystal structure of nanodomains is not cubic, this transition is similar to the induced transition from the AFE state into the FE state.

The results of [265] also strongly suggest a possible presence of the AFE states in the solid solutions from the MPB region. The crystal structure of micron-sized domains and nanodomains was investigated by transmission electron microscopy. Authors of [265] attributed the complex domain structures observed at $x = 0.5$ and $0.95$ (the last solid solution is located inside the FE-AFE morphotropic region) to a physical manifestation of the frustration between and Zr- and Ti-rich clusters.

Results pointing to the possible existence of the AFE phase domains in the PZT-based solid solutions with compositions from the MPB region have been obtained by inelastic X-ray scattering [266]. The measurement revealed that upon cooling from the paraelectric phase, the spectral response of the soft $M_5^/$ zone boundary phonon mode associated with the antiferroelectric vibrations of lead ions existed and is progressively transformed into a broad central mode. The authors believed that the behavior observed in their experiments was directly related to the nanoscale inhomogeneity of the structure.

It is necessary to draw ones attention to one more peculiarity of properties of the PZT-based solid solutions belonging to the MPB region. Neutron diffraction, spectroscopic, as well as electron diffraction experiments indicated the presence of displacements of ions that correspond to the rotational modes of the oxygen octahedra with wave vector at the Brillouin zone boundary. As a rule, researches do not relate these modes and ion displacements to the presence of the AFE phase in these solid solutions. The authors of [267] carried out density functional calculations in a Hedin-Lundqvist local density approximation using a general potential linearized augmented plane-wave method, with local orbital extensions to relax linearization errors and treat semi-core states. The purpose of this paper was to investigate the competitions between ferroelectric and rotational instabilities in the rhombohedral PZT from the MPB region. They found a ferroelectric instability along with a substantial $R$-point rotational instability, close to the ferroelectric one. The existence of both these instabilities close to one another is similar to the situation in pure $PbZrO_3$. As it was also noted in Ref. 30 these two instabilities are both strongly pressure dependent, but in opposite directions so that lattice compression of less than 1% is sufficient to change their ordering. Authors of [267] concluded that such pressure dependence and the presence of the local stress fields caused by the $B$-site cation disorder may lead to the coexistence of both types of instabilities which are likely present in the MPB region. It is again quite similar to the situation in pure $PbZrO_3$.

Here we should mention results of [268]. The high-resolution powder diffraction measurements were performed on solid solutions from the MPB region of the PZT phase diagram as well as on the AFE lead zirconate in this article. The AFE $PbZrO_3$ was used for comparison of results obtained on this compound with the ones for the PZT solid solutions form the MPB region. The result of this comparison allowed concluding that for $T > T_C$ the cubic phase is far from having a simple crystal structure, exhibits a more complex local structure than had previously been thought and has been overlooked by all but a few researchers.

Results of our studies of the PZT-based solid solutions at the temperatures above the Curie temperature (given in Chapter 3 and also in Chapters 6, 7 and presented in Fig.11.12 and 11.13) are in complete agreement with the results of [268]. Our results can be easily explained if the complex (FE+AFE) domains existed in the paraelectric phase in the vicinity of the $Rh_{FE} − T_{FE} − PE_C$ triple point in the phase diagram. To our knowledge, until now there has not been any other explanation of the fact that the local structure of the paraelectric phase in the PZT solid solutions belonging to the MPB region is different from the cubic crystal structure in such a wide temperature interval.

In our opinion, the results of [266] support our approach. Dispersion curves of the lowest frequency transverse optic and the transverse acoustic phonon modes propagating along the [100] direction have been determined in the high-temperature paraelectric phase as well as at the room temperature. Recall that the spectral response of the zone boundary phonon modes obtained in this paper is associated with the antiferroelectric vibrations of lead ions.

We have noted earlier and want to stress one more time that we are not trying to prove that the monoclinic phases do not exist in the PZT-based solid solutions with compositions belonging to the MPB region of the phase diagram. We are simply trying to draw ones attention to the fact that not all the factors were taken into account during the interpretation of experimental results. It is primarily relevant to the results obtained by the treatment of the profiles of the lines in the scattering spectra of different nature (X-ray, neutron, and optic).

Let us give one more argument supporting our point of view. A large number of publications substantiating the possibility of development of piezoelectric materials with large values of piezoelectric parameters appeared after the publication of the first papers about monoclinic phase. In this respect one can ask a question about the results that have been achieved in this direction during last 15 − 16 years. PZT-based ceramic materials had the largest value of the piezoelectric coefficient $d_{33}$ at the level of 560 − 600 pC/N before the appearance of the idea about the monoclinic phase. Using the monoclinic phase idea one could improve the value of $d_{33}$ up to ~ 630 pC/N. However, in the framework of the idea about the local decomposition of the solid solutions near the coherent interphase boundaries between the rhombohedral and tetragonal phases and the idea about the formation of nanodomains of the AFE phase the value of more than 1100 pC/N (see Fig. 3) for the piezoelectric coefficient has been achieved. As one can see, the results are disparate. However, it is not the most interesting circumstance. We have obtained the value of the $d_{33}$ piezoelectric modulus equal to 1200 pC/N in the hard ferroelectric PZT-based ceramics with composition from the MPB region and a coercive field of 1900 V/mm. These results demonstrate one more time that not all the physical effects have been taken into account while interpreting experiments.